\documentclass[a4paper,10pt]{fullarticle}
\usepackage{sciencestuff}
\usepackage[hang,centerlast]{subfigure}

\title{HyperPCA: a Powerful Tool to Extract Elemental Maps from Noisy Data Obtained in \libs Mapping of Materials}

\author[1,2]{Riccardo Finotello\emailfoot{riccardo.finotello@cea.fr}}
\author[2]{Mohamed Tamaazousti\emailfoot{mohamed.tamaazousti@cea.fr}}
\author[1]{Jean-Baptiste Sirven\emailfoot{jean-baptiste.sirven@cea.fr}}
\affil[1]{%
  Université Paris-Saclay, CEA, \protect\\
  Service d'études analytiques et de réactivité des surfaces (SEARS), \protect\\
  Gif sur Yvette, F-91191, France
}
\affil[2]{%
  Université Paris-Saclay, CEA, \protect\\
  LIST, \protect\\
  Palaiseau, F-91120, France
}

\date{}

\hypersetup{%
  pdfkeywords={machine learning, unsupervised, libs, mapping, pca, kernel},
  pdfsubject={kernel sparse \pca for \libs mapping}
}

\newcommand{\kspca}{\textsc{kspca}\xspace}
\newcommand{\hyperpca}{\textsc{HyperPCA}\xspace}
\newcommand{\hardt}{\textsc{ht}\xspace}

\newcommand{\dcs}{\textsc{dc}s\xspace}

\newcommand{\acs}{\textsc{ac}s\xspace}
\newcommand{\mva}{\textsc{mva}\xspace}
\newcommand{\sem}{\textsc{sem}\xspace}
\newcommand{\eds}{\textsc{eds}\xspace}
\newcommand{\pc}{\textsc{pc}\xspace}
\newcommand{\pcs}{\textsc{pc}s\xspace}
\newcommand{\pcn}[1]{\textsc{pc}-\oldstylenums{#1}\xspace}
\newcommand{\rmt}{\textsc{rmt}\xspace}
\newcommand{\snr}{\textsc{snr}\xspace}

\begin{document}

\maketitle

\begin{abstract}
Laser-induced breakdown spectroscopy (\libs) is a preferred technique for fast and direct multi-elemental mapping of samples under ambient pressure, without any limitation on the targeted element.
However, \libs mapping data have two peculiarities: an intrinsically low signal-to-noise ratio due to single-shot measurements, and a high dimensionality due to the high number of spectra acquired for imaging.
This is all the truer as lateral resolution gets higher: in this case, the ablation spot diameter is reduced, as well as the ablated mass and the emission signal, while the number of spectra for a given surface increases.
Therefore, efficient extraction of physico-chemical information from a noisy and large dataset is a major issue.
Multivariate approaches were introduced by several authors as a means to cope with such data, particularly Principal Component Analysis (\pca).
This technique is useful to analyse correlations between different elements, but it is limited to low signal-to-noise ratios.

In this paper, we introduce \hyperpca, a new analysis tool for hyperspectral images based on a sparse representation of the data using Discrete Wavelet Transform and kernel-based sparse \pca to reduce the impact of noise on the data and to consistently extract the spectroscopic signal, with a particular emphasis on \libs data.
The method is first illustrated using simulated \libs mapping datasets to emphasise its performances with an extremely low shot-to-shot signal-to-noise ratio, and with a variable degree of spectral interference.
Comparisons to standard \pca and to traditional univariate data analyses are provided.
Finally, it is used to process real data in two cases that clearly illustrate the potential of the proposed algorithm.
We show that the method presents advantages both in quantity and quality of the information recovered, thus improving the physico-chemical characterization of analysed surfaces.
\end{abstract}

\highlights{we introduce \hyperpca as a tool to efficiently recover physico-chemical information from noisy hyperspectral mapping data.}


\clearpage

{\small\tableofcontents}

\clearpage

\section{Introduction}

Multi-elemental mapping of surfaces on a microscopic scale enables the study of morphology, structure and composition of heterogeneous materials.
Techniques based on electron microprobe are commonly used for that purpose in different fields such as materials science~\cite{llovet_electron_2021}, geology~\cite{ning_electron_2019} or biology~\cite{marshall_quantitative_2017}.
Other approaches can be found, particularly based on mass spectrometry coupled to laser ablation~\cite{rodionov_spatial_2019} or to an ion beam~\cite{yang_application_2015}, on X-ray spectrometry induced by synchrotron radiation~\cite{crean_expanding_2015} or by a charged particle beam~\cite{hartnell_review_2020}.
Alternatively, Laser-Induced Breakdown Spectroscopy (\libs) is the only technique that enables to carry out rapid measurements in ambient air, and to detect all elements, including the lightest ones up to hydrogen, with a typical lateral resolution of \SIrange{10}{20}{\micro\meter}.
In the best case scenario, this resolution can reach \SIrange{1}{10}{\micro\meter}.
Its applications in materials mapping have been growing rapidly over the past years~\cite{jolivet_review_2019}.

\libs mapping consists in focusing a laser pulse on the sample surface.
The resulting plasma emission spectrum is recorded, then the sample is moved and the operation is repeated on the desired surface.
The typical lateral resolution in \libs mapping extends from \SI{1}{\micro\meter} to several tens to hundreds of \SI{}{\micro\meter}, while the depth resolution in a single laser shot is of the order of a fraction of \SI{}{\micro\meter} up to a few \SI{}{\micro\meter}, depending on the material and on laser ablation conditions.
High resolution mapping is sometimes required for different applications.
Several works reaching even higher resolutions can be found in the literature, for example mapping of ceramics with a \SI{3}{\micro\meter} resolution~\cite{menut_micro-laser-induced_2003} or measurement of hydrogen in zirconium-based materials with a \SI{1}[\sim]{\micro\meter} spatial resolution~\cite{brachet_study_2017}.

Yet, this particular configuration is a challenge for several reasons.
Indeed, in order to increase the lateral resolution, the size of the laser spot is minimised, and consequently the ablated mass is reduced.
Thus, the signal-to-noise ratio of single-shot spectra is inherently low.
In addition, for a given analysed surface, the smaller the laser spot size, the higher the number of collected spectra, which can reach up to several tens or thousands of spectra per mapping.
Efficient extraction of physico-chemical information from a noisy and large dataset is therefore a major issue, intrinsic to the high resolution \libs mapping technique.

Currently, data processing usually consists in extracting the intensity of the line of the element of interest from each spectrum and then reconstructing the associated intensity map.
This approach is based on only a small fraction of the total information, and it is subject to spectral interference and matrix effects.
Multivariate (\mva) approaches were used by several authors in order to exploit the data more efficiently and to facilitate the interpretation.
In particular, Principal Component Analysis (\pca), probably the most widespread multivariate technique in \libs~\cite{porizka_utilization_2018}, was successfully implemented to process mapping data and identify mineral phases in geological samples from \pca scores~\cite{moncayo_exploration_2018}, or to measure the distribution of elements of interest in seeds~\cite{gamela_hyperspectral_2019}.
The authors in~\cite{rifai_emergences_2020} developed a supervised approach to identify minerals in rocks with a \SI{50}{\micro\meter} spatial resolution: the quantitative training data were measured by means of Scanning Electron Microscopy (\sem) and Energy Dispersive X-ray Spectroscopy (\eds).
The machine learning approach was based on \pca scores.
As an alternative to \pca, k-means clustering was also used to map major and minor elements in a rock sample~\cite{nardecchia_detection_2020}: the approach proved effective at displaying correlations of chemical species even in the case of high spectral interference.

\begin{figure*}[t]
    \centering

    \includegraphics[width=\linewidth]{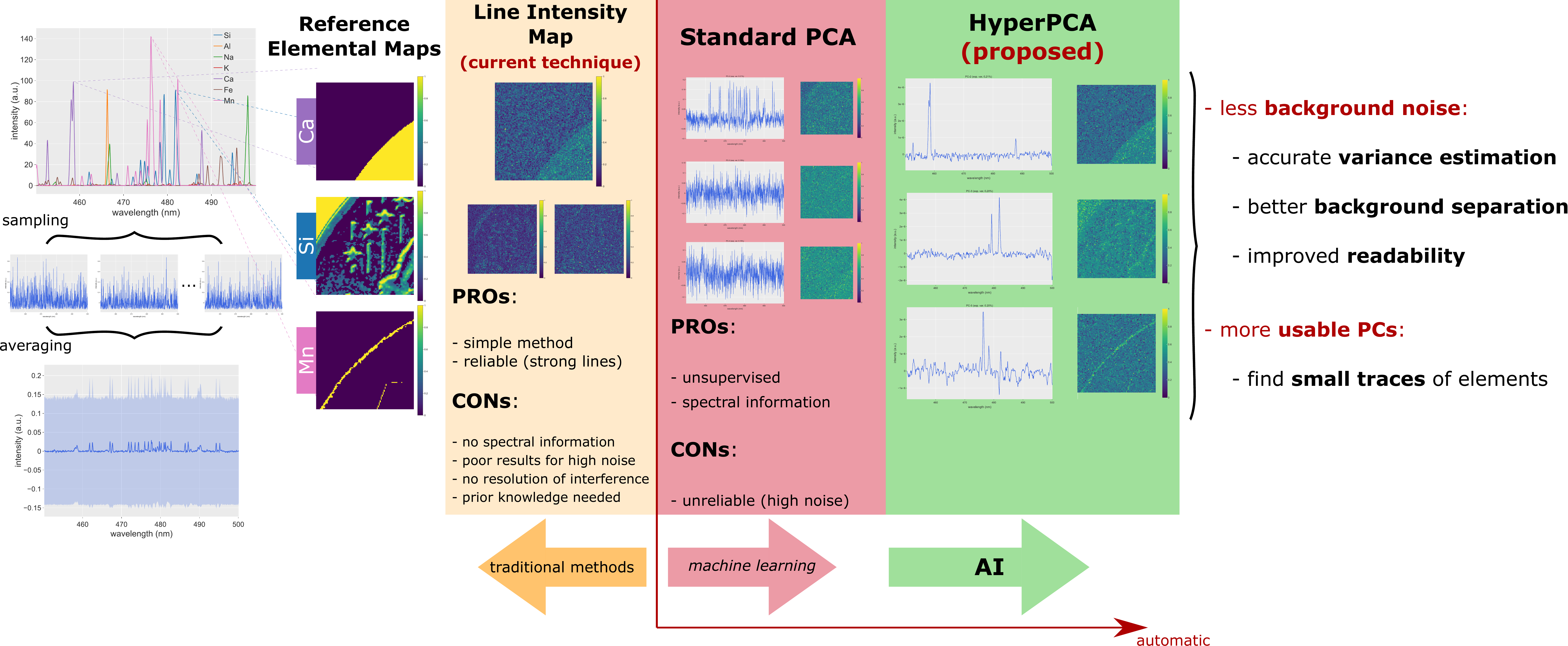}
    \caption{%
        \emph{Illustration of the results.}
        We consider three instances in the ``granite'' dataset, introduced later in the paper, to compare the information extraction ability of our proposed method (\hyperpca), the line intensity map and standard \pca.
        In the presence of noise and spectral interference, the extraction of the \pcs can be problematic, and sometimes impossible.
        Moreover, complete readability of the spectra is not assured, as the estimation of the \pc is affected by the background and spectral interference, thus the effective contribution of the elements can be underestimated.
        Differently, our proposal focuses on the ability to suppress noise and to distinguish the different sources of the signal, in such a way that better variance estimation and resolution of the spectra can be achieved.
        The algorithm can extract more \pcs, both in quality and in number.
        Separation of the different elements is also improved, as the presence of an intermediate sparse representation helps to distinguish the different profiles of the lines.
    }
    \label{fig:results}
\end{figure*}

\pca has the great advantage of being easy to implement and being unsupervised, hence no previous knowledge on the data specifics is required.
It is a useful tool to put into evidence correlations and anti-correlations among different elements.
However, as \pca computes orthogonal components, it is not designed to extract pure elemental contributions~\cite{karhunen_representation_1994}, which would be easier to interpret from a physical-chemical point of view, or at least that would yield some complementary information.
In addition, on the one hand, high resolution \libs data intrinsically present low signal-to-noise ratio, and on the other hand, the number of spectra is not significantly larger than the number of spectral channels in each spectrum.
This is, in general, even more problematic if the rank of the data matrix is lower than maximal: in this case, less realisations of the measurement, or spectra, are available with respect to the number of variables, the wavelength channels, probed by the experiment.
In these conditions, \pca is known to present theoretical limits to the consistent extraction of the \pcs~\cite{baik_phase_2005, montanari_statistical_2014, perry_optimality_2018, johnstone_sparse_2009}.
As the dimensionality of \libs mapping datasets lies at the threshold of a phase transition in the detectability of the \pcs~\cite{johnstone_distribution_2001,  paul_asymptotics_2007, baik_phase_2005, perry_optimality_2018}, other approaches may result in improvements in the extraction of the information.
In particular, \emph{sparse \pca} has been proposed to overcome the issues~\cite{johnstone_sparse_2009}: when the signal data can be constructed using a basis of functions whose coefficient vectors present null entries (i.e.\ the signal is sparse), it is possible to reduce the initial number of variables to consistently apply \pca.
The issues related to high dimensionality have already being tackled in \libs~\cite{yi_laser_2017}, by proposing lower rank approximations of the spectral data via the introduction of sparsity filters.
Moreover, the extraction of salient features by means of alternatives to the standard \pca is a standard practice in the analysis of hyperspectral images~\cite{6819824, 8447427, 6648433, 7974741}.
As the \libs mapping data are not in a sparse representation (see for instance~\cite{kepes_addressing_2021} where the authors show that enforcing sparsity on the \pcs alone is not enough to improve mapping or clustering results), we introduce a new approach to recover the \pcs.

In this paper, we propose \hyperpca: a new approach based on sparse \pca and sparse representations to process hyperspectral images, and specifically high resolution \libs mapping data.
Namely, we propose to couple a specific Discrete Wavelet Transform (\dwt), in order to build the sparse representation of the \emph{dense} data by extending standard and recent techniques in signal processing~\cite{elad_sparse_2010, xiong_sparse_2021}, and a kernel-based approach to noise reduction and to the estimation of the \pcs (\emph{kernel-based sparse \pca}, or \kspca henceforth)~\cite{seddik_kernel_2019}.
The goal is to develop a utility capable of consistently reducing the size of the information contained in the \libs or hyperspectral datasets, in order to be able to extract a clean signal, even when the low rank of the datasets limits the use of more standard approaches like \pca~\cite{johnstone_sparse_2009, montanari_statistical_2014}.
First, \hyperpca is described in~\Cref{sec:methodology}, then it is applied to simulated datasets in~\Cref{sec:synthetic} to emphasise its performances with highly noisy and/or highly interfered spectra, and to compare them to a standard \pca algorithm.
Finally, it is used to process real data in three cases that clearly illustrate the potential of \hyperpca in~\Cref{sec:experimental}.
The main improvements introduced in the \libs analysis are shown in~\Cref{fig:results}, where we compare the current technique, based on standard \pca, with our \hyperpca.
We found that \hyperpca is capable of extracting a larger number of \pcs, and to provide more readable loadings and score maps.
As shown, the proposed algorithm is able to both reduce the background noise and to extract more \pcs, thus enhancing the quality of the elemental maps and improving the detection of weak or strongly interfered emission lines.


\section{Methodology}\label{sec:methodology}

\begin{figure*}[t]
    \centering
    \includegraphics[width=\linewidth]{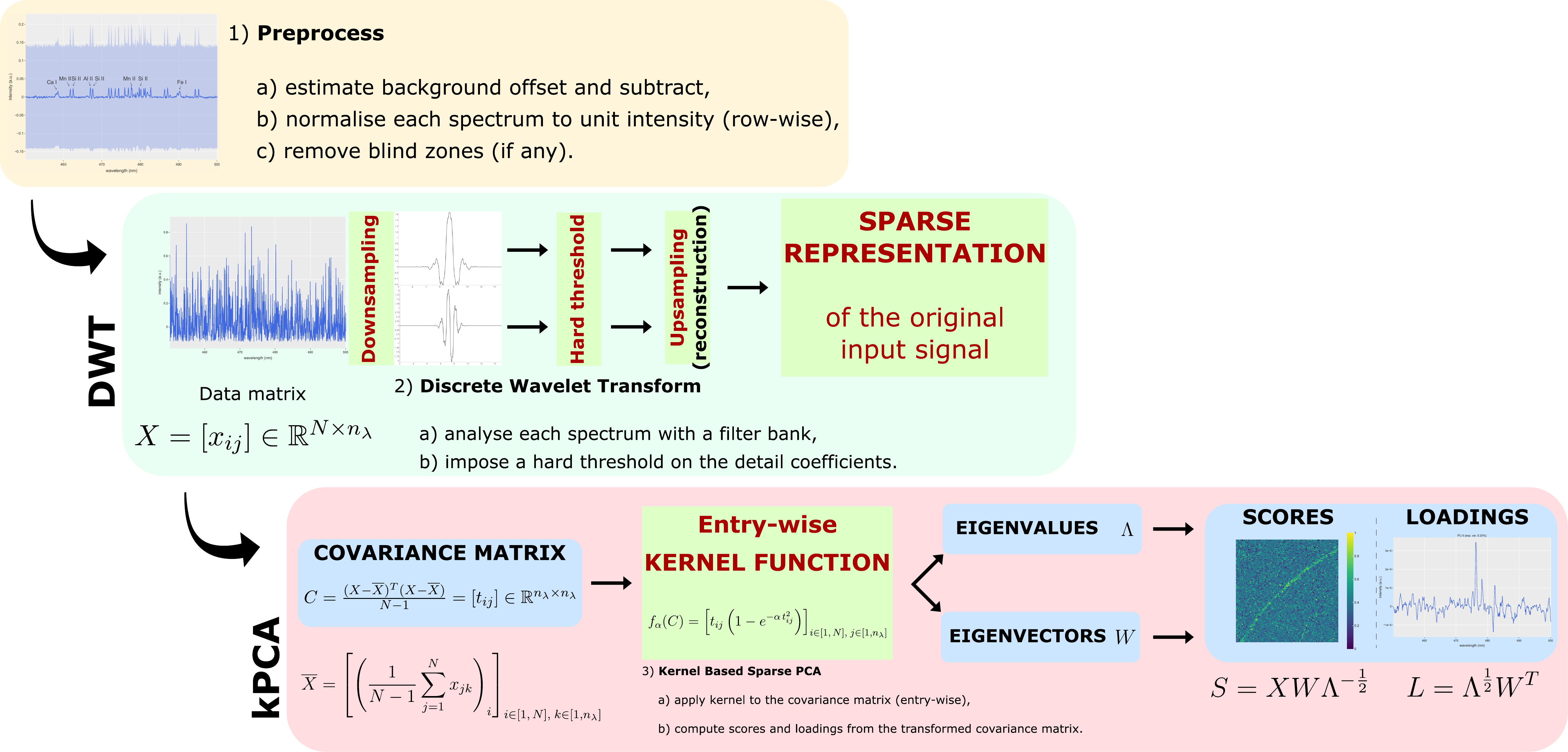}
    \caption{\emph{Methodology.} The proposed algorithm is composed of multiple building blocks. The preprocessing step aims at highlighting the emission lines with respect to a common offset, and at reducing the zone of interest. The application of sparse kernel-based methods needs a sparse representation of the input signal. We build such representation using a \dwt transform and imposing a \hardt on the detail coefficients. Finally, the kernel-based method can be applied in order to recover an improved reconstruction of the scores and loadings.}
    \label{fig:pipeline}
\end{figure*}

We consider \libs maps of $n$ pixels in width and $m$ pixels in height from both synthetic datasets and real world cases.
For each dataset, the number of wavelength channels per spectrum $n_{\lambda}$ is variable and in the order of \num{e3}.
The original data are organized in real-valued tensors $\bfT \in \R^{n \times m \times n_{\lambda}}$.
The \pca is performed considering the unfolded data cube, i.e.\ the data matrix $X = \mathrm{vec}\left( \bfT \right) \in \R^{p \times n_{\lambda}}$, where $p = n\, m$, constructed by concatenating the second axis of $\mathbf{T}$ along the first.
Its row components are vectors $x_{(i)} \in \R^{n_{\lambda}}$, for $i = 1, 2, \dots, p$, identifying the spectra.
Details of the technical background on \pca and \dwt are reported in~\Cref{app:tech}.

The analysis is carried out using the Python modules \texttt{PyWavelets}~\cite{lee_pywavelets_2019} and \texttt{NumPy}~\cite{harris_array_2020}.
Datasets are processed in \texttt{Pandas}~\cite{mckinney_data_2010, reback_pandas-devpandas_2021}.
Plots were created using \texttt{Plotly}~\cite{plotly_technologies_inc_collaborative_2015}.
Development took place on a Dell \texttt{Precision 7550} laptop with \SI{32}{\giga\byte} of \ram and an Intel\textsuperscript{\textcopyright}~Core\texttrademark~i7-10875H \cpu.


\subsection{General Pipeline}

In most \libs mapping analyses involving \pca, the unsupervised technique is directly applied to the experimental data, after being preprocessed~\cite{costa_calibration_2020}.
In our analysis, we propose a different approach to \pca by introducing the concept of \kspca to \libs mapping data~\cite{seddik_kernel_2019}.
This unsupervised statistical tool relies on the sparsity of the input, i.e.\ on input vectors containing mostly null elements, to extract the \pcs.
We therefore introduce an intermediate step aimed at finding such sparse representation, with the additional effect of providing an initial denoising filter for the spectra.
We utilise a \dwt to decompose the original signal.
The sparse representation is then given by imposing a hard threshold (\hardt) on the spectral energy density of the high frequency noise, and by upsampling the transformed data to the original signal.
The adopted algorithm, \hyperpca, is defined as follows (see~\Cref{fig:pipeline}):
\begin{enumerate}
    \item data preprocess;

    \item data filtering with the \dwt using the \hardt on the detail coefficients;

    \item spectra reconstruction;

    \item application of \kspca and study of the \pcs and score maps.
\end{enumerate}

Before showing experimental results, we dissect the process and analyse in detail each introduced aspect to ensure reproducibility.
The qualitative improvements with respect to standard \pca will be highlighted directly on simulated and real data of \libs mapping experiments.
The newly introduced methodologies are object of detailed analysis of their separate contributions in the supplementary material.


\subsection{Preprocessing}

We preprocess each dataset by normalising the spectra, and by removing offsets.
Namely, we proceed as follows:
\begin{enumerate}
    \item each spectrum in the sample matrix is normalised to unit maximum;

    \item a common electronic background offset is identified, and its average contribution, typically computed across a window of \SIrange{10}{20}{} wavelength channels, is subtracted from all spectra;

    \item for real data, flat zones on the non-intensified edges of the CCD sensor are identified and cropped from the dataset.
\end{enumerate}
These operations aim at putting in better evidence the emission lines with respect to a common value, typically zero, and at reducing the zone of interest to only the necessary wavelengths.
In general, the removal of the offset only slightly affects methods such as \pca or \kspca, as its purpose is only to have the emission lines visibly stand out.
Since no additional baseline corrections are necessary for data treatment, no discontinuities are introduced by this preprocessing technique.


\subsection{Data Representation}

\begin{figure*}[t]
    \centering

    \includegraphics[width=0.75\linewidth]{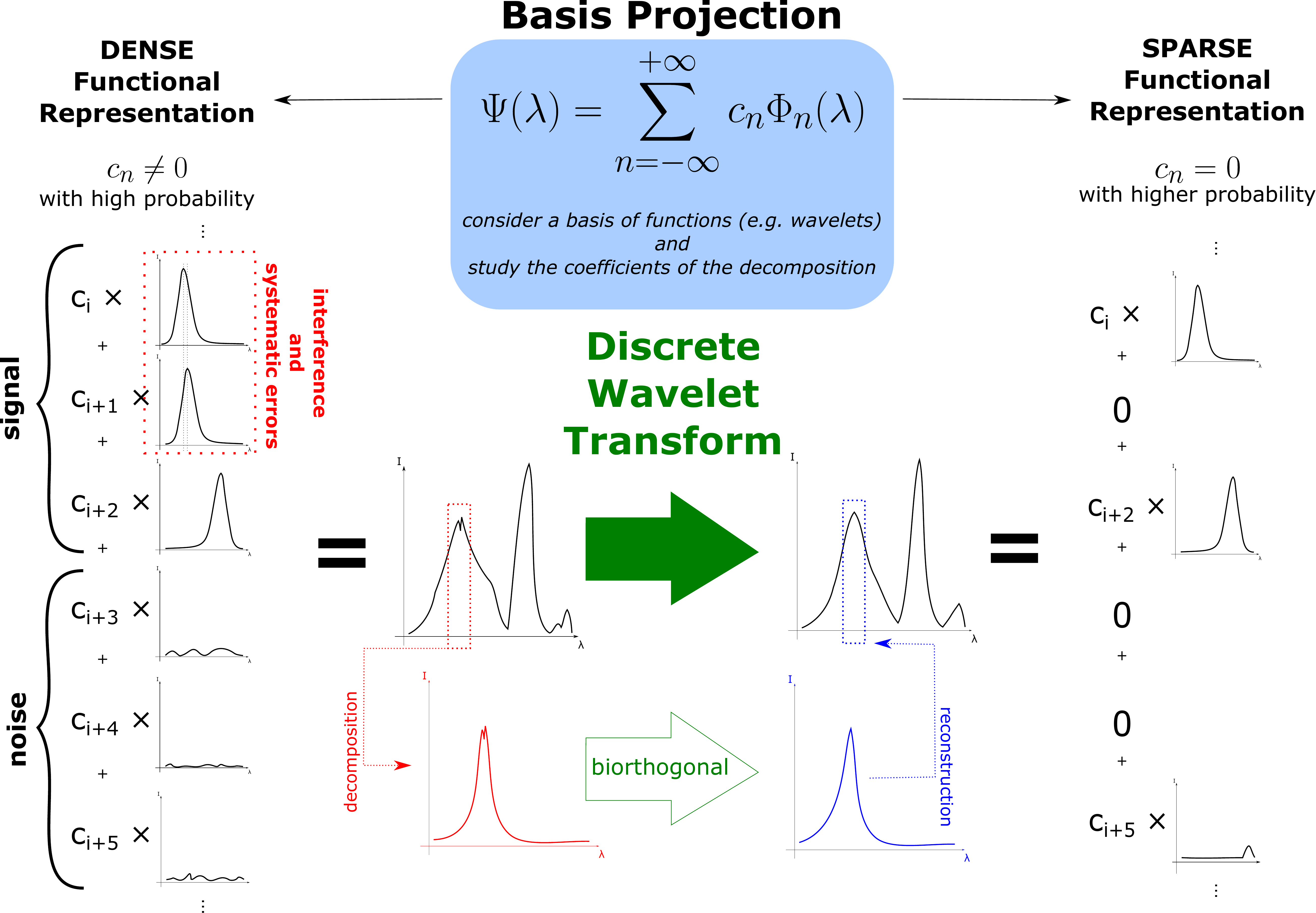}
    \caption{%
        \emph{Sparse representation.}
        The \dwt is used to process the input signal in order to obtain a new representation of the data where the information on the chemical elements are contained only in a finite number of coefficients in the decomposition.
        Biorthogonal wavelets have been chosen to directly resolve phenomena such as systematic errors leading to poor resolution (e.g.\ the emission of the same element captured by multiple neighbouring pixels of the detector) and spectral interference through the identical reconstruction of similar profiles (corresponding to same element) in a given spectral range.
    }
    \label{fig:sparse_rep}
\end{figure*}

\libs mapping data can present high degrees of noise and spectral interference.
As such, the underlying signal in spectra is in a \emph{dense} representation.
The application of \hyperpca requires the input signal, in the ideal case, to be in a \emph{sparse} representation.
That is, the signal has to be expressible as linear combination of functions in a basis where most of the coefficients vanish, or are at least suppressed with respect to the variance of the dataset.

We use wavelets to scan the input spectra at different scales (see~\Cref{app:wavelets} for the formal details).
They ensure the representation of the input to be approximately sparse.
That is, a finite number of coefficients in the functional decomposition approximately vanish, as only the coefficients corresponding to certain wavelengths at given positions differ from zero (see~\Cref{fig:sparse_rep}).
However, high frequency noise can spoil the sparsity in real-world applications.
We mainly focus on \dwts as they represent a natural way to produce the sparse basis, while they also provide good results as a denoising filter, as already observed in many \libs applications~\cite{ZHANG2015939, wiens_pre-flight_2013, maurice2021supercam, ZHANG201532, Zou:14, YUAN201429, Cisewski:2012}.
They are organised as high and low-pass filters, the first producing the \emph{detail} coefficients (\dcs), and the latter giving the \emph{approximation} coefficients (\acs), usually containing the signal.

We use \dwts to process the experimental data and produce a new sparse representation.
Specifically, we proceed as follows:
\begin{enumerate}
    \item we compute \dcs and \acs for all spectra;
    \item we apply a sparsity filter on the \dcs;
    \item we reconstruct the spectra by inverting the \dwt.
\end{enumerate}

To create the sparsity filter, we compute the energy density spectrum of the \dcs and impose a \hardt on it, thus creating a boolean mask to apply to the \dcs themselves.
As coefficients are real numbers, this corresponds to simply squaring the \dcs, without the need for an absolute value.
We then reconstruct the original spectra.
Notice that the choice of the \hardt and downsampling levels can lead to different results.
As only one \libs map is available for each scenario, we studied different values of these parameters.
We then compared the \pcs between different runs of \hyperpca and with the \pcs extracted by the \pca to get better insights on the quantity and quality of information extracted by \hyperpca.
The optimal values of \hardt and downsampling levels enable the best possible identification of the spectral signatures of the chemical elements present in the sample.
Empirically, we noticed that high \hardt often help in removing the unnecessary information, though the \dwt alone is not sufficient to lead to the cleanest possible \pcs (see~\Cref{app:ablation} for additional information).
Downsampling the signal beyond the first filter pass can help for very noisy \libs datasets, though the risk of losing relevant information on the spectral signal is a trade-off to take into account.

For completeness, we analysed both \dwts and continuous wavelet transforms (\cwts) with different profiles.
In particular, the profile needs to approximate the physical line shapes in the spectra, i.e.\ Voigt profiles.
This leaves, in general, a limited number of choices, such as symlets \dwts, or Mexican hat \cwts.
The usage of \cwts is addressed in~\Cref{app:ablation}, though it presents a few disadvantages and reconstruction artefacts.
In this analysis, we use discrete \emph{biorthogonal} wavelets as they represent the best option to approximate the physical profile, while also accounting for deformations due to interference.
Specifically, we focus on \texttt{bior3.*} in the \texttt{PyWavelets} module~\cite{lee_pywavelets_2019}.


\subsection{Kernel-based Sparse \pca}

\kspca is based on the entry-wise application of a real-valued kernel function $f$ to the correlation matrix $C = Y^T Y / (p - 1)$, where we assume the data matrix $Y \in \R^{p \times n_{\lambda}}$ to be centred, that is $Y = X - \barX$, where $\barX \in \R^{p \times n_{\lambda}}$ is a matrix containing the column-wise average contribution as rows.
It is possible to show that, under constraining conditions, there exists a class of such functions able to disentangle the stochastic, not necessarily normally distributed, noise component from the signal.
Restrictions mainly concern the size of the data matrix $Y$ which has to be large, though with a finite ratio between columns and rows: ideally, $p \to \infty$ and $n_{\lambda} \to \infty$ such that $n_{\lambda} / p \to q$.
Such request is usually met by \libs mapping datasets.
We choose the kernel function
\begin{equation}
    f_{\alpha}(t) = t \qty( 1 - e^{-\alpha t^2} ),
    \label{eq:kernel}
\end{equation}
as in the original paper in which the method was first presented~\cite{seddik_kernel_2019}, where $t$ can be any entry of the covariance matrix $C = \qty(t_{ij})_{i, j \in \qty[ 1,\, n_{\lambda}]}$.

The approach emerges by leveraging the potentiality of standard \pca with Random Matrix Theory (\rmt): while \pca itself does not rely on a specific model of the data, \rmt provides a framework in which to study how the unsupervised technique works.
The action of \pca aims at building the \pcs by isolating the largest eigenvalues associated to signal, the \emph{spikes}~\cite{johnstone_distribution_2001, perry_optimality_2018}, from the bulk distribution associated to noise.
Methods such as \kspca enable a better separation of the two components by acting directly on the distribution of eigenvalues: the application of the function~\eqref{eq:kernel} to the covariance matrix actively suppresses small entries, which are usually connected to the presence of noise, while it keeps the largest eigenvalues unchanged.
In our application, the expected effect of \kspca with respect to \pca is therefore a cleaner extraction of scores and loadings leading to a better definition of the distribution of elements on the sample, and to a better estimation of their contribution to the variance.
Such result relies on the input signal being sparse in a given basis, hence the need to introduce an intermediate step in the analysis, the \dwt.
\hyperpca also improves the ability of \pca to give a quantification of the confidence and importance contained in each extracted \pc in terms of the explained variance retained: by acting as a kernel-based algorithm at the level of the covariance matrix, the extracted values of the eigenvalues better approach their true values, thus providing a more realistic overview of the extraction.

\pca in its standard definition does not contain parameters which can be set.
\kspca introduces at least two possible choices: the functional form of~\eqref{eq:kernel} and its parameter $\alpha$.
For the first aspect, we refer to the original paper for an in-depth discussion~\cite{seddik_kernel_2019}.
The choice of the value of $\alpha$ is in general related to the signal to be reconstructed.
Experimentally, datasets containing noisy data need smaller values of $\alpha$ to filter out some contributions, while other cases may employ larger values to avoid losing relevant physical information.
We also notice that, in all cases we consider, a tangible difference in behaviour is shown for very different orders of magnitude of $\alpha$, while, for a fixed magnitude, its precise value does not heavily influence the outcome.
In this sense, while $\alpha$ is indeed a free parameter, it does not need fine tuning, and it can be optimised simply over different trials by comparing the quality of the extracted mappings and loading vectors.

Notice that \pca decomposes the original data matrix as $Y = S\, L$, where the \emph{scores} $S \in \R^{p \times n_{\lambda}}$ contain the rotated data, and the \emph{loadings} $L \in \R^{n_{\lambda} \times n_{\lambda}}$ contain the \pcs as row vectors (see~\Cref{app:pca} for the analytical derivation).
We conventionally adjust the signs of the loadings such that the most intense emission lines have positive values.
The sign of the scores are modified accordingly, to preserve the original data.
We also rescale both scores and loadings by the square root of the corresponding eigenvalues, that is by the standard deviation of the \pcs, to restore the information on signal intensity to the \pcs.
The columns of $S$ are then folded into $n \times m$ matrices and rescaled in the interval $[0,\, 1]$ to allow comparison between different score maps.


\section{Results and Discussion}

In this section, we use the pipeline presented in~\Cref{sec:methodology} to analyze different \libs mapping datasets.
In all cases, we provide a comparison with the standard \pca, recently reviewed for instance in~\cite{jolivet_review_2019}, and the line intensity maps, obtained in the traditional way.


\subsection{Synthetic Data}\label{sec:synthetic}

We first consider three synthetic datasets with different characteristics:
\begin{enumerate}
    \item a sample containing Ni and Au, where there are only two main emission lines, but in the presence of very high degree of stochastic noise and spatial superposition of the elements;

    \item a sample simulating a basalt rock, containing many interfering lines, but a very low noise contribution;

    \item a granite-like sample, with many interfering lines, strong spatial superposition and large noise component.
\end{enumerate}

Data were generated by first simulating a mixture of the elements in local thermodynamical equilibrium (\lte) using the \emph{NIST Atomic Spectra Database}~\cite{kramida_nist_2021}.
The spatial distribution was created by taking $100 \times 100$ portions of images (in this case, the front face of Euro coins), and thresholding it based on the relative concentrations of the elements.\footnotemark
\footnotetext{Images on Wikipedia, drawn by Luc Luycx, under a \emph{CC‑BY‑SA‑3.0} licence}
Experimental distributions are then created by extracting Poissonian distributed spectra $x_{(i)} \in \R^{n_{\lambda}}$, for $i = 1, 2, \dots, p$, from the \lte distributions, and adding stochastic noise as
\begin{equation}
    {x'}_{(i)}^{[n]} =
    \begin{cases}
        \abs{x_{(i)}^{[n]} + \beta\, N\, g_{a}\qty( x_{(i)}^{[n]} )} &\qif*{x_{(i)}^{[n]} > 0}
        \\
        \abs{\beta\, N} &\qif*{x_{(i)}^{[n]} = 0}
    \end{cases},
    \label{eq:synthesis}
\end{equation}
where $x_{(i)}^{[n]}$ represents an input intensity at wavelength channel $n \in \qty[ 1, n_{\lambda} ]$, $\beta$ is uniformly distributed in the interval $\qty[0, \beta_{\text{max}}]$ where $\beta_{\text{max}}$ differs for each dataset and $N$ is distributed according to a Gaussian with zero mean and unit variance.
The signal-to-noise ratio is of the order of $2.40 / \beta_{\text{max}}$ (see~\Cref{app:snr} for details on the computation).
In this section we show scenarios involving different values of $\beta_{\text{max}}$.
The sampling function $g_a$ has an exponentially suppressed behaviour $g_a( x ) = 1 - a\, e^{-x^2}$, where $a \in \R$ is defined on a case basis to control the intensity of the suppression term.
For all synthetic datasets, we simulate wavelengths in the region \SIrange{450}{500}{\nano\meter} with a resolution of $\lambda / \Updelta \lambda = 2048$.
Even though real-world cases do not present normally distributed noise, we first deal with Gaussian priors in order to better highlight the properties of \hyperpca in a controlled \rmt environment.
Even though some differences are present in the case of photonic (Poissonian) noise, the outcome does not strongly differ, and the conclusions stand also in the real-world datasets.

\subsubsection{Case 1: High Noise and Moderate Spectral Interference}\label{sec:niau}

\begin{figure}[t]
    \centering
    \includegraphics[width=0.5\linewidth]{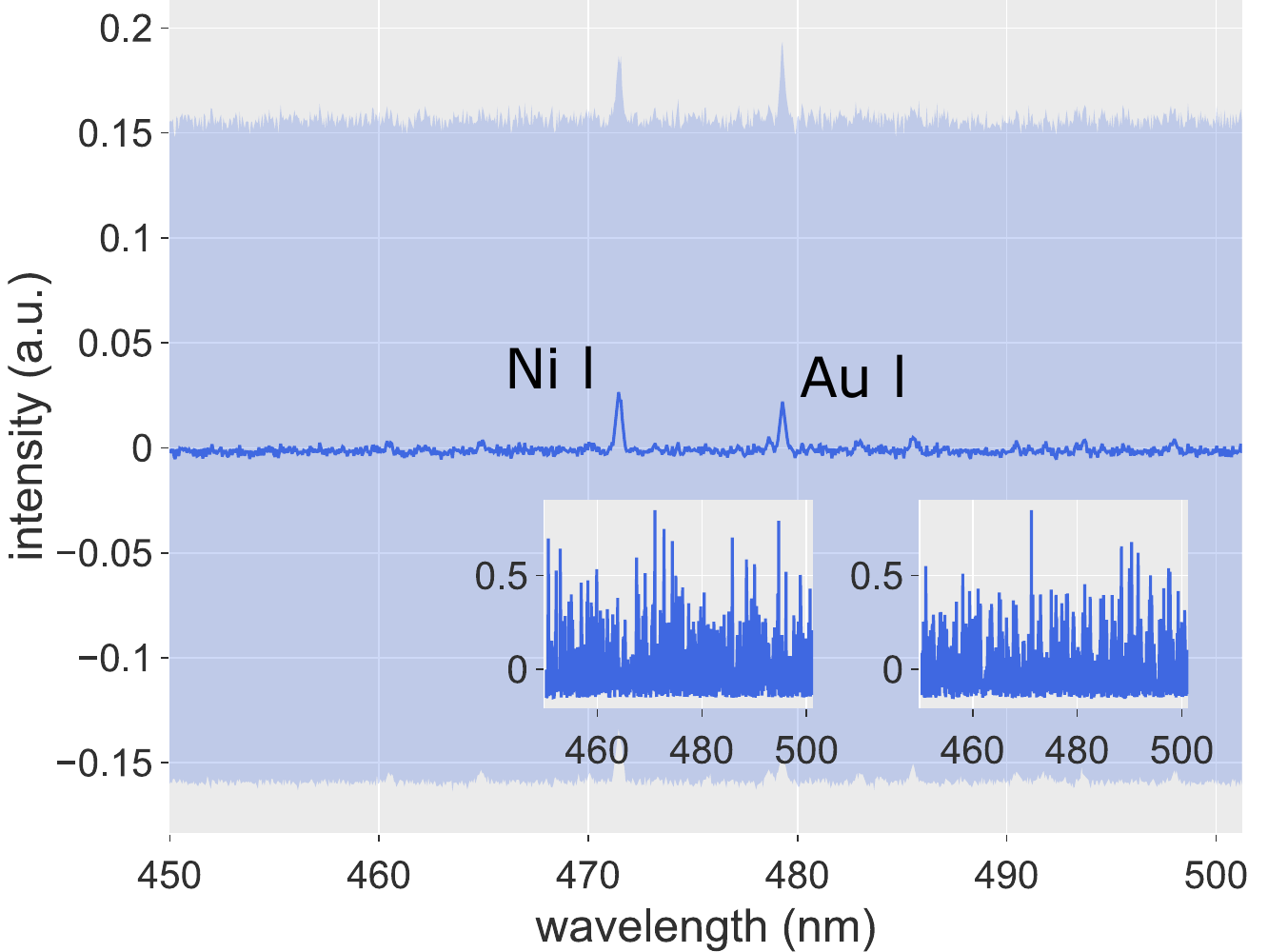}
    \caption{Preprocessed average and $1\upsigma$ spectra of the NiAu dataset with examples of single-shot spectra.}
    \label{fig:niau_spectra}
\end{figure}

\begin{figure}[t]
    \centering

    \subfigure[Ni]{\includegraphics[width=0.3\linewidth]{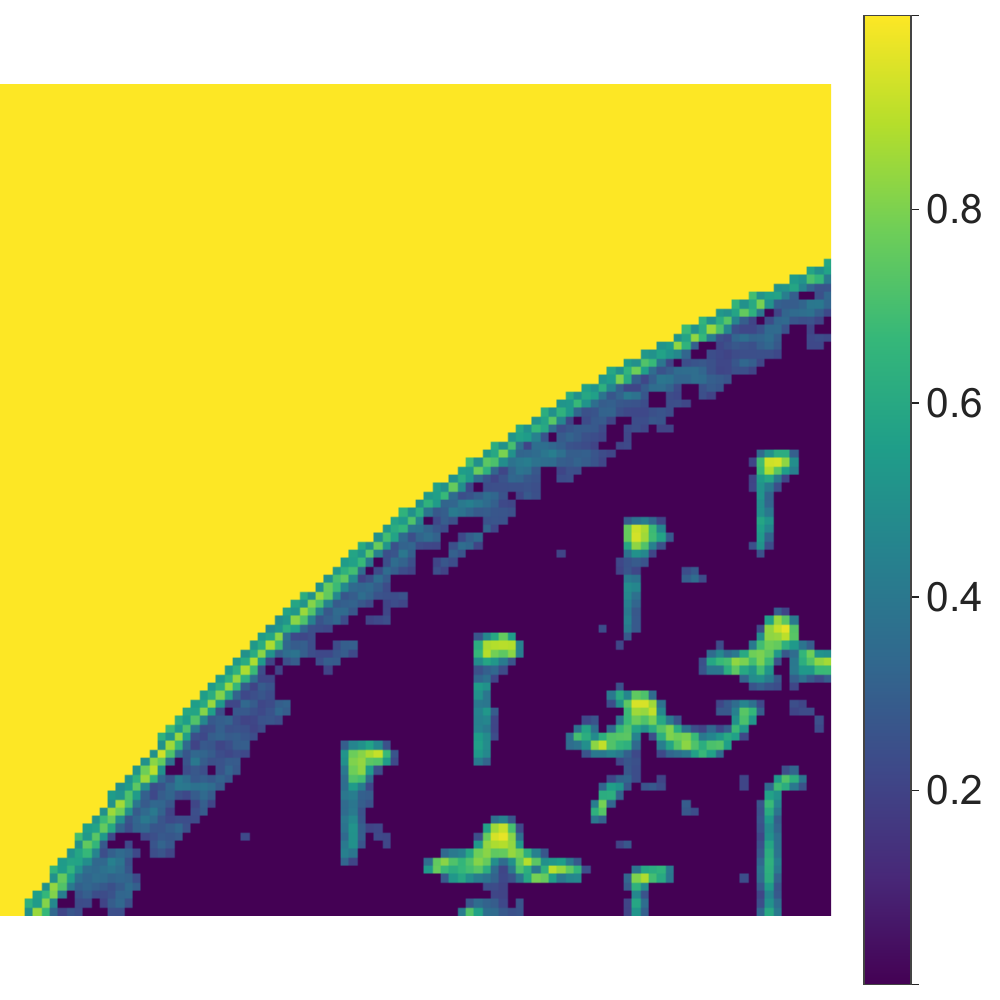}}
    \qquad
    \subfigure[Au]{\includegraphics[width=0.3\linewidth]{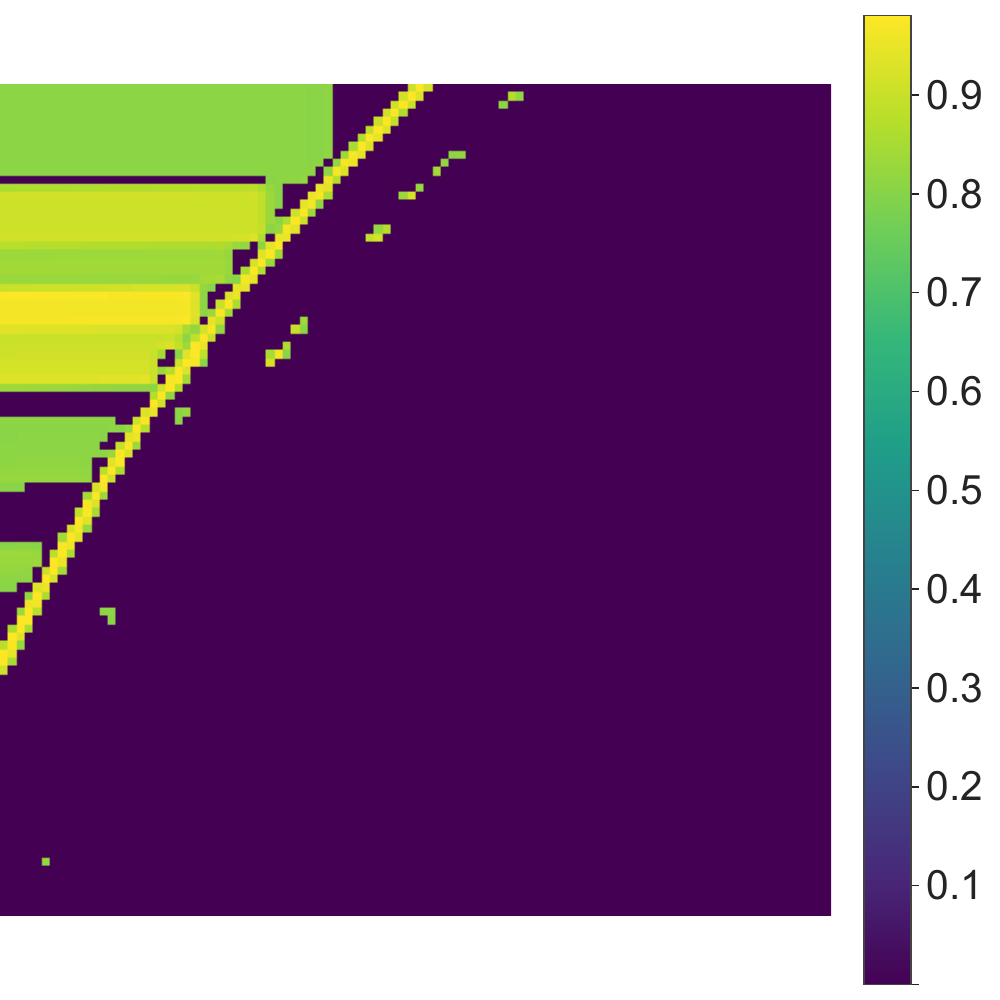}}
    \\
    \rule{0.85\linewidth}{1pt}
    \\
    \subfigure[\SI{471.4}{\nano\meter} (Ni)]{\includegraphics[width=0.3\linewidth]{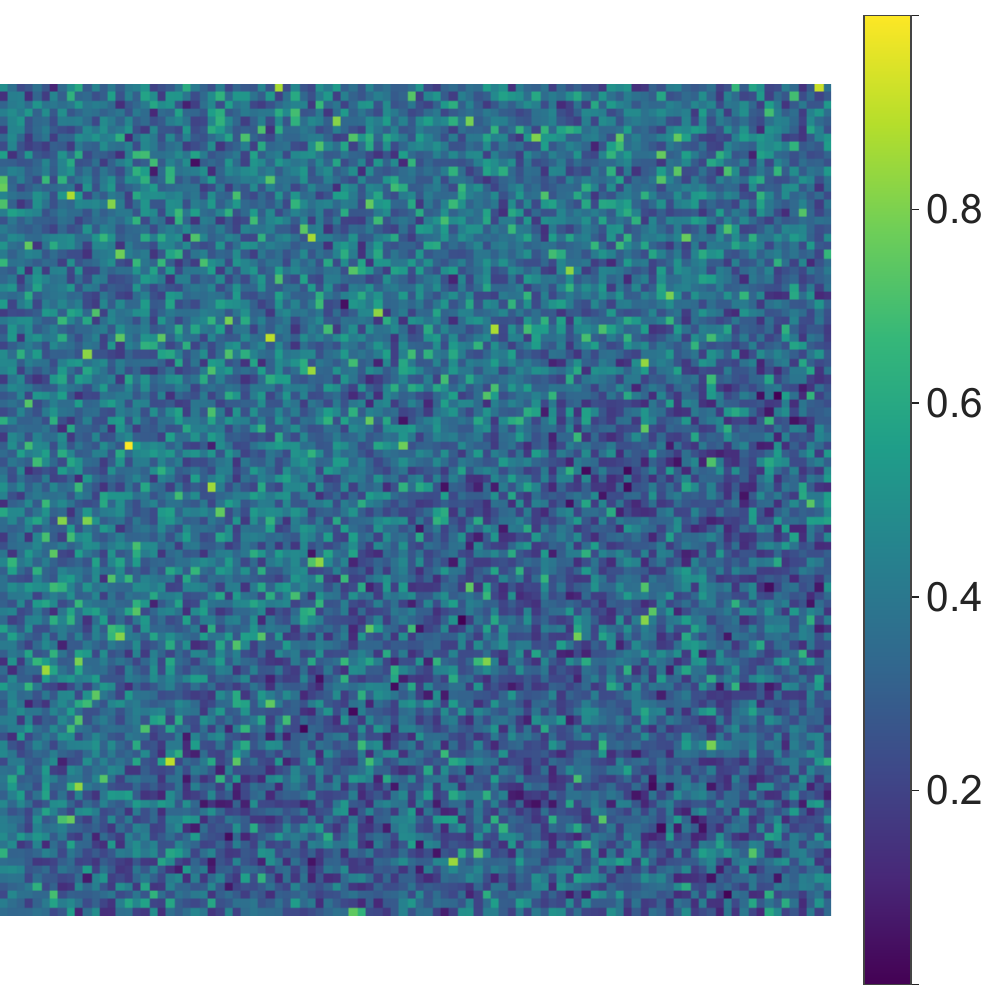}}
    \qquad
    \subfigure[\SI{479.3}{\nano\meter} (Au)]{\includegraphics[width=0.3\linewidth]{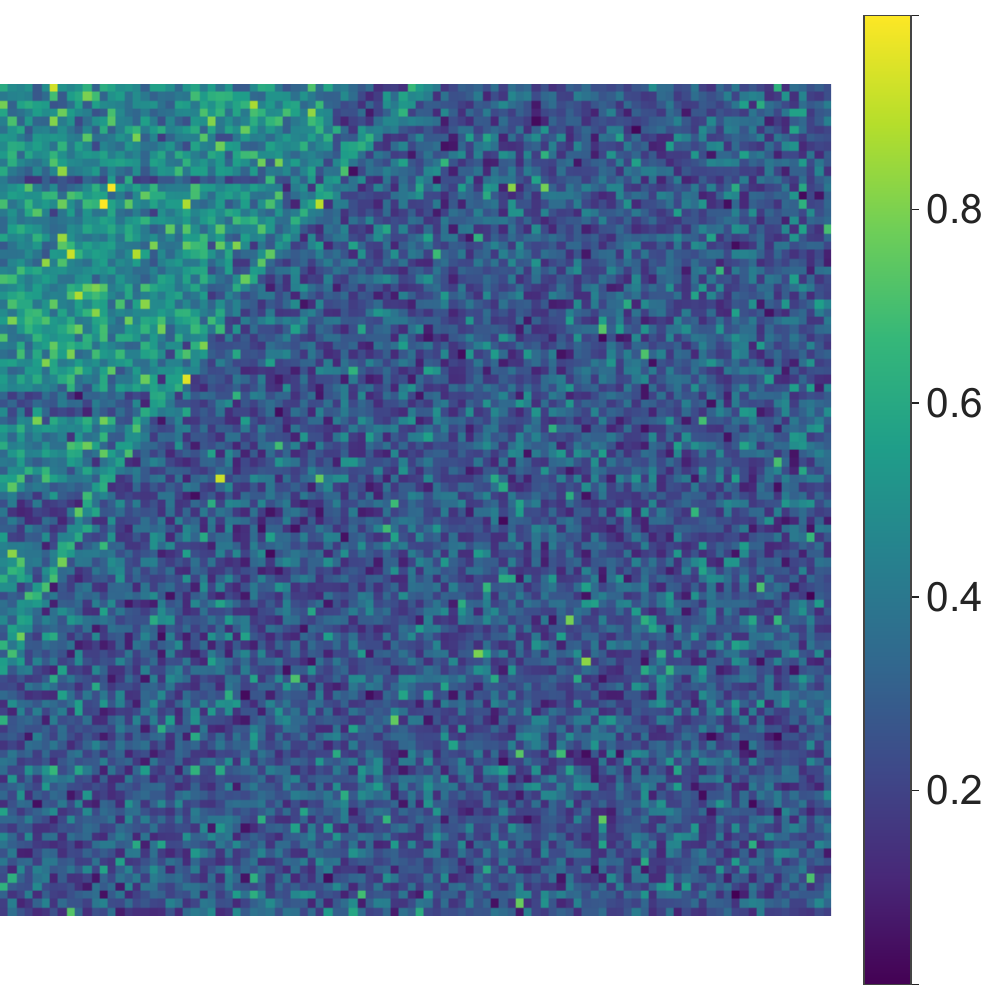}}

    \caption{NiAu reference elemental maps and line intensity maps. The scanned wavelength is reported. The element in parenthesis corresponds to the element whose intensity is maximal at the given value.}
    \label{fig:niau_ground_truth}
\end{figure}

The first dataset is a mixture of Ni (\SI{80}{\percent}) and Au (\SI{20}{\percent}).
The simulated plasma temperature is \SI{1}{\electronvolt}, and the electron density is \SI{2e17}{\per\centi\meter\cubed}.
In this case, the noise parameters are $a = \num{e-3}$ and $\beta_{\text{max}} = \num{70}$, i.e.\ a very low shot-to-shot signal-to-noise ratio, of the order of \num{3.4e-2}.
We use two layers downsampling of the data using the \dwt (\texttt{bior3.9}), and the kernel parameter $\alpha = \num{e-4}$.
The \hardt value is set to \SI{99}{\percent} of the maximum value of the power spectrum of the \dcs.

The preprocessed spectrum is shown in~\Cref{fig:niau_spectra}, while the spectra in \lte are in~\Cref{fig:niau_spectra_lte} in the appendix.
In this simulation we add a large noise base in order to test the extraction abilities of \pca and \hyperpca.
The main emission lines of Ni at \SI{471.44}{\nano\meter} and Au at \SI{479.26}{\nano\meter} are visible in the average spectrum, even though they may be hidden by the noise in single spectra.
Reference elemental distributions are shown in~\Cref{fig:niau_ground_truth}: the sample describes a substrate with a diffused and intense distribution of Ni and a quite concentrated insertion of Au in the top left part.
Notice that it completely overlaps with the Ni distribution, even though the two main emission lines should be easily resolvable.
The bottom of~\Cref{fig:niau_ground_truth} shows the line intensity maps obtained over a window of \num{10} sensor pixels across the most intense wavelength channels: due to noise, only the reconstruction of the Au distribution is significant.

\begin{figure}[t]
    \centering
    \begin{tabular}{c|c}
        {\LARGE \textsc{standard} \pca}                          & {\LARGE \hyperpca}
        \\[1em]
        \includegraphics[width=0.47\linewidth]{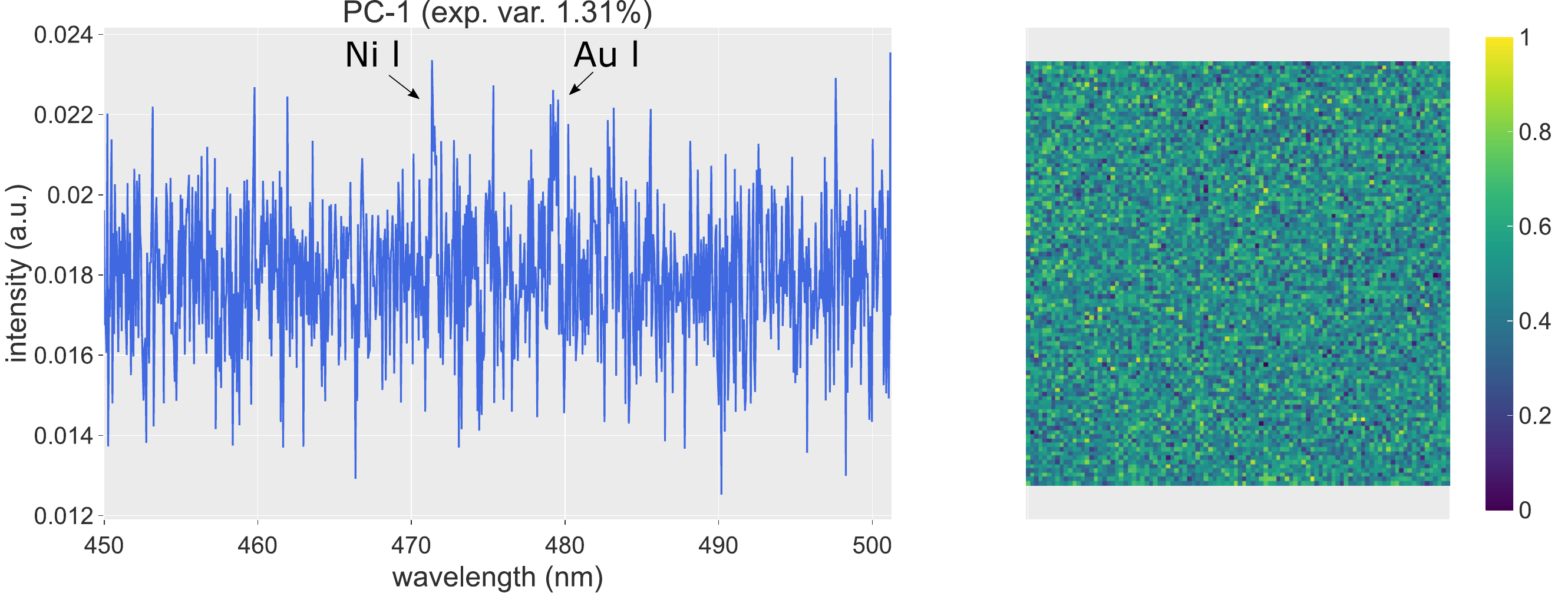} & \includegraphics[width=0.47\linewidth]{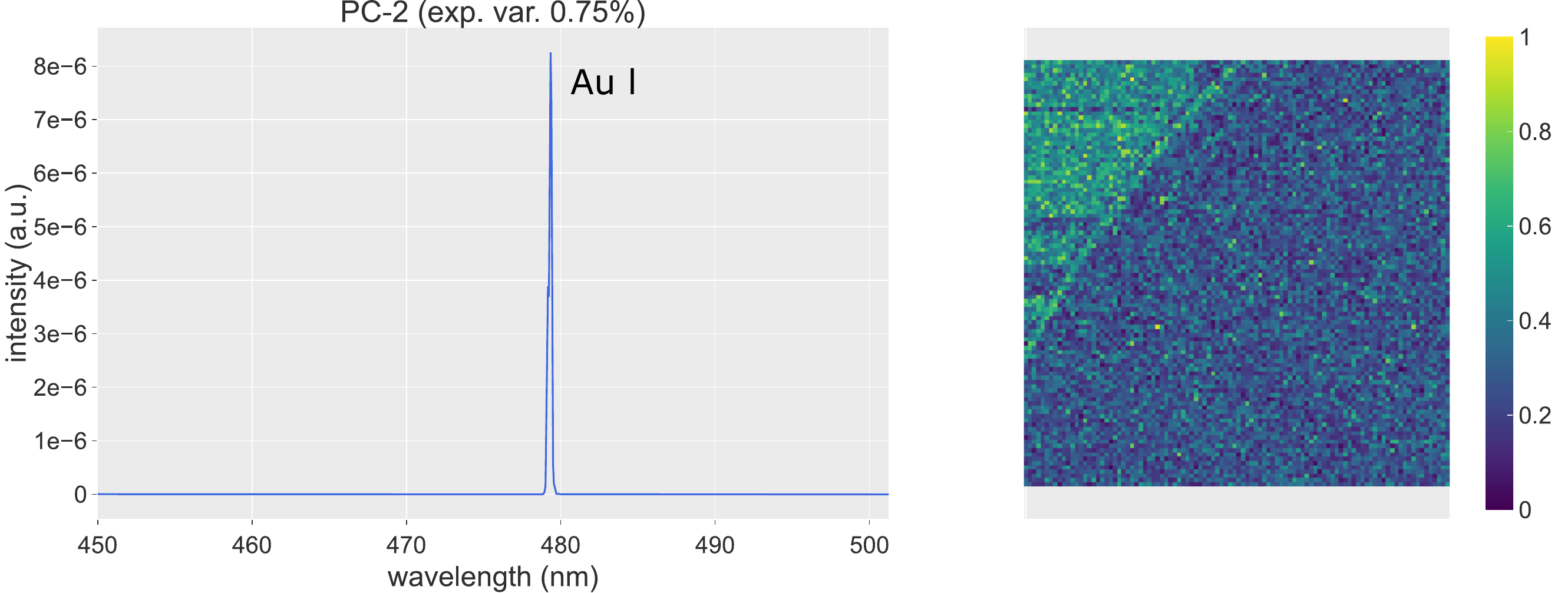}
        \\
        \includegraphics[width=0.47\linewidth]{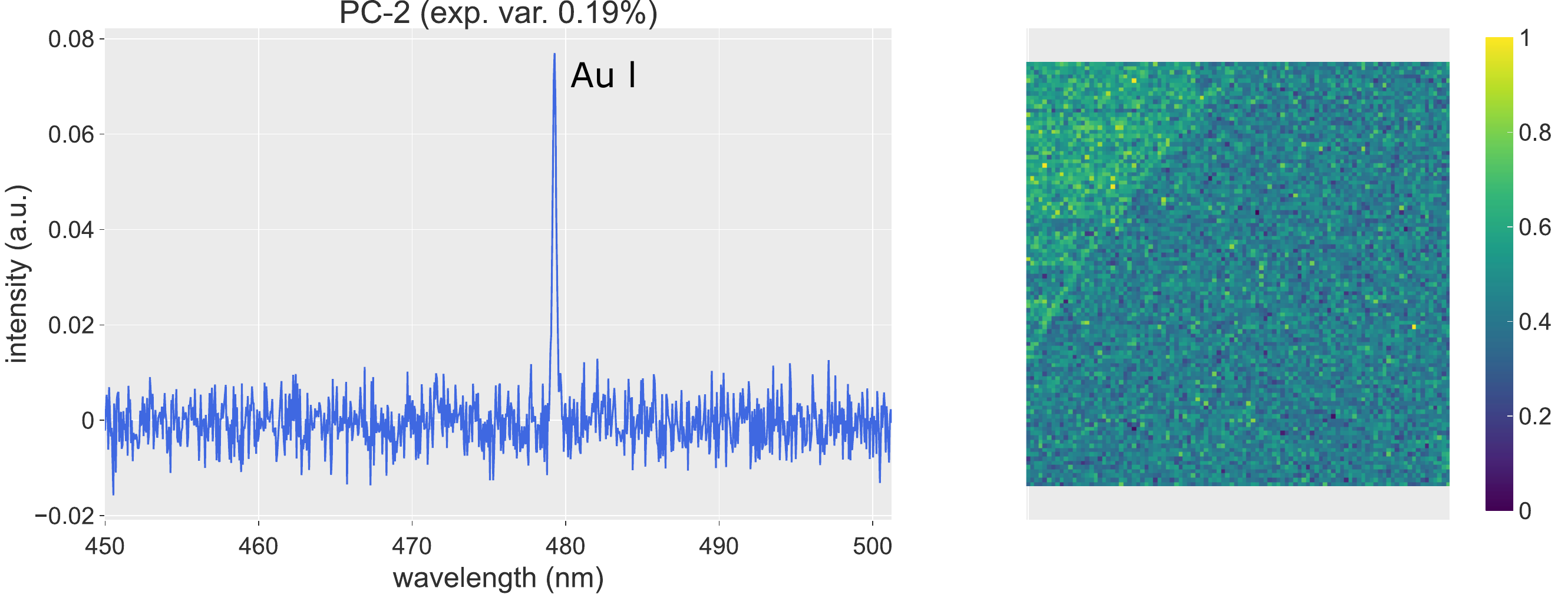} & \includegraphics[width=0.47\linewidth]{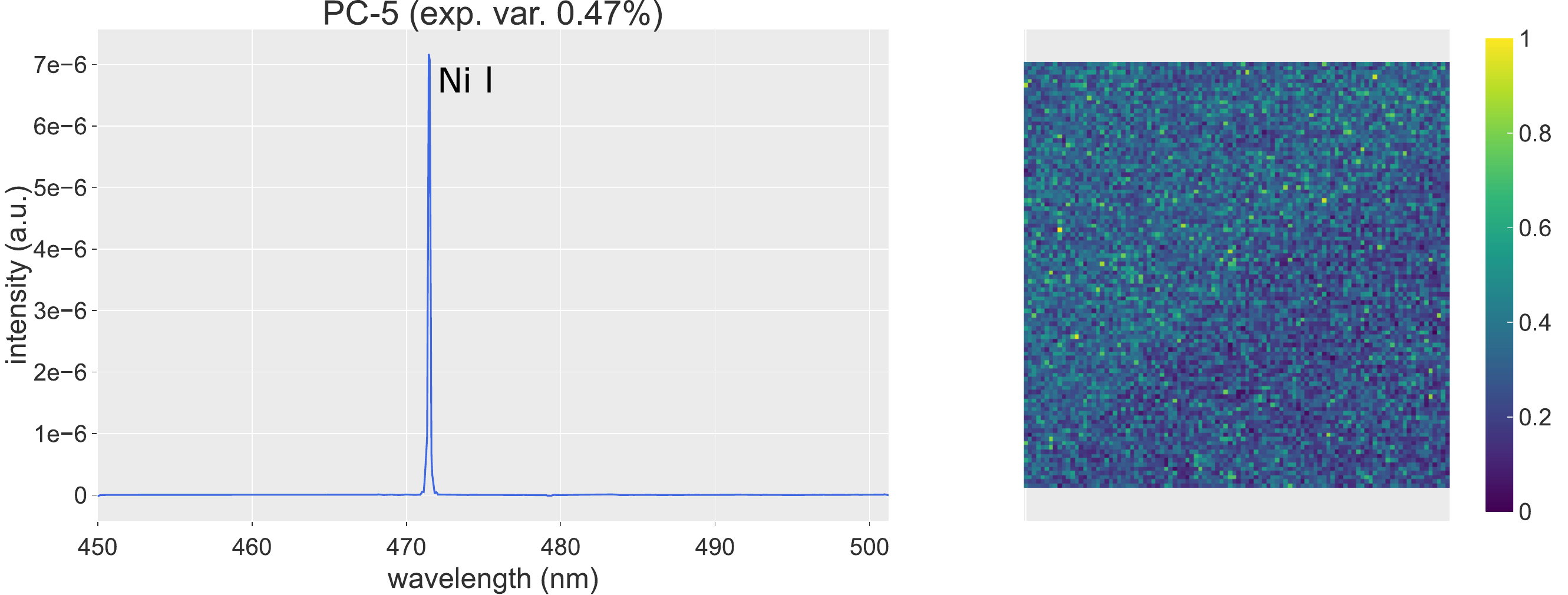}
        \\
        \includegraphics[width=0.47\linewidth]{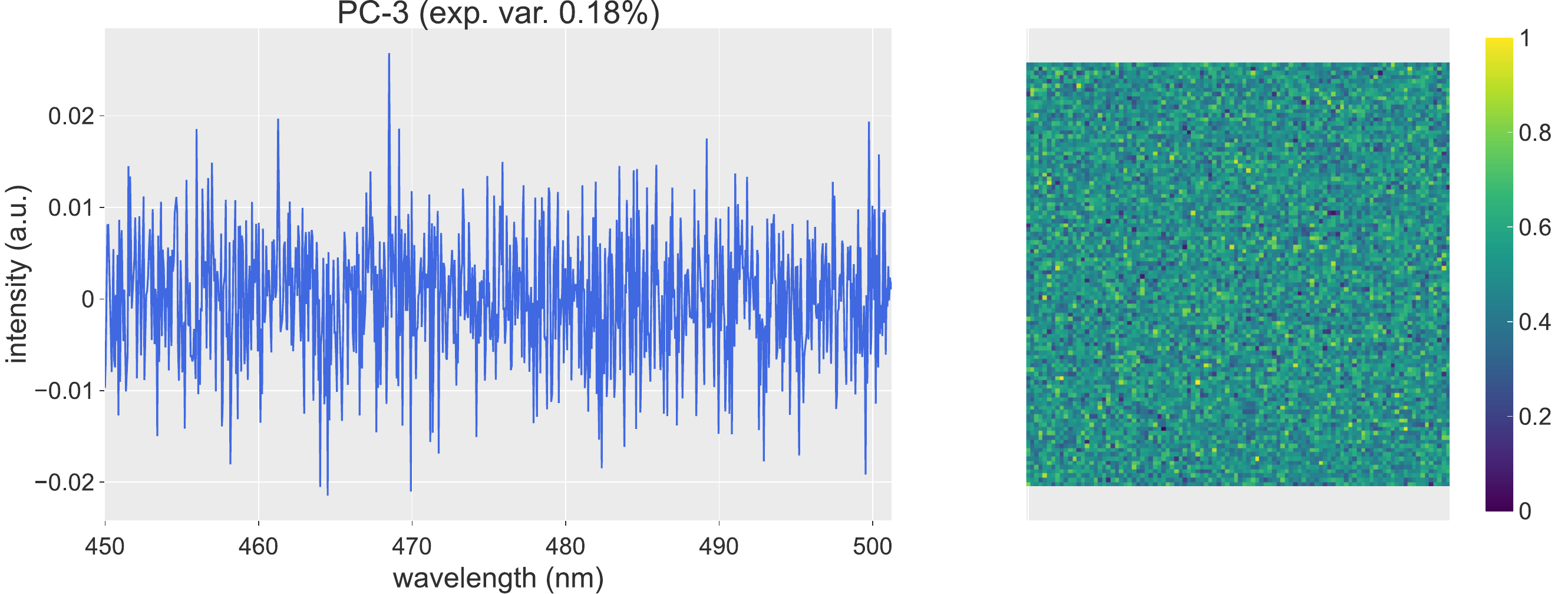} &
    \end{tabular}
    \caption{NiAu reconstruction using the standard \pca and \hyperpca.}
    \label{fig:niau_pca}
\end{figure}

In the first column of~\Cref{fig:niau_pca} we show the extraction of the first \pcs using the current approach with standard \pca.
As for \pca, the addition of a large noise component inevitably leads to a very noisy first \pc: by definition, the average spectrum could be used as an approximation up to a global rescaling factor of such component, which is therefore informative, though not easy to interpret.
As the noise is easily extracted by this component, the information retained is not entirely useful for the extraction of the maps.
The Au principal line is extracted in \pcn{2}: the mapping starts to acquire some physical significance, but the underlying random noise distribution spoils the result.
Information on the Ni distribution cannot be retrieved from subsequent \pcs.

With respect to \pca, \hyperpca suppresses the noise components, thus extracting the main emission lines of Ni and Au as separate contributions.
The second columns of \Cref{fig:niau_pca} shows \pcn{2} and \pcn{5} containing Ni and Au separately.
Other \pcs between those present effects due to experimental conditions: noise cancellation and wavelet filtering expose uncertainties due to photons being captured by different pixels in the detection system
Though, chemically, these \pcs may represent the same element, they must be extracted to reflect the correct geometrical rank of the signal matrix.
\hyperpca leads to a more precise representation of the signal, at the cost of a greater number of \pcs being extracted.
For instance, \Cref{fig:niau_dwt-kpca_derivative} in the appendix shows one component presenting a residual effect due to photons being detected by neighboring pixels of the CCD sensor: this was extracted as a \emph{first derivative} signal of the Au line.
The principal contribution of \hyperpca is therefore a better extraction of the loading vectors and their associated explained variance ratio: though less than \SI{1}{\percent}, it is still important to quantify correctly these values in order to extract the whole information contained in the data as accurately as possible.
Improvements are also present in the score maps as the Au map in \pcn{2}, in this case, is definitely better defined than in the case of \pca, but it also presents a neater edge segmentation with respect to the line intensity maps in~\Cref{fig:niau_pca}.
The mapping of the Ni distribution, not extracted by \pca, is here visible in a lighter nuance of the colour map in the top left part of \pcn{5} on the right of~\Cref{fig:niau_pca}.
The increase in quality is also visible with respect to the line intensity maps in~\Cref{fig:niau_pca} as the edge of the Ni distribution is neater in \hyperpca, and the almost total removal of the background noise leads to a better definition of the Ni distribution in the bottom right part of the mapping, as shown in the reference maps in~\Cref{fig:niau_ground_truth}.

The proposed method increases the quality and readability of the loadings by removing the noise components, thus showing good resolution of the elements.
The improved differentiation of the emission lines as separate signals enables the resolution of the interference created by the superposition of the Au distribution to the Ni-dominated pixels, leading to more accurate elemental maps.
Moreover, as in the case of \pca, no previous knowledge of the sample is necessary for \hyperpca.
In fact, the elemental maps extracted with \hyperpca in~\Cref{fig:niau_pca} are the result of an unsupervised analysis, differently from the extraction using the line intensities, which makes the interpretation of the results straightforward through the loadings.

\subsubsection{Case 2: Low Noise and Strong Spectral Interference}

\begin{figure}[t]
    \centering
    \includegraphics[width=0.5\linewidth]{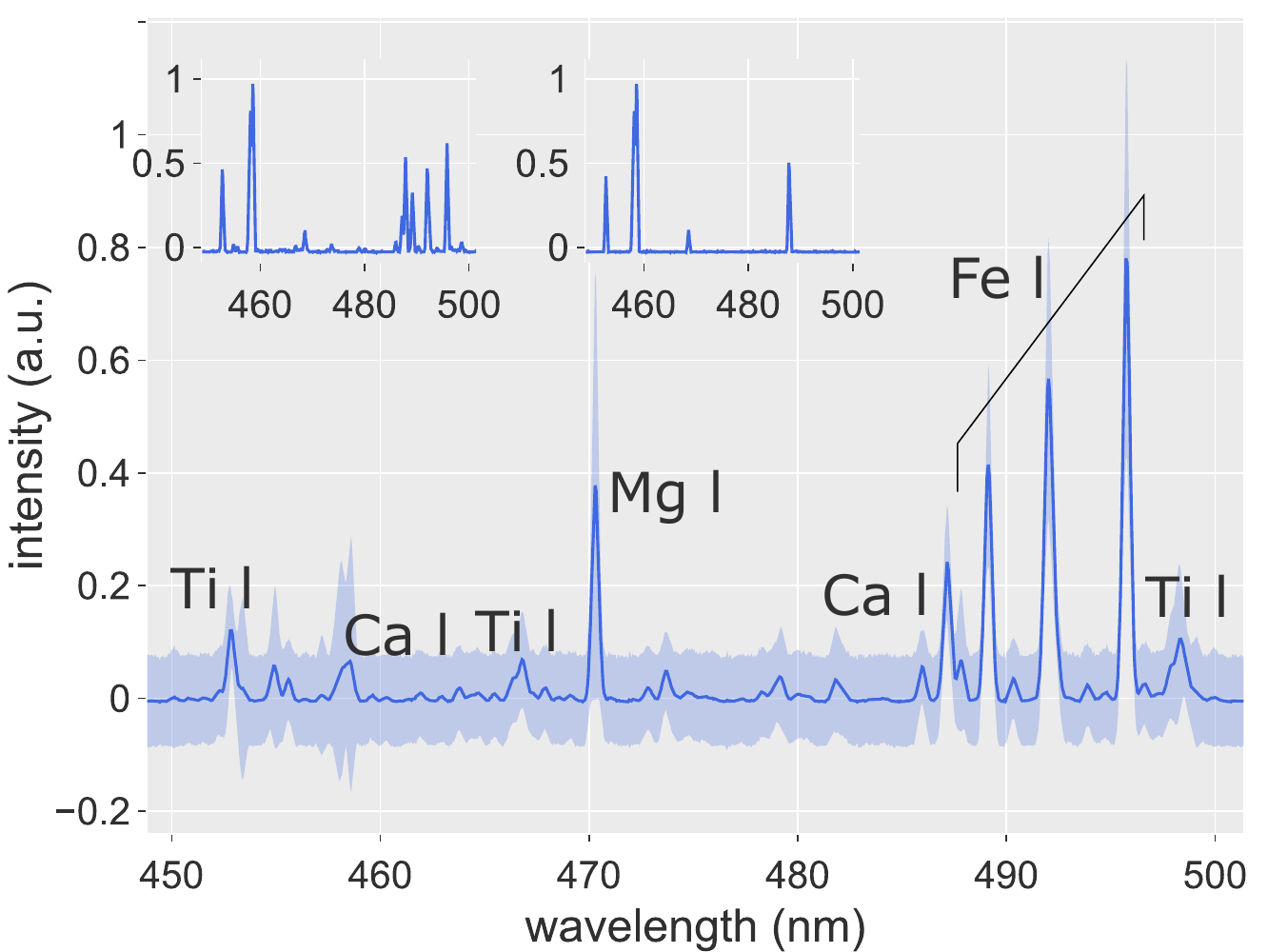}
    \caption{Preprocessed average and $1\upsigma$ spectra of the ``basalt'' dataset with examples of single-shot spectra.}
    \label{fig:basalt_spectra}
\end{figure}

We then consider a basalt-like mixture of Si (\SI{50}{\percent}), Al (\SI{16}{\percent}), Fe (\SI{10}{\percent}), Ca (\SI{9}{\percent}), Mg (\SI{7}{\percent}), Na (\SI{4}{\percent}) and Ti (\SI{4}{\percent}).
The plasma temperature is \SI{1}{\electronvolt}, and the electron density \SI{5e17}{\per\centi\meter\cubed}.
In this dataset, $a = 5$ and $\beta_{\text{max}} = 10$, i.e.\ an average shot-to-shot signal-to-noise ratio of \num{2.4e-1}.
For \kspca, we choose the kernel parameter $\alpha = \num{e-4}$.
We compute the \dwt (\texttt{bior3.5}) at the first downsampling layer, with a \hardt of \SI{99}{\percent} of the maximum value of the power spectrum of the \dcs.
We show the preprocessed average spectrum in~\Cref{fig:basalt_spectra} and the theoretical \lte distributions in~\Cref{fig:basalt_spectra_lte} of the appendix.

\begin{figure}[t]
    \centering

    \subfigure[Si]{\includegraphics[width=0.24\linewidth]{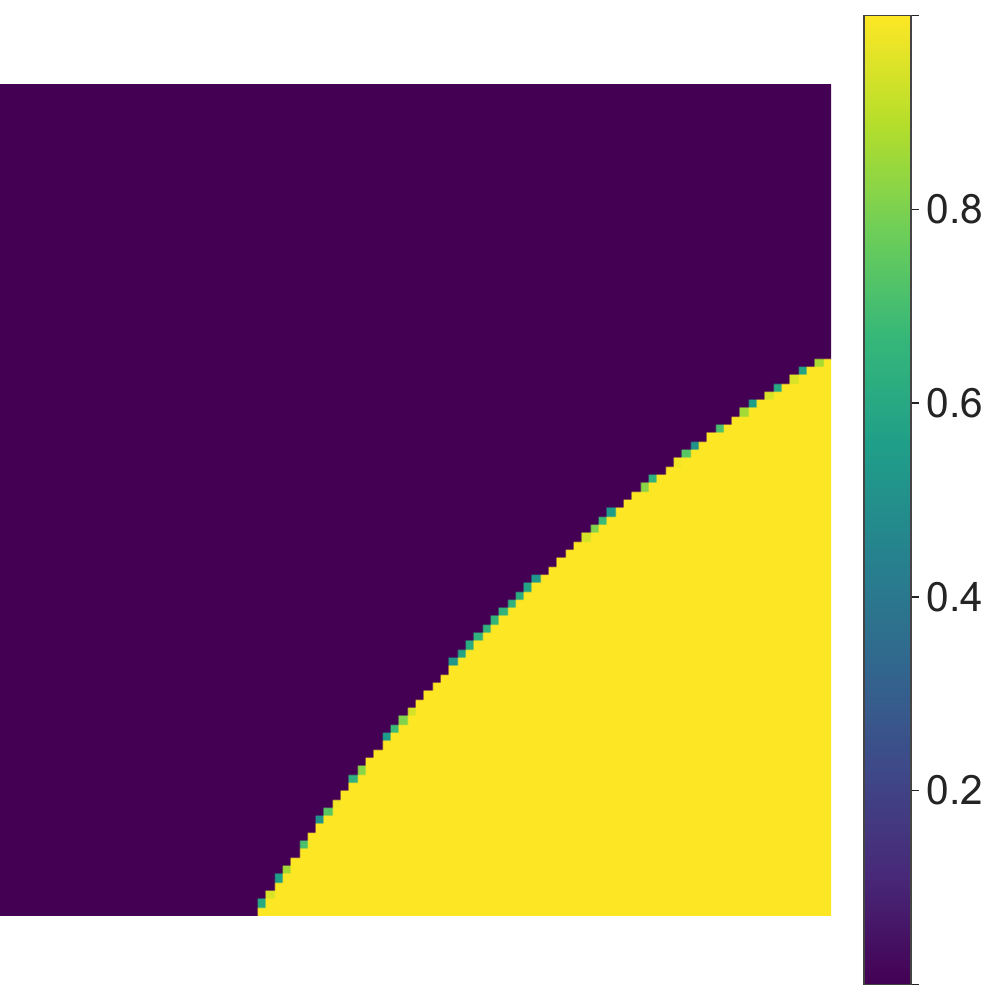}}
    \hfill
    \subfigure[Al]{\includegraphics[width=0.24\linewidth]{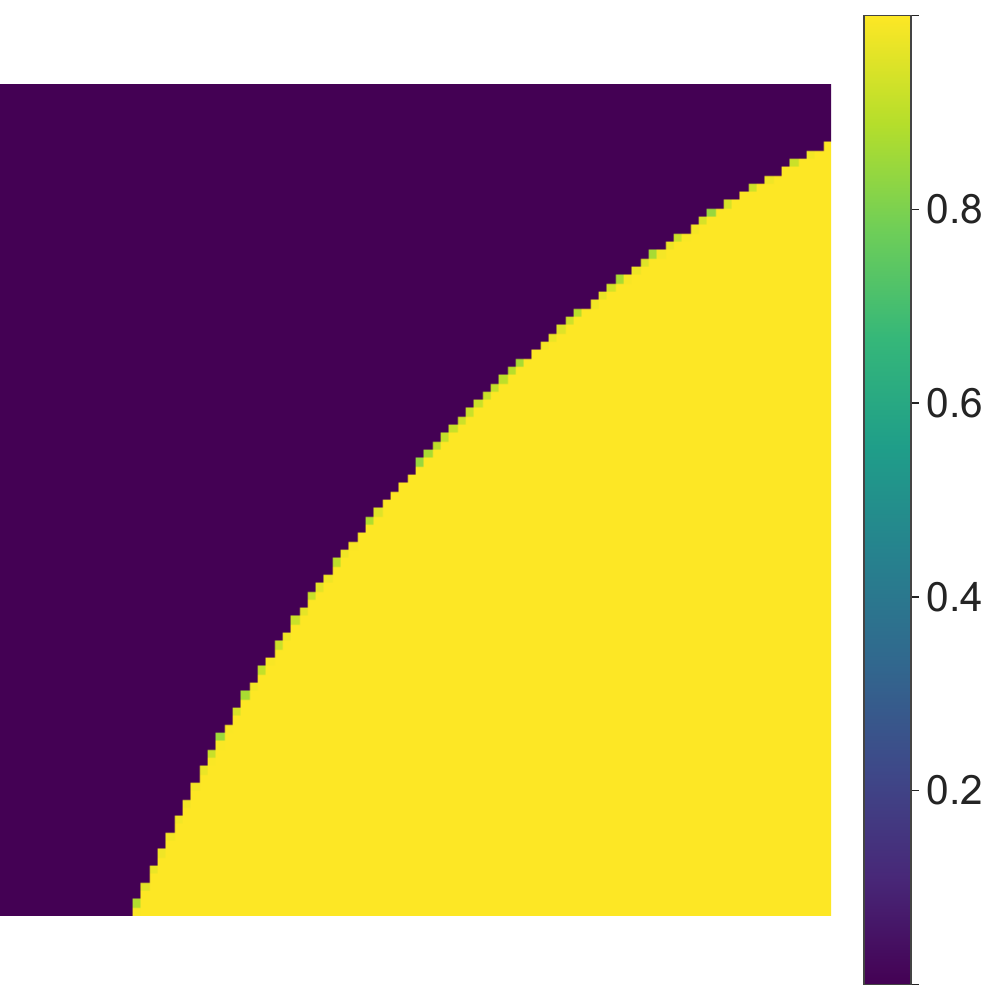}}
    \hfill
    \subfigure[Fe]{\includegraphics[width=0.24\linewidth]{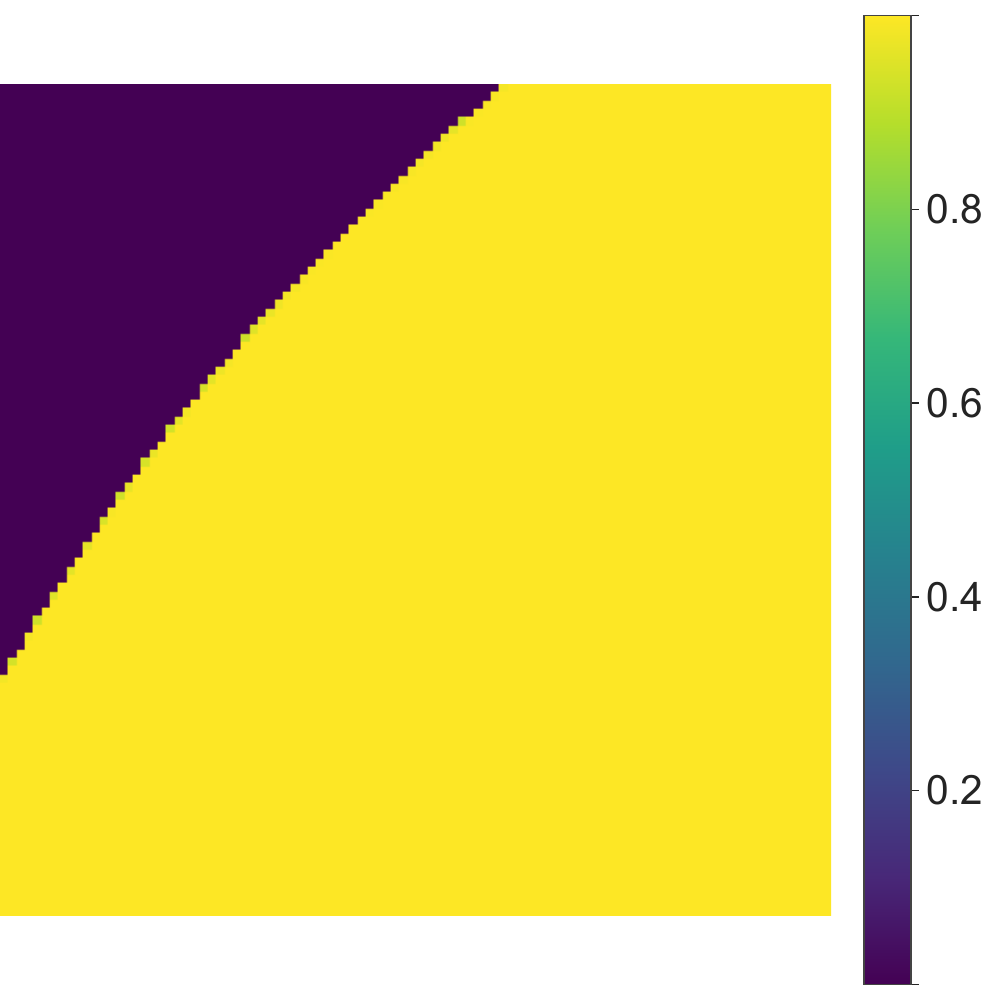}}
    \hfill
    \subfigure[Ca]{\includegraphics[width=0.24\linewidth]{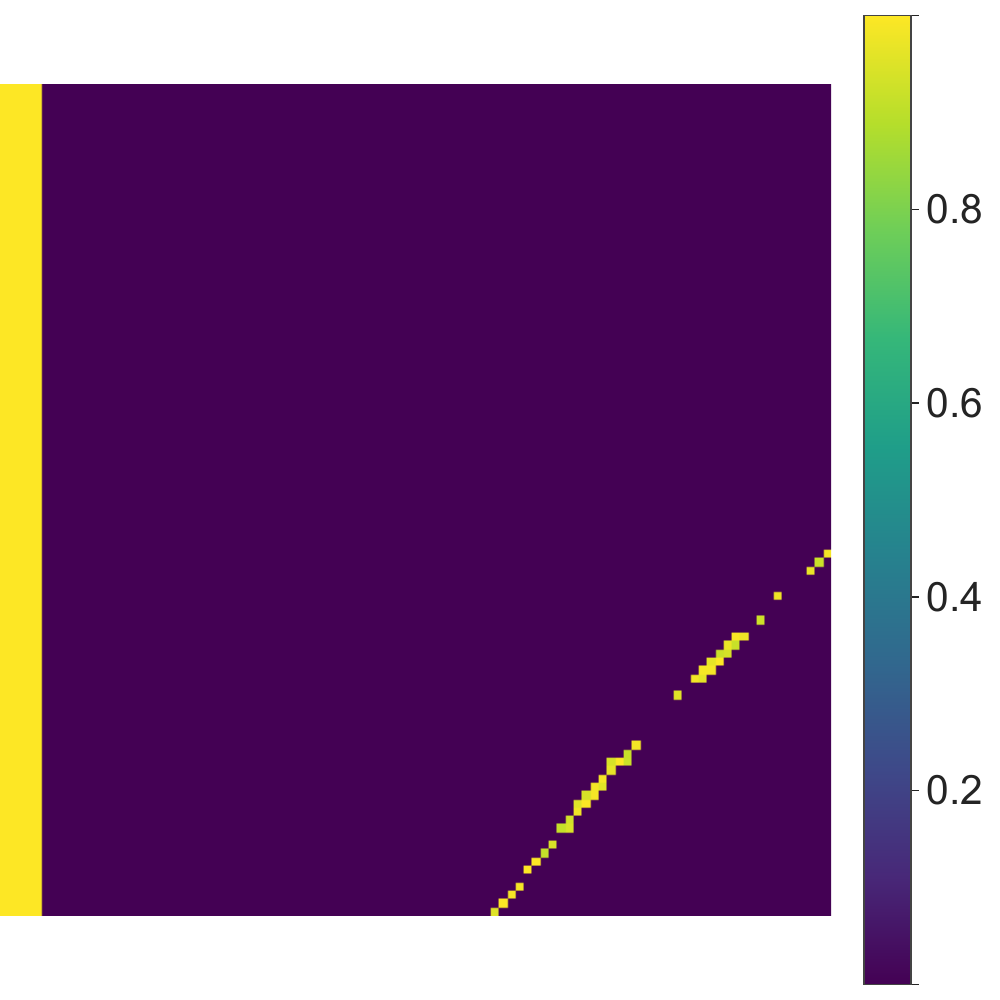}}
    \\
    \subfigure[Mg]{\includegraphics[width=0.24\linewidth]{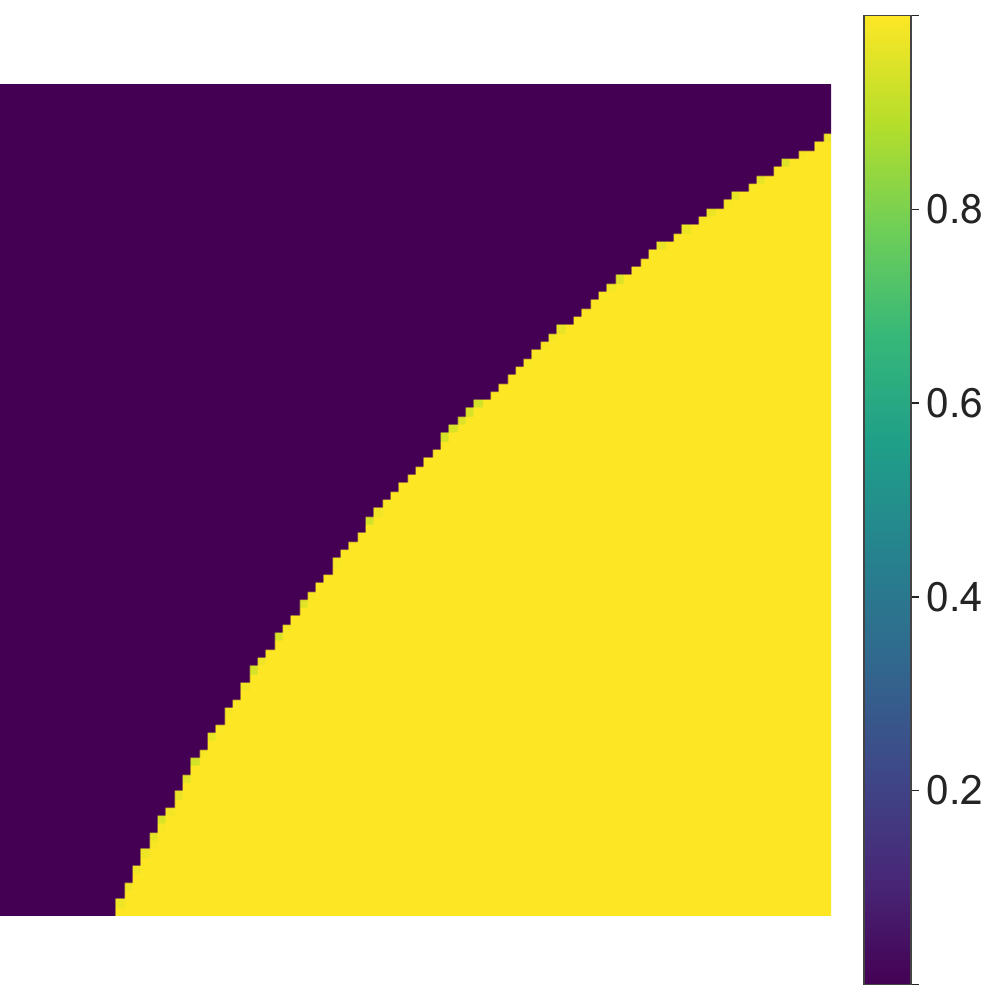}}
    \qquad
    \subfigure[Na]{\includegraphics[width=0.24\linewidth]{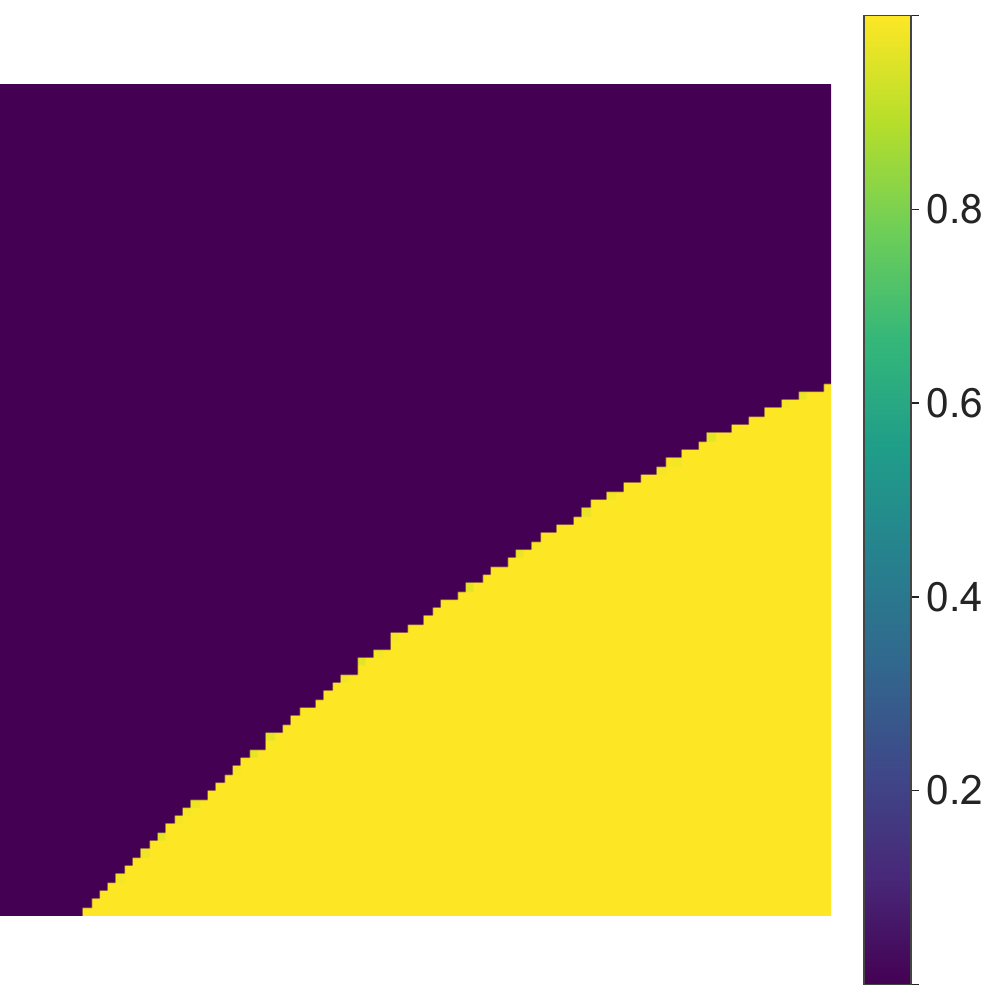}}
    \qquad
    \subfigure[Ti]{\includegraphics[width=0.24\linewidth]{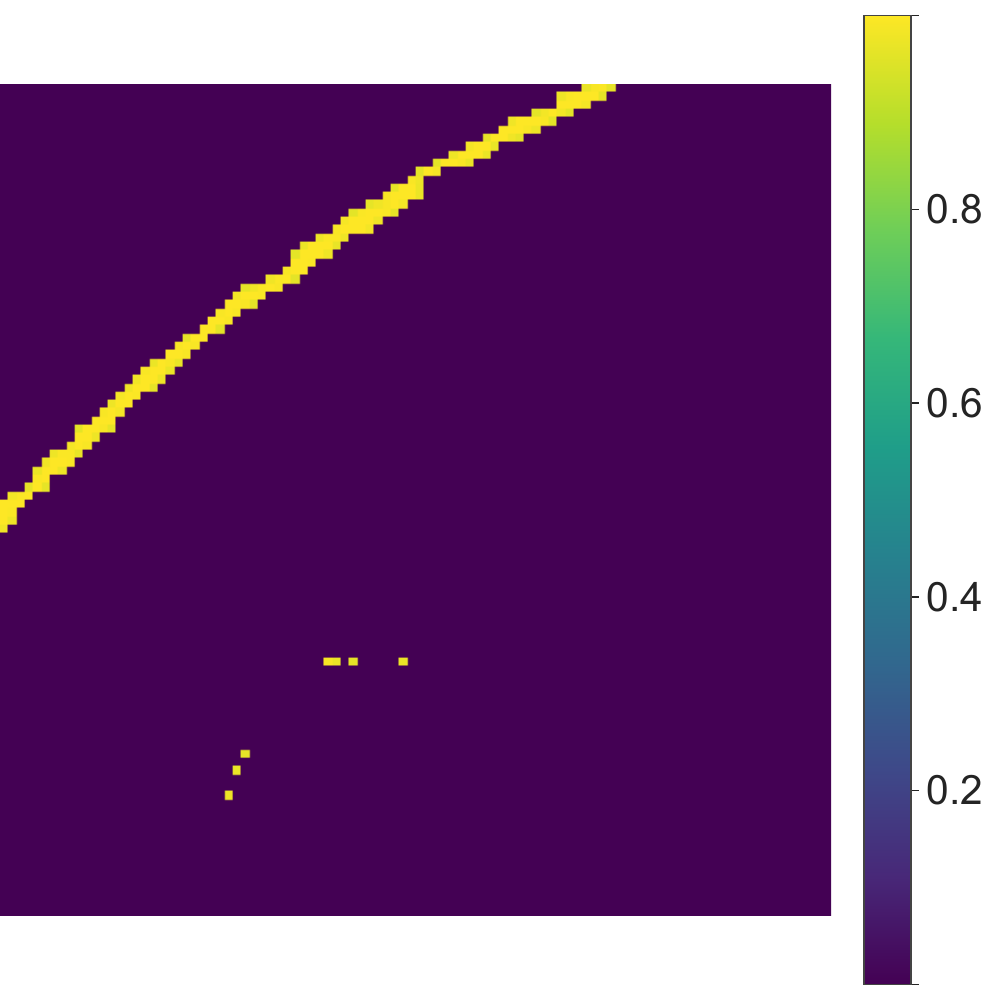}}
    \\
    \rule{0.85\linewidth}{1pt}
    \subfigure[\SI{452.8}{\nano\meter}, \SI{454.9}{\nano\meter} and \SI{498.4}{\nano\meter} (Ti)]{%
        \includegraphics[width=0.185\linewidth]{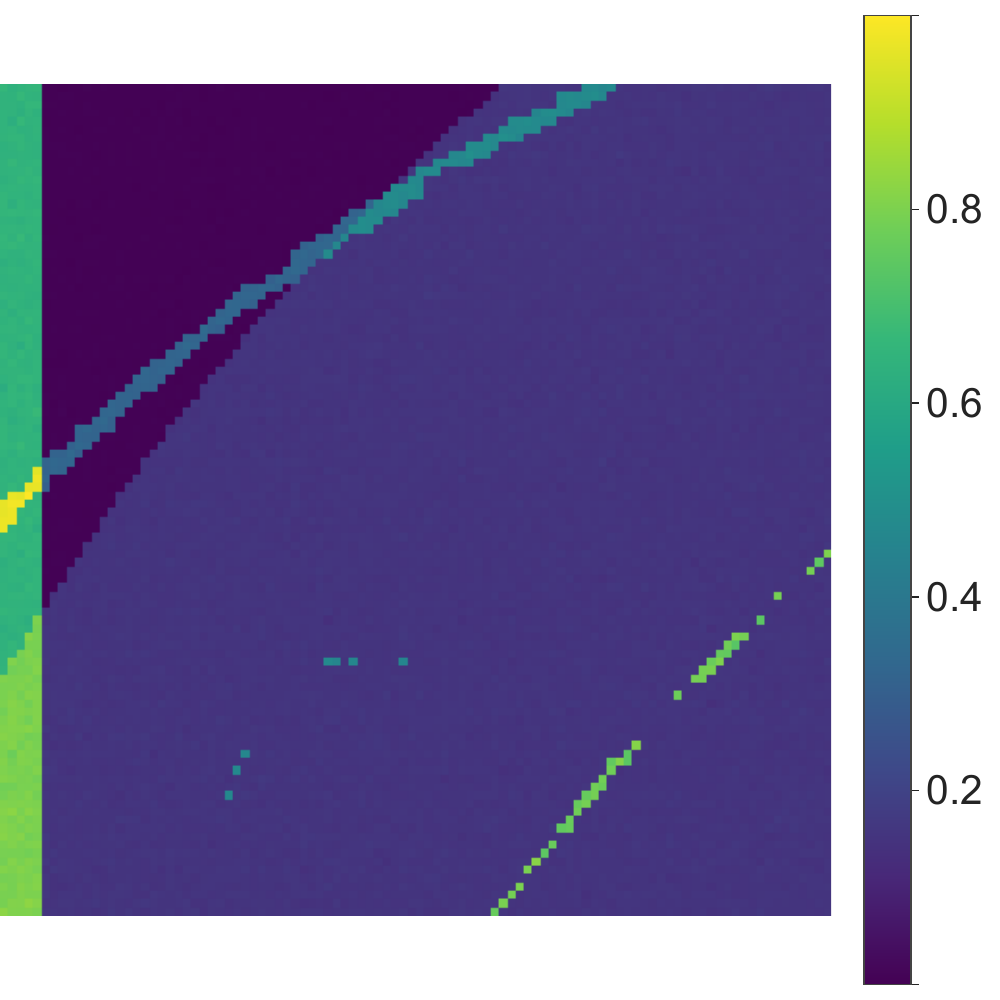}
        \includegraphics[width=0.185\linewidth]{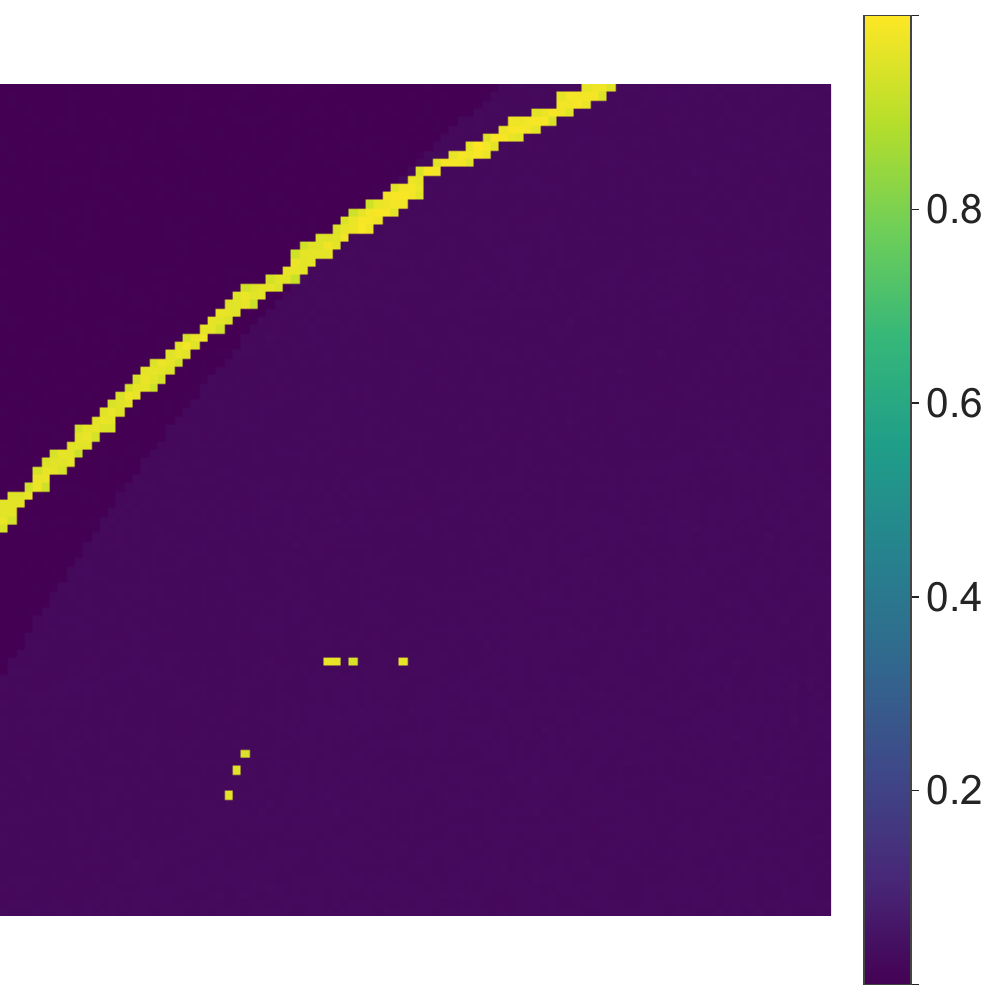}
        \includegraphics[width=0.185\linewidth]{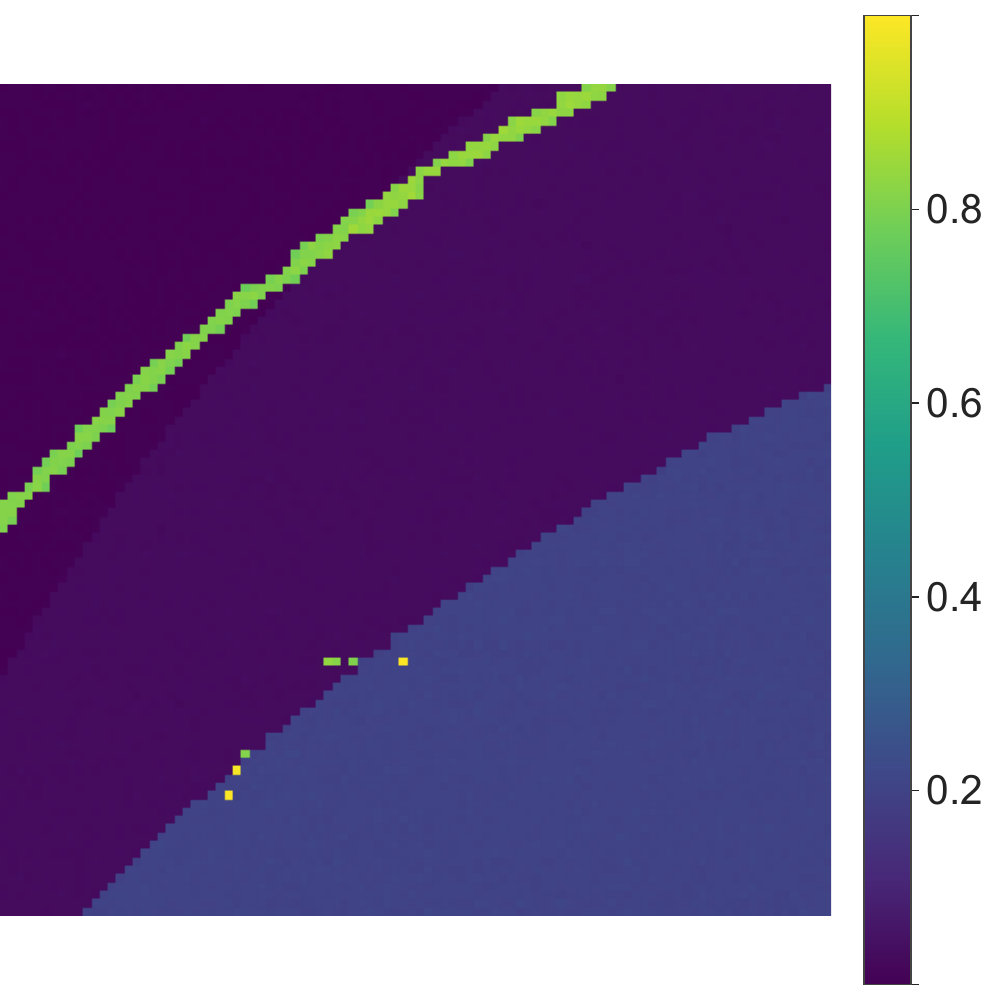}
    }
    \hfill
    \subfigure[\SI{458.6}{\nano\meter} and \SI{487.2}{\nano\meter} (Ca)]{%
        \includegraphics[width=0.185\linewidth]{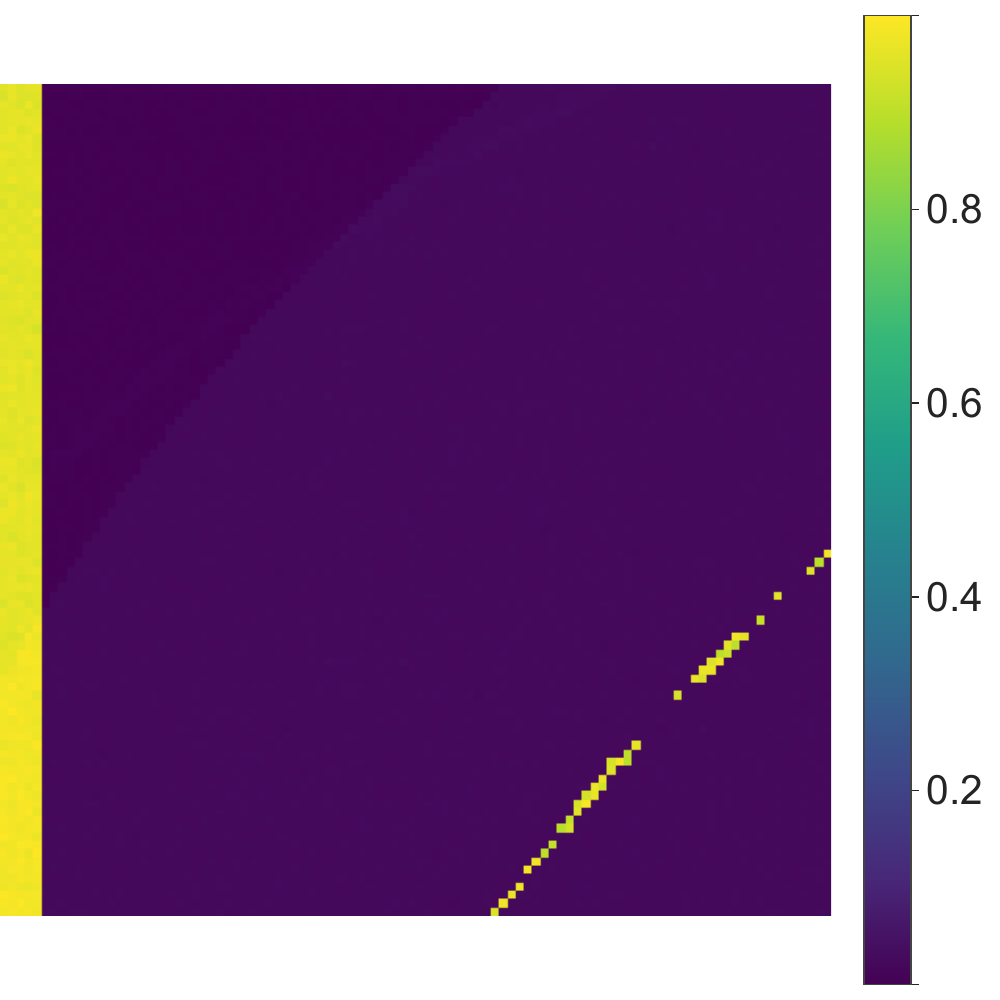}
        \includegraphics[width=0.185\linewidth]{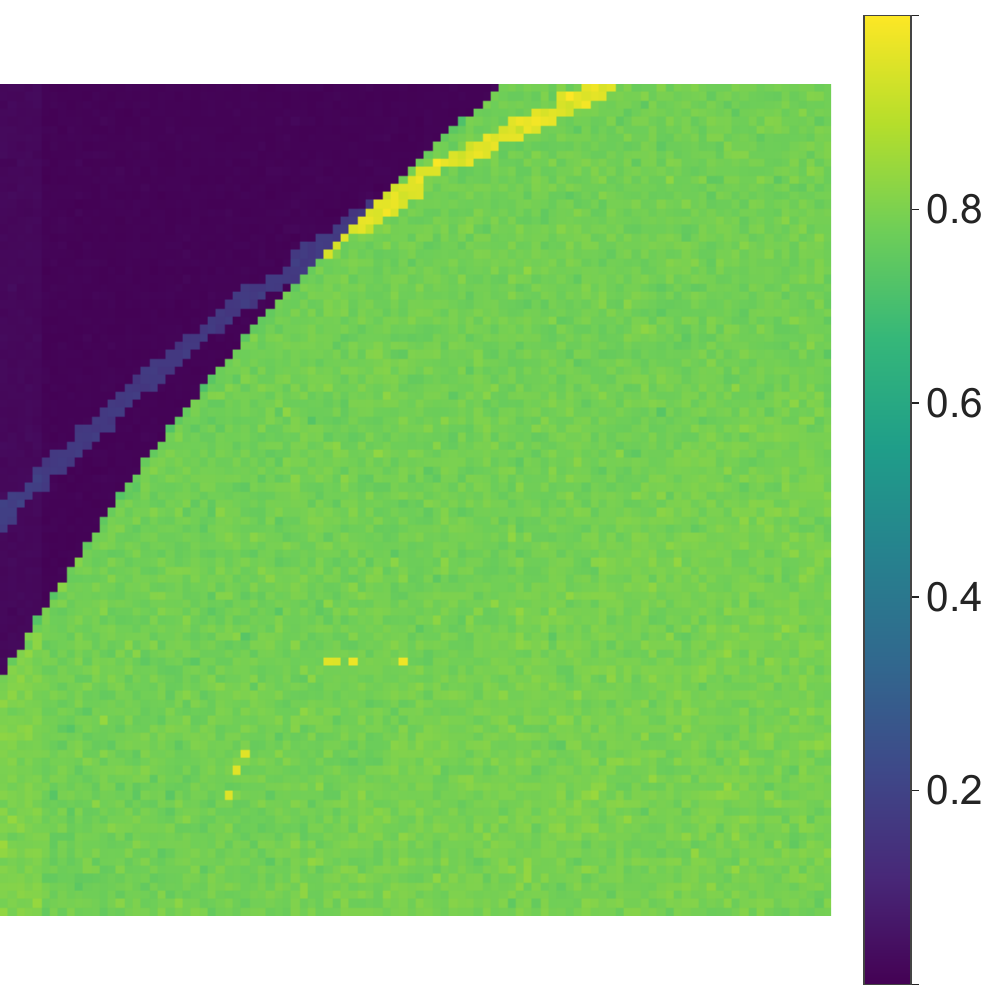}
    }
    \\

    \caption{``Basalt'' reference elemental maps and line intensity maps. The scanned wavelength is reported. The element in parenthesis corresponds to the element whose intensity is maximal at the given value.}
    \label{fig:basalt_ground_truth}
\end{figure}

In~\Cref{fig:basalt_ground_truth}, we show the  reference elemental maps of the elements.
The sample describes a basalt-like rock formation with an intense and superimposed presence of Si, Fe, Al, Mg and Na as main components, and concentrated insertions of Ca and Ti across the surface.
As the elements overlap frequently and for extended portions of the sample, we expect a high degree of spectral interference in the reconstruction, especially from emission lines which have similar characteristic wavelengths.
In this scenario, we simulate a clean measurement, in the presence of a very small noise
component in order to more directly address issues related to spatial superposition of the elements and spectral interference.

\begin{figure}[t]
    \centering
    \begin{tabular}{c|c}
        {\LARGE \textsc{standard} \pca}                            & {\LARGE \hyperpca}
        \\[1em]
        \includegraphics[width=0.47\linewidth]{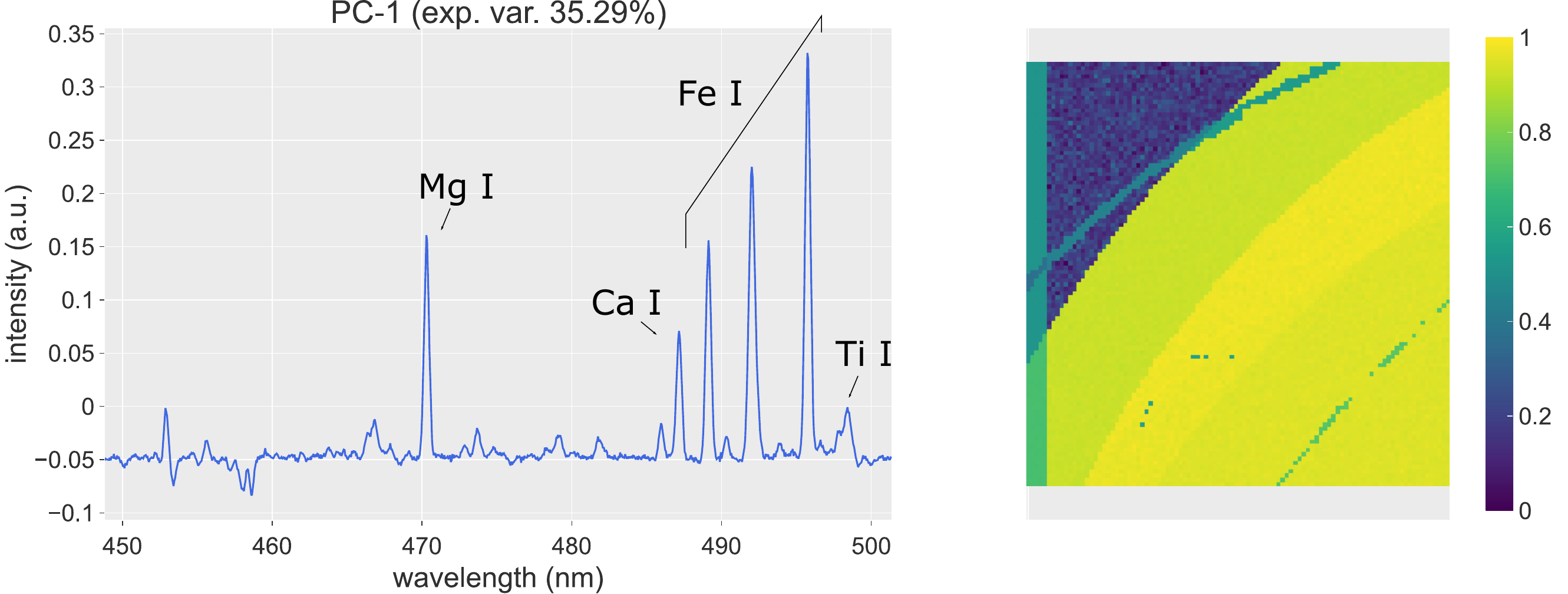} & \includegraphics[width=0.47\linewidth]{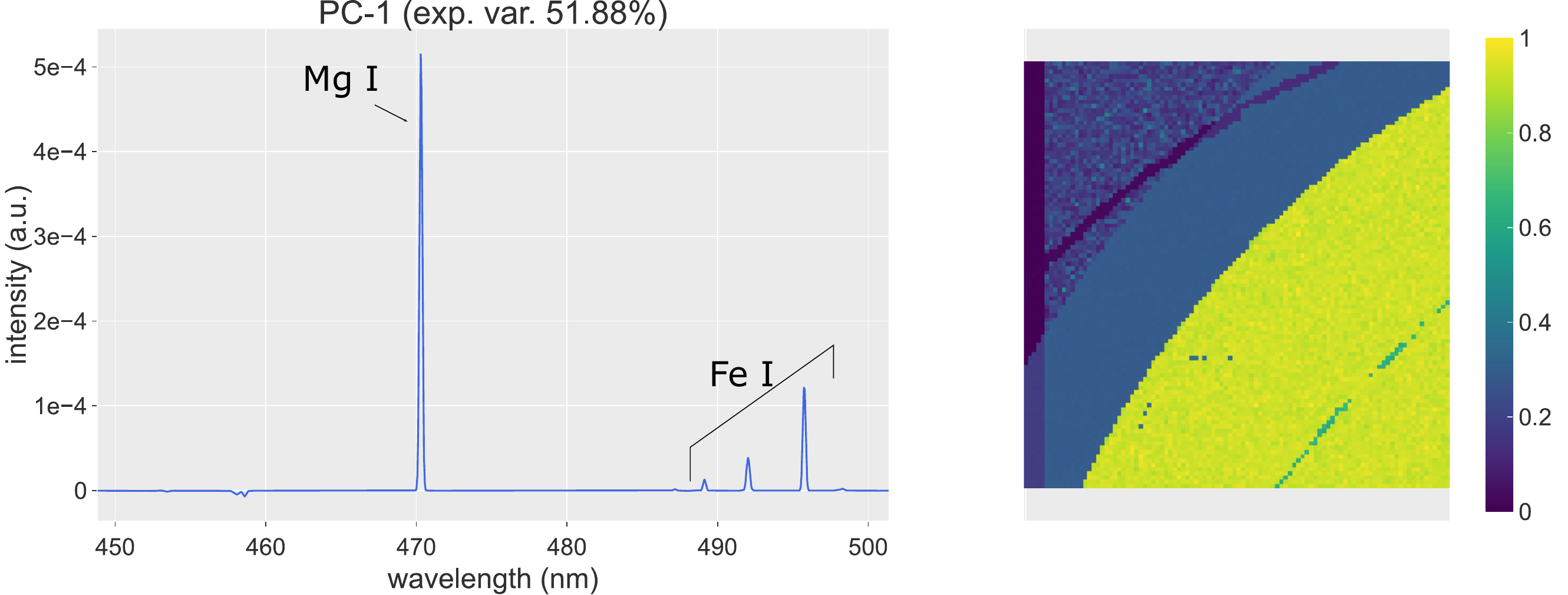}
        \\
        \includegraphics[width=0.47\linewidth]{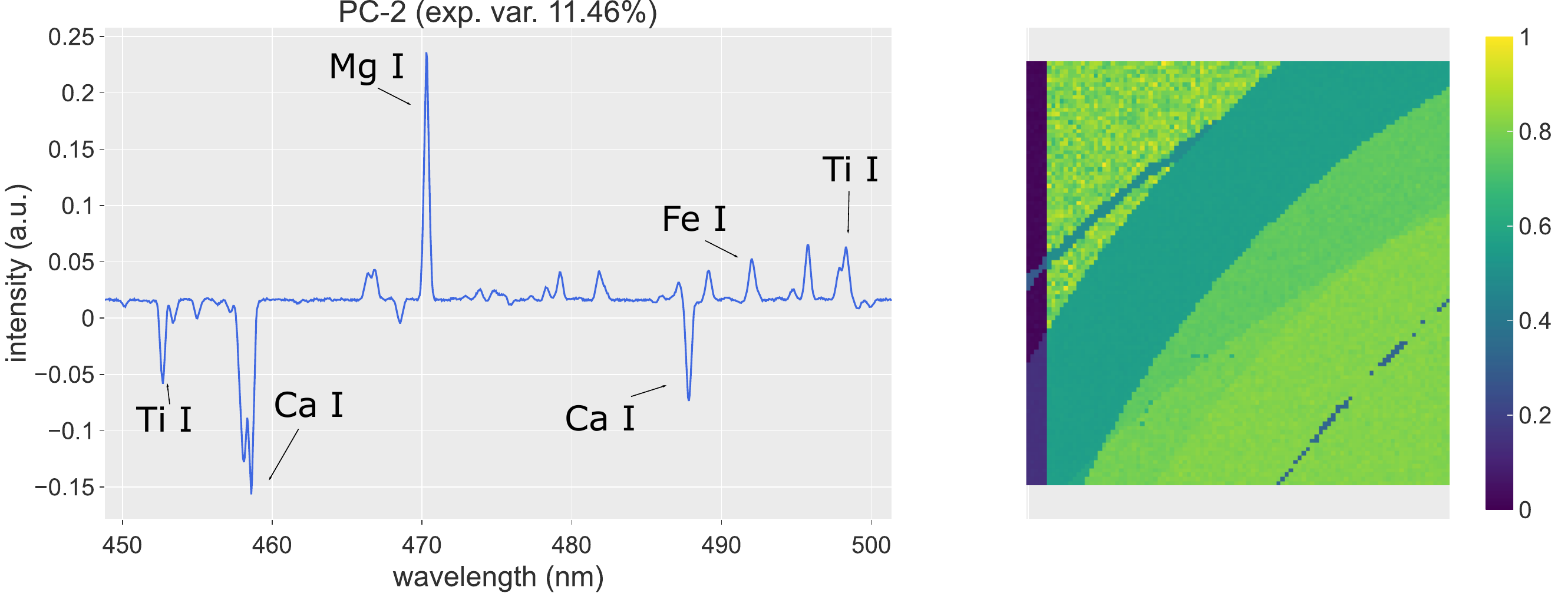} & \includegraphics[width=0.47\linewidth]{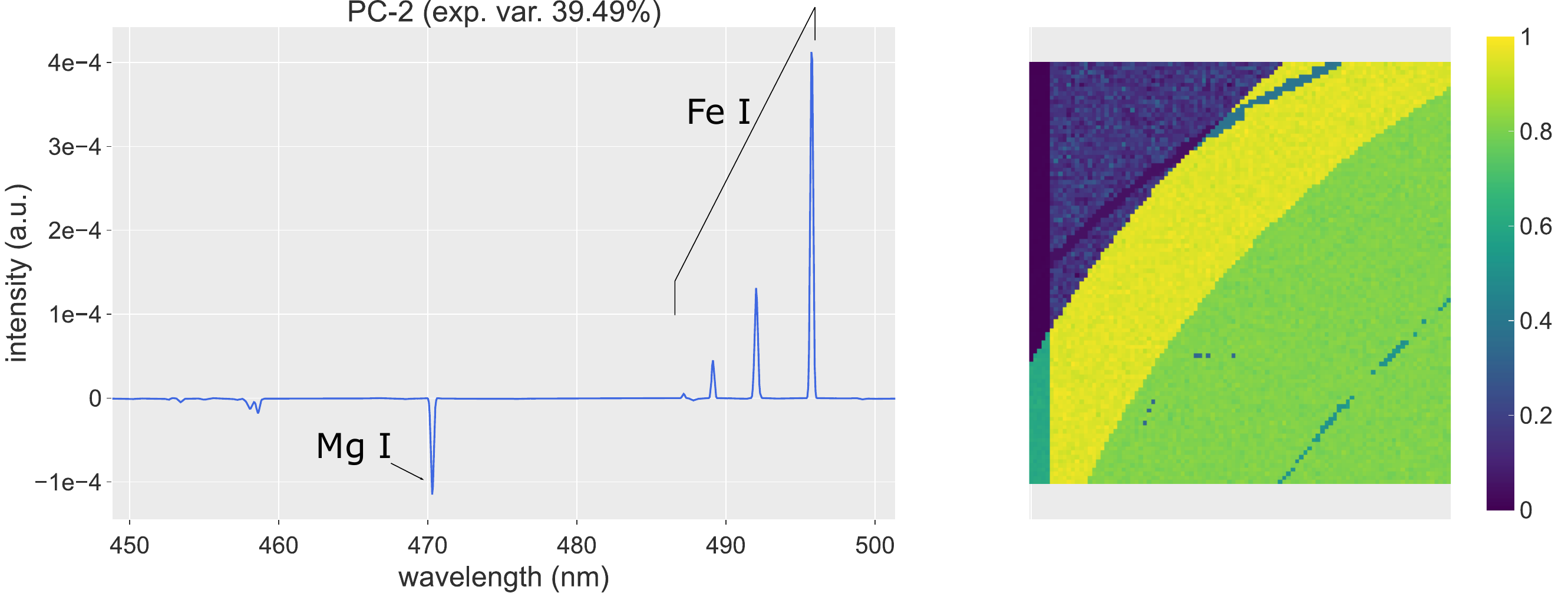}
        \\
        \includegraphics[width=0.47\linewidth]{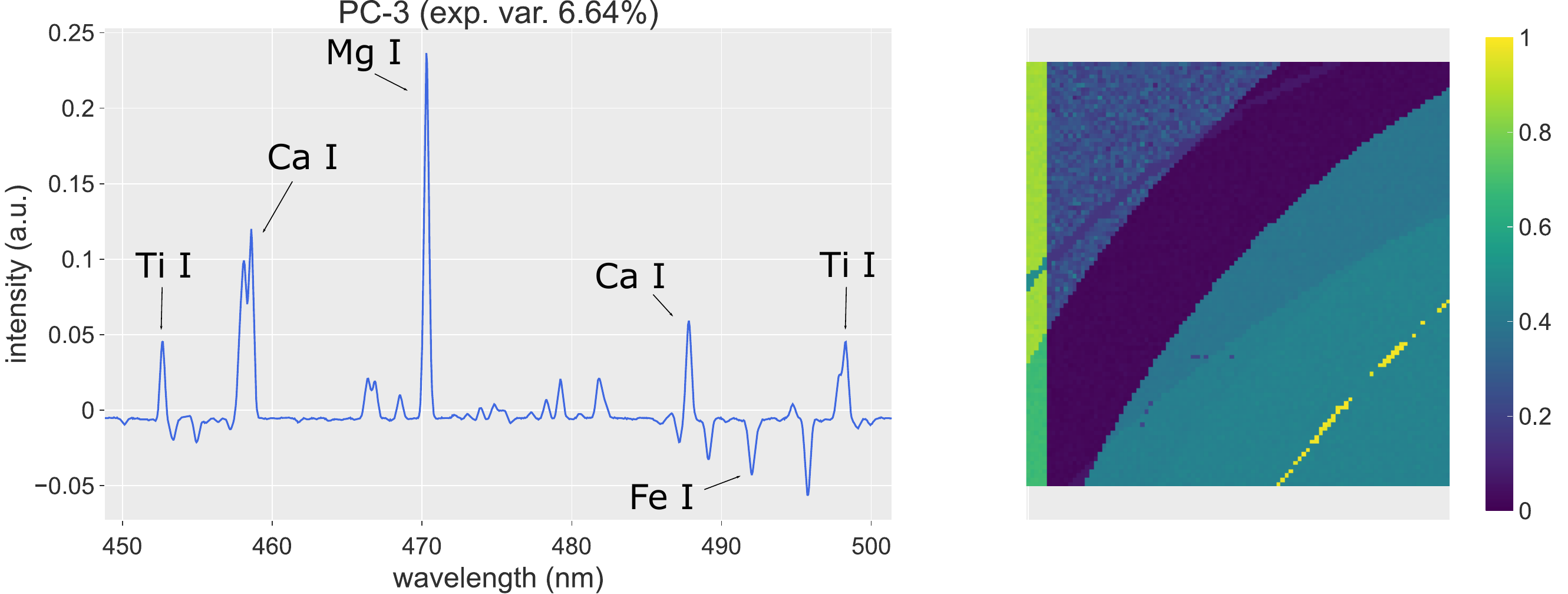} & \includegraphics[width=0.47\linewidth]{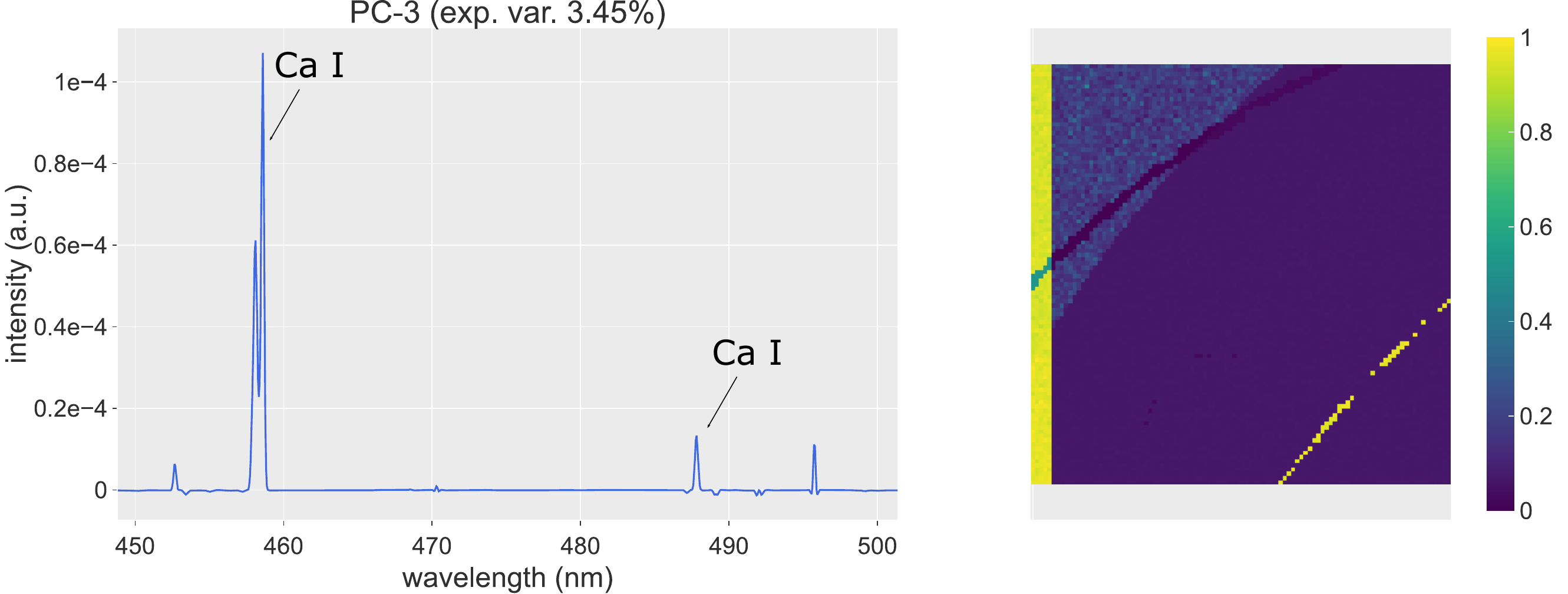}
        \\
        \includegraphics[width=0.47\linewidth]{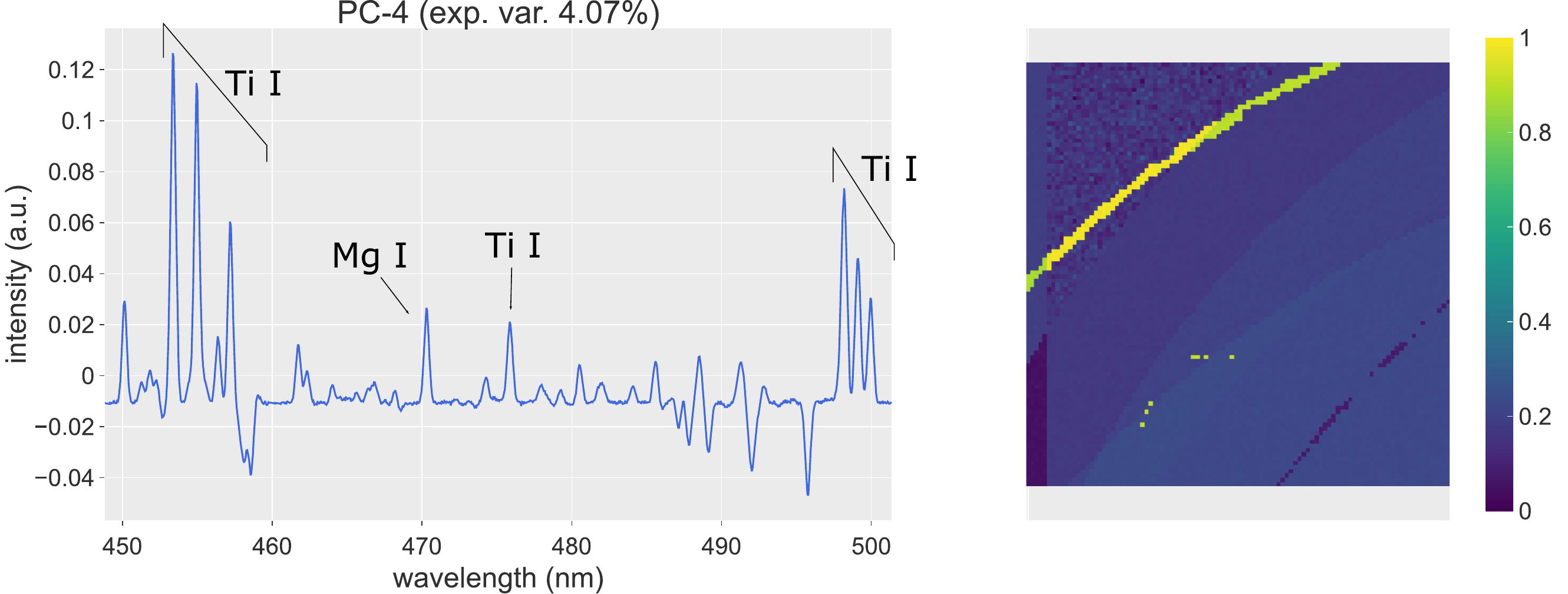} & \includegraphics[width=0.47\linewidth]{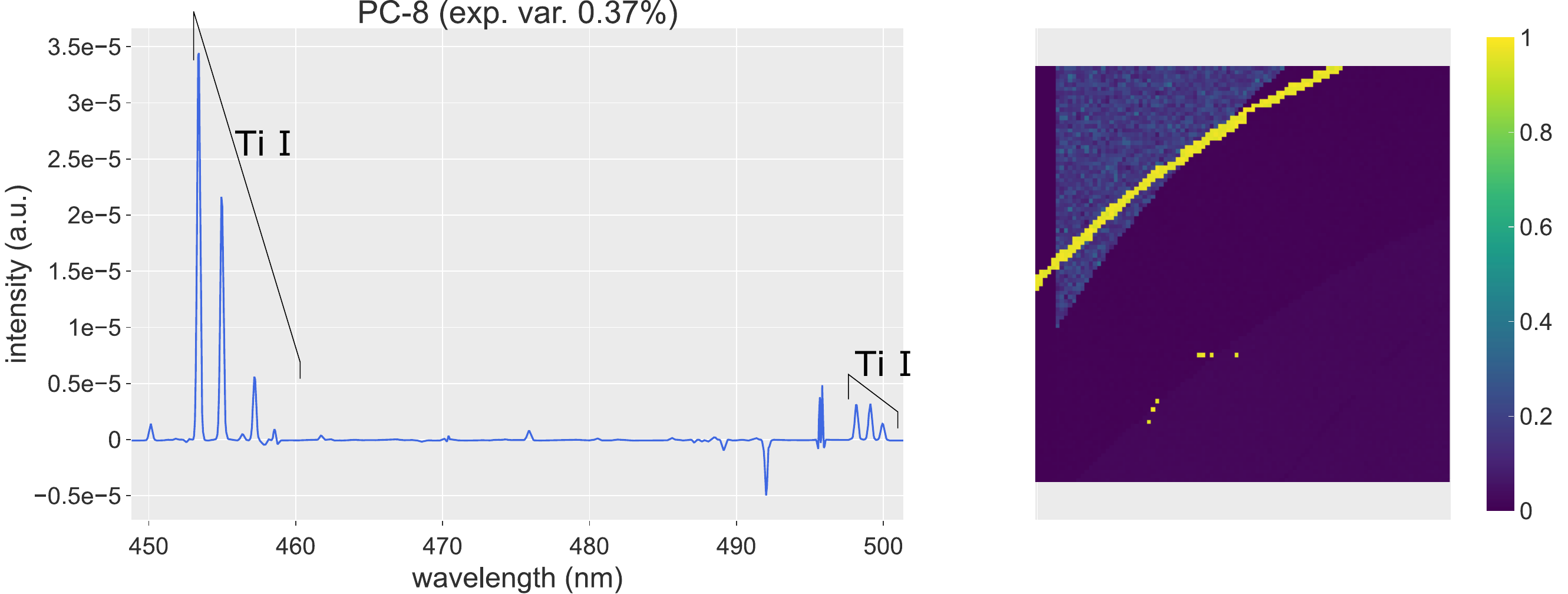}
    \end{tabular}
    \caption{``Basalt'' reconstruction using the standard \pca and \hyperpca.}
    \label{fig:basalt_pca}
\end{figure}

A first example of such problems is visible in the reconstructions obtained using the intensity of the emission lines in the average spectrum, shown in~\Cref{fig:basalt_ground_truth}.
While the Fe and Mg distributions, not shown here, are cleanly detected using the most intense lines at \SI{489.1}{\nano\meter}, \SI{492.0}{\nano\meter} and \SI{495.7}{\nano\meter}, and \SI{470.3}{\nano\meter}, the Ti distribution suffers from interference with Fe at \SI{452.8}{\nano\meter} and Na at \SI{498.4}{\nano\meter}, and with Ca at \SI{452.9}{\nano\meter}.
Interference between Ca and Fe is also present at \SI{487.2}{\nano\meter}.

After preprocessing the data as shown in~\Cref{sec:methodology}, we compute the \pcs using standard \pca.
While the highest ranked \pcs do not present large noise components and extract the distributions of the elements with good accuracy, some uncertainties are, however, present in \pcn{4}, as visible in~\Cref{fig:basalt_pca} where some minor lines start to appear in the corresponding loading vector, thus spoiling the readability of the spectrum, and the connected elemental map.
In this case, the application of \pca is, however, already sufficient to clearly and neatly reveal the presence of Ca and Ti on a substrate mainly dominated by other metallic elements.
In~\Cref{fig:basalt_pca_less_imp} in the appendix we show a few additional \pcs extracted using the standard \pca, which show also the detection of Si in the loadings, though its spatial distribution cannot be fully separated.
Even though the signal-to-noise ratio is favourable, the \pcs show only noise starting from \pcn{8}.

\begin{figure}
    \centering
    \includegraphics[width=0.9\linewidth]{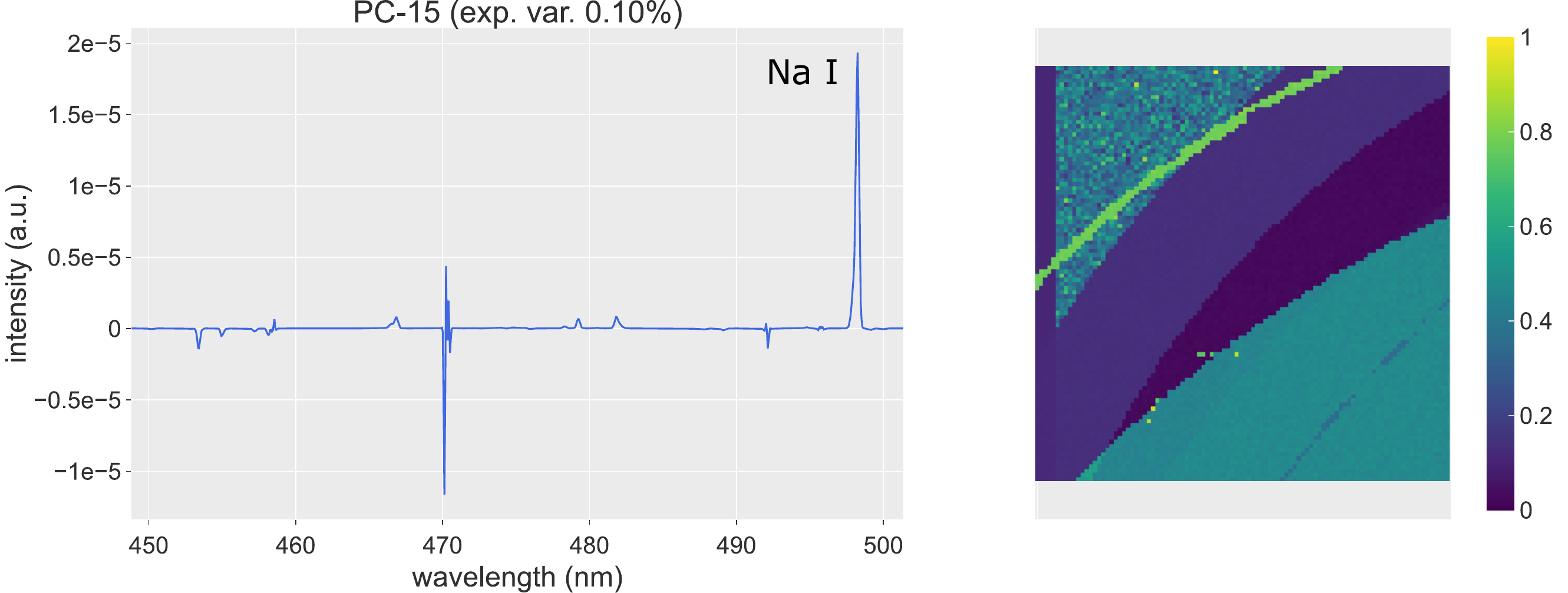}
    \caption{Extraction of the Na line using \hyperpca.}
    \label{fig:basalt_na_kpca}
\end{figure}

In the second column of~\Cref{fig:basalt_pca} we present \pcs computed using the proposed pipeline.
For the purposes of this article, we show only the loadings and score maps presenting genuine new information with respect to the previous ones.
As discussed in the previous section, \hyperpca exposes a higher number of details which account for the correct geometrical rank of the signal matrix: these lead to a greater number of \pcs being extracted.
As shown in the appendix in~\Cref{fig:basalt_kpca_less_imp_1,fig:basalt_kpca_less_imp_2,fig:basalt_kpca_less_imp_3}, the intermediate \pcs extracted by \hyperpca up to \pcn{15} display additional information on the spectral signatures already present in other \pcs.
They highlight the ability of \hyperpca to display different details on the spectral signatures of the chemical elements, and correlations between them as in the standard \pca, such as \pcn{5}, \pcn{9} and \pcn{12} to \pcn{15}.
Some components carry additional mono-elemental information, such as \pcn{6} and \pcn{10} in~\Cref{fig:basalt_kpca_less_imp_1,fig:basalt_kpca_less_imp_2}, which extract the distribution of Fe.
As in the previous scenario, residual effects due to photons being captured by neighbouring pixels of the CCD sensor are still visible in a few \pcs.
In~\Cref{fig:basalt_kpca_less_imp_3} we finally show a low-rank \pc, namely \pcn{38}, which clearly shows the presence of Si, already detectable in \pcn{13} in the same figure: by comparison of \pcn{38}, showing Si and Ti, and, for instance, \pcn{8}, showing Ti, the spatial distribution of Si can be reconstructed using \hyperpca.
With respect to standard \pca, \hyperpca provides access to more readable \pcs, which are extracted with higher confidence, since the noise components are suppressed by the kernel.
Using different types of functions in decomposition and reconstruction, the \dwt grants the possibility to distinguish different elements: the line pattern in a given spectral range depends on the element and the \dwt is able to reconstruct this characteristic.
Emission lines of a given element are therefore more likely to be extracted together in separate \pcs, as they are recognised as part of the same signal vector by \hyperpca.
For instance, the first two \pcs in the second column of~\Cref{fig:basalt_pca} can be used to better differentiate between the Fe and Mg, with respect to the same \pc computed with standard \pca.
Ca is extracted as isolated, as well as Ti in the following \pcs: \hyperpca extracts their associated loading vectors as totally separated from contributions of other elements, thus improving the quality of the elemental maps computed by standard \pca in~\Cref{fig:basalt_pca}.
They also represent a strong improvement over the line intensity maps in~\Cref{fig:basalt_ground_truth}, as the spectral interference is completely resolved by \hyperpca, contrary to the traditional approach.
Finally, notice that Na is detected in the score map of \pcn{15}, shown in~\Cref{fig:basalt_na_kpca}, since the line profile differs from the Ti line in \pcn{8}, which is totally unreadable in the case of standard \pca.
The associated score map represents thus the correct distribution of Na in the sample (even though some Ti is still detected, though it could be easily subtracted using \pcn{8} of \hyperpca): the mapping is comparable to the mapping at \SI{498.4}{nm} in~\Cref{fig:basalt_ground_truth}, which was wrongly assigned to Ti, since it displays the most intense contribution.

In this sense, not only does \hyperpca provide means to address the presence of noise, but it also solves the issue of heavy spectral interference in the samples.
In fact, differently from \pca, \hyperpca leverages the action of the kernel function, with the sparse representation in terms of wavelets.
As every element has characteristic emission lines, the sparse basis enables the reconstruction of their profiles in similar ways, thus making it possible to resolve them.
The advantage over the line intensity map is the association of elemental maps directly to loading vectors, i.e.\ spectra identifying physical elements, rather than a single wavelength, which leads almost inevitably to failure in the case of strong interference.
\hyperpca is able to provide information at the elemental level, while also dealing with correlations and anti-correlations between the elements as the standard \pca.
In this sense, it provides additional information as it also proves effective in disentangling the spectral signatures of the elements.
Having access to this kind of information translates in the ability to study both the distributions of the elements on the surface, but also to analyse independently the spatial correlations between the elements.
Notice that, in this case, both \pca and \hyperpca present strong positive and negative contributions in the loading vectors, even though, conventionally, we normalise the score maps in the interval $[0,\, 1]$: when the difference between the contributions with opposite signs are relevant, it is possible to observe the intermediate zero-level as a homogeneous colour in an intermediate nuance of the colour map, such as the top left corner in both columns of~\Cref{fig:basalt_pca}.

\subsubsection{Case 3: High Noise and Strong Spectral Interference}
\label{sec:granite}

\begin{figure}[t]
    \centering

    \includegraphics[width=0.5\linewidth]{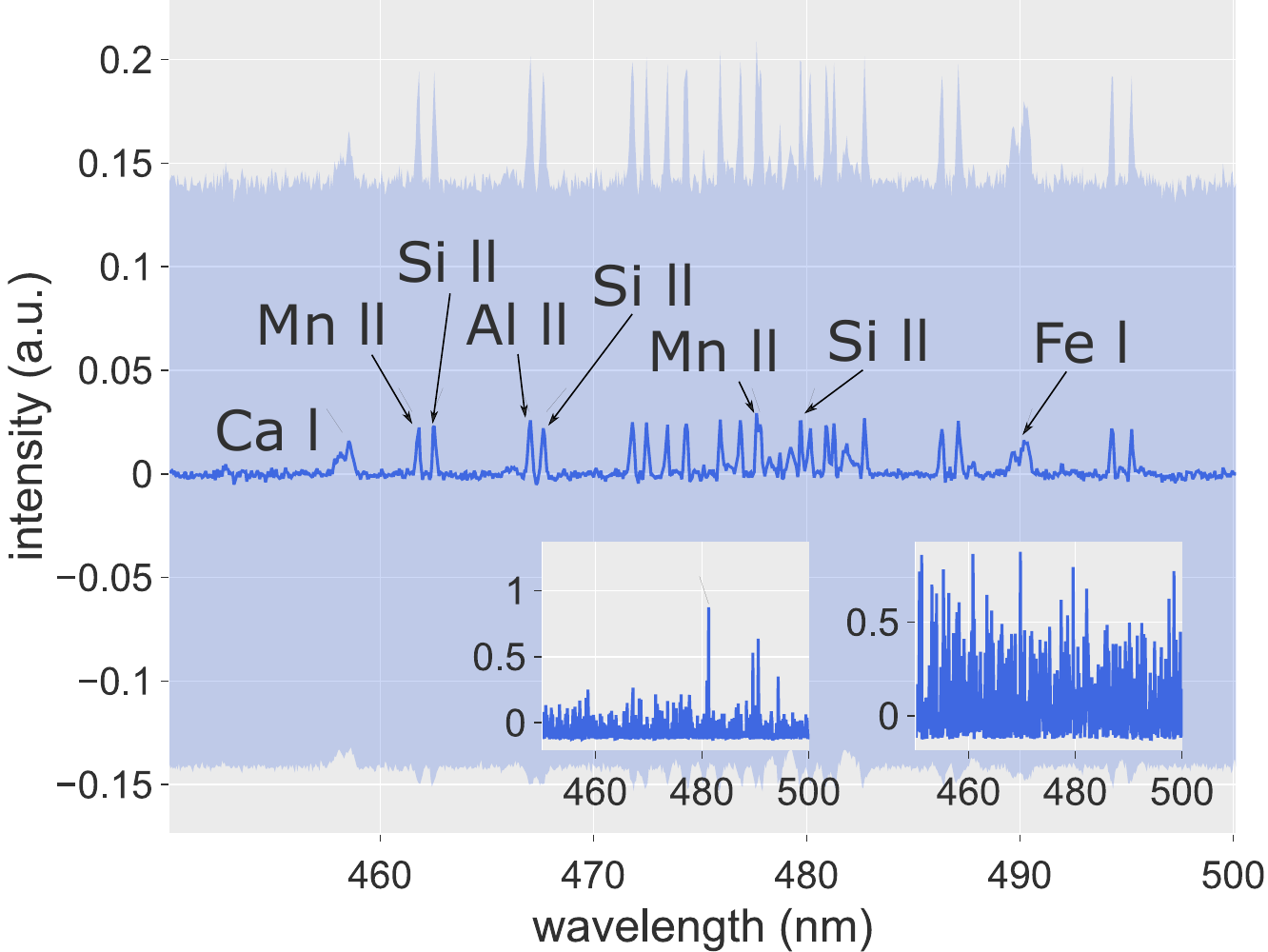}

    \caption{Preprocessed average and $1\upsigma$ spectra in the ``granite'' dataset with examples of single-shot spectra.}
    \label{fig:granite_spectra}
\end{figure}

The last synthetic dataset is a granite-like rock made from a mixture of Si (\SI{74}{\percent}), Al (\SI{14}{\percent}), Na (\SI{4}{\percent}), K (\SI{4}{\percent}), Ca (\SI{2}{\percent}), Fe (\SI{1}{\percent}) and Mn (\SI{1}{\percent}).
The plasma temperature is \SI{1}{\electronvolt}, while the electron density \SI{e17}{\per\centi\meter\cubed}.
For the dataset, $a = 5$ and $\beta_{\text{max}} = 200$, i.e.\ an extremely low shot-to-shot signal-to-noise ratio, of the order of \num{1.2e-2}.
The kernel parameter, in this case, is $\alpha = \num{e-10}$.
We compute the \dwt (\texttt{bior3.9}) using one downsampling pass, with a \hardt of \num{0.99} times the maximum value of the power spectrum of the \dcs.
Given the strong noise problem, in this case we standardise the spectrum for the \mva algorithms, that is, after centring by removing the mean contribution, we divide the spectra by the standard deviation of the dataset.
This technique, often referred to as \emph{whitening}~\cite{friedman_exploratory_1987}, can be used in cases where the signal-to-noise ratio is small, in order to better highlight the weak signal components in the data.

\begin{figure}[t]
    \centering

    \subfigure[Si]{\includegraphics[width=0.24\linewidth]{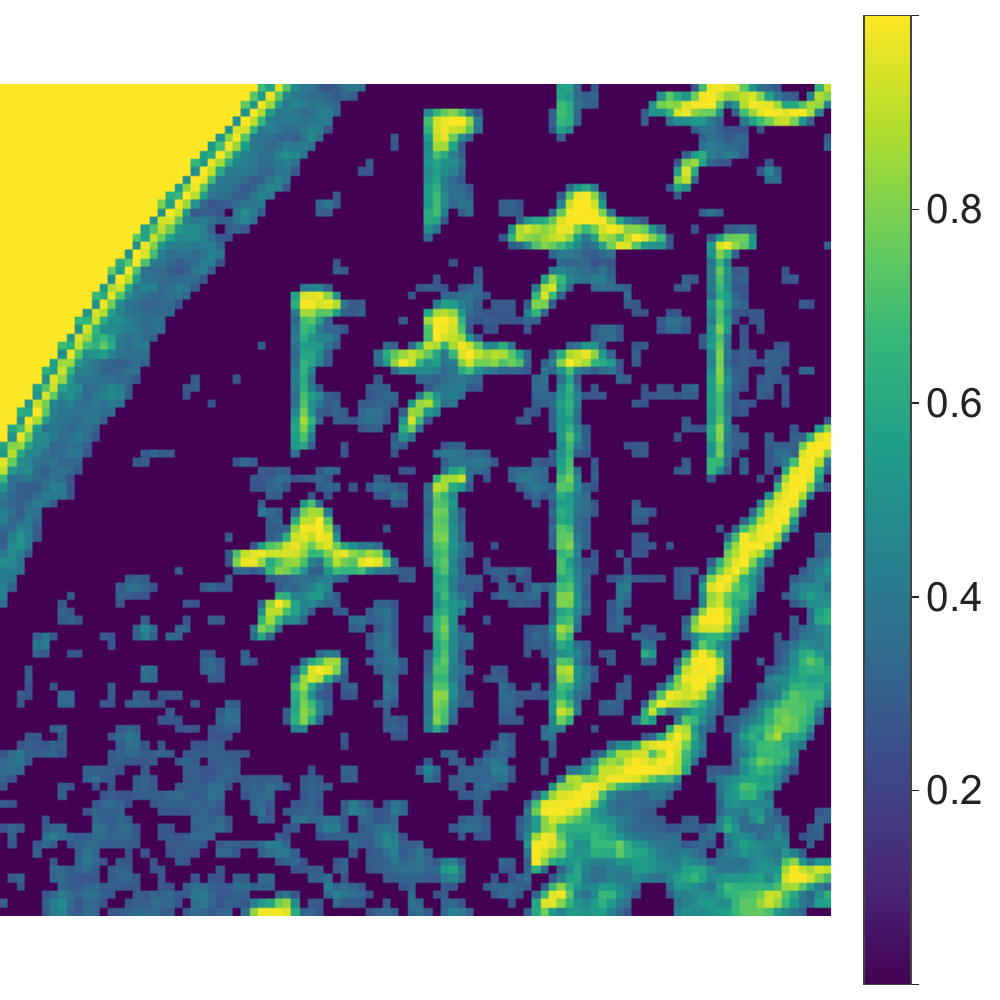}}
    \hfill
    \subfigure[Al]{\includegraphics[width=0.24\linewidth]{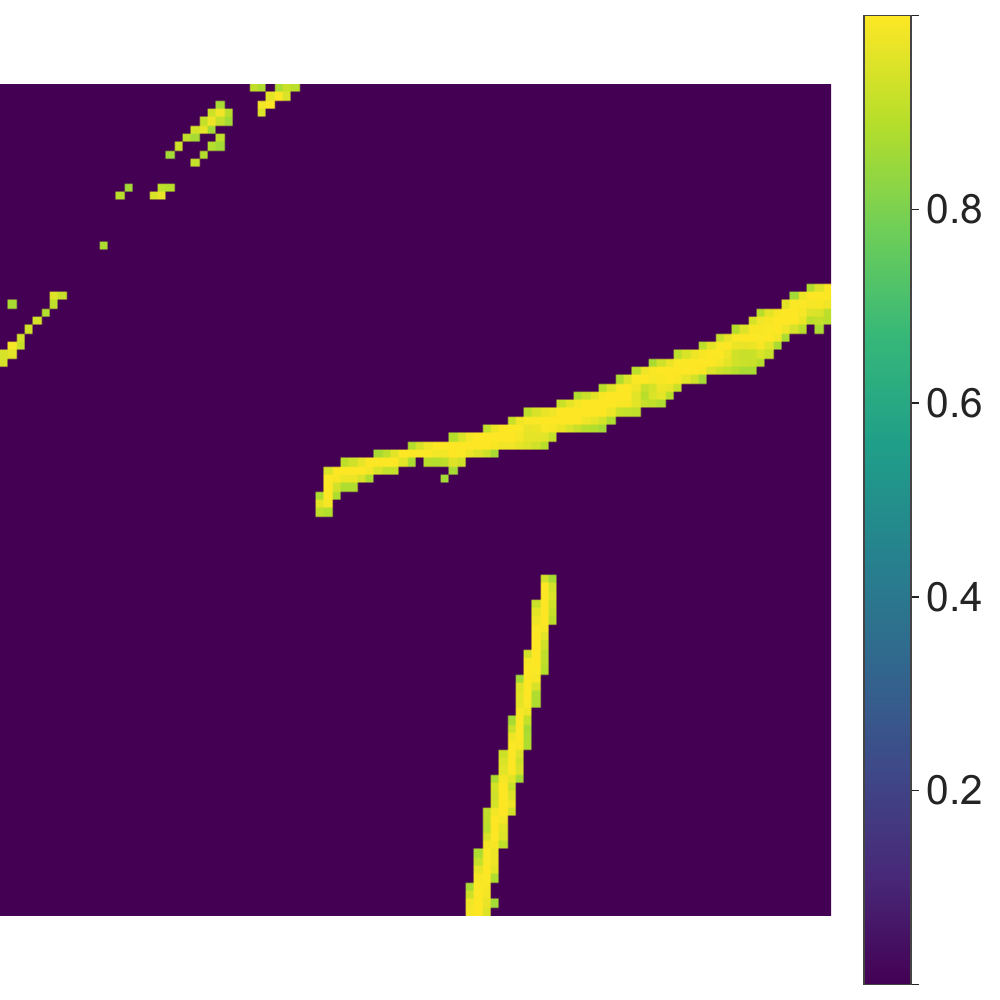}}
    \hfill
    \subfigure[Na]{\includegraphics[width=0.24\linewidth]{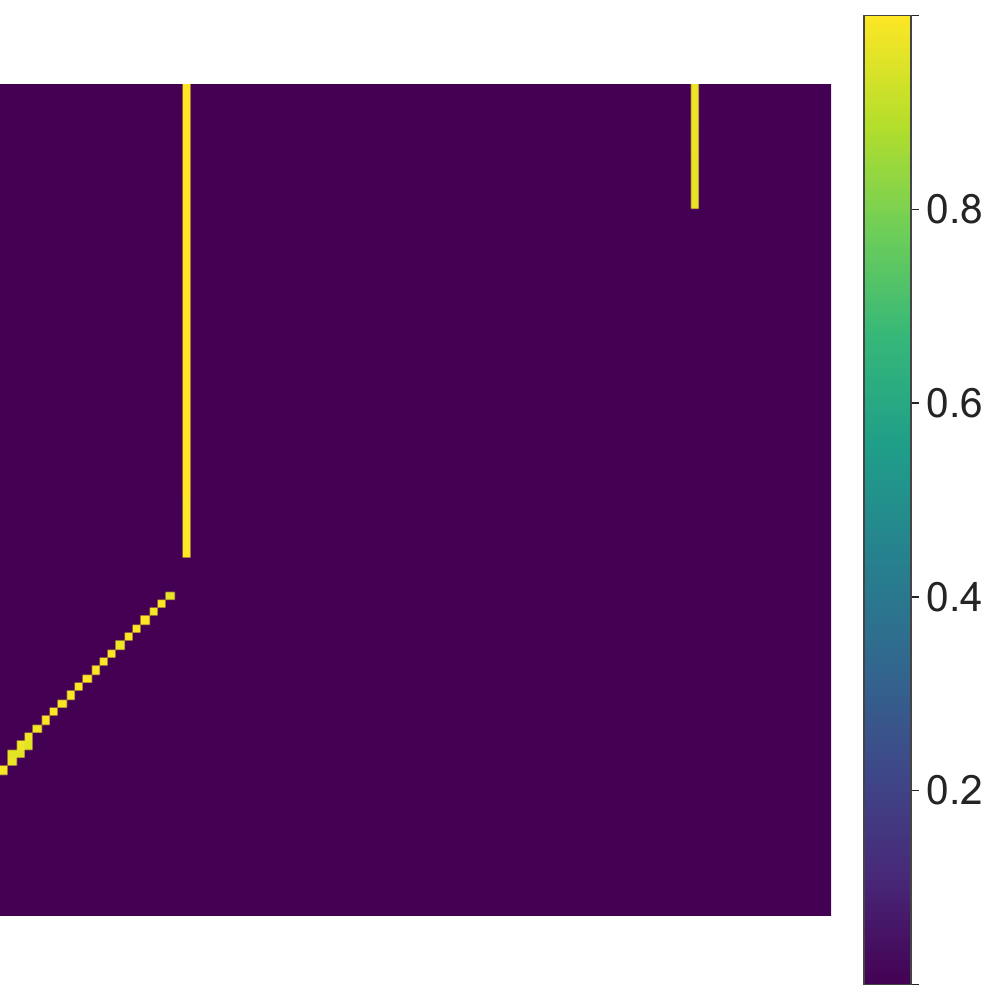}}
    \hfill
    \subfigure[K]{\includegraphics[width=0.24\linewidth]{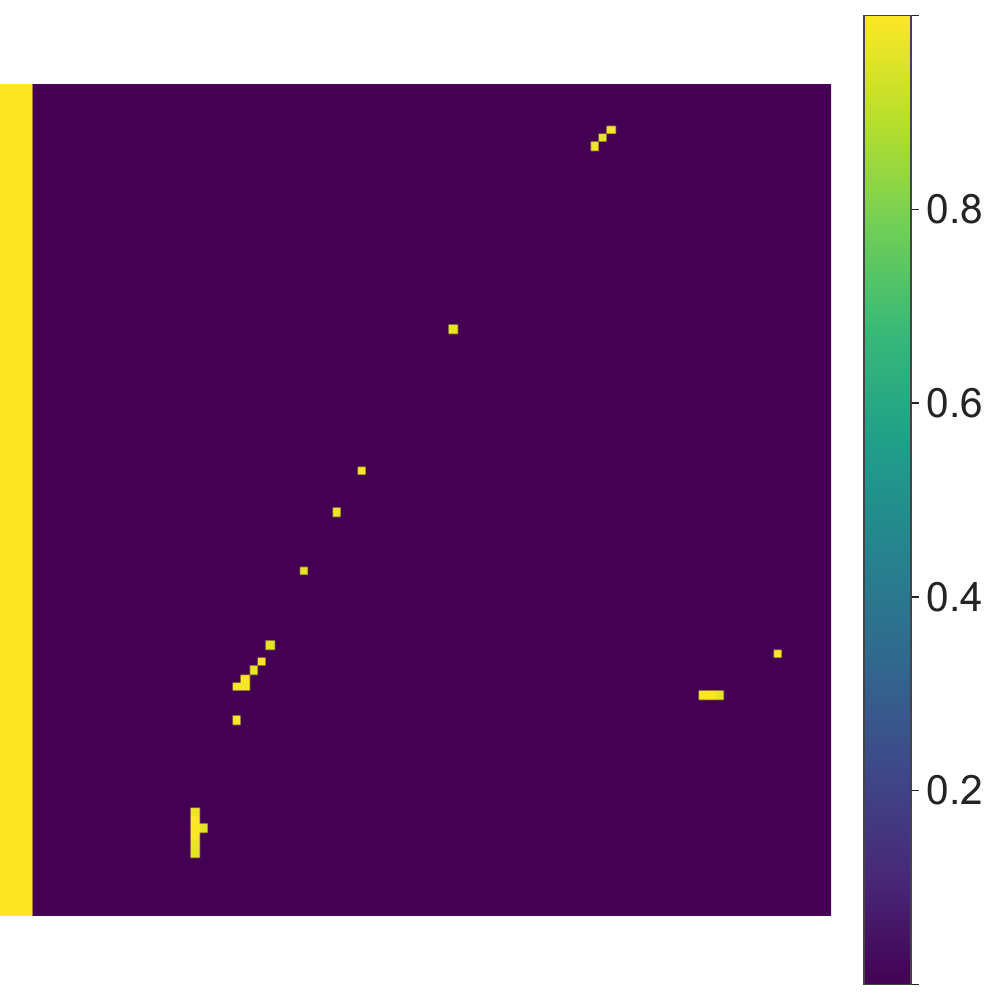}}
    \\
    \subfigure[Ca]{\includegraphics[width=0.24\linewidth]{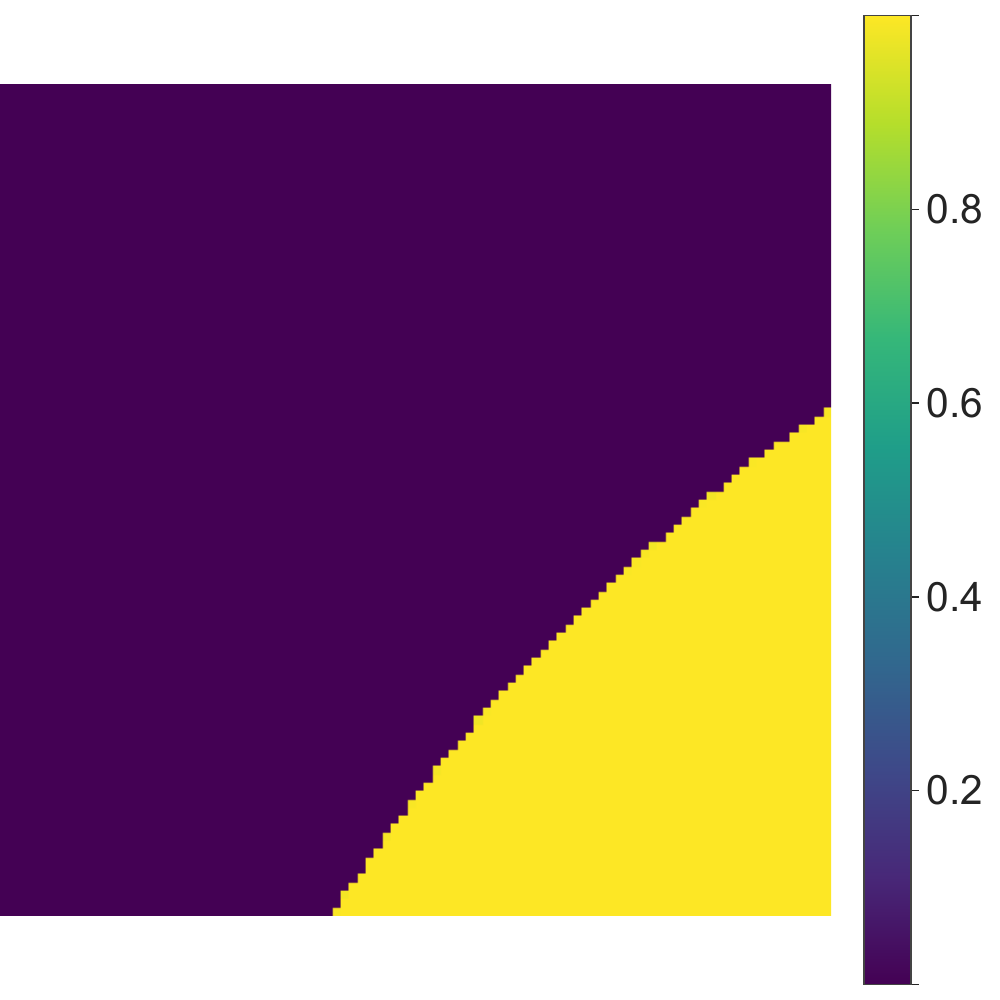}}
    \qquad
    \subfigure[Fe]{\includegraphics[width=0.24\linewidth]{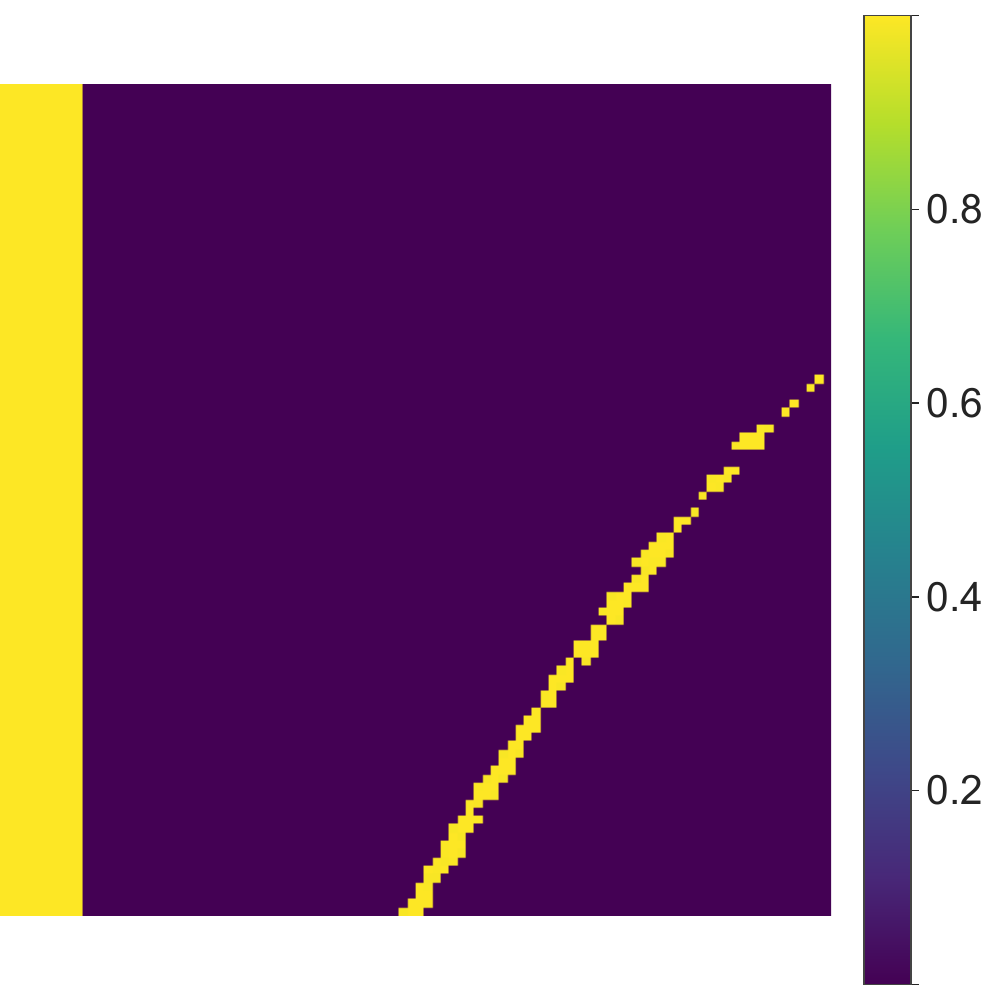}}
    \qquad
    \subfigure[Mn]{\includegraphics[width=0.24\linewidth]{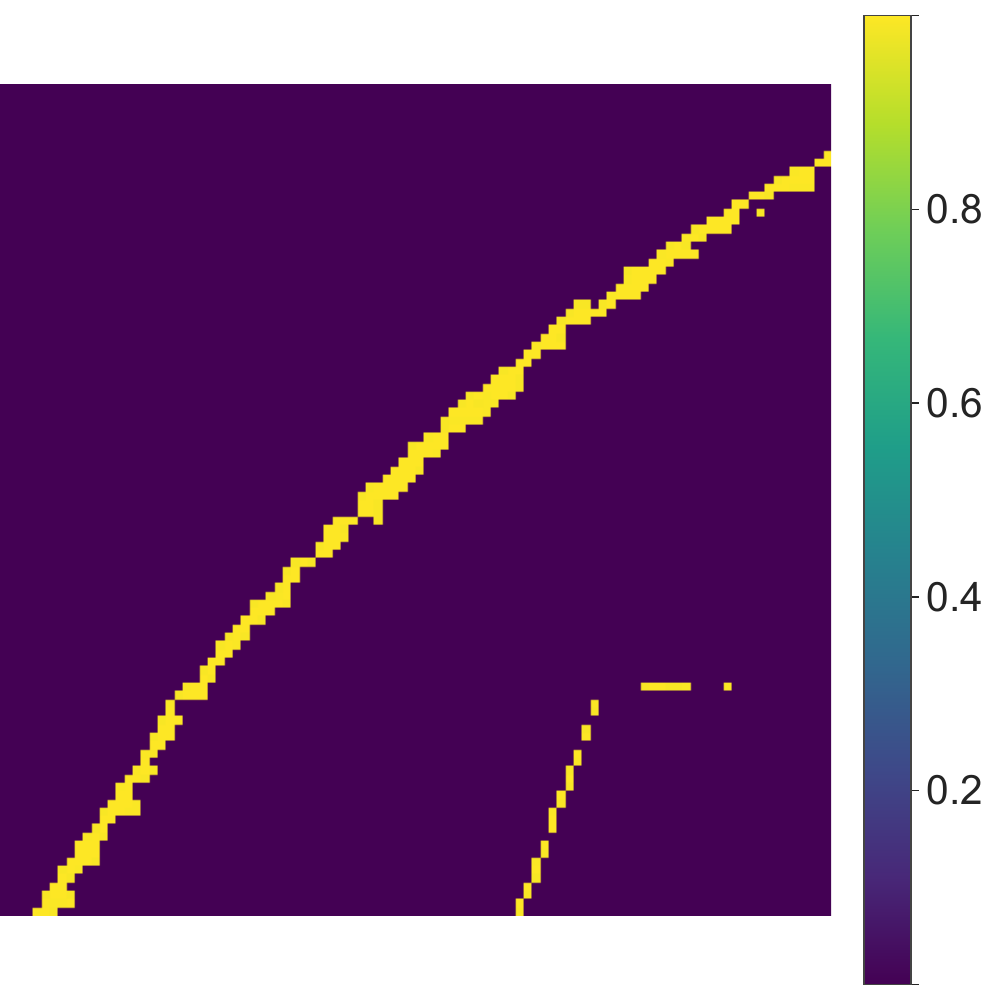}}
    \\
    \rule{0.85\linewidth}{1pt}
    \\
    \subfigure[\SI{458.53}{\nano\meter} (Ca)]{%
        \includegraphics[width=0.185\linewidth]{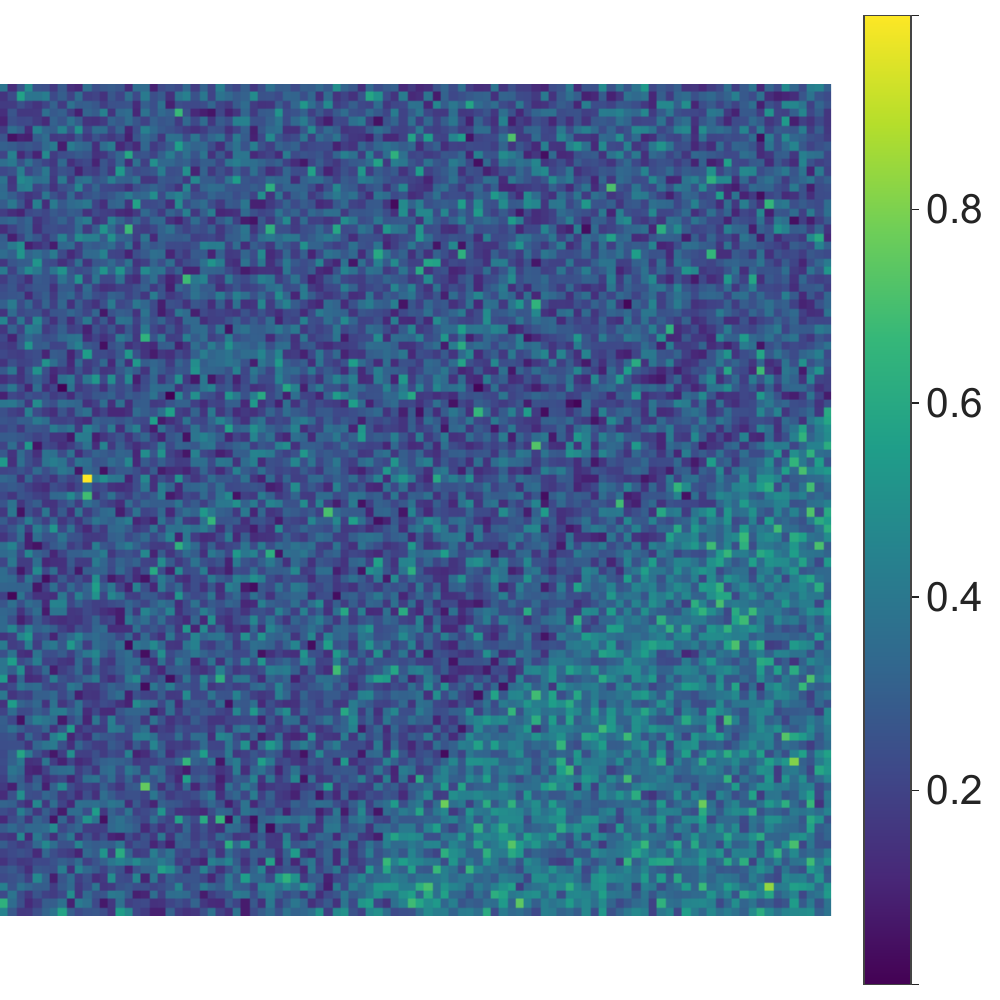}
    }
    \qquad
    \subfigure[\SI{479.69}{\nano\meter} (Si)]{%
        \includegraphics[width=0.185\linewidth]{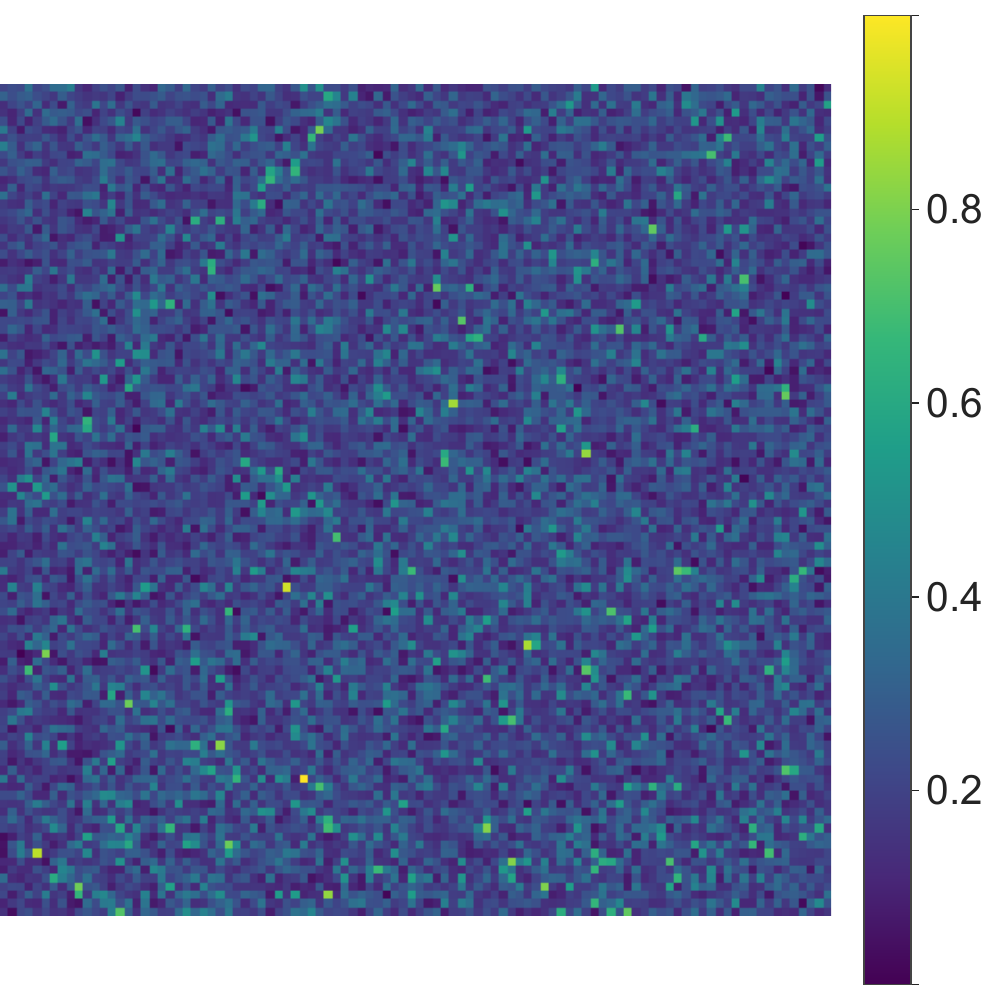}
    }
    \\

    \caption{``Granite'' reference elemental maps and line intensity maps. The scanned wavelength is reported. The element in parenthesis corresponds to the element whose intensity is maximal at the given value.}
    \label{fig:granite_ground_truth}
\end{figure}

The theoretical distributions in \lte are shown in the appendix in~\Cref{fig:granite_spectra_lte}, while in~\Cref{fig:granite_spectra} we show the preprocessed spectrum, together with some example spectra present in the dataset.
This represents a difficult case in which both interference and large noise components are present in the dataset.
\Cref{fig:granite_ground_truth} shows the reference elemental maps.

\begin{figure}[t]
    \centering
    \begin{tabular}{c|c}
        {\LARGE \textsc{standard} \pca}                            & {\LARGE \hyperpca}
        \\[1em]
        \includegraphics[width=0.47\linewidth]{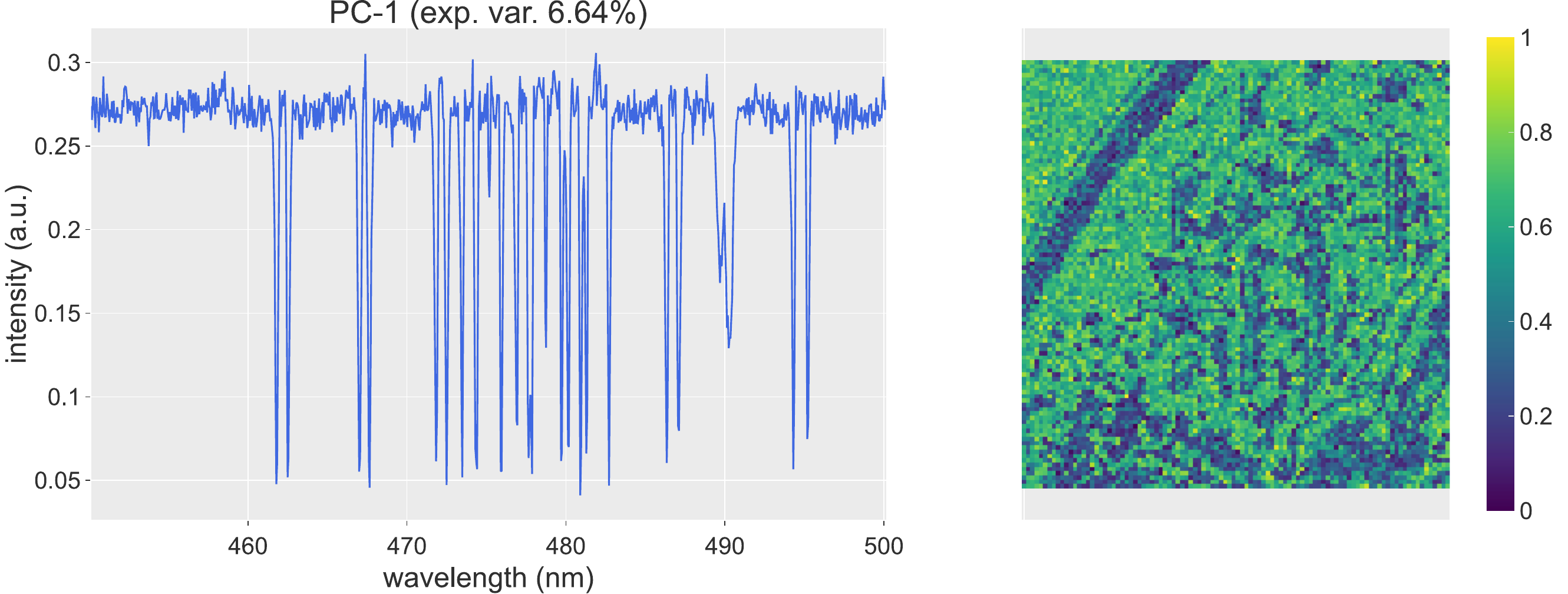} & \includegraphics[width=0.47\linewidth]{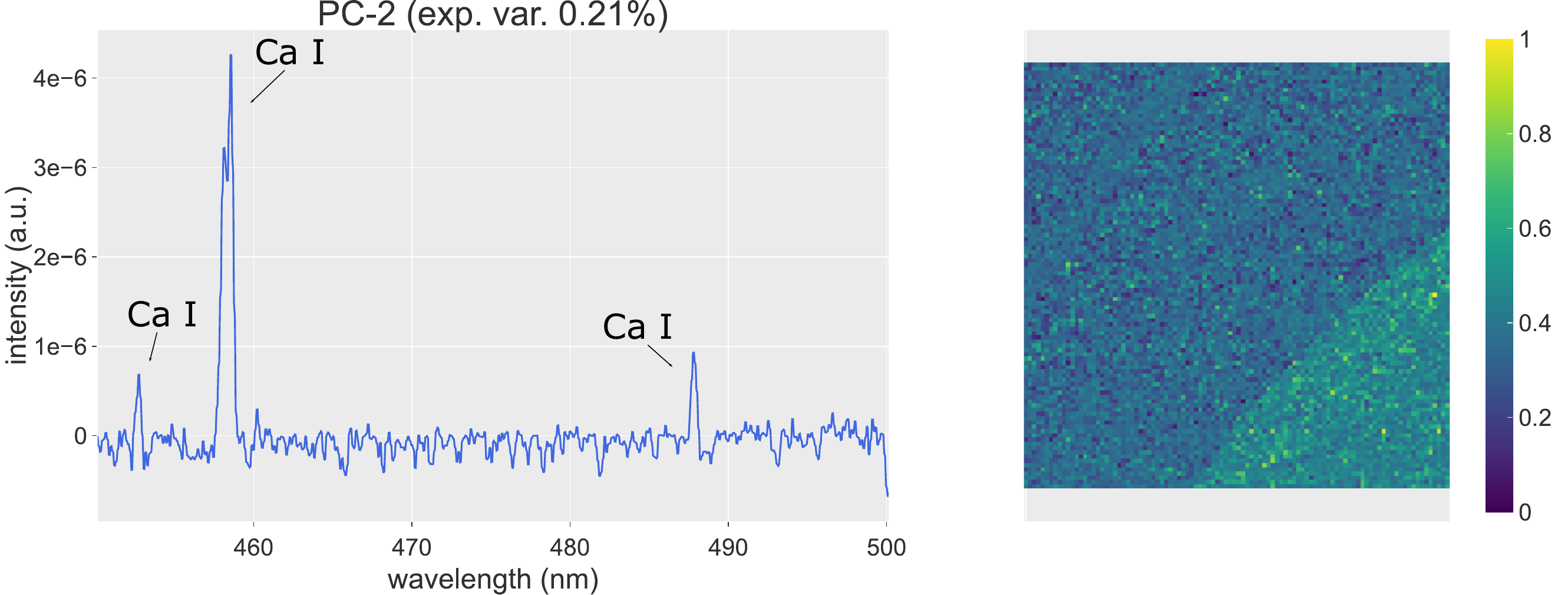}
        \\
        \includegraphics[width=0.47\linewidth]{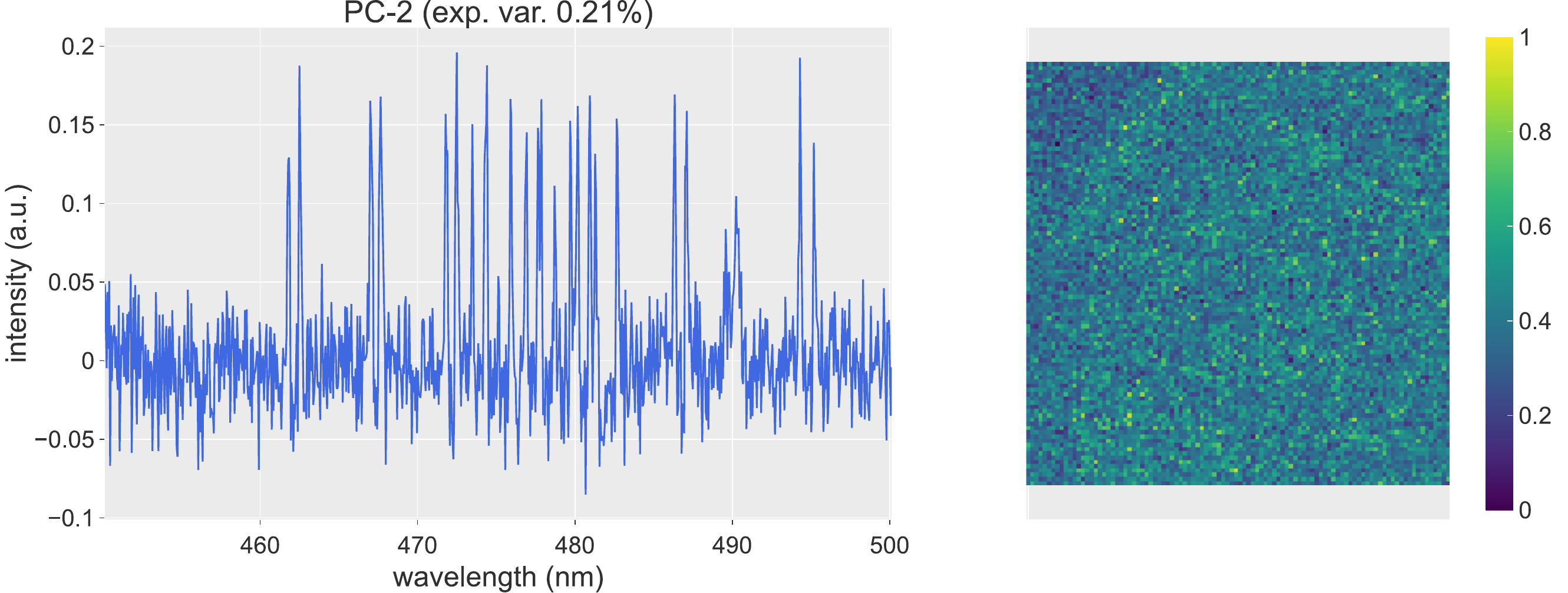} & \includegraphics[width=0.47\linewidth]{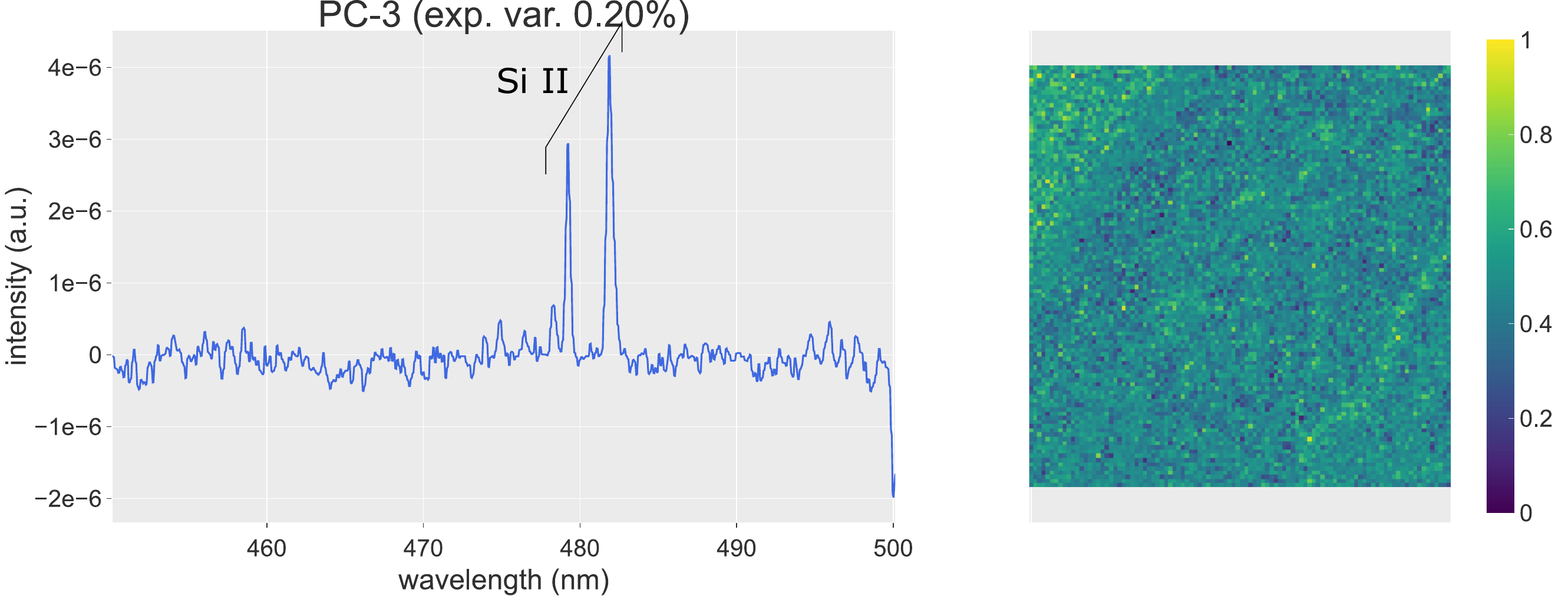}
        \\
        \includegraphics[width=0.47\linewidth]{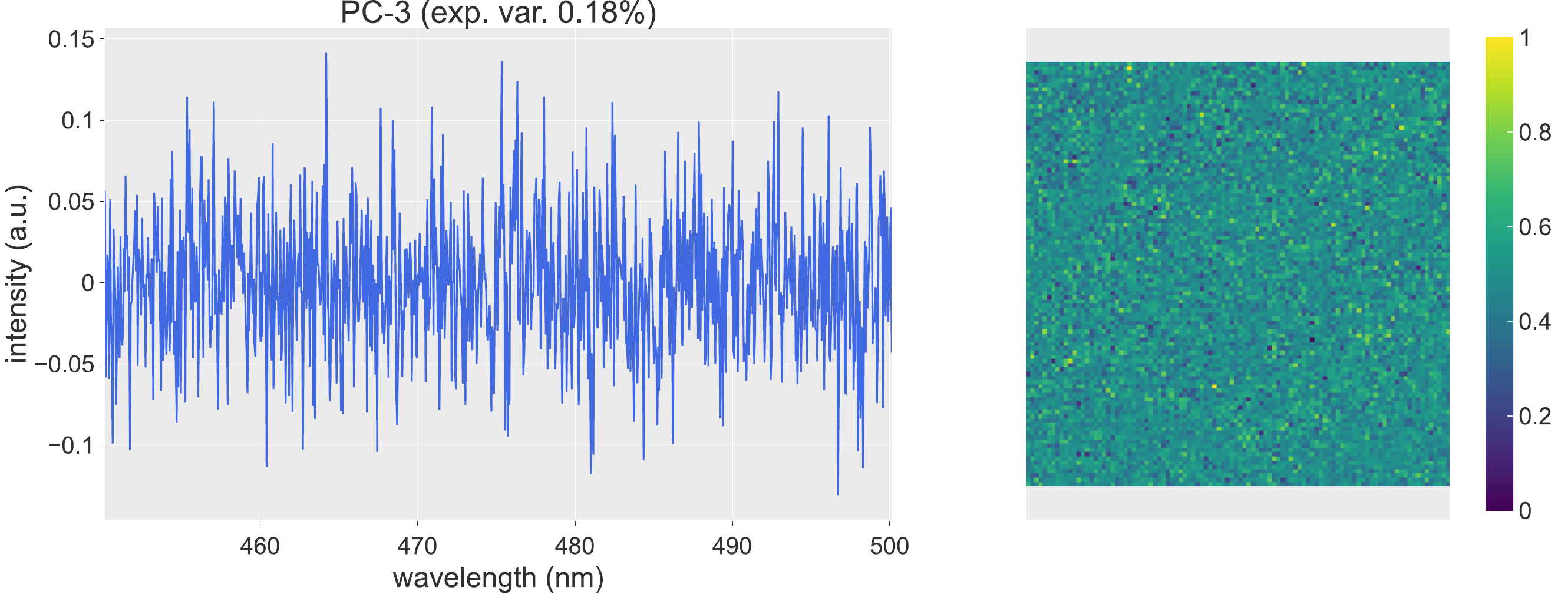} & \includegraphics[width=0.47\linewidth]{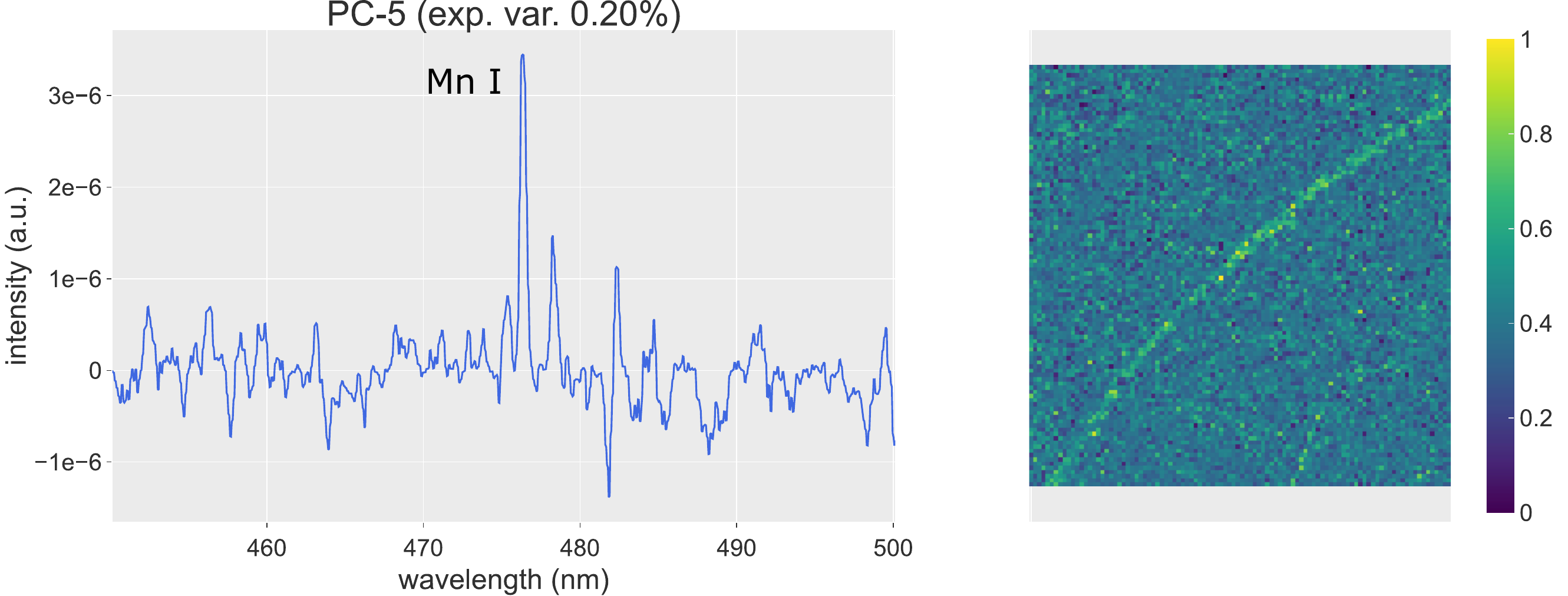}
    \end{tabular}
    \caption{``Granite'' reconstruction using the standard \pca and \hyperpca.}
    \label{fig:granite_pca}
\end{figure}

The simulated sample is a substrate containing a diffused distribution of Si, together with more concentrated insertions of the other elements.
The Fe distribution is almost entirely covered by Ca, and the Si locations basically hide elements as Al and Mn almost completely.
At the bottom of~\Cref{fig:granite_ground_truth} we present the only two readable maps obtained using the traditional approach using the line intensities: Ca and Si are sufficiently intense in the dataset to be traced and mapped, even though the quality of the maps is strongly affected by spectral interference and noise.

In the first column of~\Cref{fig:granite_pca} we show the first \pcs built with the standard \pca.
As it is clearly visible, \pca cannot extract any elemental map or loading vector properly due to very high level of noise present in the dataset.
In this case, a standard \mva approach is not recommendable as no information can be recovered.
\pcs extracted by \hyperpca are shown in the second column of~\Cref{fig:granite_pca}.
Even in this case \pcn{1} is not easy to interpret and excluded from the figure, due to the high level of noise, and it is presented in~\Cref{fig:granite_dwt_kpca_granite_noise} of the appendix.
The advantage of using a more advanced machine learning model is visible both in the noise level, reduced with respect to the standard \pca, and in the ability to detect minor elements and weak or strongly interfered emission lines via the unsupervised and automatic reconstruction of the line profiles using the \dwt.
\hyperpca is able to extract correctly the score maps of Ca, Si and Mg: the noise reduction mechanism of the kernel approach clearly increases the quality of the loading vectors, while the wavelet approach leverages it by grouping lines associated to the same element in the same \pc.
The combined action of the processes boosts the ability to build meaningful elemental maps and to extract physico-chemical information efficiently.

The refinement of the extracted \pcs and the ability to distinguish interfering lines can therefore lead to an improved analysis of the components in the samples, with respect to standard \pca and the traditional approach using the line intensity.
As a reference, in~\Cref{fig:granite_examples} of the appendix, we show spectra taken in four random spots on the sample matrix to show the information retrieval capabilities of \pca and \hyperpca in very noisy data: while the first can still recover meaningful \pcs, the unsupervised \hyperpca improves the overall quality of the \pcs and provides additional information.


\subsection{Experimental Data}\label{sec:experimental}

In this section, we apply \hyperpca to real-world datasets.
With respect to the simulated datasets, we expect several differences in the spectra, such as:
\begin{enumerate}
    \item noise in real-world scenarios follows a Poisson distribution: while the Gaussian nature was necessary to show some of the properties of \hyperpca in simulations, the outcome does not change in case of different distributions, as the effect of the application of the kernel is the same.
    The efficacy can be less intense than in a normally distributed background, thus more noise might be retained;

    \item specific physical phenomena may influence the shape and resolution of the profiles, depending on the experimental setup.
    \hyperpca needs to be versatile enough to deal with different experimental situations.
\end{enumerate}

\begin{figure}[t]
    \centering

    \subfigure[Preprocessed average and $1\upsigma$ spectra with examples of single-shot spectra.]{\includegraphics[width=0.5\linewidth]{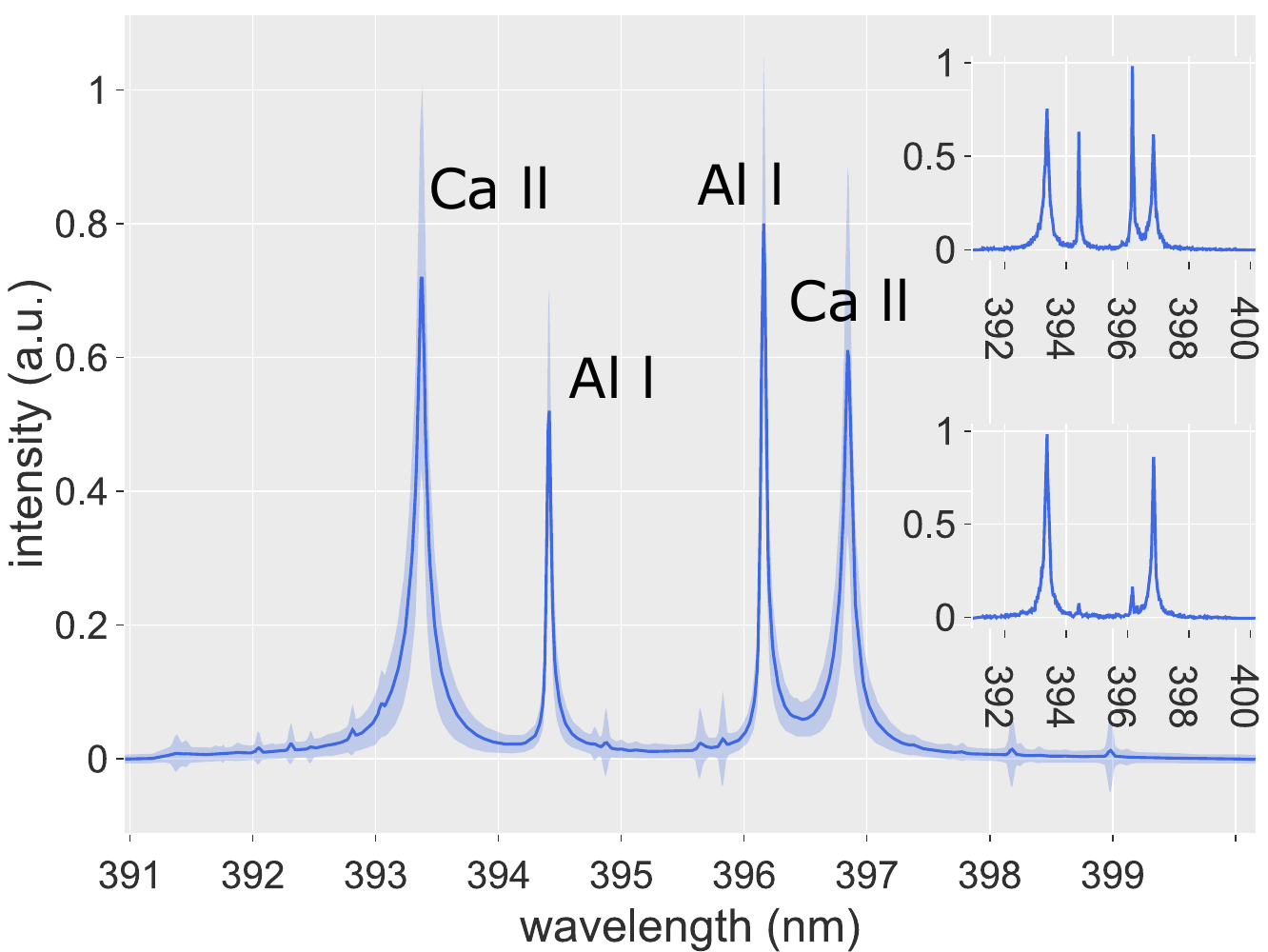}}
    \\

    \subfigure[\SI{393.37}{\nano\meter} and \SI{396.85}{\nano\meter} (Ca)]{%
        \includegraphics[width=0.23\linewidth]{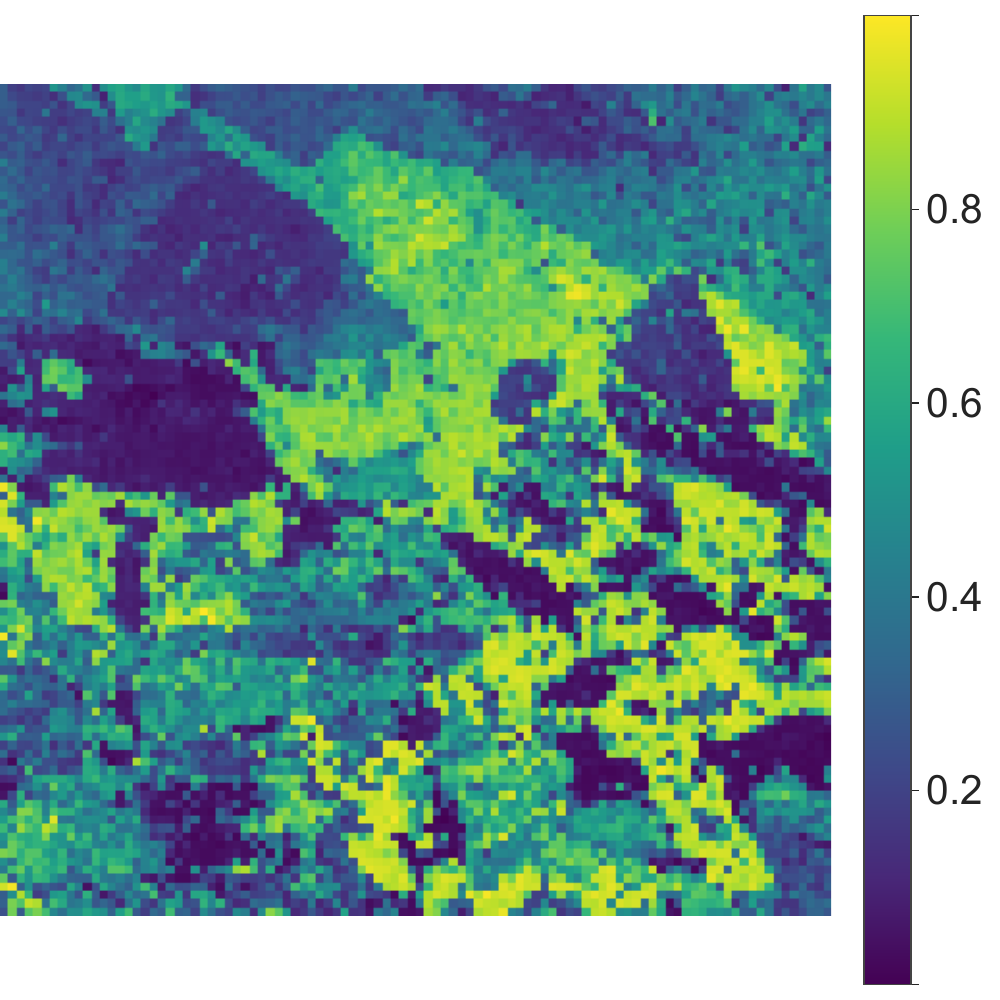}
        \includegraphics[width=0.23\linewidth]{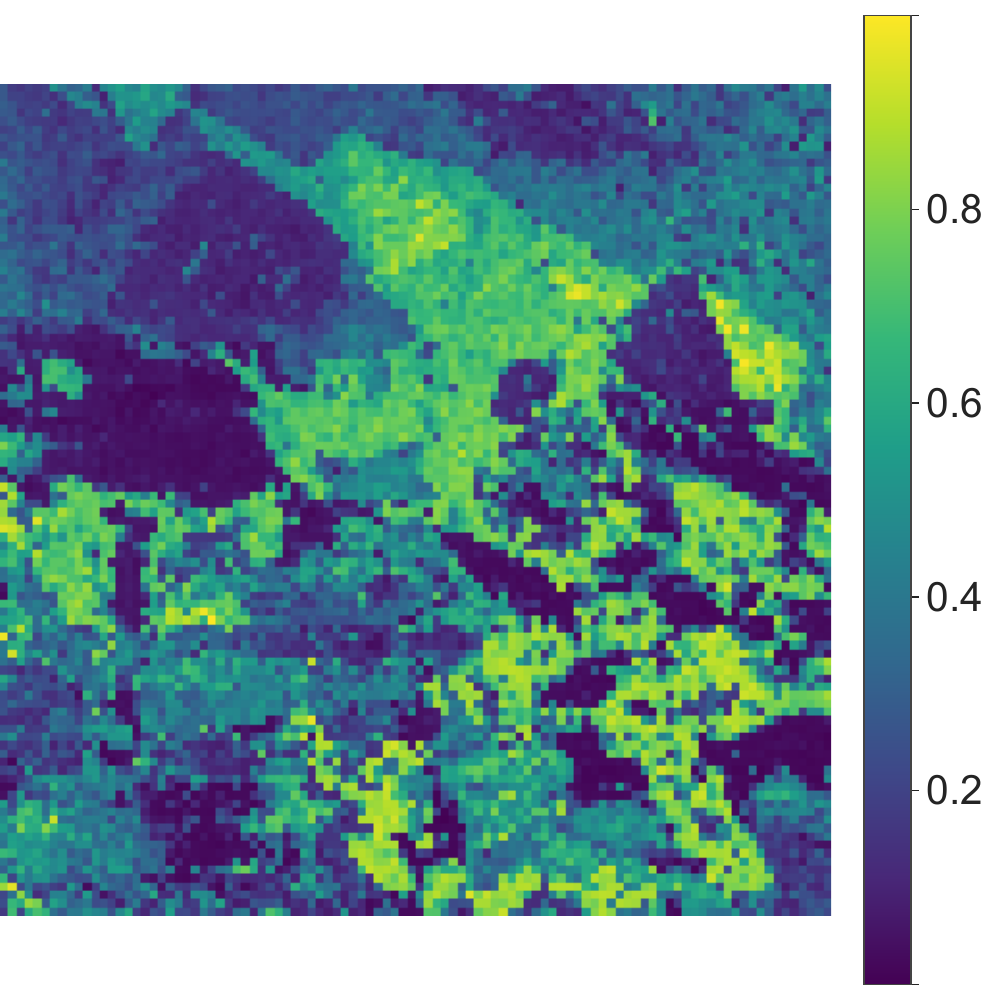}
    }
    \hfill
    \subfigure[\SI{394.40}{\nano\meter} and \SI{396.15}{\nano\meter} (Al)]{%
        \includegraphics[width=0.23\linewidth]{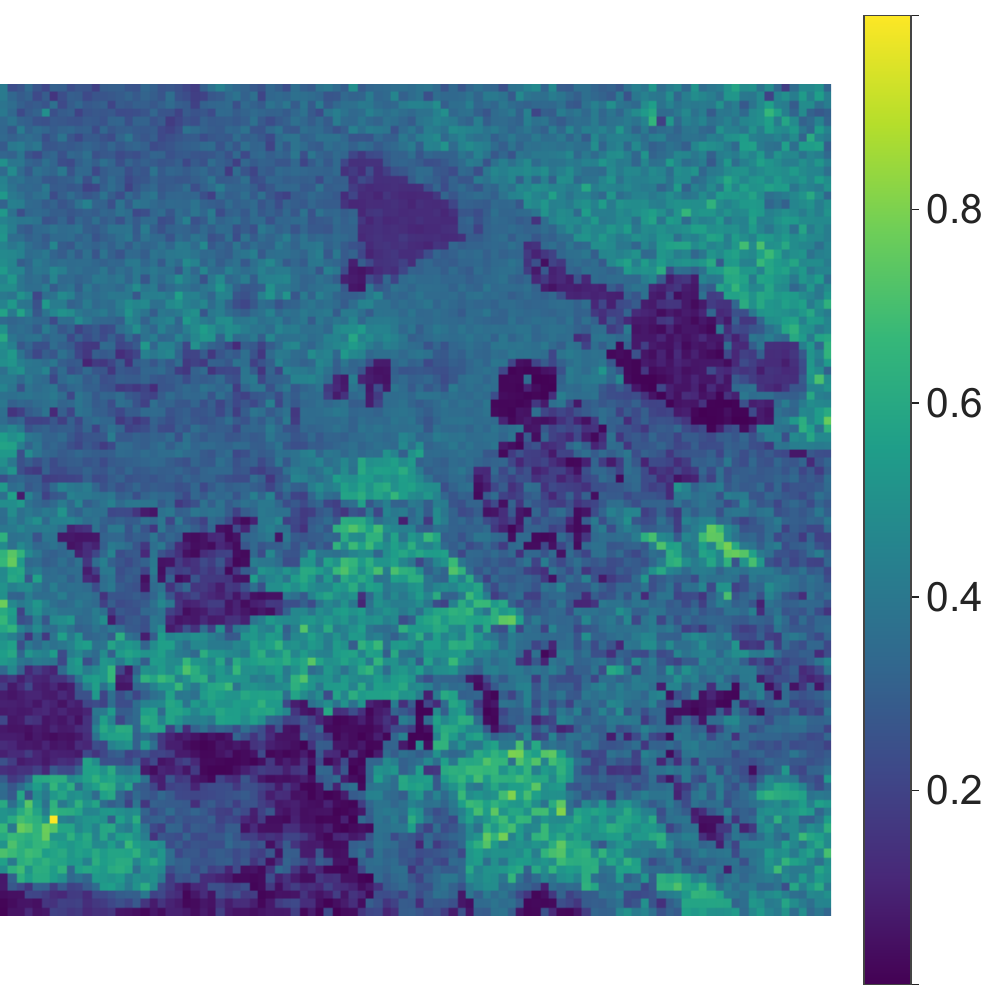}
        \includegraphics[width=0.23\linewidth]{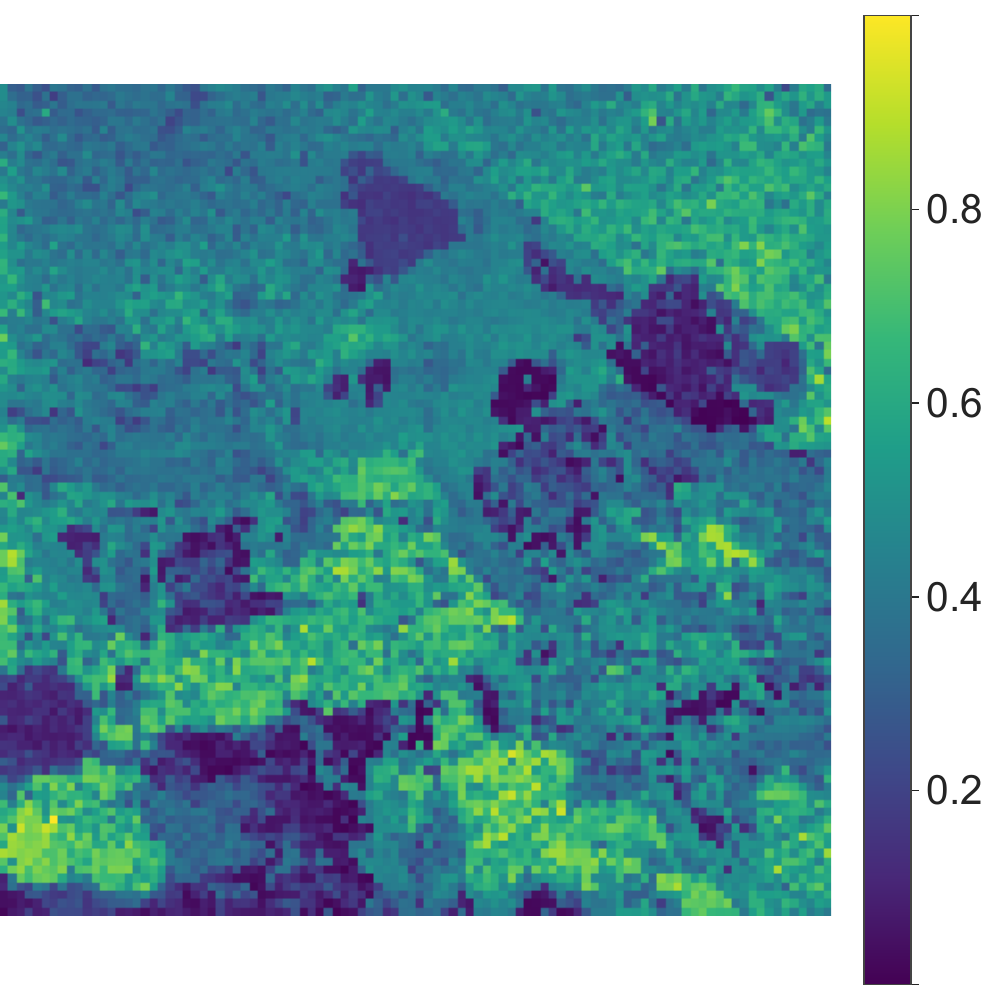}
    }
    \\

    \caption{%
        Reconstructions of the gabbro specimen using the line intensity maps.
        The scanned wavelength is reported.
        The element in parenthesis corresponds to the element whose intensity is maximal at the given value.
    }
    \label{fig:gabbro_average}
\end{figure}

In what follows, we present different scenarios involving rock and metal samples.
First, we analyse a gabbro specimen.
Finally, we show a peculiar property on an Al/Cu alloy.
These datasets have been chosen to display the properties of \hyperpca while dealing with different issues: strong spatial overlap of elements in the first case, increased \pc extraction threshold due to a small \libs map in the second, self-reversal of the Cu lines in the third.
In~\Cref{app:316L}, we also consider a sample of 316L stainless steel used in additive manufacturing to address additional issues arising in traditional analyses.

When possible, a reference measurement using \sem-\eds (\emph{Tescan Vega} with W filament) is provided to mimic the reference elemental maps available in the synthetic datasets.
The microscope was operated at \SI{15}{\kilo\volt}, at different values of magnification.
For the \libs mapping we used a Nd:YAG laser, operating at \SI{266}{\nano\meter}, with a gate delay of \SI{100}{\nano\second} and a gate width of \SI{1.0}{\micro\second}.
The spectrometer chosen for the measurements is an \emph{Acton} VM505, coupled to a \emph{Princeton Instruments} PI-MAX3 ICCD camera.
In general, we choose a slit width of \SI{50.0}{\micro\meter}, unless otherwise specified.
Measurements were performed under Ar atmosphere to increase the signal-to-noise ratio, unless otherwise stated.

\subsubsection{Case 1: Strong Spectral Interference}

\begin{table}[t]
    \centering
    \begin{tabular}{@{}lc@{}}
        \toprule
        \textbf{} & \textbf{concentration (\%)} \\ \midrule
        O         & $45.0$                      \\
        Si        & $22.3$                      \\
        Fe        & $8.6$                       \\
        Al        & $8.41$                      \\
        Ca        & $6.11$                      \\
        Mg        & $3.92$                      \\
        Na        & $2.56$                      \\
        Ti        & $1.38$                      \\
        others    & $< 1$                       \\ \bottomrule
    \end{tabular}
    \caption{Concentration of elements in the gabbro specimen~\cite{salle_comparative_2006}.}
    \label{tab:gabbro_conc}
\end{table}

In this section we show the analysis of a gabbro specimen, whose composition has been previously studied~\cite{salle_comparative_2006}.
In~\Cref{tab:gabbro_conc} we summarise the average concentration of elements in the sample, and we show the reference elemental distributions obtained from the \sem analysis in the appendix, presented in~\Cref{fig:gabbro_ground_truth}.
Notice that the \sem analysis is not performed on the same area as the \libs mapping.
The elemental maps show a diffused and heterogeneous distribution of Al and Ca on a scale of $\SI{2.5}{\milli\meter} \times \SI{1.5}{\milli\meter}$.
More concentrated insertions of Ti and Fe are present in the sample, though Fe is quite more diffused than Ti, in general.
The granular formation in the lower right is of particular interest: from the \eds analysis, the composition of this formation is almost exclusively Ti, though Fe is also present in small concentration.
Given its specific and quite different nature, it may be possible for the \pca analysis to completely disentangle this granular formation from the rest of the components.
These granularities are present in several points on the surface.
Detection of other elements, relevant to the \libs analysis, is complicated: for instance, the distributions of Si is completely homogeneous on the sample, while K and Na require very high excitation energies, in the chosen spectral band, and present a homogeneous matrix.
For the \libs imaging measurements we use a \SI{10}{\micro\joule} laser pulse on a $100 \times 100$ matrix on the surface.
Craters have generally \SI{1.5}{\micro\meter} of depth and a diameter of \SI{3.5}{\micro\meter}.
The slit width of the spectrometer is, in this case, \SI{120}{\micro\meter}.
We use a grating with \SI{2400}{grooves\per\milli\meter} centred at \SI{395.57}{\nano\meter}.
\Cref{fig:gabbro_average} shows the average spectrum obtained from the measurements.

We first approach the analysis by providing the elemental maps obtained from the traditional univariate method: we sum the emission intensity of the principal emission lines across a window of \num{10} wavelength channels centred at the peaks visible in the average spectrum.
As shown in \Cref{fig:gabbro_average}, using this approach we can reconstruct good quality cartographies of Al, using the lines at \SI{394.40}{\nano\meter} and \SI{396.15}{\nano\meter}, and Ca, with the emissions at \SI{393.37}{\nano\meter} and \SI{396.85}{\nano\meter}.
The quality of the cartographies is indeed good, though only two elements have been reconstructed.
They present good indication of the heterogeneity of the rock at the scale of interest for \libs mapping, with distinct mineral formations.

\begin{figure}[t]
    \centering
    \begin{tabular}{c|c}
        {\LARGE \textsc{standard} \pca}                           & {\LARGE \hyperpca}
        \\[1em]
        \includegraphics[width=0.47\linewidth]{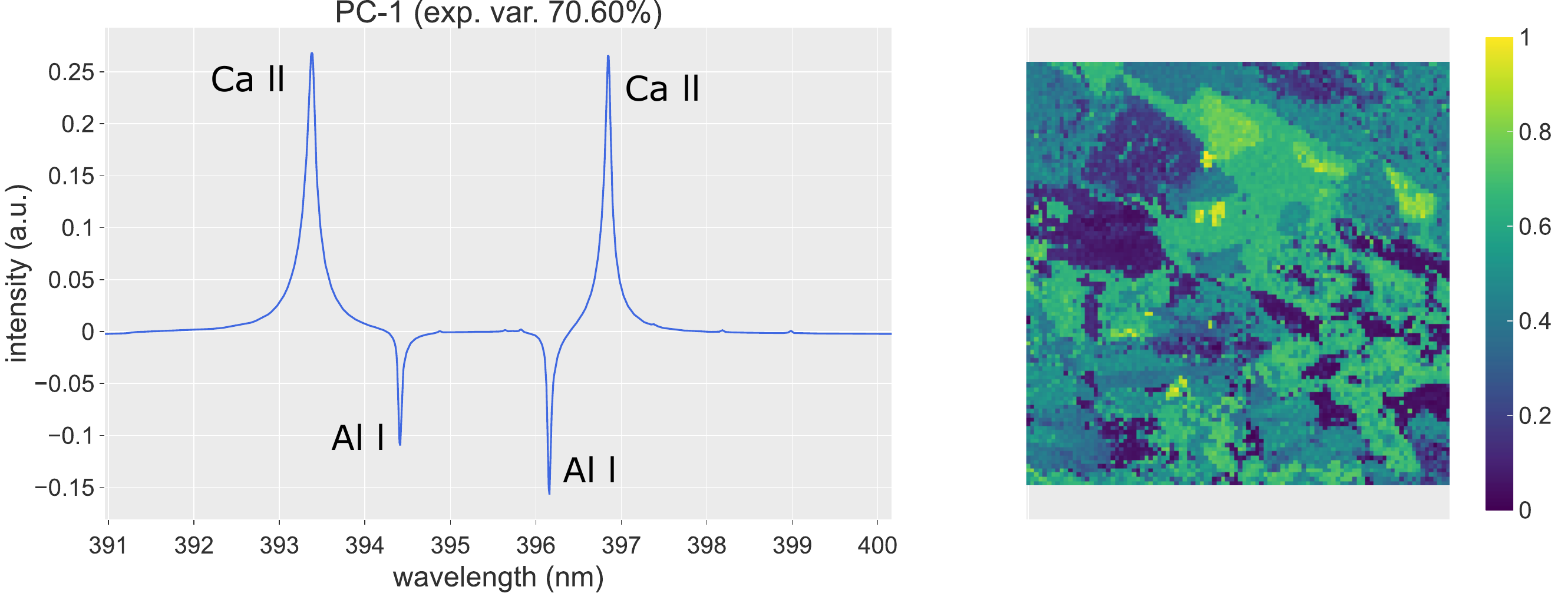} & \includegraphics[width=0.47\linewidth]{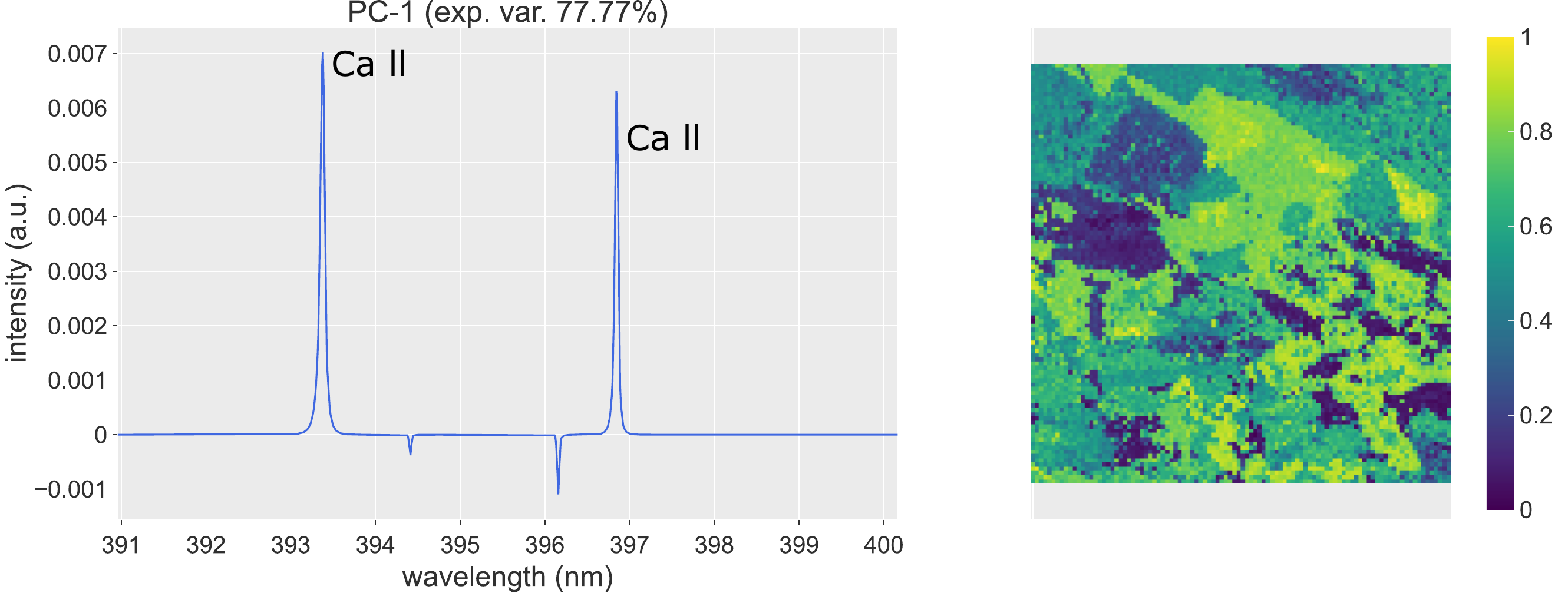}
        \\
        \includegraphics[width=0.47\linewidth]{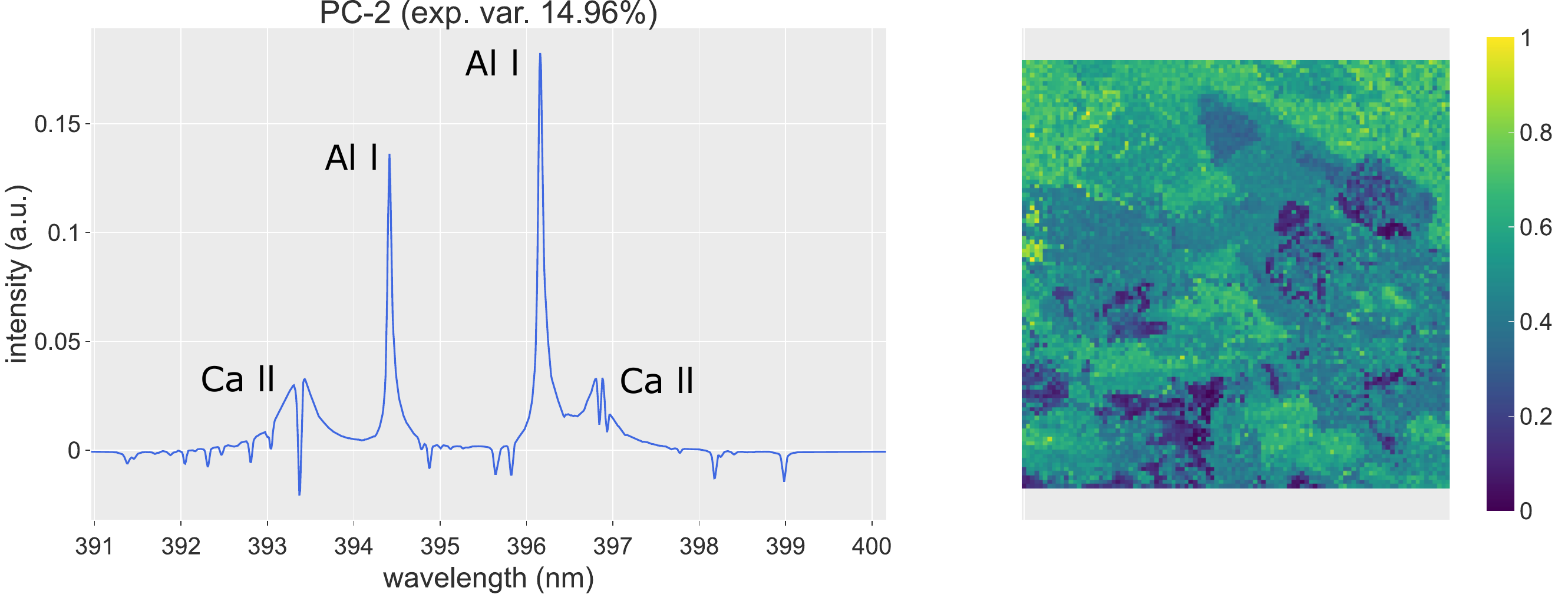} & \includegraphics[width=0.47\linewidth]{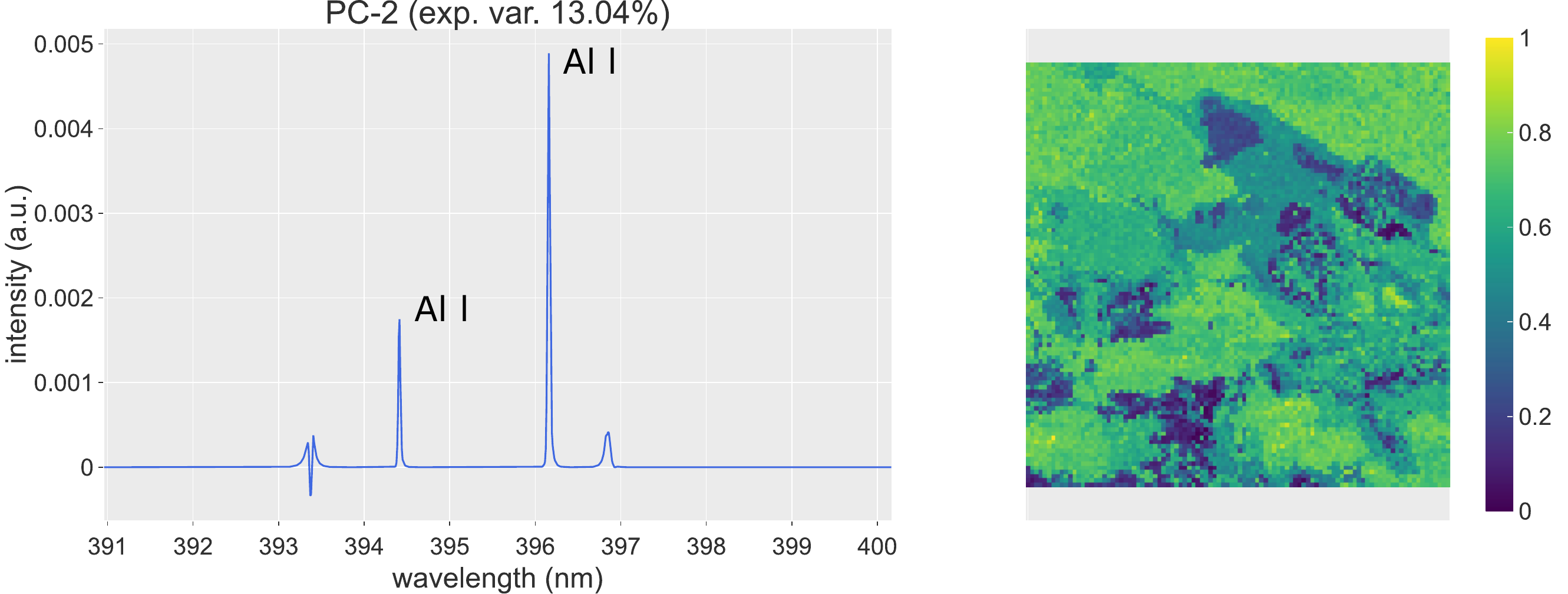}
        \\
        \includegraphics[width=0.47\linewidth]{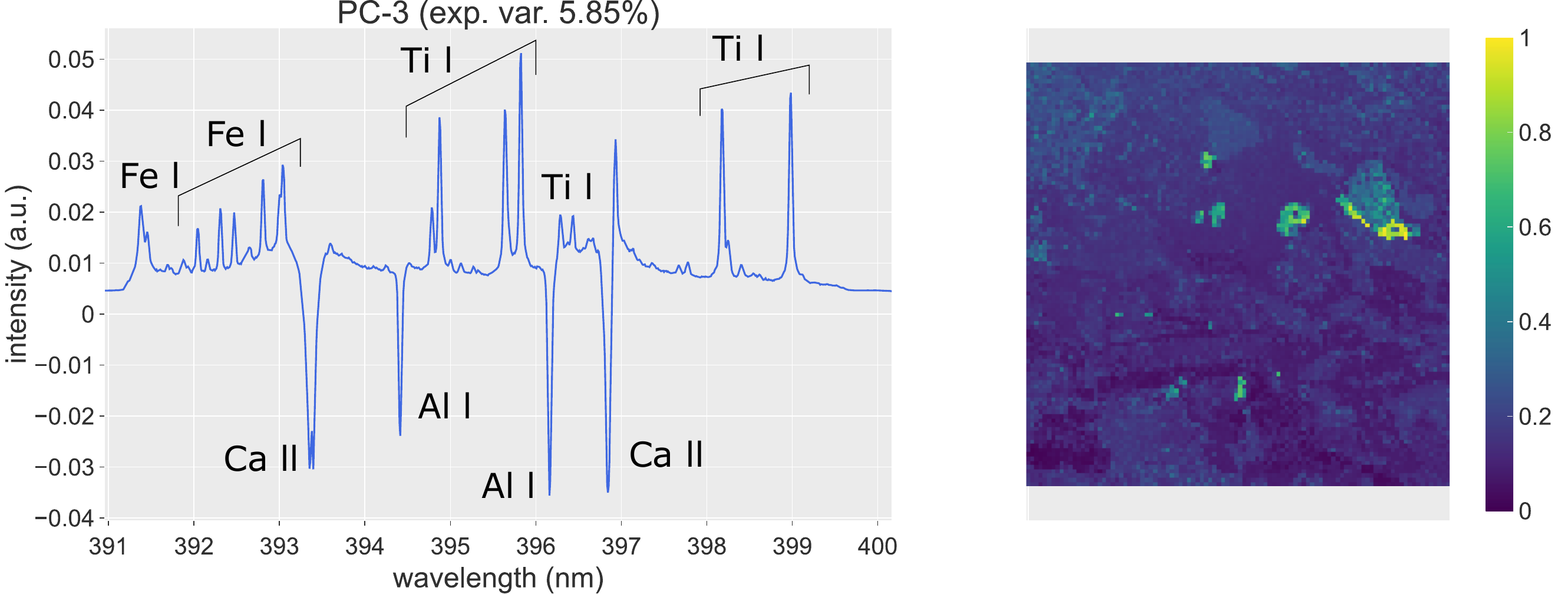} & \includegraphics[width=0.47\linewidth]{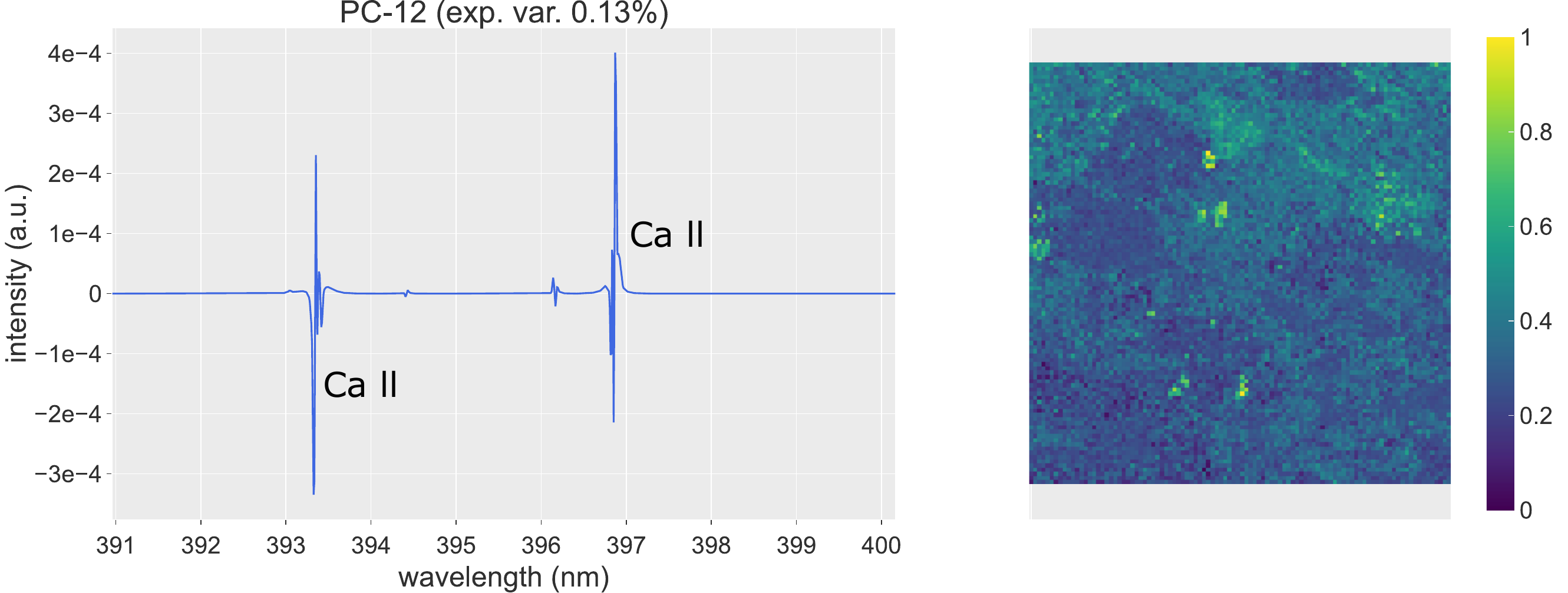}
        \\
                                                                  & \includegraphics[width=0.47\linewidth]{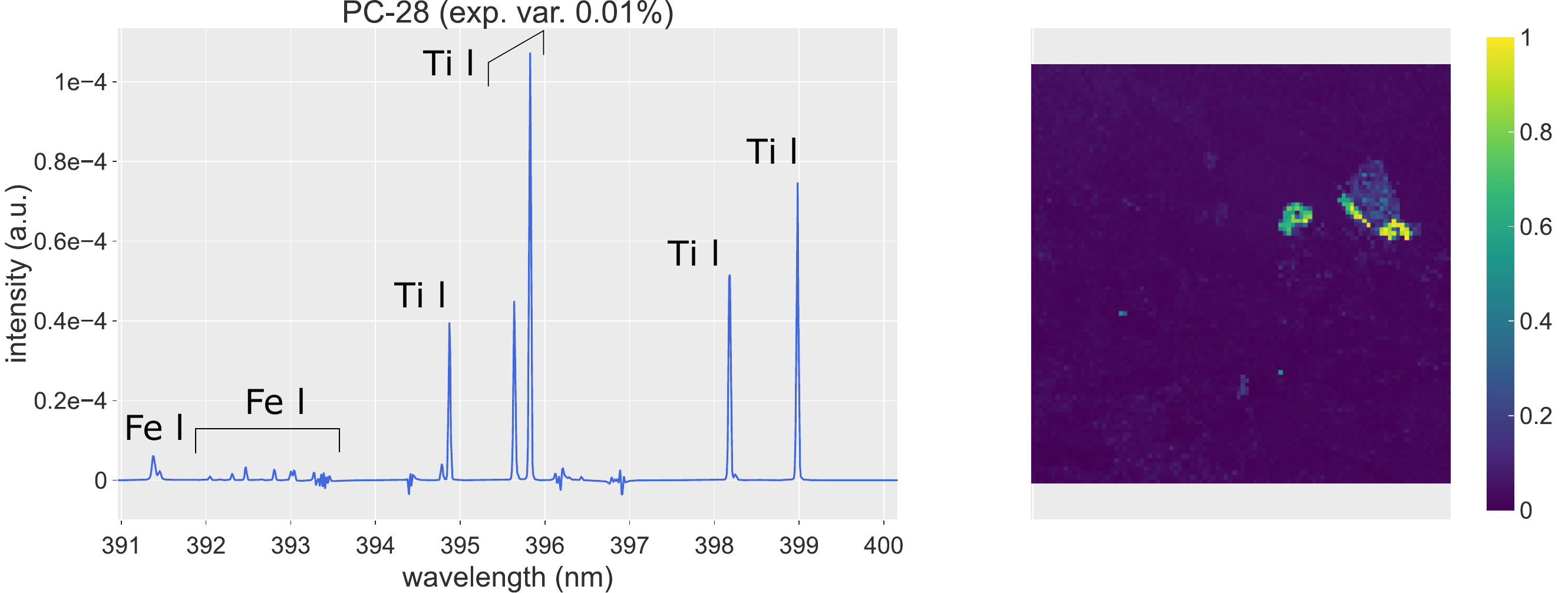}
    \end{tabular}
    \caption{Reconstruction of the gabbro specimen using the standard \pca and \hyperpca.}
    \label{fig:gabbro_pca}
\end{figure}

In the first column of~\Cref{fig:gabbro_pca} we show the retrieval of highest ranking \pcs of the gabbro specimen using the standard \pca.
The first \pc shows a perfect distinction of the lines associated to Al and Ca present in the average spectrum, representing the former as negative contributions to the loading vectors, associated to the darker zones of the map, and the latter as positive contributions, associated to the lighter colours.
The principal component thus represents a map of the differential distribution of Al and Ca in the sample.
\pcn{2} highlights the presence of Al in the sample, though in the presence of other elements which spoil the quality of the associated elemental map.
The loading vector also shows the self-reversal phenomenon of the Ca lines, though its visualisation on the elemental map is mostly unfeasible.
Differently from the univariate approach, \pcn{3} is able to efficiently extract the emission lines of Fe and Ti, scattered in the entire spectral range.
Again, the \pc represents a differential distribution of such element with respect to Al and Ca, as these are represented as negative contributions in the loading vector.
Notice that part of the self-reversal phenomenon of the Ca lines is represented by a positive contribution, which makes it difficult to disentangle completely Fe and Ti lines from the most concentrated Ca areas.
Even a simple \mva approach such as the standard \pca is therefore able to provide a good quality spatial information on the distribution of the elements.
In this case, the analysis does not disentangle completely the Ti and Fe granularities from the other mineral formations.

\begin{figure}[t]
    \centering

    \subfigure[Preprocessed average and $1\upsigma$ spectra with examples of single-shot spectra.]{\includegraphics[width=0.5\linewidth]{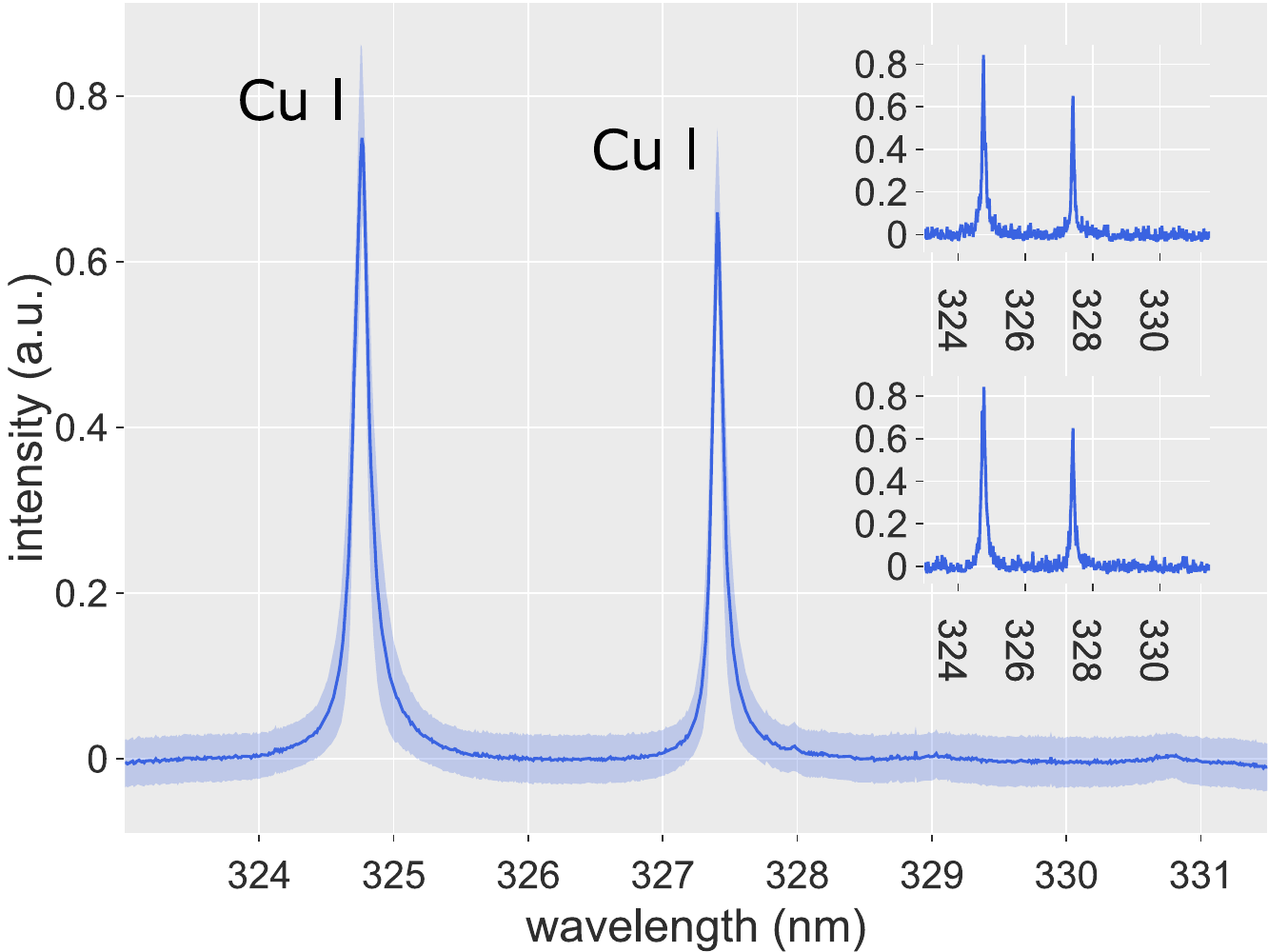}}
    \\
    \subfigure[\SI{324.75}{\nano\meter} and \SI{327.40}{\nano\meter} (Cu)]{%
        \includegraphics[width=0.25\linewidth]{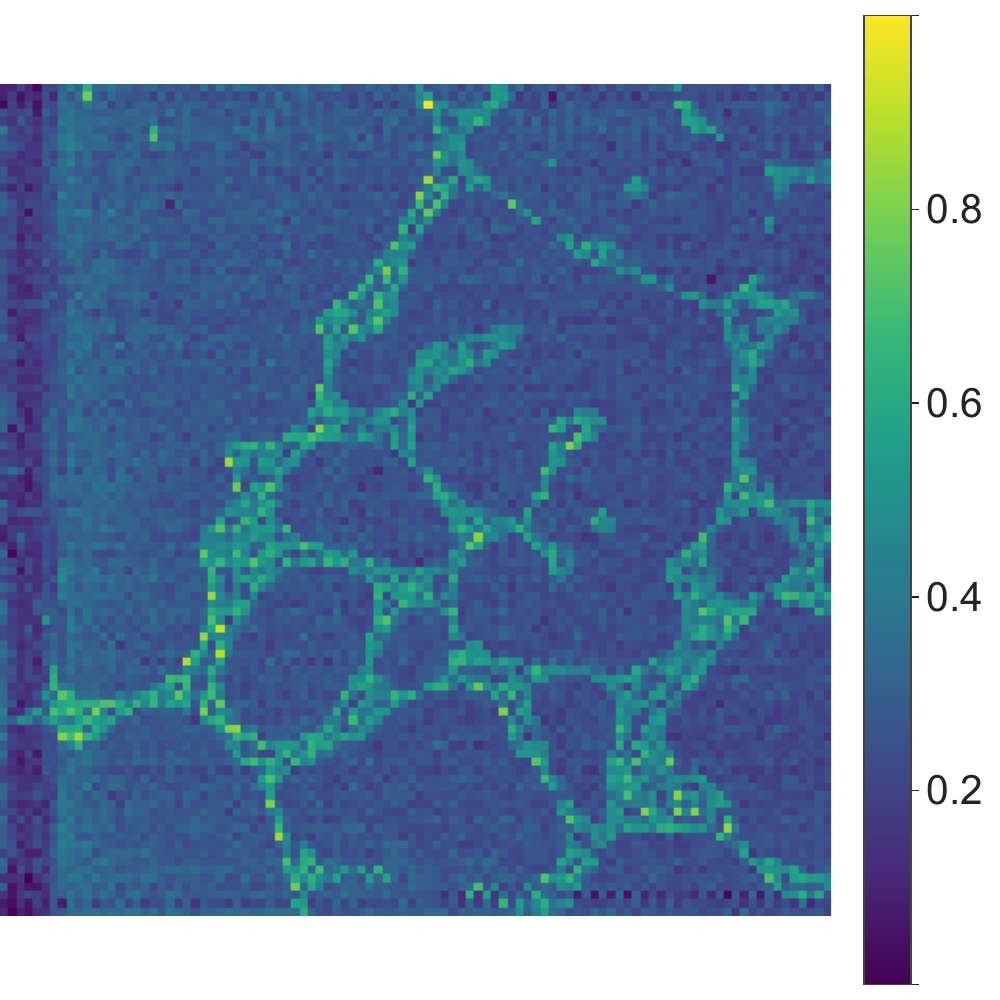}
        \qquad
        \includegraphics[width=0.25\linewidth]{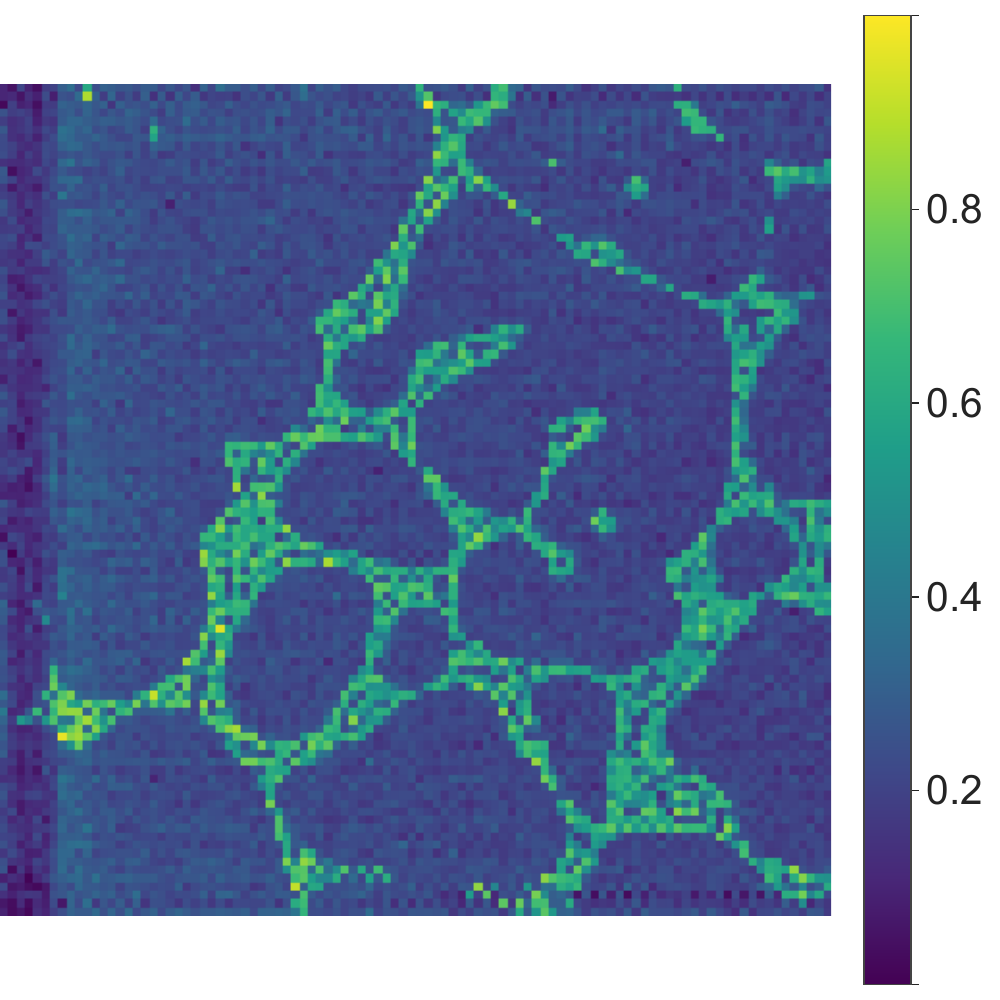}
    }

    \caption{%
        Reconstructions of the Al/Cu alloy using the line intensity maps.
        The scanned wavelength is reported.
        The element in parenthesis corresponds to the element whose intensity is maximal at the given value.
    }
    \label{fig:self_reversal_sample}
\end{figure}

In the second column of~\Cref{fig:gabbro_pca} we present a collection of the most relevant components extracted using our proposed method, \hyperpca.
For the analysis we choose \texttt{bior3.9} filters, stopping at the first downsampling layer, and $\alpha = \num{e-1}$.
The effect of the application of a kernel function to the covariance matrix obtained from the \dwt of the original data is immediately visible in the highest ranking \pcs.
\pcn{1} shows the distribution of Ca on the sample, almost deprived of any relevant Al contamination.
In the same way, \pcn{2} shows the distribution of Al, without significant Ca presence, which starts to show part of the self-reversal phenomenon.
Readability of the loadings is thus clearly enhanced, as noise is no longer present in the spectra.
As a consequence, the associated elemental maps only show the effective distribution of the elements, thus making their interpretation easier and immediate.
The effect of the \dwt is visible in the grouping of the lines associated to Al and Ca in separate \pcs: the loadings only contain strong contributions coming from a single element, thus disentangling the emission lines of interfering elements.
As in standard \pca, \hyperpca is also capable of extracting information which would otherwise be difficult, if not impossible, to highlight using a univariate method.
These properties are visible in \pcn{12} which shows additional information on the self-reversed Ca lines: the corresponding elemental map presents clearly the distribution where the presence of Ca is most intense, without Fe and Ti interference.
By comparison with the reconstruction of the standard \pca in~\Cref{fig:gabbro_pca}, it is thus already possible to distinguish the Ca contribution from the Ti formation.
\hyperpca is also capable of completely separating the Ti emission lines present in the granularities in \pcn{28} at \SIlist{394.87;395.63;395.82;398.18;398.98}{\nano\meter}.
Less intense Fe lines, notably at \SIlist{391.36;392.29;392.79;393.03}{\nano\meter} are also recovered in the same \pc.
Given its properties, \hyperpca is thus capable of providing more readable loading vectors, and, consequently, more relevant elemental maps (e.g.: isolated elemental contributions, specific mineral formations), with higher quality than standard \pca or other univariate approaches.
\hyperpca also grants the ability to recover other properties of the analysed sample, shown in~\Cref{fig:gabbro_kpca_less_imp}: \pcn{3} and \pcn{4} show the self-reversal phenomenon of the Ca lines, represented as brighter colours in the score maps due to the sign convention, by singling out the Ca contributions and the reversed peaks.
Minor lines of Fe can also be recovered in lower ranking \pcs, such as the \pcn{62} shown in the same figure in the supplementary material.
This component also shows the correlation between Ti and Fe, the latter being present in less concentrated quantities on the entire surface of the sample, as confirmed also by the \sem images in~\Cref{fig:gabbro_ground_truth} in the appendix.
Other \pcs contain additional information on the line profiles at different detail levels and are not presented here for brevity.

Our proposed method is able to extract higher quality information with respect to both to traditional approaches and standard \pca.
The kernel-based method and the creation of a sparse representation enable the resolution of spectral interference and the elimination of most noise contributions, granting better quality for both loadings and score maps, and providing a larger number of readable and interpretable \pcs.

\subsubsection{Case 2: Self-reversal Phenomenon}

We show the case of an Al sample in the presence of Cu insertions.
In~\Cref{fig:self_reversal_sample} we show its average spectrum, after preprocessing the data as in~\Cref{sec:methodology}.
The two distinguishable peaks are the characteristic most intense emission lines of Cu at \SI{324.75}{\nano\meter} and \SI{327.40}{\nano\meter}.
We use a \SI{3}{\micro\joule} laser pulse on a $100 \times 100$ matrix with a resolution of \SI{5}{\micro\meter} in both axes.
Craters are \SI{3}{\micro\meter} both in diameter and depth.
We use a grating with \SI{2400}{grooves \per\milli\meter}, centred at \SI{327.46}{\nano\meter}.
Measurements are not performed in Ar atmosphere, as the Cu signal is already quite intense.
In this case, we choose a kernel parameter $\alpha = \num{5e-3}$.
We use the \texttt{bior3.5} filter bank at the first downsampling layer, with a \hardt of \num{0.99} times the maximal entry in the power spectrum of the \dcs.
Due to the physical size of the sample, \sem reference maps are not provided.
Moreover, at the scale probed by the \sem, the images would not display the same heterogeneous structures present in the \libs map.
The line intensity maps are shown in~\Cref{fig:self_reversal_sample}.
This dataset presents a few peculiar aspects, such as the presence of self-reversal in the brightest patterns of~\Cref{fig:self_reversal_sample} and a homogeneous Cu distribution elsewhere.

\begin{figure}[t]
    \centering
    \begin{tabular}{c|c}
        {\LARGE \textsc{standard} \pca}                         & {\LARGE \hyperpca}
        \\[1em]
        \includegraphics[width=0.47\linewidth]{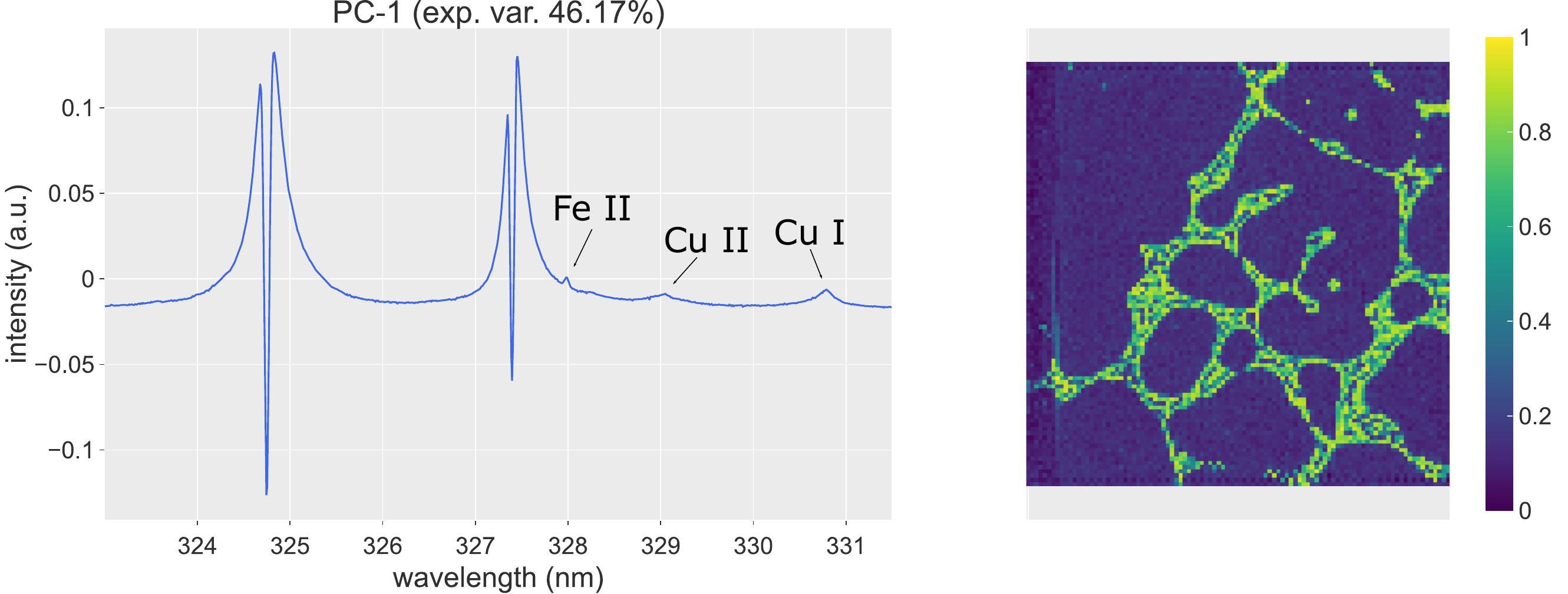} & \includegraphics[width=0.47\linewidth]{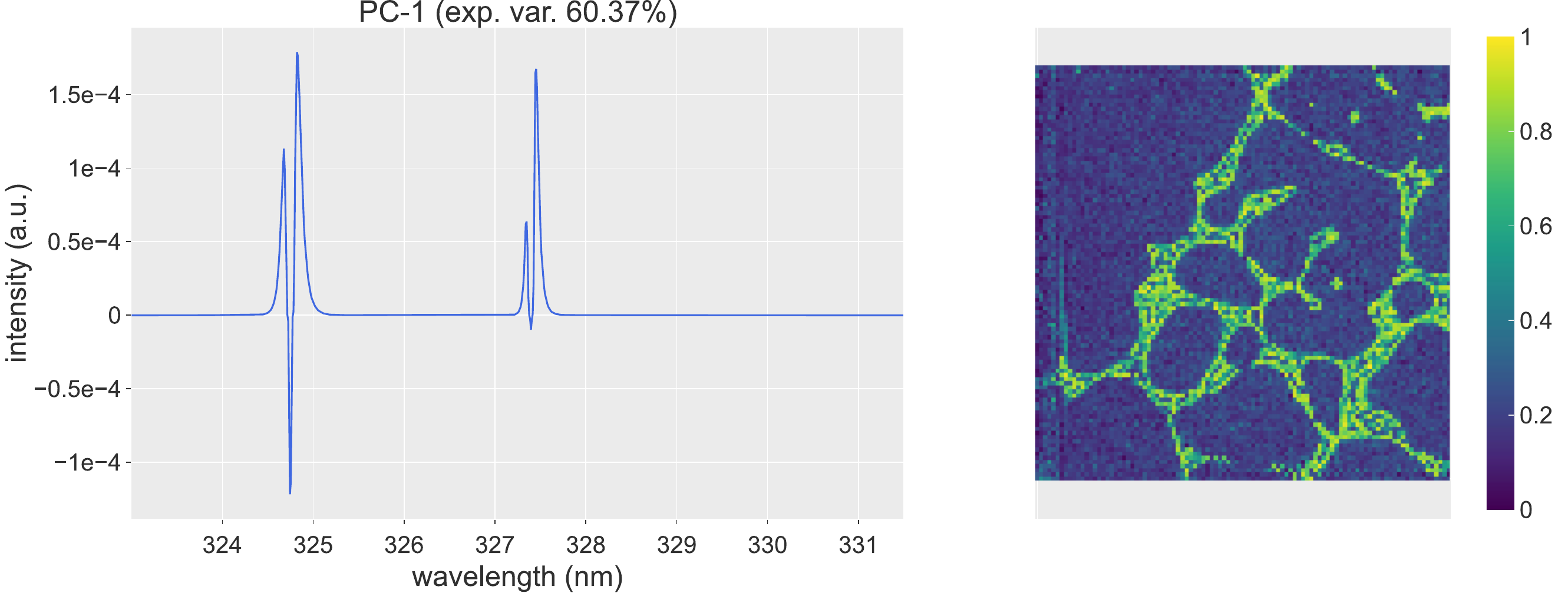}
        \\
        \includegraphics[width=0.47\linewidth]{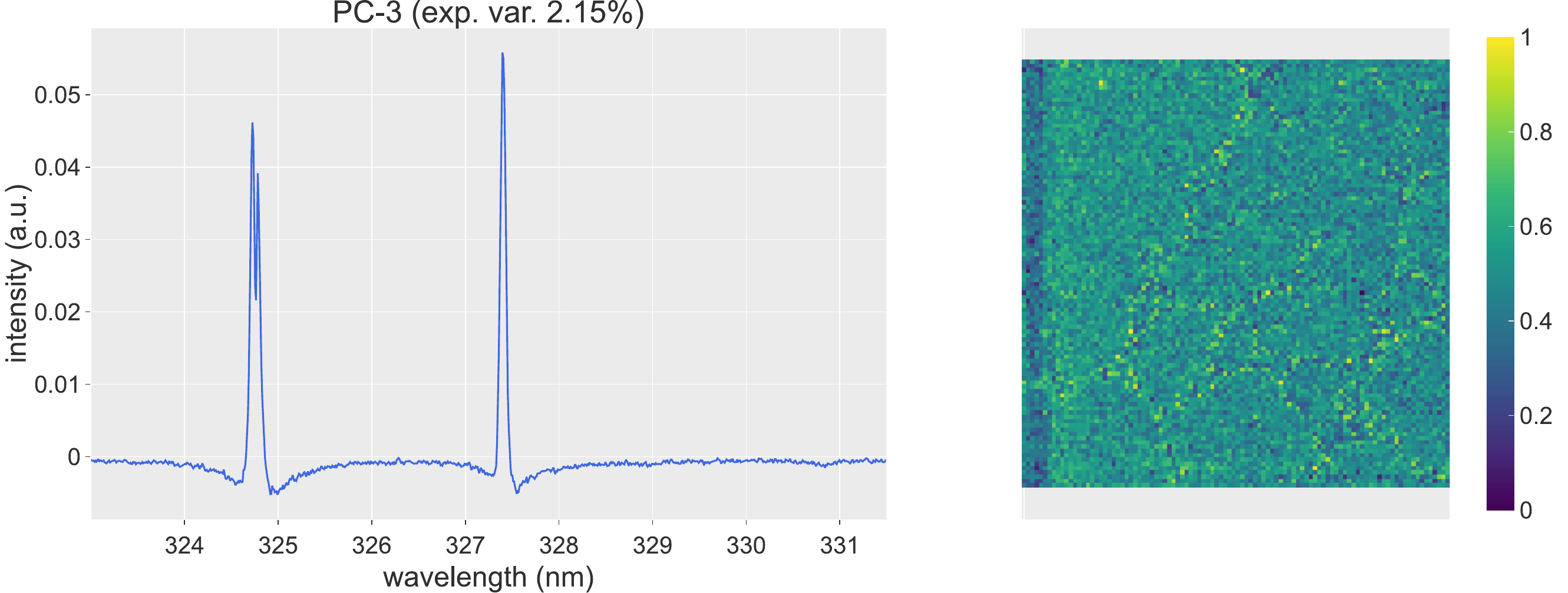} & \includegraphics[width=0.47\linewidth]{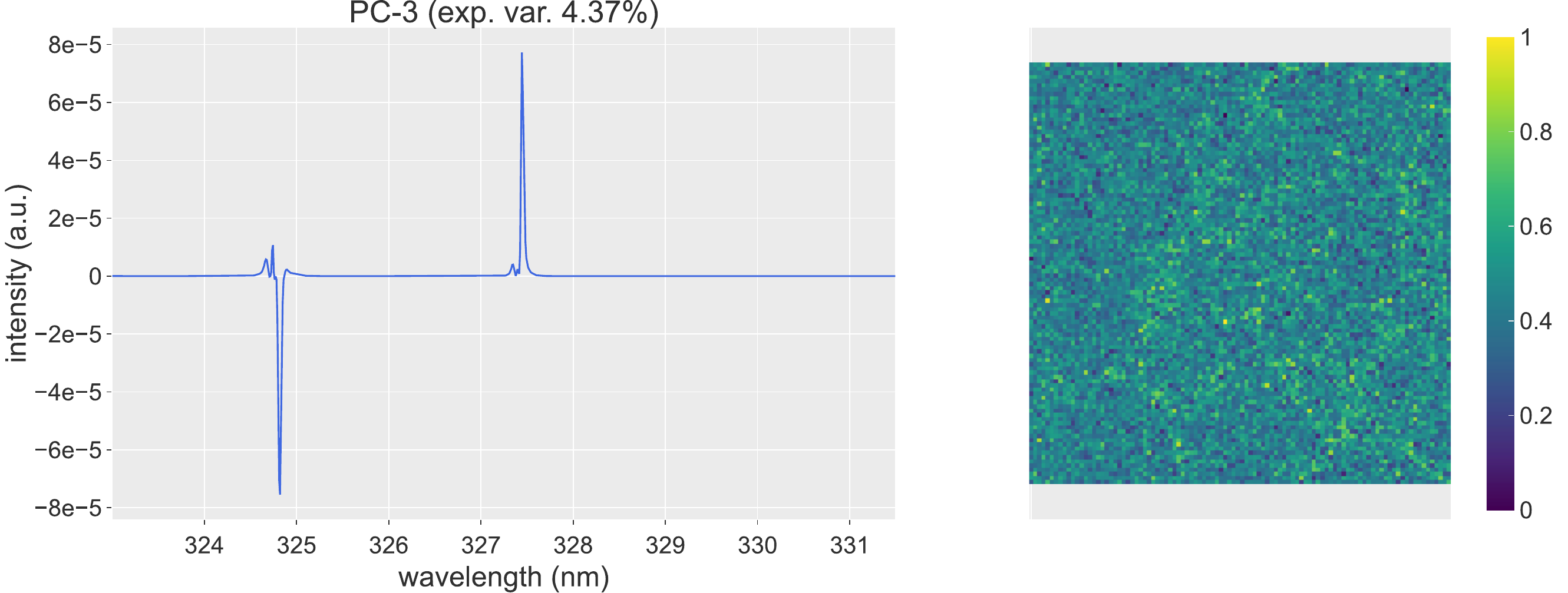}
        \\
                                                                & \includegraphics[width=0.47\linewidth]{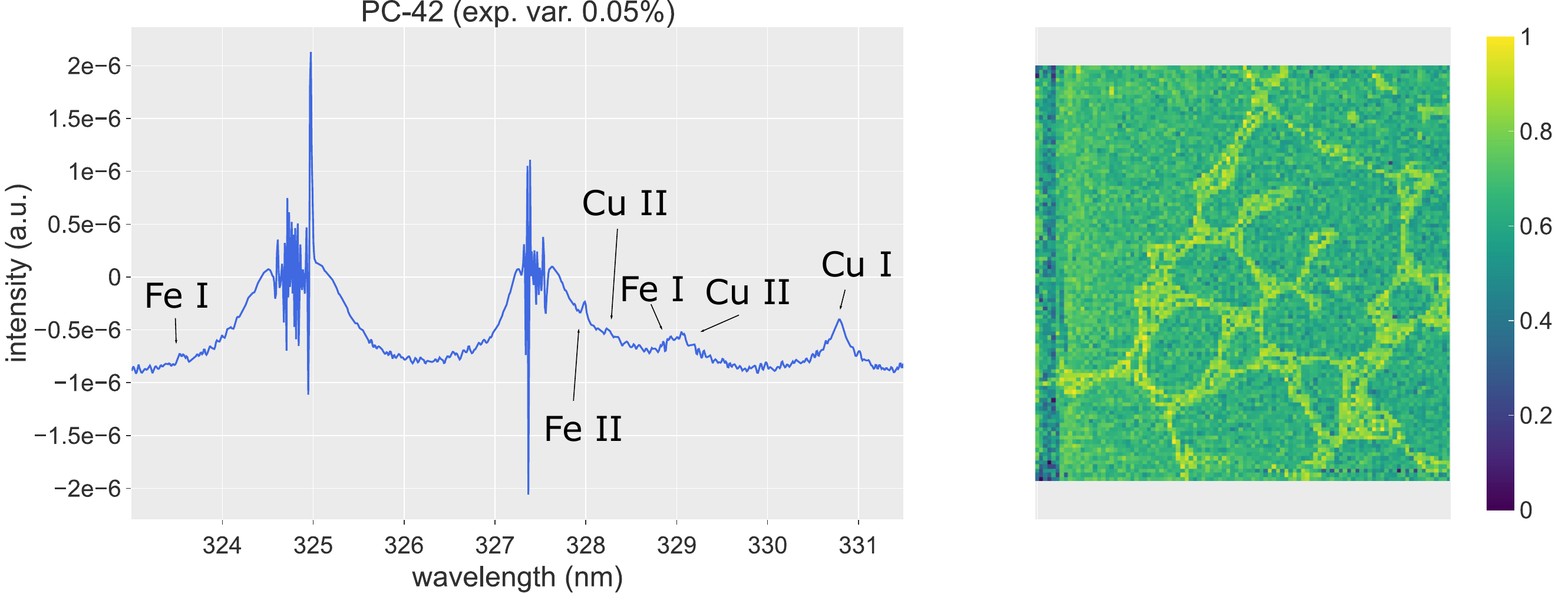}
    \end{tabular}
    \caption{Reconstruction using the standard \pca and \hyperpca.}
    \label{fig:self_reversal_pca}
\end{figure}

As visible in the first column of~\Cref{fig:self_reversal_pca}, the first \pc extracted using the standard \pca approach highlights the presence of self-reversed lines: the brightest spots in the score maps are in correspondence to the strong positive contributions of the loading vector, present only in self-reversed lines, which are usually processed as two separate peaks by \mva algorithms.
The centre peak, shown as a negative contribution by \pca, identifies Cu in general (shown in darker colour on the map).
\pcn{3} on the left of~\Cref{fig:self_reversal_pca} shows the homogeneous presence of Cu on the sample by accentuating the self-reversed peaks, shown as positive contributions by our sign convention, which identify both self-reversed and non self-reversed lines of Cu.
The peak at \SI{324.75}{\nano\meter} presents some reconstruction artefacts, differently from the line at \SI{327.40}{\nano\meter}.

As \pca already reaches good results, \hyperpca does not drastically improve the quality of the first \pc, as it is visible on the right of~\Cref{fig:self_reversal_pca}, though it shows the Cu presence over a perfectly clean background, totally deprived of noise, making it easier to recognise the spectral signature of the element.
By singling out entirely the Cu contribution, \pcn{1} also enables the localisation of the element on the entire surface of the sample, confirmed by \pcn{3} of both \pca and \hyperpca.
However, the scan with the \dwt enables a finer characterisation of the self-reversal phenomenon by providing access to a larger number of \pcs showing the reversed lines in details.
The first \pc, in fact, does not show the minor peaks above \SI{327.5}{\nano\meter}, thus the amount of explained variance in the \pc is totally related to the presence of the self-reversed Cu lines.
As for standard \pca, \pcn{3} in the second column of~\Cref{fig:self_reversal_pca} displays the homogeneous distribution of Cu on the sample, by almost totally removing the self-reversal effects from the data.
The opposite sign of the emission lines are due to the effect of self-reversal: according to its intensity, the line width and profile change, thus being reconstructed differently by \dwts.
Such information can be therefore used, in this particular case, to better characterise the phenomenon.
Moreover, as the stochastic noise distribution is almost completely reduced by the kernel of \kspca, it is possible to have access to more readable \pcs, and more information accordingly, with respect to standard \pca.
In lowest ranked \pcs, such as \pcn{42} in~\Cref{fig:self_reversal_pca}, it is in fact possible to recognise the presence of minor Cu peaks at \SI{329.04}{\nano\meter} and \SI{330.80}{\nano\meter}, and a Fe line at \SI{327.98}{\nano\meter}, already extracted in \pcn{1} using standard \pca but with a larger profile width than \hyperpca.
In addition to the information of standard \pca, it is also possible to recognise another Cu line at \SI{328.17}{\nano\meter} and Fe lines at \SI{323.62}{\nano\meter} and \SI{328.94}{\nano\meter}: as \hyperpca better extracts the hierarchy of eigenvalues, their variance is better estimated, and their contribution is particularly evident.
Notice that the line at \SI{323.62}{\nano\meter} is only slightly visible in \pcn{1} on the left of~\Cref{fig:self_reversal_pca}, but it is extracted as part of the signal information by \hyperpca.
The same \pc also shows more details of the main emission lines, as it enhances experimental effects, such as the uncertainty of the measure, which is registered as a fluctuation around the Cu lines, where the intensity of the signal is stronger.
Finally, it also shows in detail the diffused distribution of Cu on the sample, since the self-reversed peaks are less intense.
The same \pc in the standard \pca is not displayed, as it resulted highly unreadable due to noise.
In the appendix in~\Cref{fig:alcu_pca_less_imp} we show two additional \pcs (\pcn{5} and \pcn{25}) extracted using the standard \pca: after the fifth \pc, the standard analysis only shows information on the Cu lines, with homogeneous score maps, which do not yield additional information on the content of the samples, or on the distributions of elements on the surface.
\hyperpca provides access to a higher number of components with respect to the standard \pca, thus making it possible to find useful information in low ranking \pcs, such as \pcn{42}.

With respect to traditional univariate methods, both \pca and \hyperpca are capable of providing more information and better reconstruction of the elemental maps: in both cases, the \libs maps are decoupled into self-reversed lines and homogeneous Cu distribution, especially in the \hyperpca case where the two behaviours are distinct.
Differently from \pca, \hyperpca grants the possibility to distinguish the different nature of a signal by reconstructing the spectra in a different and sparse basis.
This way finer details can emerge and it may be possible to distinguish different kinds of sources, in addition to correlations and anti-correlations of spectral signatures as in the standard \pca.
In some difficult cases, where specific effects are not evident, low-ranking \pcs may display additional information, generally lost when limiting the dimensionality reduction to only the first few \pcs.
\hyperpca gives such possibility, as noise no longer spoils the loading vectors.
Notice that, even though \pcn{42} can look irrelevant, dimensionality reduction is nonetheless greatly achieved, as the total number of components is approximately \num{e3} in all cases we consider.


\section{Conclusion}

In this paper we introduce \hyperpca as an alternative to the standard \pca approach for the analysis of \libs mapping spectral data, particularly with a very low ($\ll 1$) shot-to-shot signal-to-noise ratio.
The proposed algorithm shows significant improvement over univariate data analysis methods and standard \pca for the extraction of physico-chemical information, and it is backed up by theoretical background in \rmt and machine learning literature.
In particular, it provides a tool to extract elemental maps, together with correlations of chemical elements.
The novel contribution is the introduction of a method to recover a certain degree of sparsity of the input signal, as noise and interference prevent the original data from being in such representation, and a kernel-based \pca method to decouple the signal from the random distribution.
These aspects can be directly controlled through the introduction of additional parameters, contrary to the standard \pca analysis which does not require any parameter tuning.
The main advantages with respect to current approaches can be found both in the quality of the score matrices, which present better details in the element map, and in the loadings, which are almost totally deprived of background noise.
Furthermore, \hyperpca shows the ability to extract emission lines of a given element in the same \pcs, as the \dwt can transform in the same way lines which present the same pattern in a given spectral range with higher probability.
We also provide an additional performance study of the algorithm in~\Cref{app:ablation}: we show that the best results are indeed due to the combined action of the \dwt and \kspca, while the use of just one of the techniques does not reliably reproduce the same quality in the results.
This enables a better extraction of the elemental maps as the probability of reconstructing the score matrices with an increased amount of details increases.
Inevitably, with respect to the standard \pca, the enhanced quality and quantity of information comes at the cost of a greater number of meaningful \pcs, representative of the correct geometrical rank of the signal matrix, which need to be analysed.
As it has been shown both on synthetic and experimental data, \hyperpca grants the possibility to improve the elemental mapping of the content of samples even in the presence of low signal-to-noise ratio, small \libs maps and strong spectral interference.
It thus represents an unsupervised technique alternative to the standard approach based on the line intensity.
Beyond that, it also represent a good option for the extraction of information with respect to the \pca approach used in \libs mapping in the presence of very high noise levels.
As an advanced unsupervised technique, here optimised for \libs mapping, \hyperpca shows versatility when adapting to different situations, in which no supervised input is given.
Quantitative analyses may greatly profit from \hyperpca since it still provides both a reduction of the number of variables, and higher quality information with respect to the standard \pca, as it retains information on both correlations and elemental maps.
It would be interesting to test \hyperpca in other unsupervised pixel-wise segmentation tasks, with different types of data, where no \textit{a priori} knowledge of the samples is available.
This would also open the possibility to study other advanced refinements of \pca, such as tensor \pca models~\cite{montanari_statistical_2014}, which in addition take advantage of the spatial structure of the data distribution.


\section*{Author Contributions}

\textbf{Riccardo Finotello}: conceptualization, data curation, formal analysis, investigation, methodology, software, validation, visualization, writing -- original draft, writing -- review \& editing;
\textbf{Mohamed Tamaazousti}: funding acquisition, project administration, resources, supervision, writing -- review \& editing;
\textbf{Jean-Baptiste Sirven}: data curation, funding acquisition, project administration, resources, supervision, writing -- review \& editing.

\section*{Conflicts of Interests}

There are no conflicts to declare.

\section*{Acknowledgements}

R.F.\ would like to thank C.\ Qu\'{e}r\'{e} and D.\ L'Hermite for the extensive support and help with \libs equipment, and C.\ Blanc and J.\ Varlet for the help with the \sem analysis and the preparation of the samples.
We acknowledge the financial support of the \emph{Cross-Disciplinary Programme on Instrumentation and Detection} of \textsc{CEA}, the French \emph{Alternative Energies and Atomic Energy Commission}, and G.\ Gallou for proposing this collaboration between the \emph{Direction des \'{e}nergies} (\textsc{DES}) and the \emph{Direction de la recherche technologique} (\textsc{DRT}).


\printbibliography[heading=bibintoc]

\clearpage

\appendix\break\pagenumbering{roman}\renewcommand{\thepage}{\roman{page}}

\section{Technical Background}\label{app:tech}

We briefly recall the definitions and properties of the Principal Components Analysis (\pca) and the Discrete Wavelet Transform (\dwt) used in the main paper.

\subsection{Principal Components Analysis}\label{app:pca}

Let $Y \in \R^{n \times p}$ be a data matrix with zero sample mean whose vector components $y_{(i)} \in \R^{p}$ for $i = 1, 2, \dots, n$ represent the observed data.
The principal components (\pcs) are an orthonormal basis of vectors $\left\lbrace w_{(i)} \right\rbrace_{i = 1, 2, \dots, p}$ representing the directions of the lines which best fit the data.
\pca~\cite{pearson_lines_1901} is a statistical and unsupervised machine learning tool to compute the \pcs and use them to ``rotate'' the data in a suitable basis, where the sample variance is the largest.

Mathematically, \pca is a linear map realised as the right multiplication between matrices:
\begin{equation}
    \begin{tabular}{lccc}
        \pca: & $\R^{n \times p}$ & $\to$      & $\R^{n \times p}$
        \\
        & $Y$                       & $\mapsto$  & $Y' = Y W$
    \end{tabular}
\end{equation}
where $W \in \R^{p \times p}$ contains the \pcs as columns, such that $W^T W = \1$, that is $W$ is an orthogonal transformation $W \in \mathrm{O}(p)$.
By its definition, the first component $w_{(1)} \in \R^p$ of $W$ is the direction which maximises the variance of the transformed data.
Let $y'_{(i)\, (1)} = y_{(i)} \cdot w_{(1)}$ for $i = 1, 2, \dots, n$ be the rotated components of the new data vectors, then the \pca computation translates into the maximisation problem:
\begin{equation}
    \begin{split}
        w_{(1)}
        & =
        \arg \max\limits_{\left|\left| w \right|\right| = 1}
        \left\lbrace \frac{1}{n - 1} \sum\limits_{i = 1}^n \left( y'_{(i)\, (1)} \right)^2 \right\rbrace
        \\
        & =
        \arg \max\limits_{w}
        \left\lbrace \frac{w^T C w}{w^T w} \right\rbrace,
    \end{split}
\end{equation}
where $C = \frac{1}{n - 1} Y^T Y \in \R^{p \times p}$ is the sample covariance matrix of the data.
The last line is known as \emph{Rayleigh quotient}.
The solution to the extermination problem is given if $w_{(1)}$ is the eigenvector of $C$ corresponding to the largest eigenvalue (see the \emph{Rayleigh-Ritz theorem}~\cite{horn_matrix_2017}).
Other components $w_{(k)}$ for $k > 1$ can be computed through the iterated application of this procedure after the \emph{deflation} of the original matrix:
\begin{equation}
    Y^{(k)} = Y - \sum\limits_{s = 1}^{k - 1} Y w_{(s)} w_{(s)}^T.
\end{equation}
The $k$-th \pc $w_{(k)} \in \R^p$ is given by the eigenvector associated to the largest eigenvalue of
\begin{equation}
    C^{(k)} = \frac{1}{n - 1} \left( Y^{(k)} \right)^T Y^{(k)}.
\end{equation}
In general, the \pcs are represented by the orthonormal basis of the eigenvectors of $C$, ordered by explained variance, that is the value of the corresponding eigenvalues.

In the main paper, we start from individual spectra $x_{(i)} \in \R^p$  for $i = 1,\, 2\, \dots, n$. The procedure is then as follows:
\begin{enumerate}
    \item the mean intensity $\barx = \frac{1}{n - 1} \sum_{i = 1}^n x_{(i)} \in \R^p$ is computed across all spectra, i.e.\ column-wise, and it is subtracted to the original data vectors.
    Vectors $y_{(i)} = x_{(i)} - \barx \in \R^p$ for $i = 1,\, 2,\, \dots,\, n$ are defined, and the corresponding centred matrix $Y = \qty( y_{(i)} )_{i \in [1, n]}$ is constructed;

    \item in order to estimate the directions of maximal variance, eigenvalues and eigenvectors of the sample covariance matrix $C = \frac{1}{n - 1} Y^T\,  Y \in \R^{p \times p}$ are then computed and collected into the diagonal matrix of the eigenvalues $\Lambda \in \R^{p \times p}$ and the eigenvectors matrix $W \in \R^{p \times p}$.
    Each column vector $w_{(i)}$ of the orthonormal matrix $W$ is the eigenvector associated to $i$-th eigenvalue;

    \item the data matrix is projected onto its loadings, $L = \Lambda^{\frac{1}{2}}~W^T \in \R^{p \times p}$, and scores, $S = Y~W~\Lambda^{-\frac{1}{2}} \in \R^{n \times p}$, such that $Y = S\, L$.
\end{enumerate}
The newly constructed representation contains the \pcs as columns of the scores, ordered by the fraction of total variance explained, and the projection coefficients as rows of the loadings.

\subsection{Wavelet Transforms}\label{app:wavelets}

In signal processing, a Fourier Transform (\textsc{ft}) is able to extract the frequency content of a signal, but it loses information related to the localisation of the frequency in the original domain.
A Short-Time \textsc{ft} (\textsc{stft}) can instead be used to partially reinstate the ability to localise the frequency components by scanning the signal with a window function $w$:
\begin{equation}
    \mathcal{F}_{\text{STFT}}\left[ f\, |\, w \right](\upomega, t)
    =
    \int\limits_{\R}
    \dd{\uptau}
    f(\uptau)
    w(t - \uptau)
    e^{- i \upomega \uptau}.
\end{equation}
The window considered is such to always consider intervals of the same length in the original domain: the number of sampled data points used to scan higher and lower frequencies is the same, thus losing the ability of a multi-level resolution.

Wavelets are square-integrable real function of real variables $\uppsi \in \mathrm{L}^2( \R )$ which have been introduced to adapt the width of the slice of the domain to the frequencies which are analysed (see~\cite{daubechies_ten_1992, williams_introduction_1994} for more details):
\begin{equation}
    \mathcal{F}_{\text{WT}}\left[ f\, |\, \uppsi \right](s, t)
    =
    \int\limits_{\R}
    \dd{\uptau}
    f(\uptau)
    \uppsi\left( \frac{t - \uptau}{s} \right),
\end{equation}
where $s$ is the scale factor to adjust the extension of the window, and $t$ represents the localisation parameter.
Given their ability to adjust the size of the scanned portion of the domain, wavelets can be used to construct multi-level representations of a signal $f \in \mathrm{L}^2(\R)$ at various levels of resolution.

A multi-level basis of wavelets is built by considering a sequence of vector spaces:
\begin{equation}
    \left\lbrace 0 \right\rbrace
    \subset
    \dots
    \subset
    V_{-1}
    \subset
    V_0
    \subset
    V_1
    \subset
    \dots
    \subset
    \mathrm{L}^2(\R),
\end{equation}
such that $\bigcup_{i \in \Z} V_i$ is dense in $\mathrm{L}^2(\R)$, and $\bigcap_{i \in \Z} V_i = \left\lbrace 0 \right\rbrace$.
Moreover, $g(x) \in V_i \Leftrightarrow g(2x) \in V_{i+1}$. Thus, $V_m = \mathrm{span}\overline{\left\lbrace g(2^m x - k), k\in \Z \right\rbrace}$, where the overline indicates the closure of the set.
An orthonormal basis of $V_m$ can be defined as the family of functions
\begin{equation}
    \left\lbrace
    \uppsi_{m,k}( x ) = 2^{\frac{m}{2}} \uppsi(2^m x - k)
    \right\rbrace_{k \in \Z},
\end{equation}
where $m$ is also referred as the \emph{scale} parameter. At any given value of $m$, a square integrable function $f$ can be approximated in the space $V_m$ by projection onto the newly defined basis:
\begin{equation}
    \mathcal{P}_m \left[ f \right]( x )
    =
    \sum\limits_{k = -\infty}^{\infty} a_{m,k}\, \uppsi_{m,k}( x ),
    \label{eq:proj_p}
\end{equation}
where $\left\lbrace a_{m,k} \right\rbrace_{k \in \Z}$ are the Approximation Coefficients (\acs) of $f$.

Let now $W_m$ be the orthogonal complement of $V_m$ such that $V_m = V_{m-1} \bigoplus W_{m-1}$, and let
\begin{equation}
    \left\lbrace \upphi_{m,k}( x ) = 2^{\frac{m}{2}} \upphi(2^m x - k) \right\rbrace_{k \in \Z}
\end{equation}
be a basis of $W_{m}$.
We then have the relation
\begin{equation}
    \mathcal{P}_m \left[ f \right]
    =
    \mathcal{P}_{m-1} \left[ f \right]
    +
    \mathcal{Q}_{m-1} \left[ f \right],
    \label{eq:complement}
\end{equation}
where $Q_m$ is the projection operator onto $W_m$:
\begin{equation}
    \mathcal{Q}_m \left[ f \right]( x )
    =
    \sum\limits_{k = -\infty}^{\infty} d_{m,k} \upphi_{m,k}( x ).
    \label{eq:proj_q}
\end{equation}
Coefficients $\left\lbrace d_{m,k} \right\rbrace_{k \in \Z}$ are the Detail Coefficients (\dcs) needed in order to move from one level of decomposition to the next.
Relation~\eqref{eq:complement} can be chained leading to:
\begin{equation}
    \eval{\mathcal{P}_m \left[ f \right]}_t
    =
    \mathcal{P}_{m - t} \left[ f \right]
    +
    \sum\limits_{n = 1}^k
    \mathcal{Q}_{m-t} \left[ f \right].
    \label{eq:chain}
\end{equation}
The equation shows the decomposition of the original signal at a given scale $m$ in terms of lower scales, indexed by $t$: the original decomposition at scale $m$ can be entirely recovered by considering the decomposition at level $m - t$ together with all the \dcs from scale $m - t$ to $m$.
In this sense, no information is lost.

A Discrete Wavelet Transform (\dwt) is a particular wavelet transform in which the signal $S$ and the basis functions are discretely sampled, and possibly finite.
At a given scale $m$, they effectively act as pass-band filters on the original signal with the \dcs, representing the projection~\eqref{eq:proj_p}, being constructed by a high-pass filter $h_{m,k}[n] = h_m[n - k]$ related to $\upphi_{m,k}$, and the \acs, representing~\eqref{eq:proj_q}, being the result of the convolution with a low-pass filter $\ell_{m,k}[n] = \ell_m[n - k]$ connected to $\uppsi_{m,k}$:
\begin{equation}
    \begin{split}
        d_{m,k}
        & =
        d_m[k]
        =
        \sum\limits_{q = -L}^{+L}
        S[q]\, h_{m,q}[k],
        \\
        a_{m,k}
        & =
        a_m[k]
        =
        \sum\limits_{q = -L}^{+L}
        S[q]\, \ell_{m,q}[k],
    \end{split}
\end{equation}
where $L$ may be theoretically very large.
The procedure can then be repeated on the \acs to further increase the frequency resolution as in~\eqref{eq:chain}.
As for each successive iteration of the filter bank the frequency domain is halved as well as the original domain, the effect of the \dwt is to subsample the signal at each scale $m$.
According to the length of the support of $S$ and the wavelet profile considered, the number of decomposition layers which can be computed changes.


\clearpage
\section{Determination of the Signal-to-noise Ratio in Synthetic Datasets}\label{app:snr}

As shown in~\Cref{sec:synthetic} of the main text, synthetic spectra are generated from the theoretical distribution in local thermodynamical equilibrium, with the addition of stochastic noise.
Wavelength channels which do not contain any signal, i.e.\ pure noise channels, are defined as:
\begin{equation}
    {x'}_{(i)}^{[n]} = \abs{\beta\, N},
    \qquad{i = 1, 2, \dots, n},
    \qquad{n = 1, 2, \dots, n_{\lambda}},
\end{equation}
where $\beta$ is uniformly distributed in the interval $\qty[0, \beta_{\text{max}}]$, and $N$ follows a standard normal distribution.
In order to determine the signal-to-noise (\snr), we need to compute the expected value and the population variance of ${x'}_{(i)}^{[n]}$ in the pure noise channels of the synthetic spectra.
This is possible through the computation of the moments of $\abs{\beta}$ and $\abs{N}$:
\begin{eqnarray}
    \Ev{\abs{\beta}}
    & = &
    \Ev{\beta}
    =
    \frac{1}{\beta_{\text{max}}}
    \int\limits_{0}^{\beta_{\text{max}}} \dd{t} t
    =
    \frac{\beta_{\text{max}}}{2},
    \\
    \mathrm{Var}\qty(\abs{\beta})
    & = &
    \mathrm{Var}\qty(\beta)
    =
    \frac{1}{\beta_{\text{max}}}
    \int\limits_{0}^{\beta_{\text{max}}} \dd{t} \qty(t - \Ev{\beta})^2
    =
    \frac{\beta_{\text{max}}^2}{12},
\end{eqnarray}
and
\begin{eqnarray}
    \Ev{\abs{N}}
    & = &
    \frac{1}{\sqrt{2 \uppi}}
    \int\limits_{-\infty}^{+\infty} \dd{t} \abs{t}\, e^{- \frac{t^2}{2}}
    =
    \sqrt{\frac{2}{\uppi}} \int\limits_{0}^{+\infty} \dd{t} t\, e^{- \frac{t^2}{2}}
    =
    \sqrt{\frac{2}{\uppi}} \int\limits_{0}^{+\infty} \dd{u} e^{- u}
    =
    \sqrt{\frac{2}{\uppi}},
    \\
    \mathrm{Var}\qty(\abs{N})
    & = &
    \frac{1}{\sqrt{2 \uppi}}
    \int\limits_{-\infty}^{+\infty} \dd{t} \qty(\abs{t} - \Ev{\abs{N}})^2\, e^{- \frac{t^2}{2}}
    =
    \mathrm{Var}\qty(N) - \Ev{\abs{N}}^2
    =
    1 - \frac{2}{\uppi}.
\end{eqnarray}

Finally, given the fact that the probability density functions of $\abs{\beta}$ and $\abs{N}$ are respectively independent, we can compute the moments of the distribution of $\abs{\beta\, N}$:
\begin{equation}
    \begin{split}
        \Ev{\abs{\beta\, N}}
        & =
        \Ev{\abs{\beta}}\, \Ev{\abs{N}}
        =
        \frac{\beta_{\text{max}}}{\sqrt{2 \uppi}},
        \\
        \mathrm{Var}\qty(\abs{\beta\, N})
        &=
        \Ev{\beta^2 N^2 - \Ev{\abs{\beta\, N}}^2}
        \\
        &=
        \Ev{\qty(\beta^2 - \Ev{\beta}^2) + \Ev{\beta}^2}
        \Ev{\qty(N^2 - \Ev{N}^2) + \Ev{N}^2}
        -
        \Ev{\abs{\beta\, N}}^2
        \\
        &=
        \qty(\mathrm{Var}\qty(\beta) + \Ev{\beta}^2) \mathrm{Var}\qty(N)
        -
        \frac{\beta_{\text{max}}^2}{2 \uppi}
        \\
        &=
        \beta_{\text{max}}^2 \qty(\frac{1}{3} - \frac{1}{2 \uppi}).
    \end{split}
\end{equation}

The \snr is defined as $\mathrm{Var}\qty(\abs{\beta\, N})^{-\frac{1}{2}}$, thus:
\begin{equation}
    \mathrm{\snr} = \frac{1}{\beta_{\text{max}}} \sqrt{\frac{6 \uppi}{2 \uppi - 3}} \simeq \frac{2.40}{\beta_{\text{max}}}.
\end{equation}


\clearpage
\section{Performance Review of the Proposed Algorithm}\label{app:ablation}

In the principal text, in~\Cref{sec:synthetic,sec:experimental}, we showed that \hyperpca performs better than standard \pca.
The application of simply one of the two techniques, either \dwt or Sparse Kernel-\pca (\kspca), to the data does not achieve the same results.
In this additional section, we show that the combined action of the \dwt and the application of \kspca delivers the best outcome: we study separately the contributions of the two techniques we introduce, and we show how they compare to the current methods and the proposed technique.
Namely, we show results using:
\begin{enumerate}
    \item standard \pca applied to the \dwt of the original data,

    \item \kspca applied directly to the original data,

    \item \kspca and \pca applied to the Continuous Wavelet Transform (\cwt) of the spectra.
\end{enumerate}
As the comparison with standard \pca applied to the data was already provided, showing these results enables a better comparison with alternative treatments.

\begin{figure}[t]
    \centering

    \subfigure[\pca on the \dwt of the ``granite'' dataset.]{\includegraphics[width=0.9\linewidth]{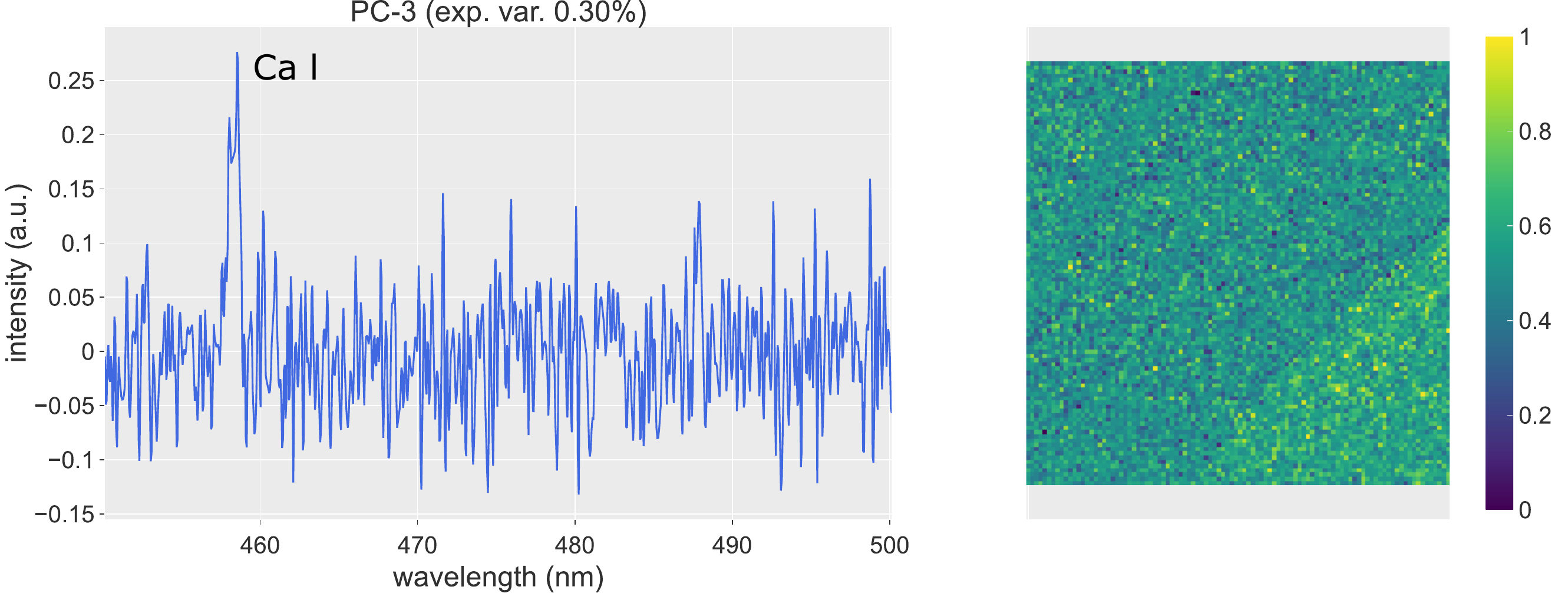}}
    \\
    \subfigure[\kspca on the ``granite'' dataset (Ca emission line).]{\includegraphics[width=0.9\linewidth]{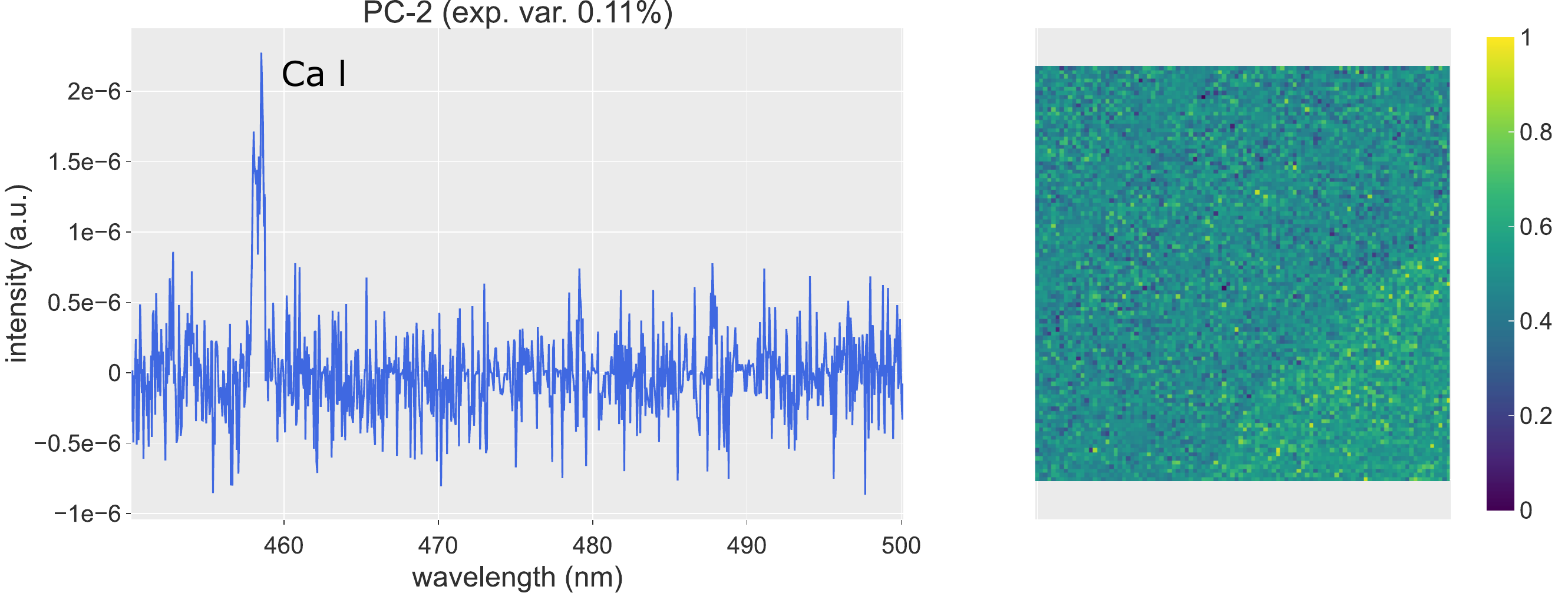}}

    \caption{Different analyses of the ``granite'' dataset.}
    \label{fig:ablation_pca}
\end{figure}

The expected natural outcome of the application of \pca on the \dwt of the original data is the retention of a high degree of noise which might prevent the complete separation of the elements in the dataset: the \dwt indeed acts as an initial, weak denoising filter, but the lack of the kernel suppressing small entries in the covariance matrix retains most of the background noise.
At the top of~\Cref{fig:ablation_pca} we show the \pc showing Ca in the ``granite'' dataset of~\Cref{sec:granite} in the main text.
Noise is indeed evidently present in the loading vector, and it prevents the complete separation of Ca from the background distribution.
This is a feature which is visible also in other datasets, as it is a general trait of the underlying physical nature of the data.
The \dwt of the original data is not enough to lead to an improvement of the standard \pca.

\begin{figure}[t]
    \centering

    \includegraphics[width=0.9\linewidth]{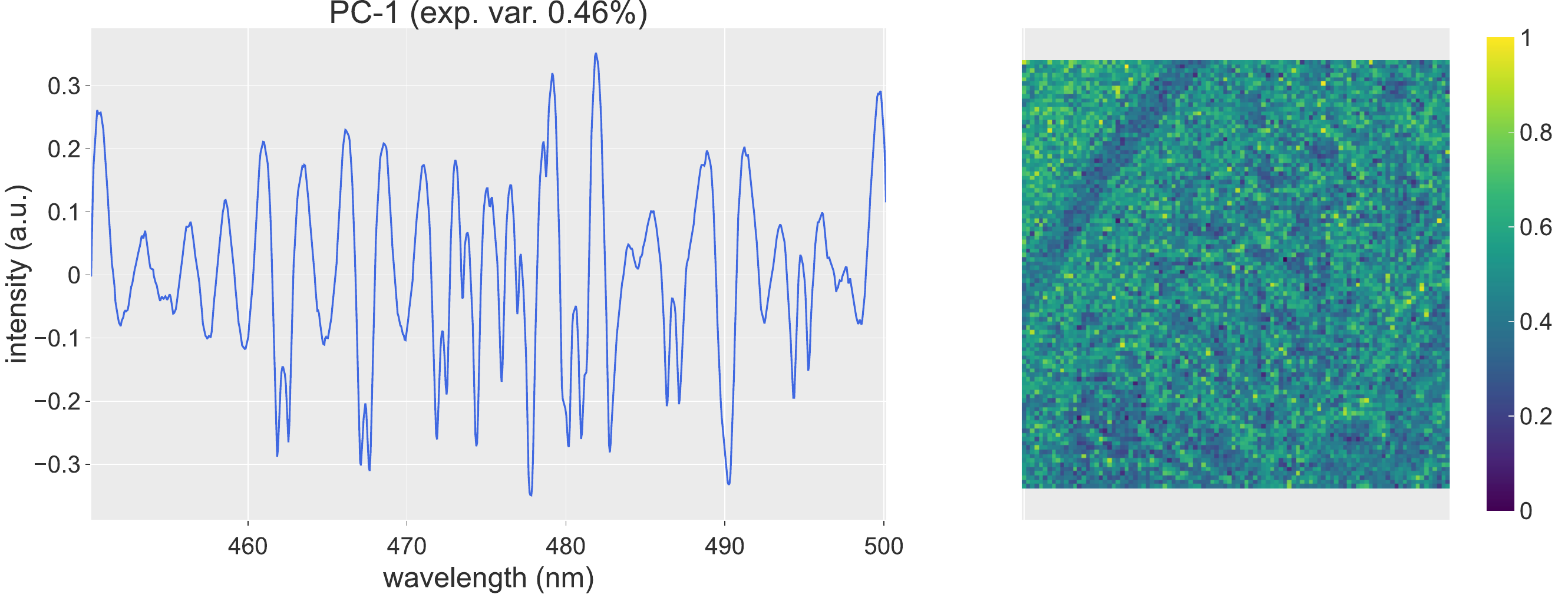}
    \caption{\pca on the \cwt of the ``granite'' data.}

    \label{fig:ablation_cwt_pca}
\end{figure}

We then consider \kspca applied directly to the data, without any kind of sparsity filter.
At the bottom of~\Cref{fig:ablation_pca}, we show the fifth \pc computed on the ``granite'' dataset.
By comparison with~\Cref{fig:granite_ground_truth} in the main text, the \pc shows the Al contribution correctly, but the quality of the score matrix is worsened by the remaining noise component.
The detection of the element is possible, as the application of the kernel lowers the detection threshold of signal.
The lack of a sparse basis to appropriately express the signal, does not allow the \kspca to reach the same quality as \hyperpca.

The comparison between scores and loadings in~\Cref{fig:ablation_pca} shows that the action of the \dwt is mostly aimed at modifying the line profile and to highlight the presence of physical signal with respect to the noise.
The modification of the eigenvalue distribution puts in evidence the physical line by acting directly on the signal-to-noise ratio, which in the case of \kspca is almost doubled with respect to the the application of \dwt only, which nonetheless has a noise filtering effect, too.

\begin{figure}[t]
    \centering

    \includegraphics[width=0.9\linewidth]{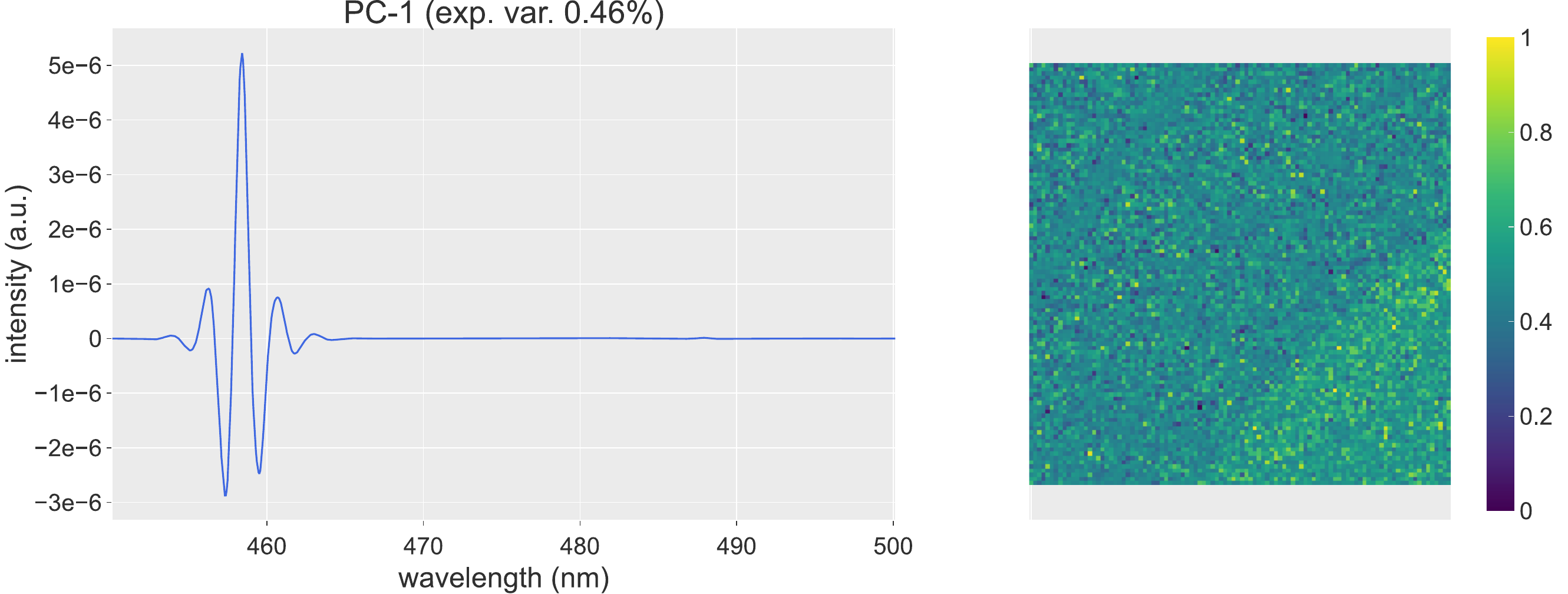}
    \\
    \includegraphics[width=0.9\linewidth]{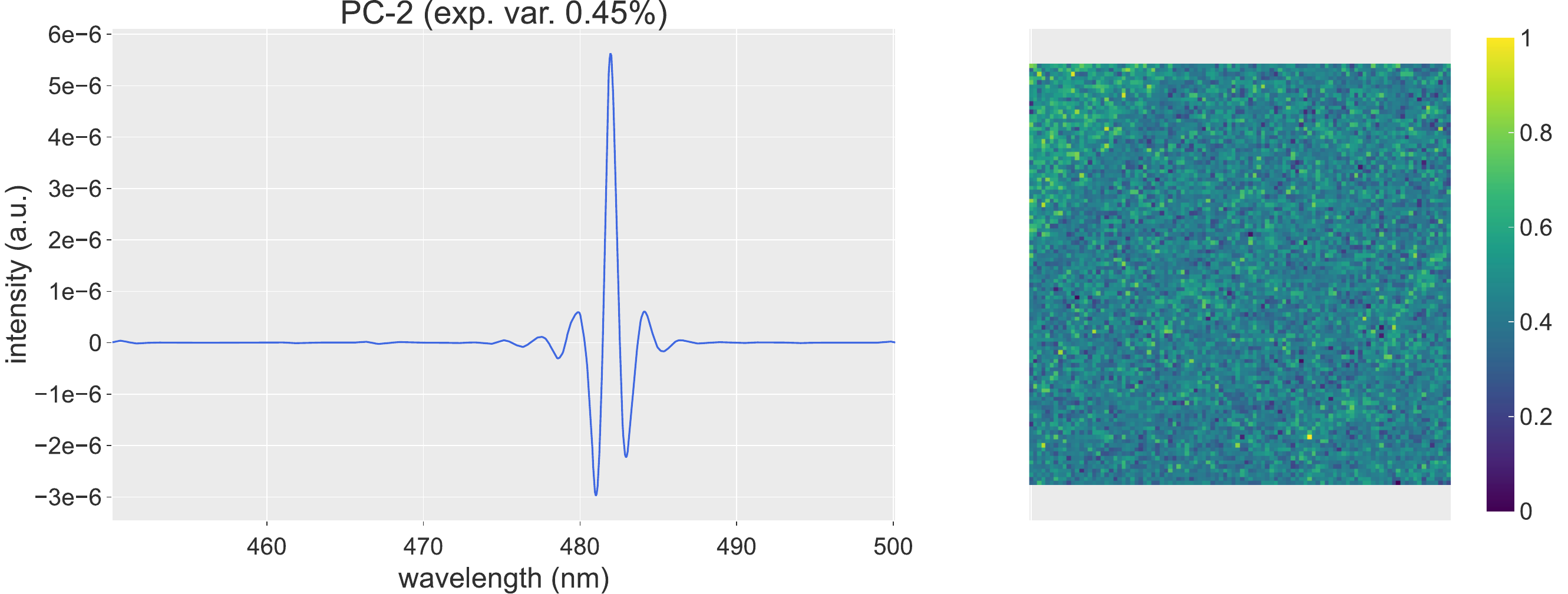}
    \\
    \includegraphics[width=0.9\linewidth]{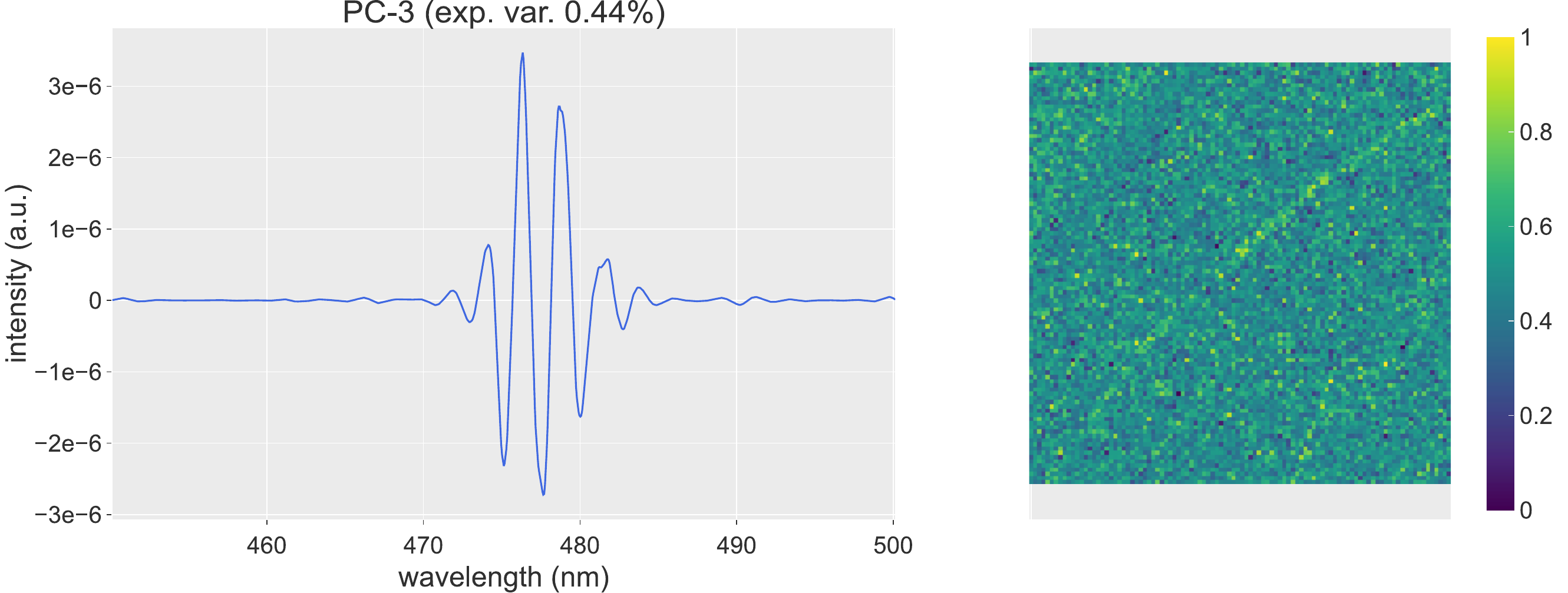}
    \\

    \caption{\kspca on the \cwt of the ``granite'' data.}
    \label{fig:ablation_cwt_kpca}
\end{figure}

The application of \pca or \kspca over the \cwt of the data opens a different kind of discussion.
In this section, we show the results obtained using the best fitting wavelet profile, a Mexican hat continuous wavelet, on the ``granite'' data.
As a general trait, we notice two shortcomings to this approach:
\begin{enumerate}
    \item the creation of strong artefacts, not present in the original dataset,

    \item the strong modification of the profiles of the emission lines.
\end{enumerate}
In the presence of strong spectral interference, imposing a continuous wavelet profile, negatively impacts neighbouring emission lines, creating strong artefacts, which completely spoil the readability of the loading vectors, as standard \pca on the \cwt shows in~\Cref{fig:ablation_cwt_pca}.
Imposing the profiles of the continuous wavelet has greatly modified the shape, which becomes rather difficult to interpret as emission spectra.
Notice, for instance the characteristic ``wells'' forming at the base of the lines.
Even the coupling of \cwt with the \kspca approach, shown in~\Cref{fig:ablation_cwt_kpca}, suffers from the same drawbacks: in this case, though the score maps present good spatial information, the extracted loading vectors are difficult to read and present too many artefacts to be considered physical.

As shown, the various components of \hyperpca play a complementary role in the outcome.
The \dwts are able to provide a sparse representation of the input and, at the same time, reduce part of the high frequency noise, as byproduct of the procedure, while retaining the physical information present in the datasets.
At the same time, the kernel-based sparse \pca method helps to distinguish noise components from genuine signal, when the latter is sparse in a certain basis.
Only the combination of both aspects in \hyperpca has the ability to automatically distinguish the sources of different signals and to disentangle the background noise, thus behaving as an unsupervised Artificial Intelligence, rather than a simple geometrical transformation of the input.


\clearpage
\section{Small Elemental Map}\label{app:316L}

In this section we highlight a specific effect of \hyperpca addressing the high dimensionality of \libs mapping datasets.
We show that the use of the kernel function is able to solve the theoretical phase transition issue in cases where the number of variables (wavelength channels) is greater than the number of available samples (spectra).
In this case, it can be shown that \pca does not correctly reconstruct the directions of maximal variance~\cite{baik_phase_2005}, which ultimately leads to meaningless \pcs.

\begin{figure}[t]
    \centering
    \subfigure[Fe]{\includegraphics[width=0.475\linewidth]{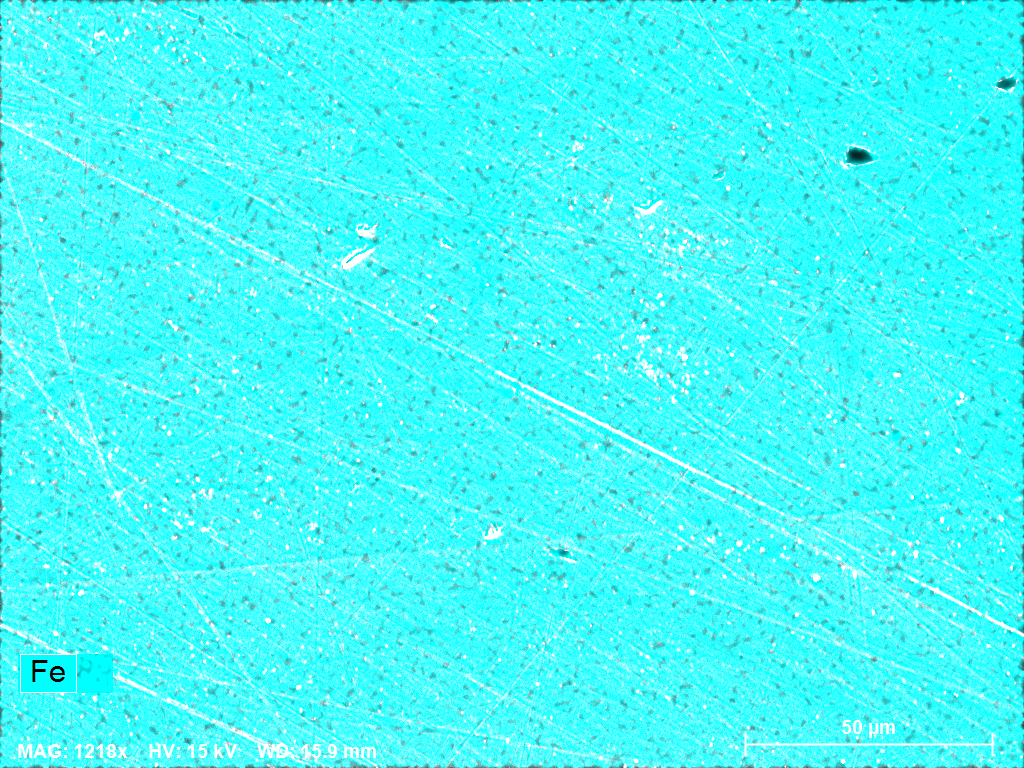}}
    \hfill
    \subfigure[Cr]{\includegraphics[width=0.475\linewidth]{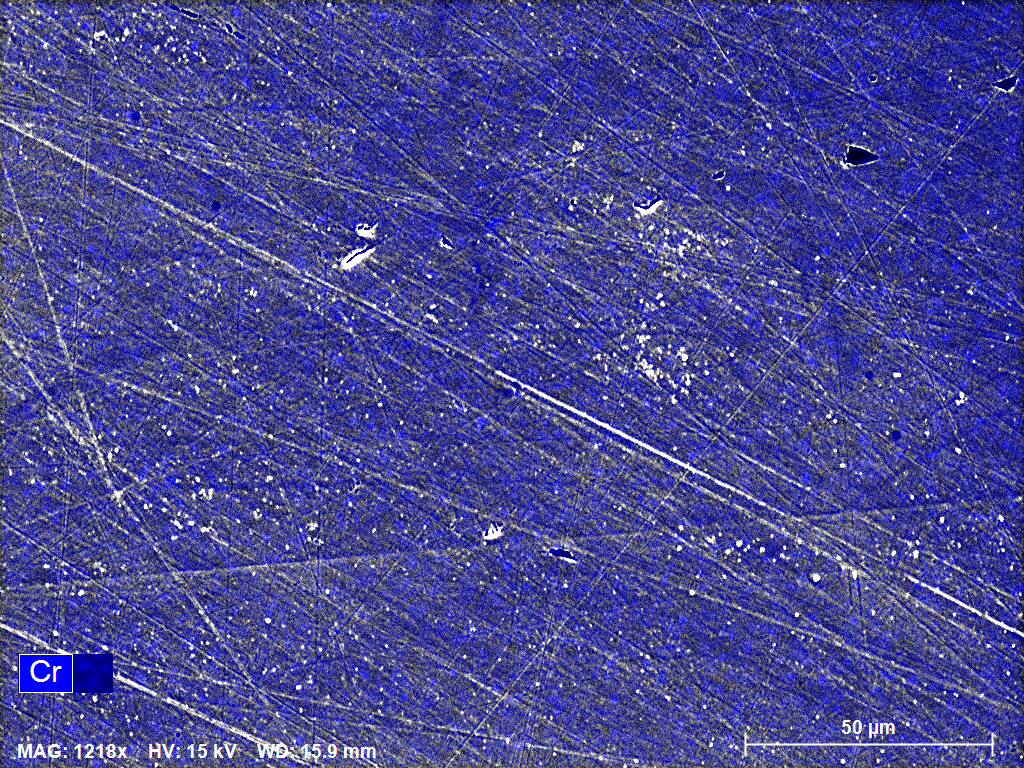}}
    \\

    \caption{Distributions of the main elements in the 316L stainless steel.}
    \label{fig:316L_ground_truth}
\end{figure}

\begin{table}[t]
    \centering
    \begin{tabular}{@{}cc@{}}
        \toprule
        \textbf{} & \textbf{concentration} \\ \midrule
        Fe        & 0.68                   \\
        Cr        & 0.17                   \\
        Ni        & 0.10                   \\
        Mo        & 0.02                   \\
        C         & 0.02                   \\
        Si        & 0.01                   \\ \bottomrule
    \end{tabular}
    \caption{Concentration of elements in the stainless steel.}
    \label{tab:316L_conc}
\end{table}

\begin{figure}[t]
    \centering

    \includegraphics[width=0.5\linewidth]{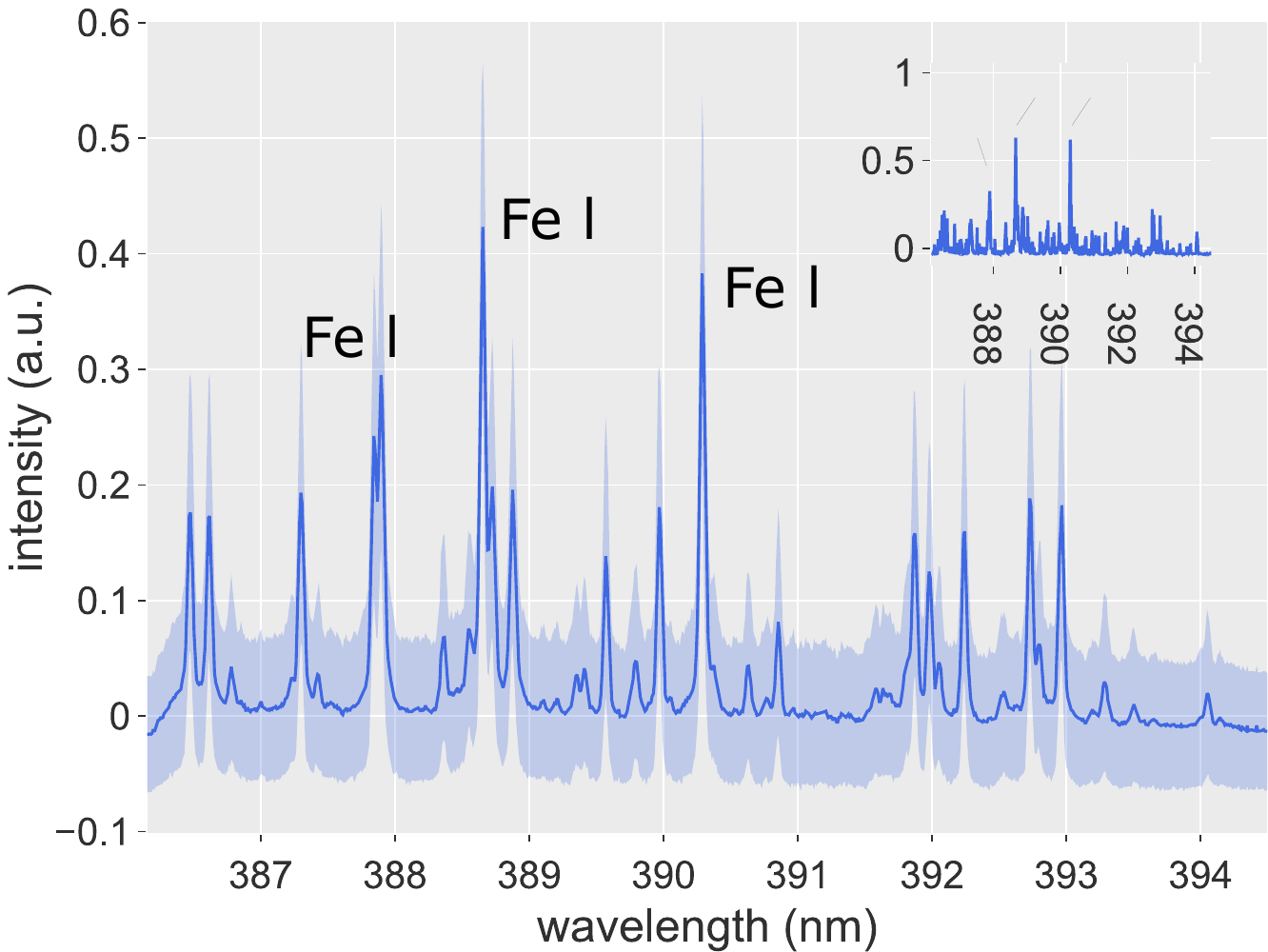}

    \caption{Preprocessed average and $1\sigma$ spectra with an example of single-shot spectrum of the 316L stainless steel sample.}
    \label{fig:316L_average}
\end{figure}

The proposed dataset is a small \libs map taken on a 316L stainless steel sample.
In~\Cref{fig:316L_ground_truth} we show the reference elemental distributions as $\SI{200}{\micro\meter} \times \SI{150}{\micro\meter}$ images, taken at the \sem in a different location with respect to the \libs map location, for the main elements.
The average spectrum recorded during the \libs mapping experiment is shown in~\Cref{fig:316L_average}.
The maps are mostly homogeneous on very small scales and different positions, and thus it is fair to assume the same composition all over the surface.
\Cref{tab:316L_conc} shows the relative concentration of the elements in the sample, estimated by \eds.

As the \libs analysis of the sample is unable to provide the same resolution to detect possible sub-structures in the sample, we focus on a different aspect of the data analysis.
We also do not provide the line intensities charts as they are not relevant to \libs mapping in this case.
We use a \SI{3}{\micro\joule} laser pulse on a $25 \times 25$ matrix on the surface, with a resolution of \SI{10}{\micro\meter} in both directions.
Craters are \SI{3}{\micro\meter} both in diameter and depth on average.
We use a grating with \SI{2400}{grooves \per\milli\meter}, centred at \SI{391.914}{\nano\meter}.
The goal of the analysis is to show that \hyperpca can extract meaningful information even below the theoretical threshold for extraction associated to \pca: in this case, the data matrix $X \in \R^{p \times n_{\lambda}}$ is such that $n_{\lambda} > p$, that is the number of columns in the dataset exceeds the number of rows.
By \rmt arguments, it is possible to show that the \pca reconstruction of the \pcs is not reliable, and the distance between the true and extracted eigenvectors increases, when the signal-to-noise ratio $\beta^{-1} \lesssim \sqrt{q} = \sqrt{\frac{n_{\lambda}}{p}}$~\cite{paul_asymptotics_2007, baik_phase_2005}.
As in this case $q > 1$, we may expect issues in the reconstruction of the loading vectors.

\begin{figure}[t]
    \centering
    \begin{tabular}{c|c}
        {\LARGE \textsc{standard} \pca}                          & {\LARGE \hyperpca}
        \\[1em]
        \includegraphics[width=0.47\linewidth]{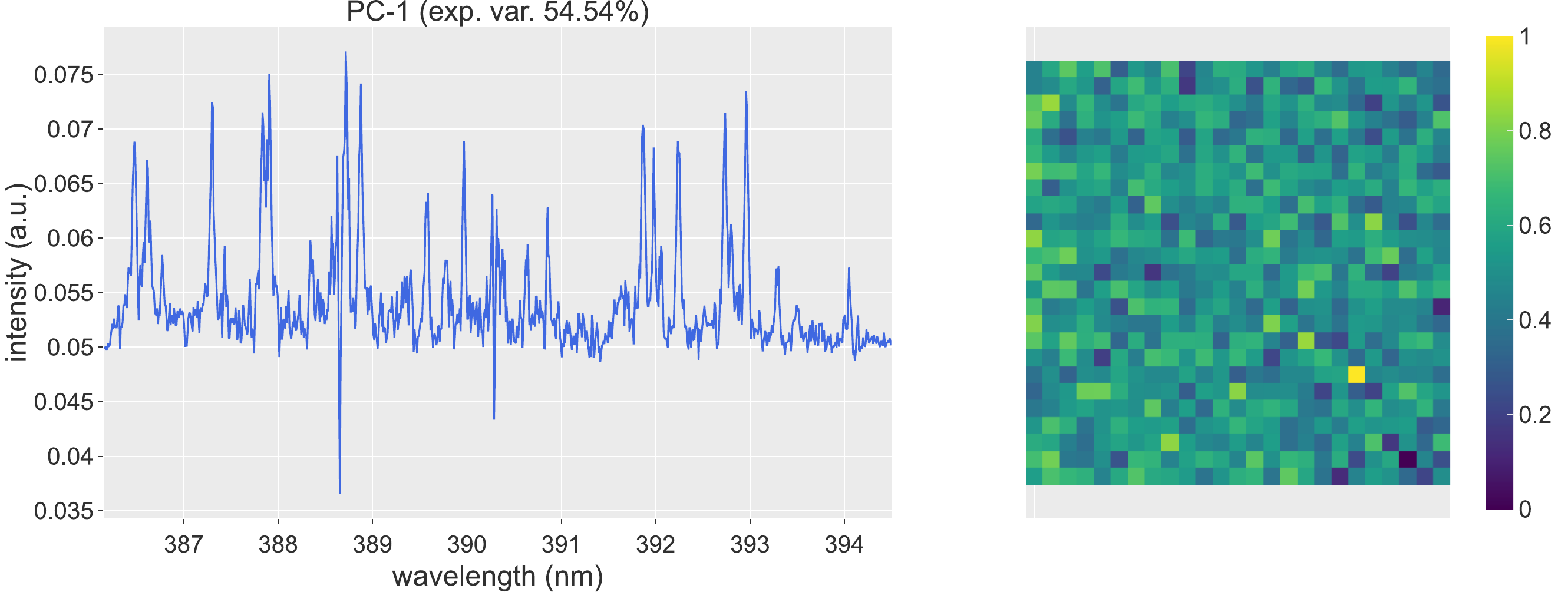} & \includegraphics[width=0.47\linewidth]{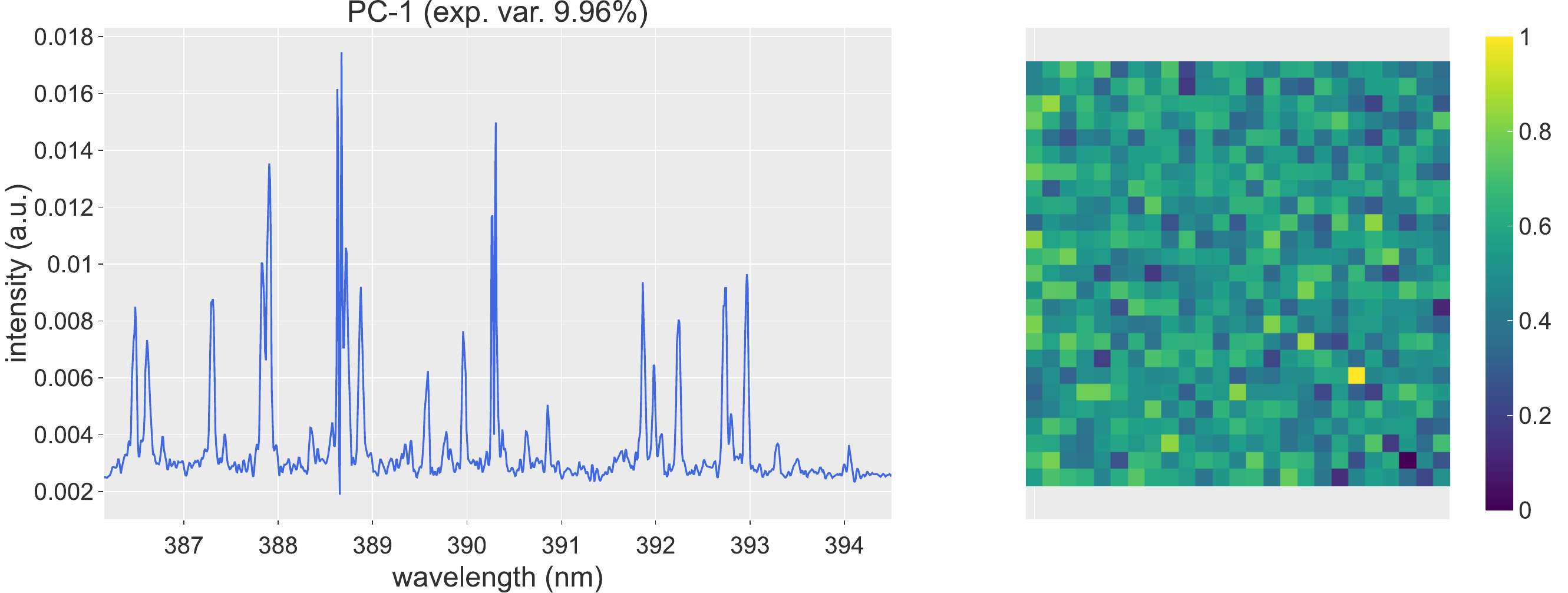}
        \\
        \includegraphics[width=0.47\linewidth]{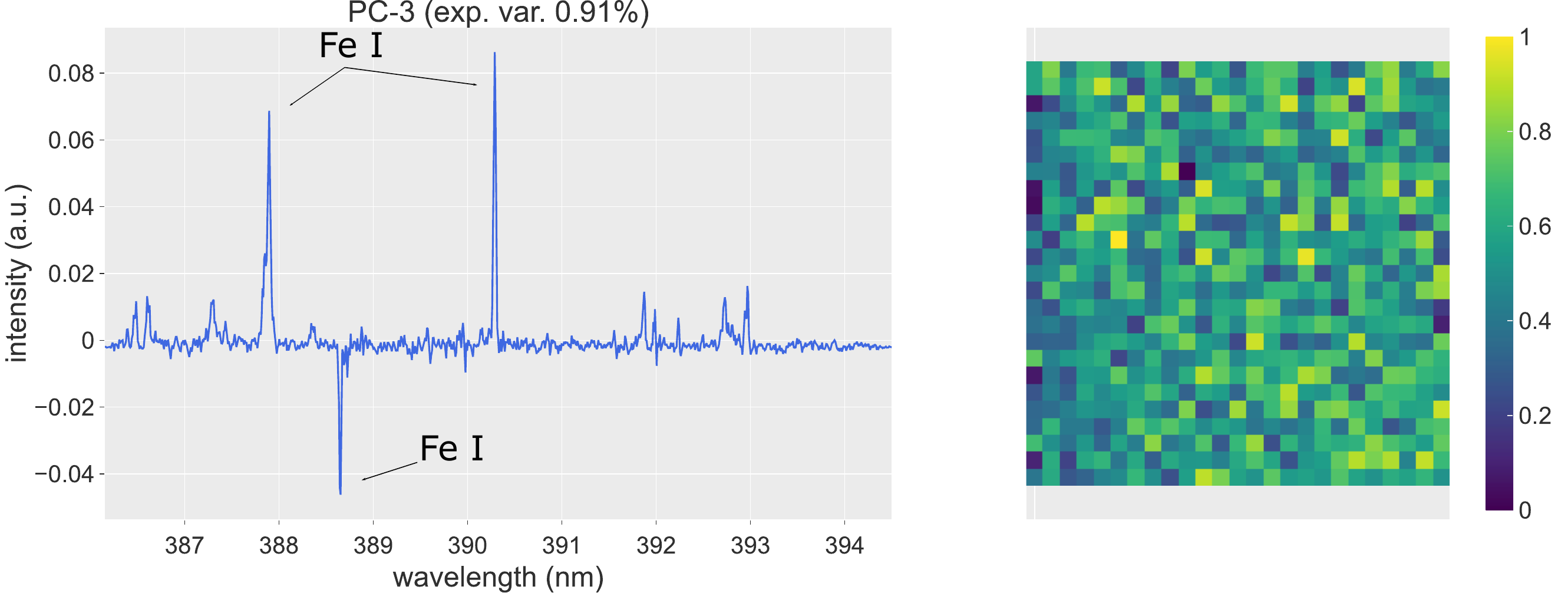} & \includegraphics[width=0.47\linewidth]{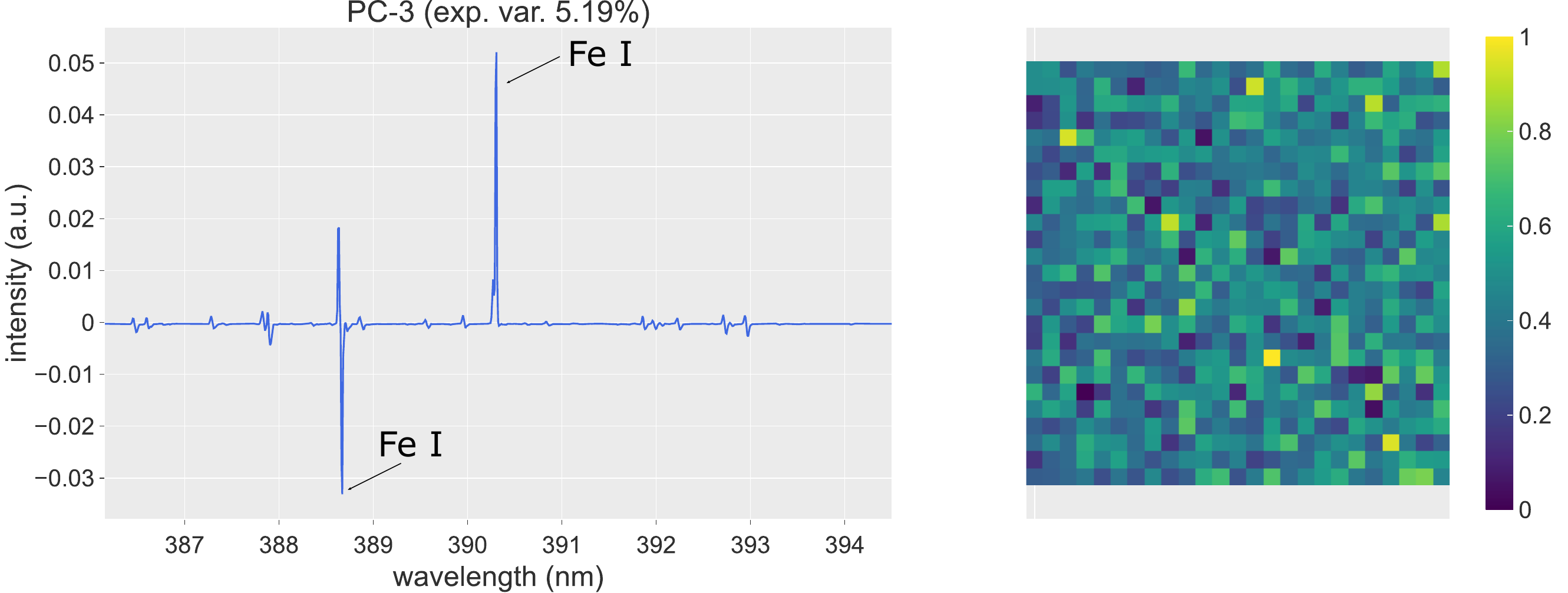}
        \\
        \includegraphics[width=0.47\linewidth]{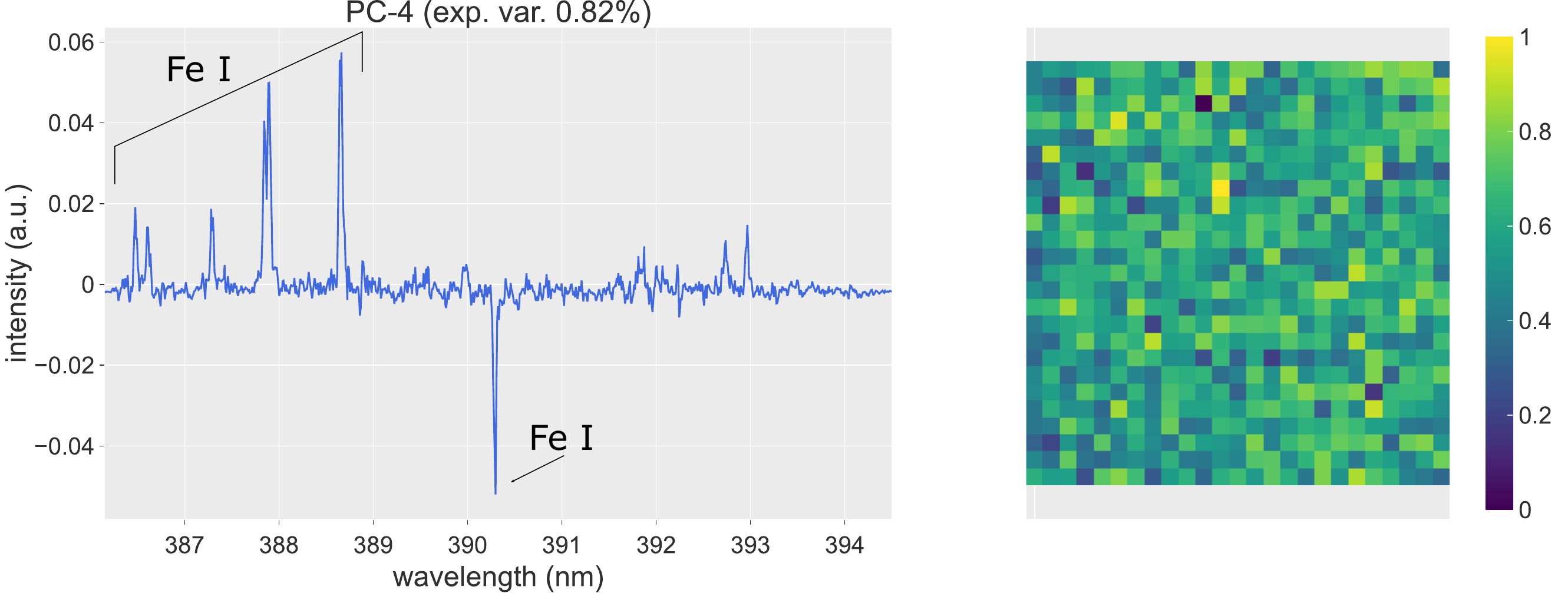} & \includegraphics[width=0.47\linewidth]{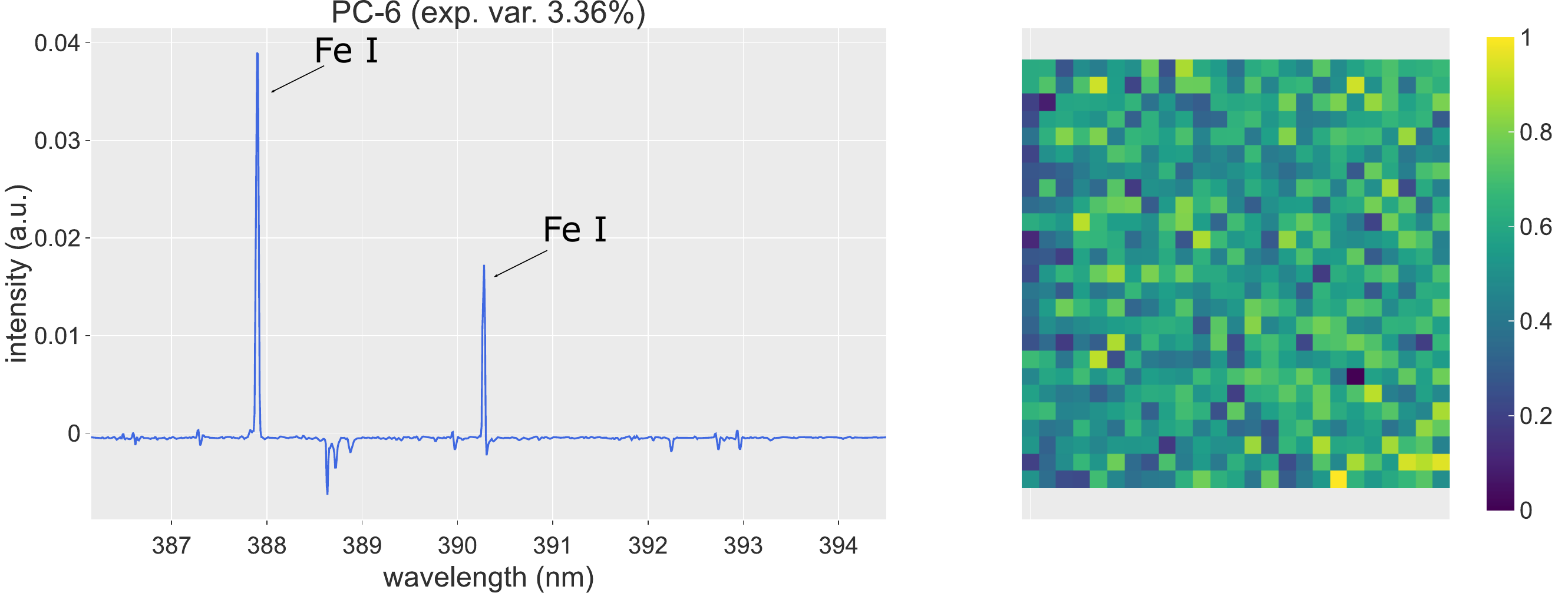}
        \\
        \includegraphics[width=0.47\linewidth]{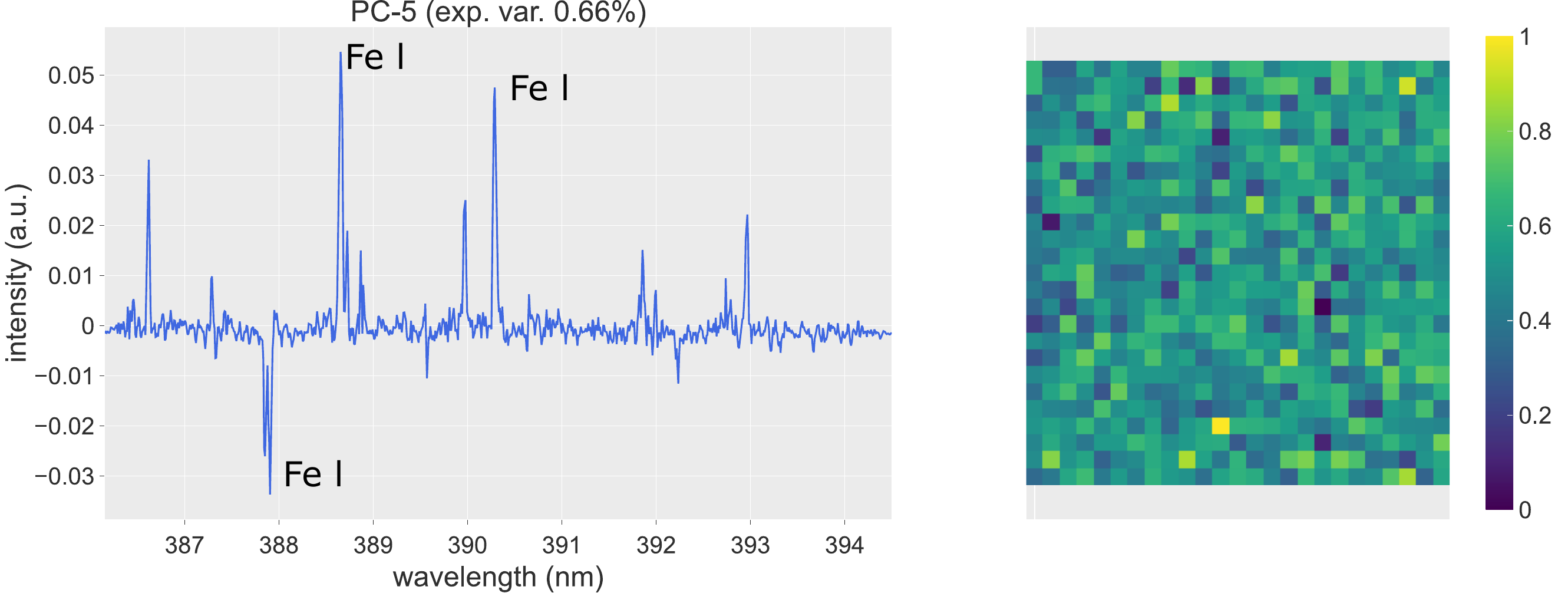} & \includegraphics[width=0.47\linewidth]{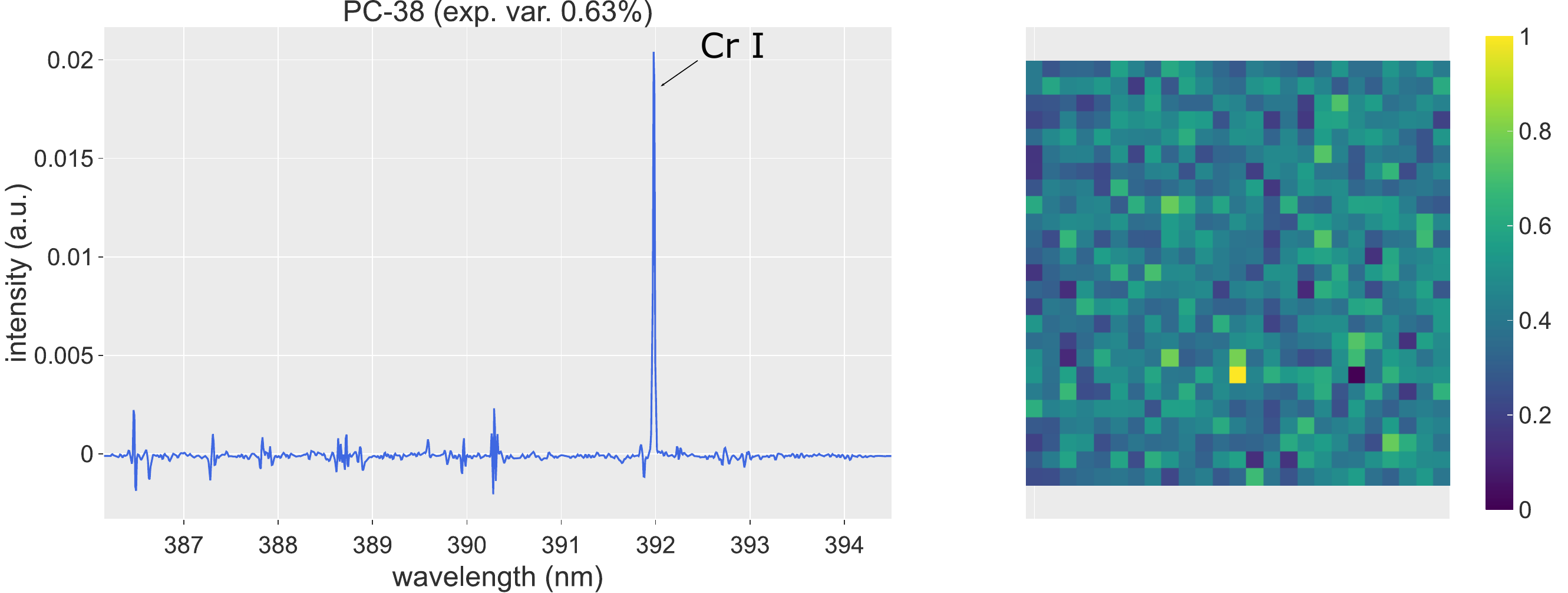}
    \end{tabular}
    \caption{Reconstruction of the 316L stainless steel sample using the standard \pca and \hyperpca.}
    \label{fig:316L_pca}
\end{figure}

On the right of~\Cref{fig:316L_pca} we show the most readable \pcs extracted using the standard \pca.
The loadings are all affected by a higher noise rate, even though the average spectrum in~\Cref{fig:316L_average} shows a strong signal presence: \pcn{1} is mostly unusable for element recognition tasks, while the line profiles in \pcn{3} and \pcn{4} present some positively reconstructed lines together with poorly resolved profiles.
Although most of the lines in this wavelength range correspond to either Fe or Cr, some of them are almost indistinguishable and hardly resolved.
As anticipated, the score matrices do not provide any information on the map, due to the homogeneity of the sample with \libs mapping resolution.

In the second column of~\Cref{fig:316L_pca} we show \pcs extracted with \hyperpca as a proof of concept of the proposed pipeline.
We use a kernel parameter $\upalpha = \num{5.0e2}$ and the first level of the \texttt{bior3.5} wavelet filter banks.
We impose a HT of \num{0.99} times the largest value of the power spectrum of the \dcs.
Even from the first ranked \pc, the impact of the noise components on the resolution of the lines is reduced, thus increasing the quality of the loading vector associated to the largest eigenvalue of the covariance matrix: the profiles of the lines are generally better resolved than in standard \pca.
Moreover, the absence of noise components in the first \pcs is such to enable the resolution of principal and subdominant emission lines which may create spectral interference, such as the Fe lines in \pcn{3} at \SI{388.63}{\nano\meter} and \SI{390.65}{\nano\meter}, and \pcn{6} at \SI{387.86}{\nano\meter} and \SI{390.65}{\nano\meter}.
The readability of the \pcs is ensured also at low rank, such as \pcn{38} in the figure, where Cr at \SI{391.92}{\nano\meter} is resolved free of Fe interference, thus also restoring some predictive power to the score map.
Even though the quality of the score matrix is reduced by its small size, as the \pc isolates Cr, the map can give an indication of the distribution of the element nonetheless.
The ratios of the explained variance are also better estimated, such that a better assignment of the reconstruction confidence can be assigned to each \pc.

\hyperpca can thus deal also with extreme cases in which the experimental framework may impose strong restrictions, such as small maps, or noisy data, usually defined in \rmt as a spiked model where the separation between the largest eigenvalue is not strongly distinct from the bulk.
Such situations are usually connected to, but not limited to, cases in which the transition threshold $n_{\lambda} = p$ is crossed, in which case the noise gains a larger importance with respect to signal.
The use of the kernel approach ensures the ability to go beyond the theoretical bounds of standard \pca, thus restoring the readability of the \pcs.
In addition, beyond the case of \libs mapping, it is interesting to note that the situation where the number of channels exceeds the number of spectra ($n_{\lambda} > p$) is very frequent in conventional \libs, where \pca is often applied to explore small datasets. As shown in this section, standard \pca is therefore limited if the signal-to-noise ratio is low. \hyperpca provides a relevant alternative to extract more meaningful information.


\clearpage
\section{Complementary Figures}\label{app:figures}

\begin{figure}[h]
    \centering

    \includegraphics[width=0.5\linewidth]{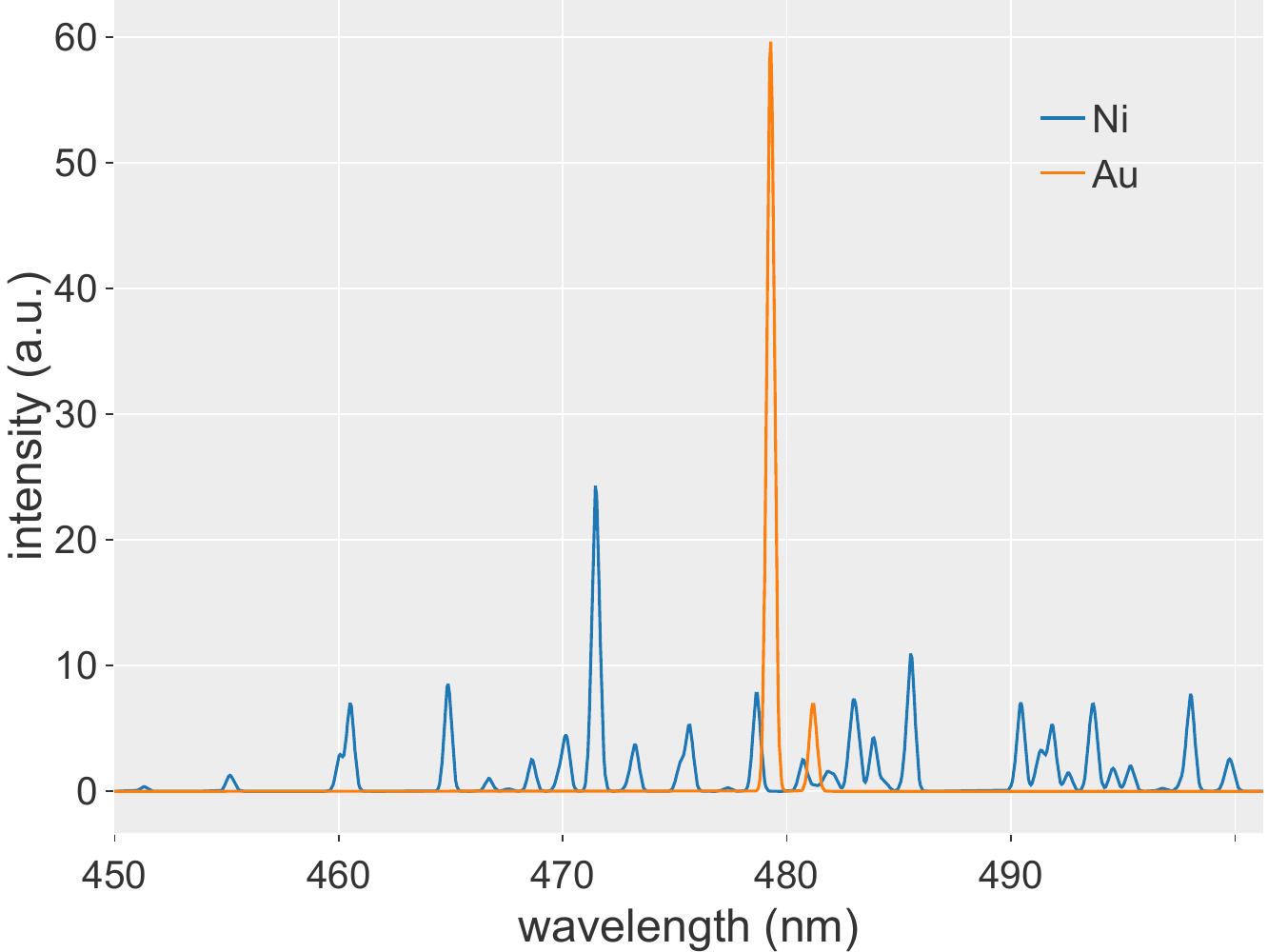}

    \caption{NiAu spectra in LTE.}
    \label{fig:niau_spectra_lte}
\end{figure}

\begin{figure}[h]
    \centering

    \includegraphics[width=0.9\linewidth]{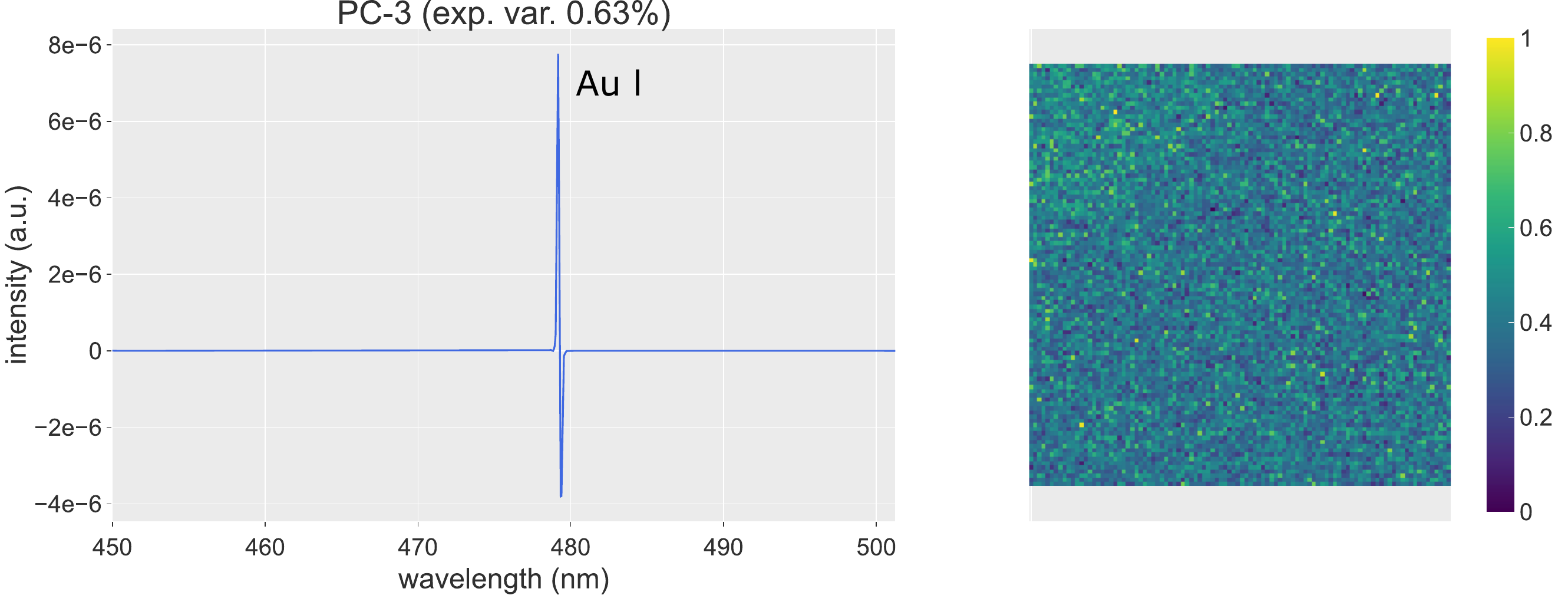}
    \\

    \caption{Due to experimental noise, the most intense photon emission can be captured by the CCD sensor in different neighbouring pixels, thus leading to a slightly different shape of the line profile. This is sometimes extracted as a \emph{first derivative} signal by \pca or, in this case, by \hyperpca.}
    \label{fig:niau_dwt-kpca_derivative}
\end{figure}

\begin{figure}[h]
    \centering

    \includegraphics[width=0.5\linewidth]{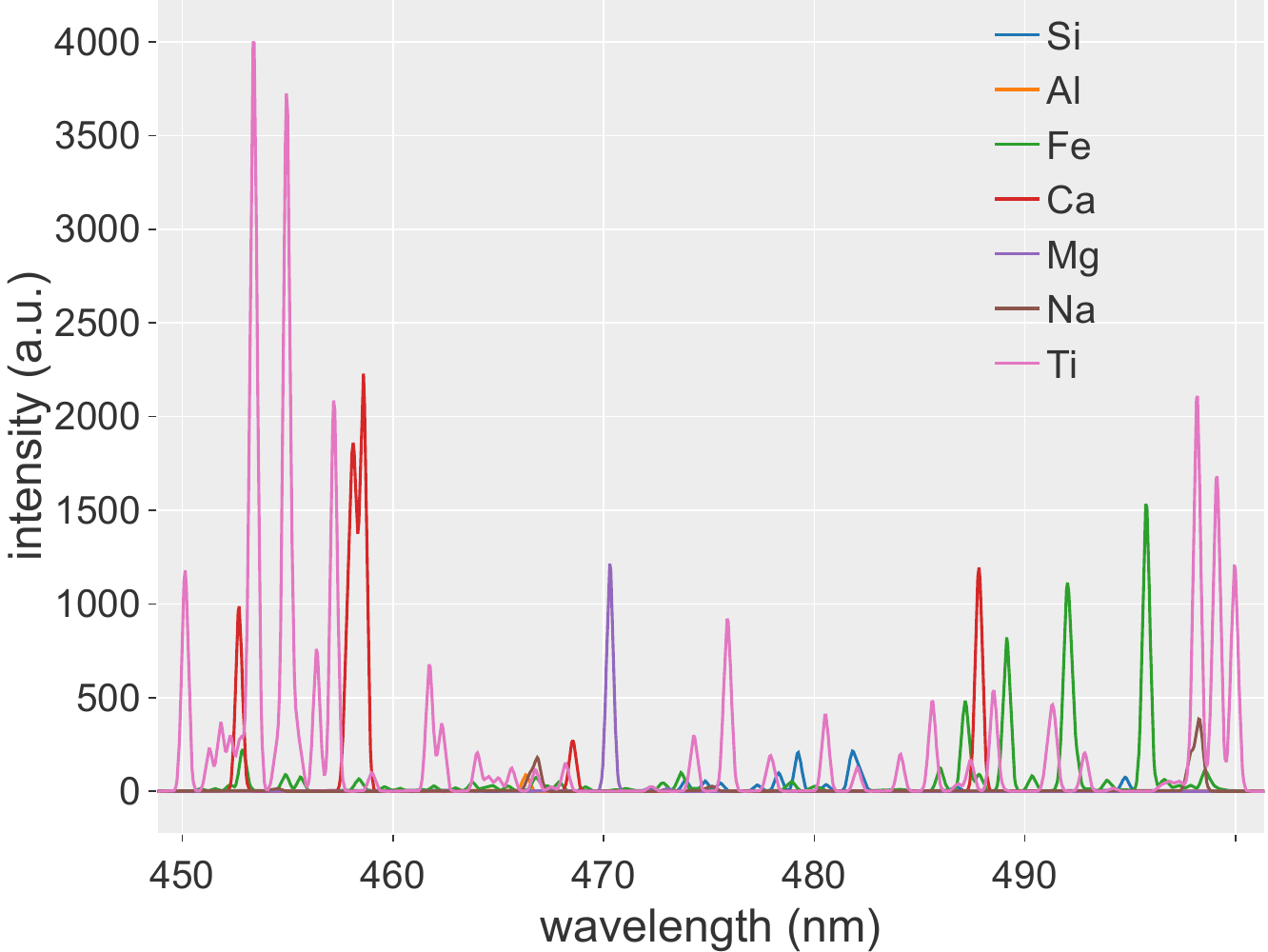}

    \caption{``Basalt'' spectra in LTE.}
    \label{fig:basalt_spectra_lte}
\end{figure}

\begin{figure}[h]
    \centering

    \includegraphics[width=0.85\linewidth]{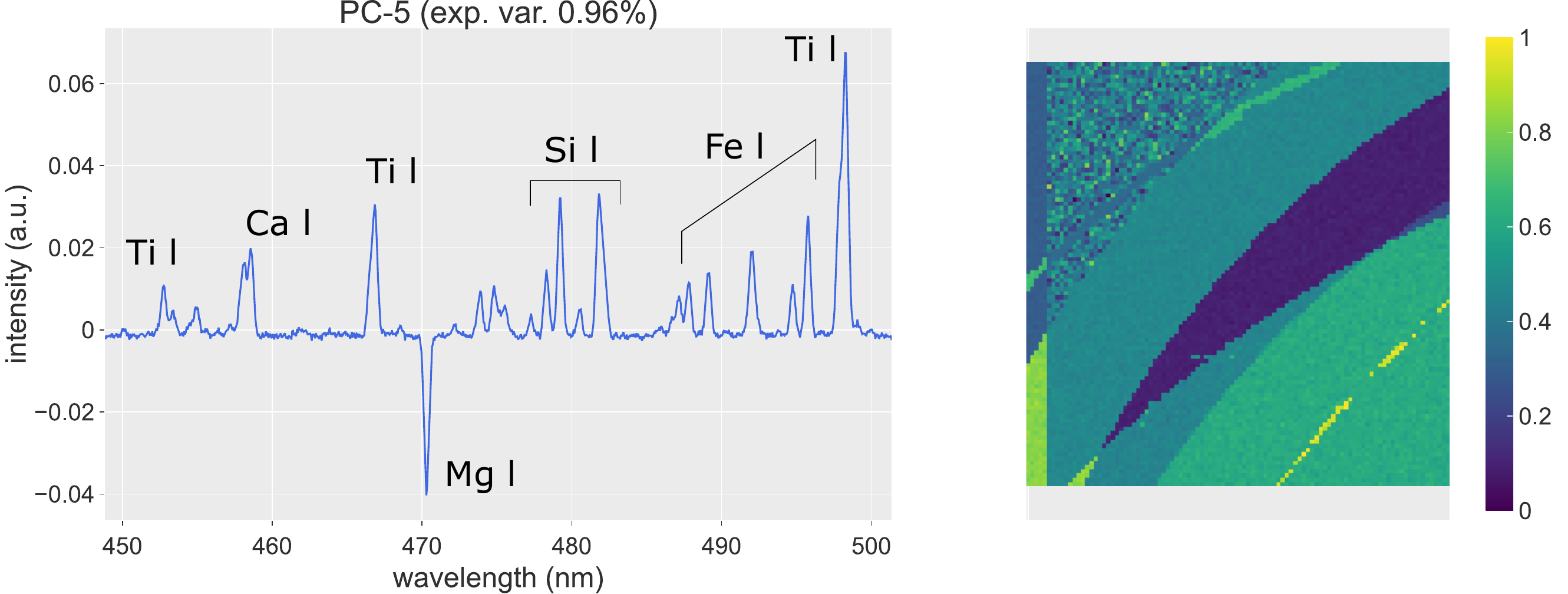}
    \\
    \includegraphics[width=0.85\linewidth]{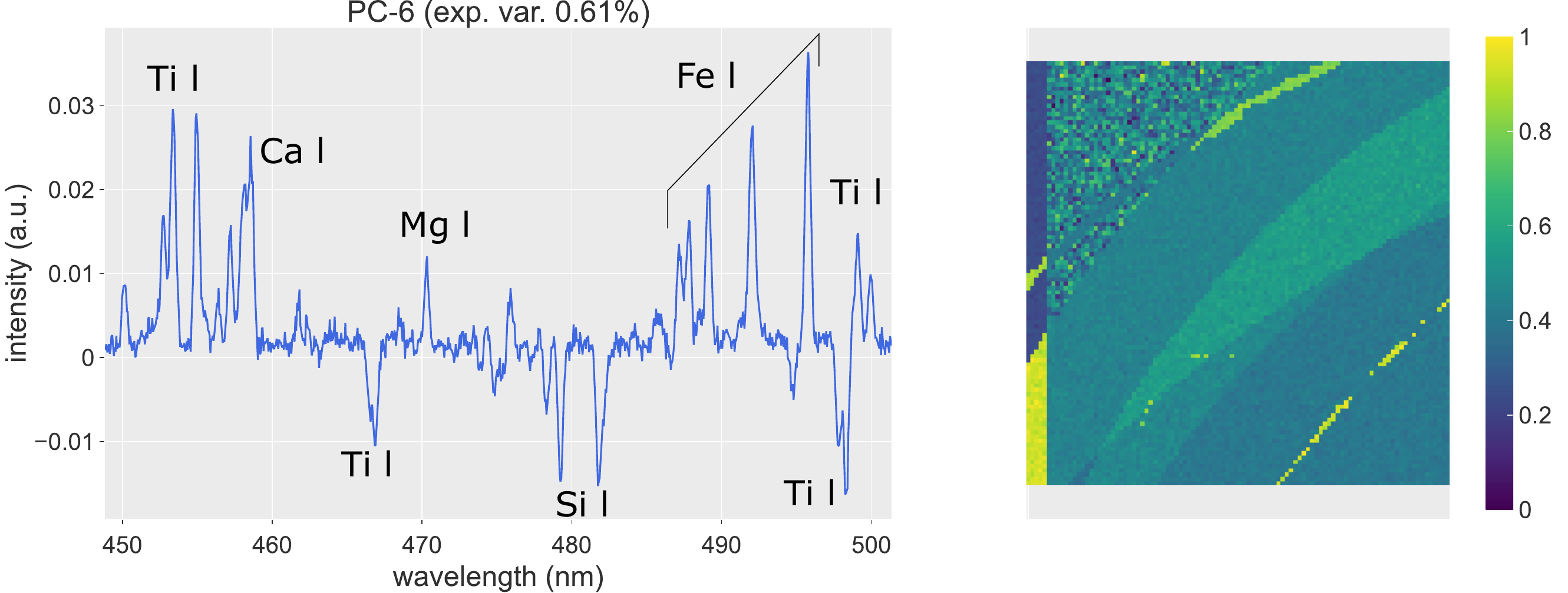}
    \\
    \includegraphics[width=0.85\linewidth]{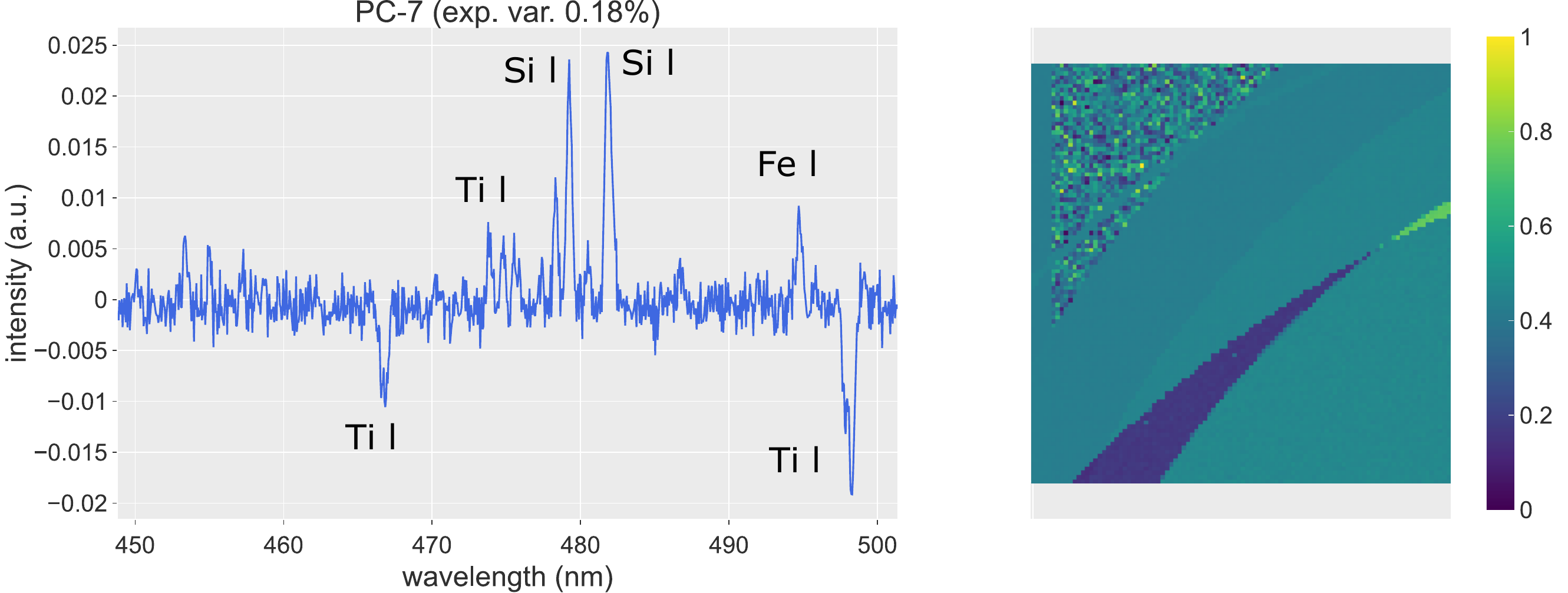}
    \\
    \includegraphics[width=0.85\linewidth]{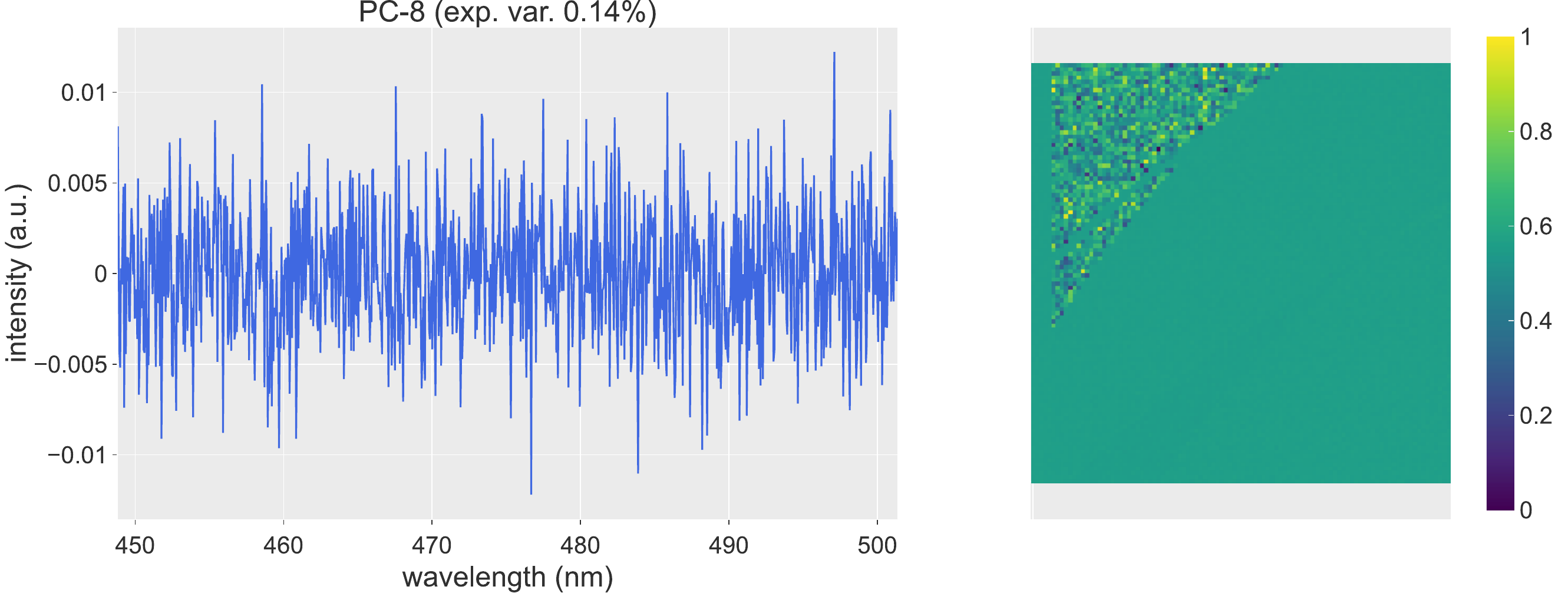}
    \\

    \caption{Complementary \pcs of the ``basalt'' specimen using the standard \pca.}
    \label{fig:basalt_pca_less_imp}
\end{figure}

\begin{figure}[h]
    \centering

    \includegraphics[width=0.9\linewidth]{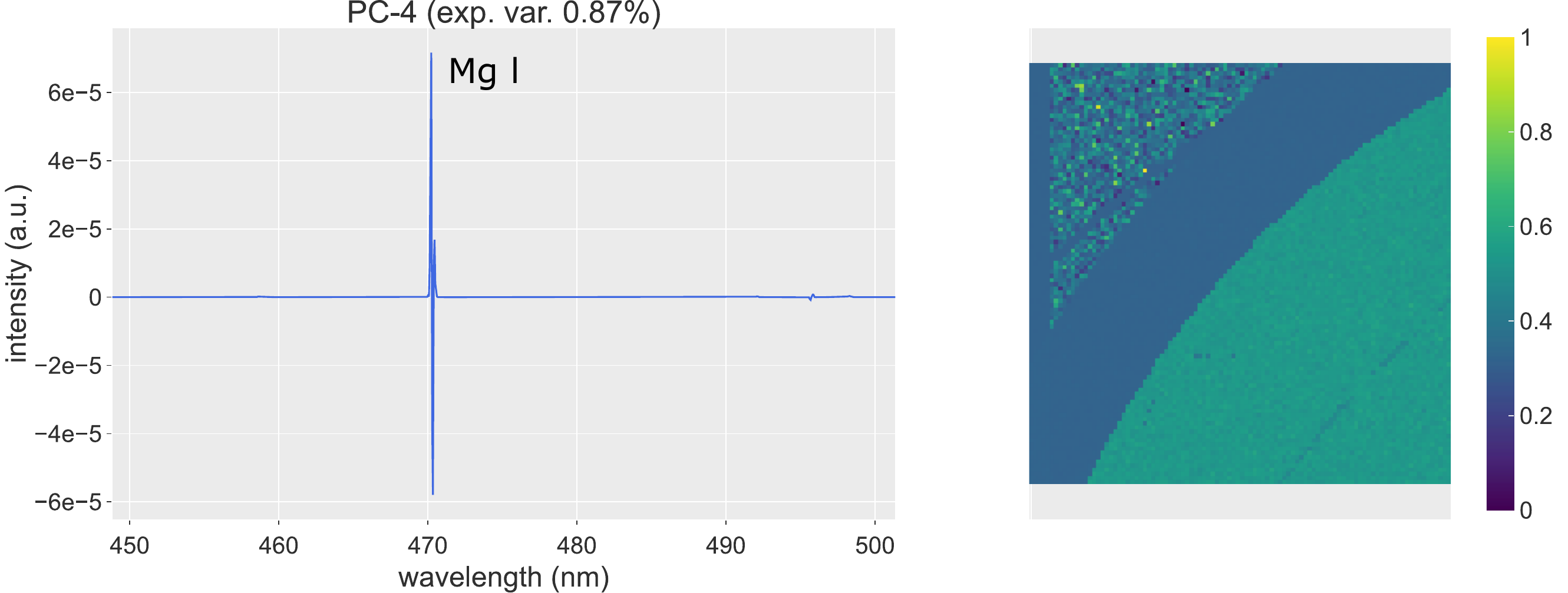}
    \\
    \includegraphics[width=0.9\linewidth]{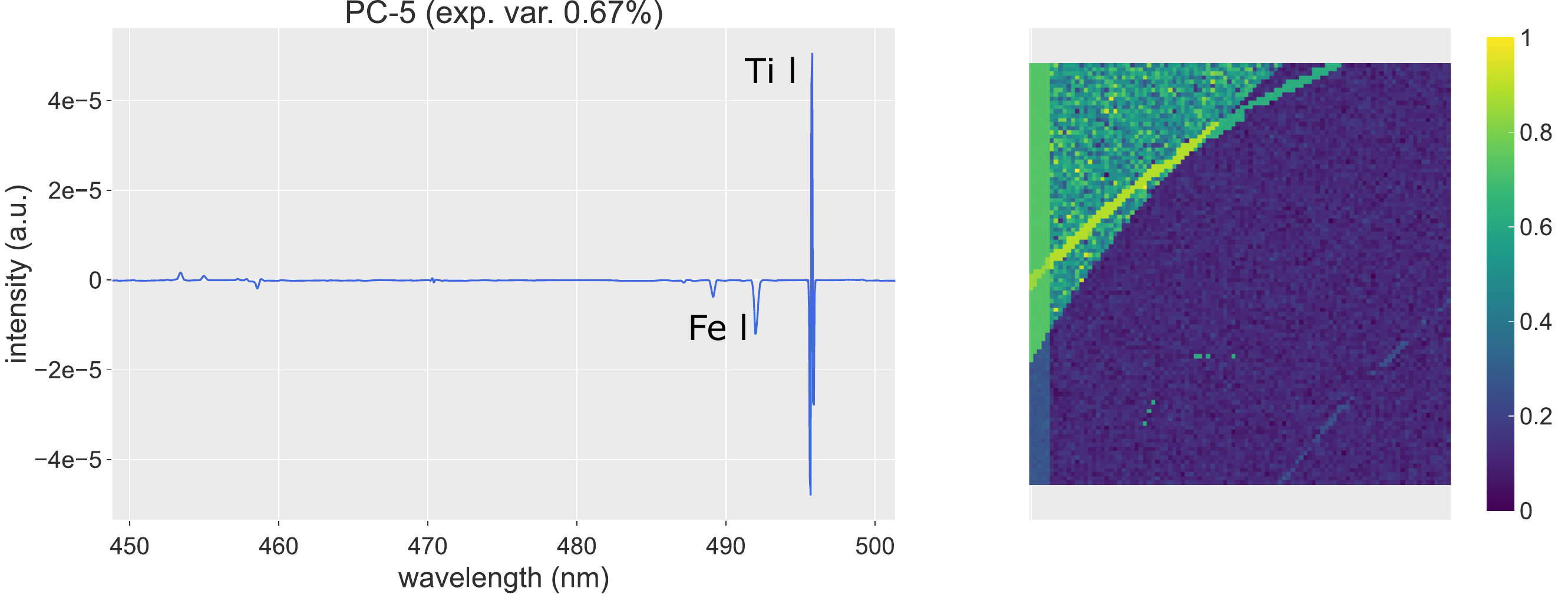}
    \\
    \includegraphics[width=0.9\linewidth]{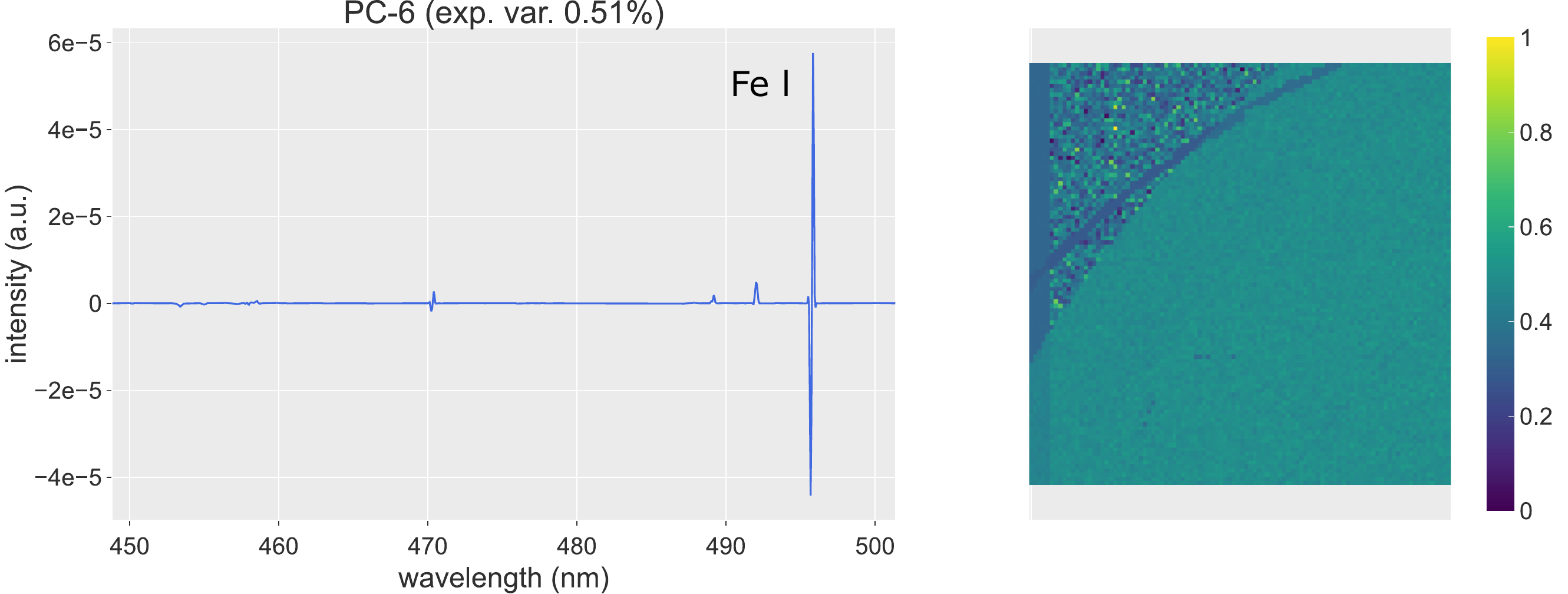}
    \\

    \caption{Complementary \pcs of the ``basalt'' specimen using \hyperpca (part 1).}
    \label{fig:basalt_kpca_less_imp_1}
\end{figure}

\begin{figure}[h]
    \centering

    \includegraphics[width=0.9\linewidth]{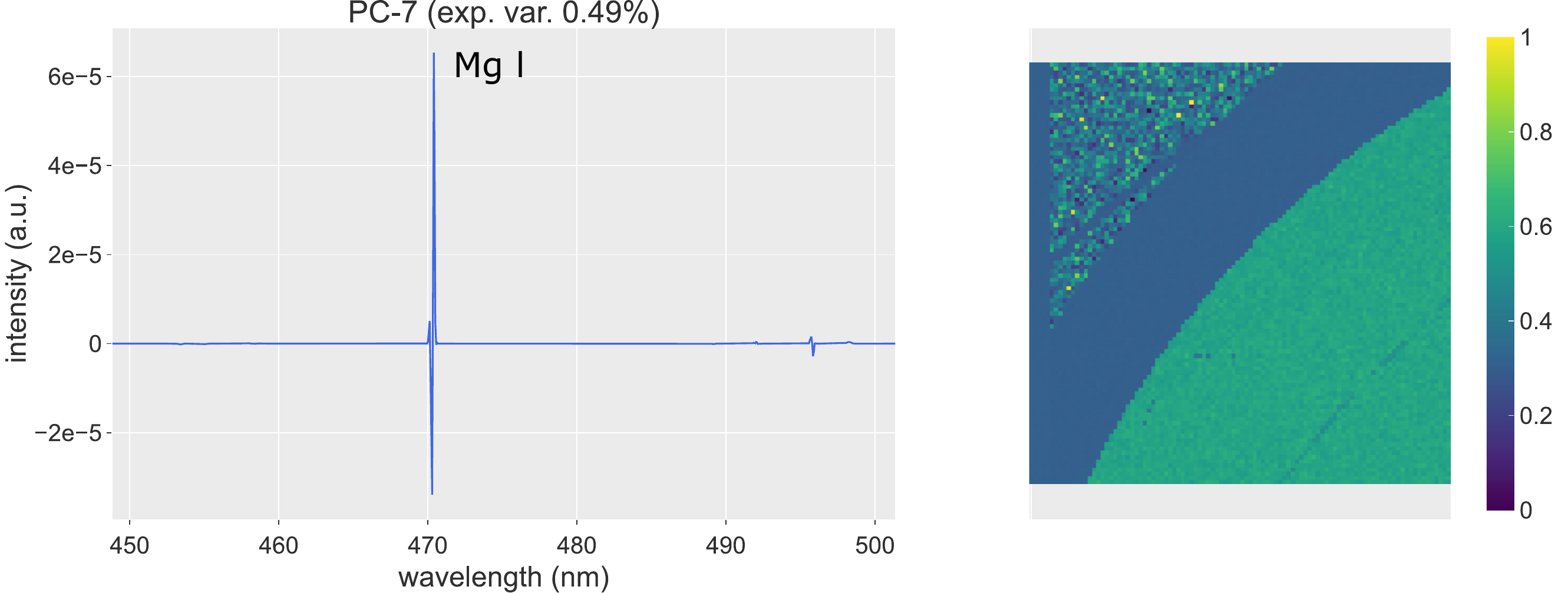}
    \\
    \includegraphics[width=0.9\linewidth]{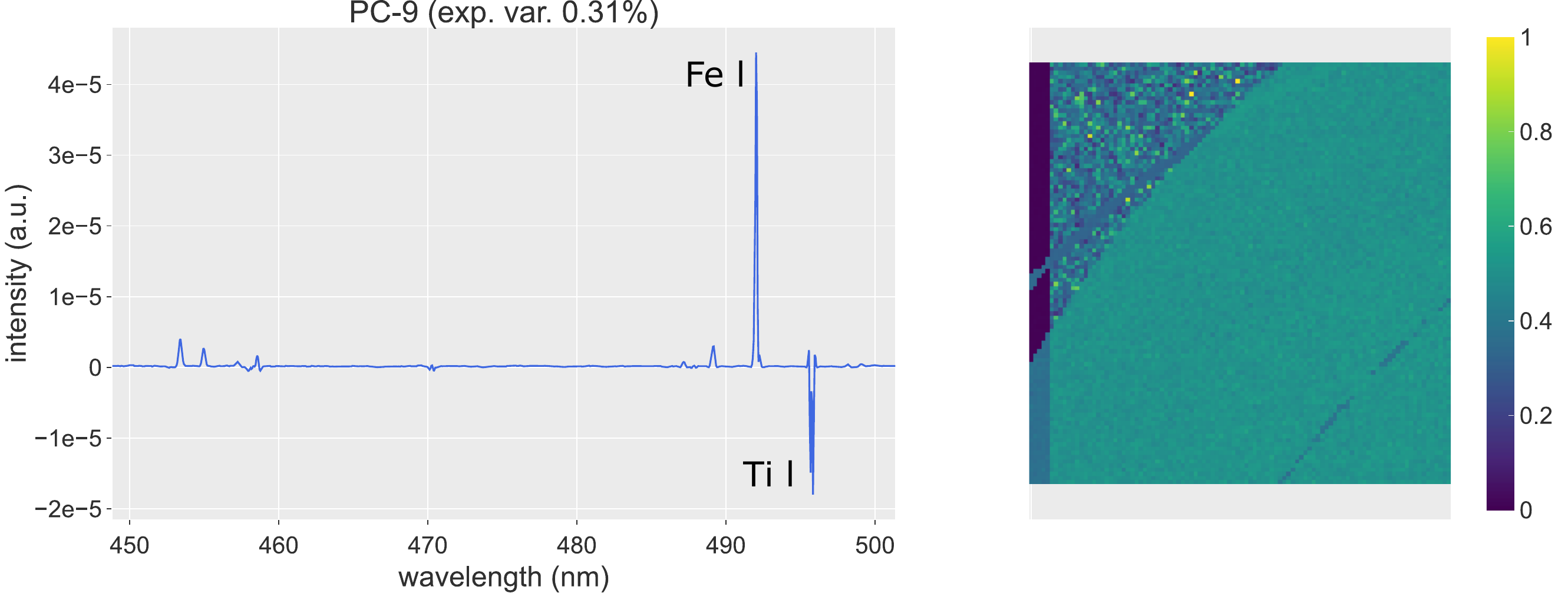}
    \\
    \includegraphics[width=0.9\linewidth]{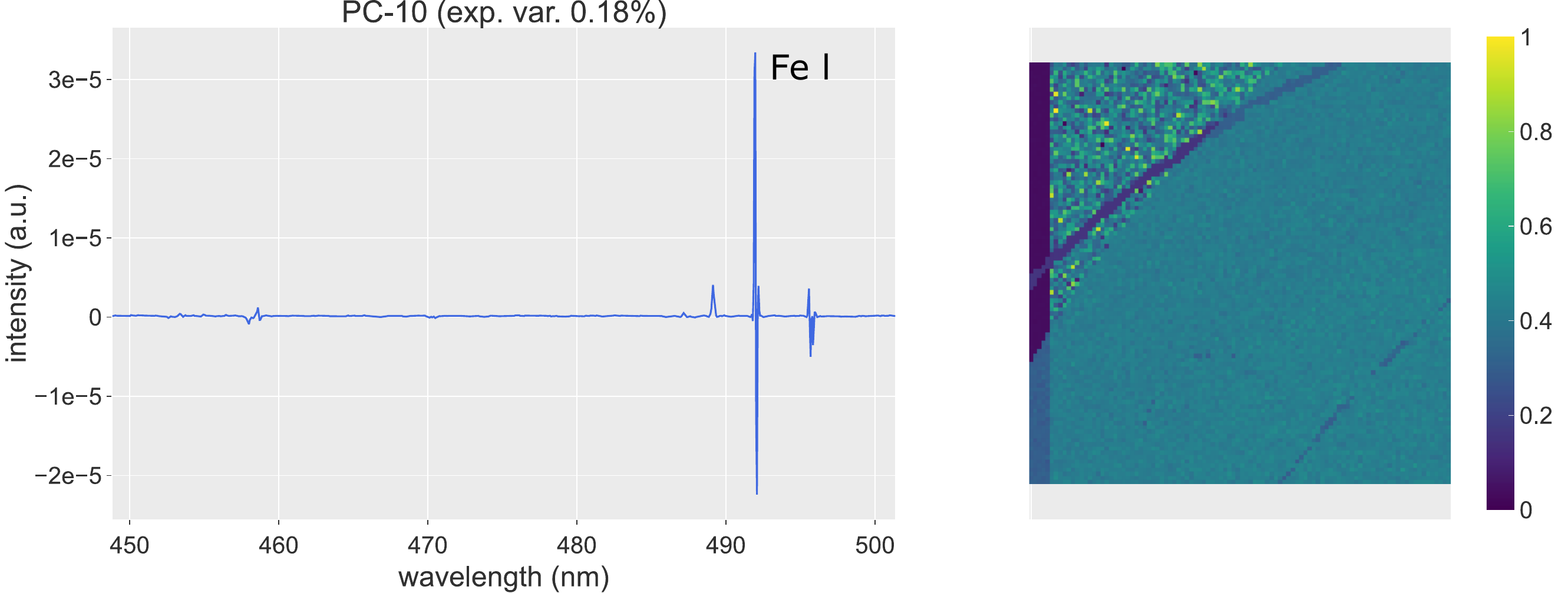}
    \\
    \includegraphics[width=0.7\linewidth]{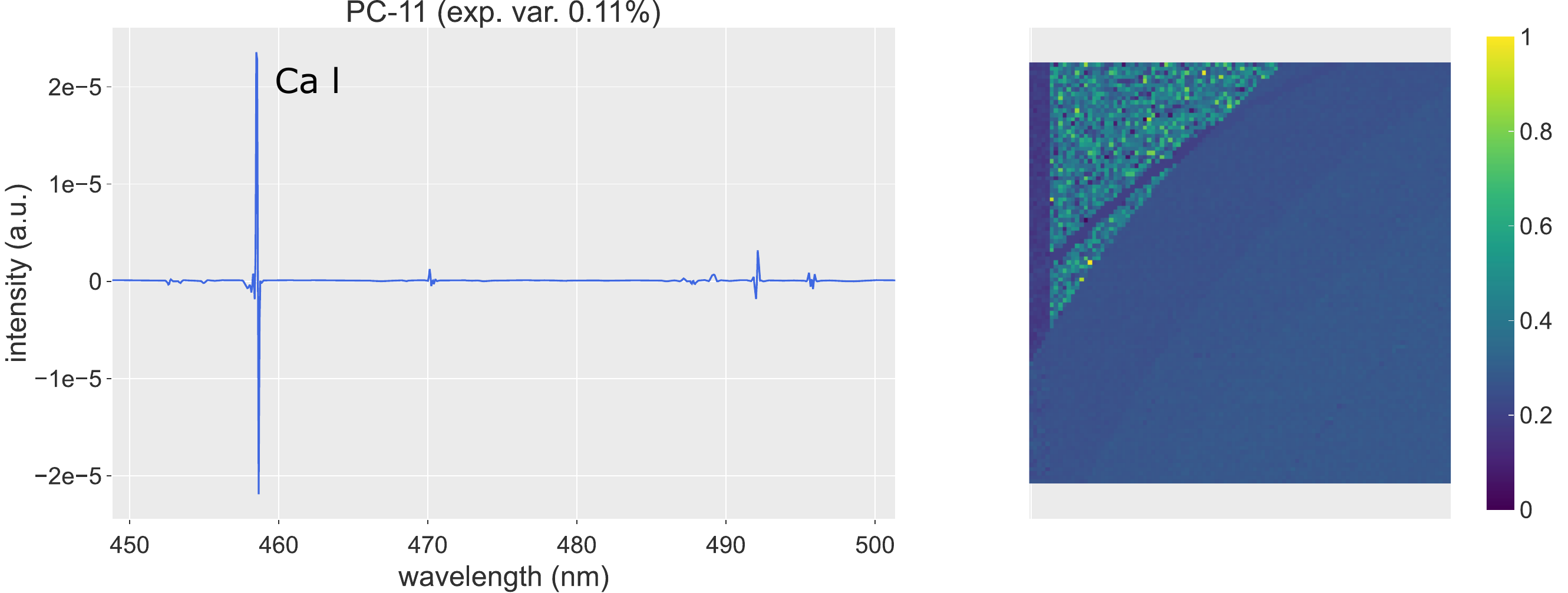}
    \\
    
    \caption{Complementary \pcs of the ``basalt'' specimen using \hyperpca (part 2).}
    \label{fig:basalt_kpca_less_imp_2}
\end{figure}

\begin{figure}[h]
    \centering

    \includegraphics[width=0.85\linewidth]{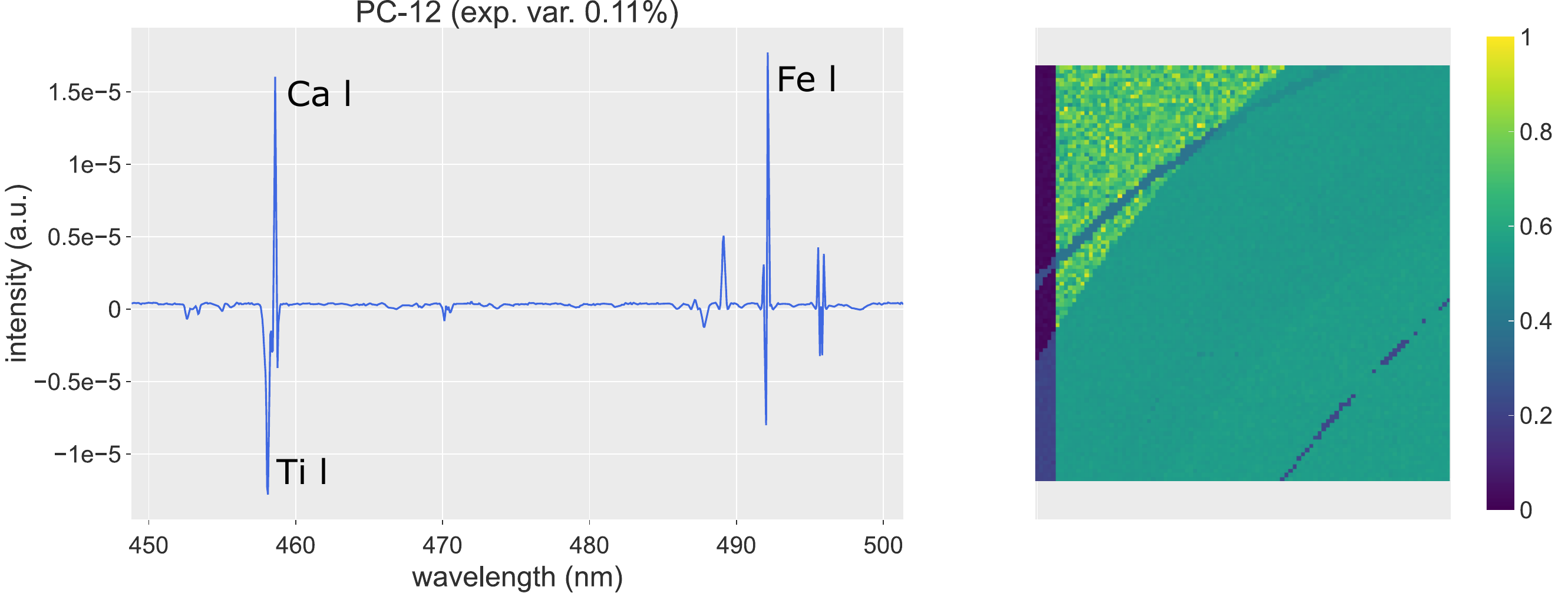}
    \\
    \includegraphics[width=0.85\linewidth]{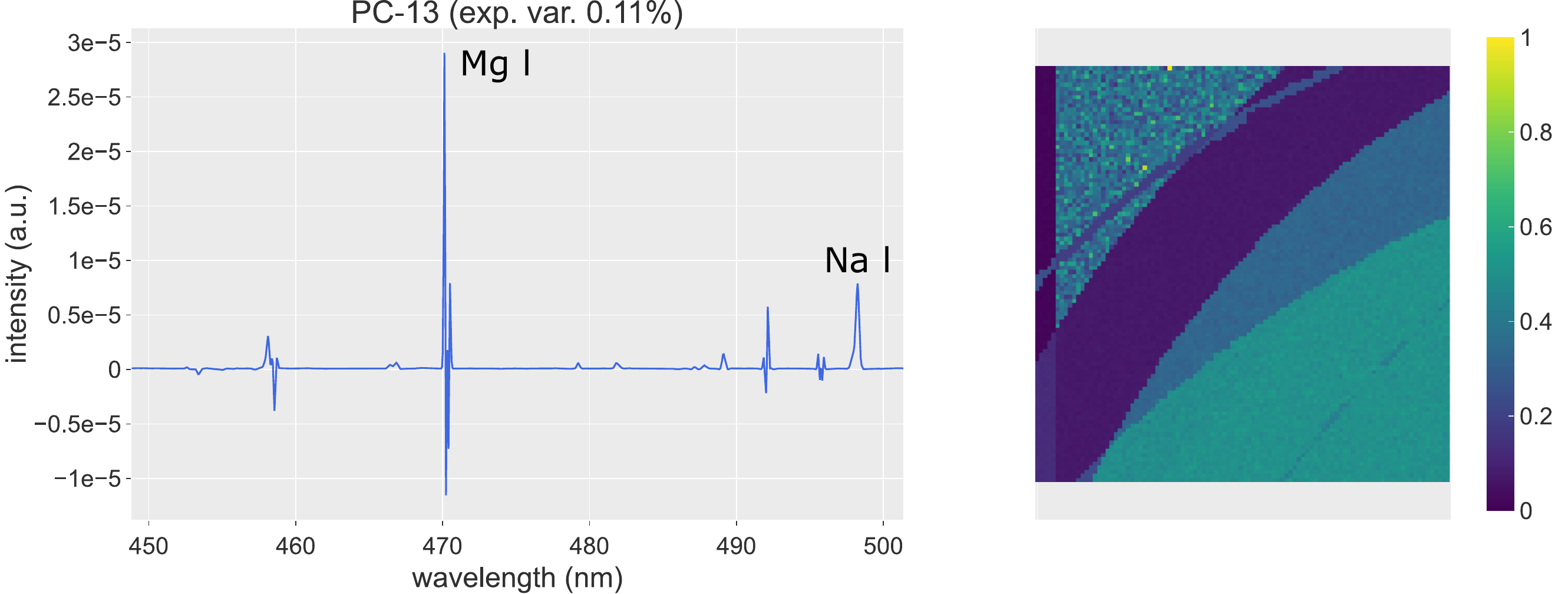}
    \\
    \includegraphics[width=0.85\linewidth]{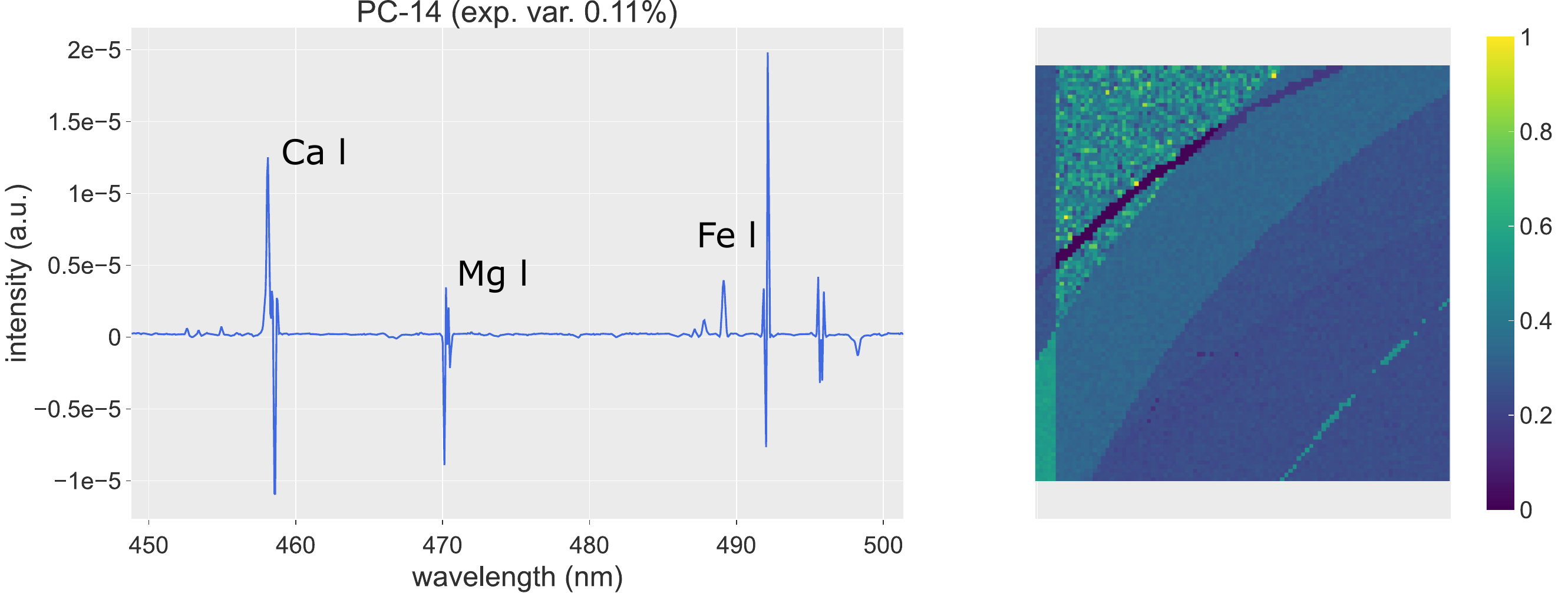}
    \\
    \includegraphics[width=0.85\linewidth]{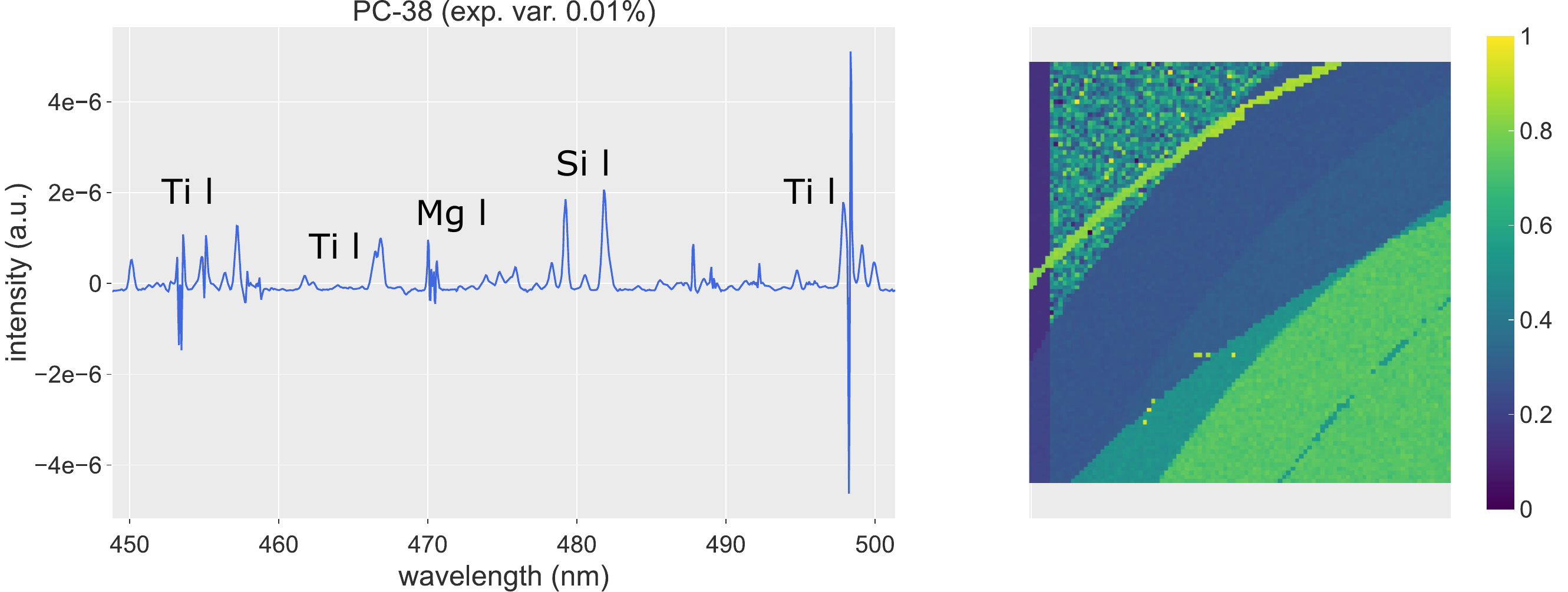}
    \\

    \caption{Complementary \pcs of the ``basalt'' specimen using \hyperpca (part 3).}
    \label{fig:basalt_kpca_less_imp_3}
\end{figure}

\begin{figure}[h]
    \centering

    \includegraphics[width=0.5\linewidth]{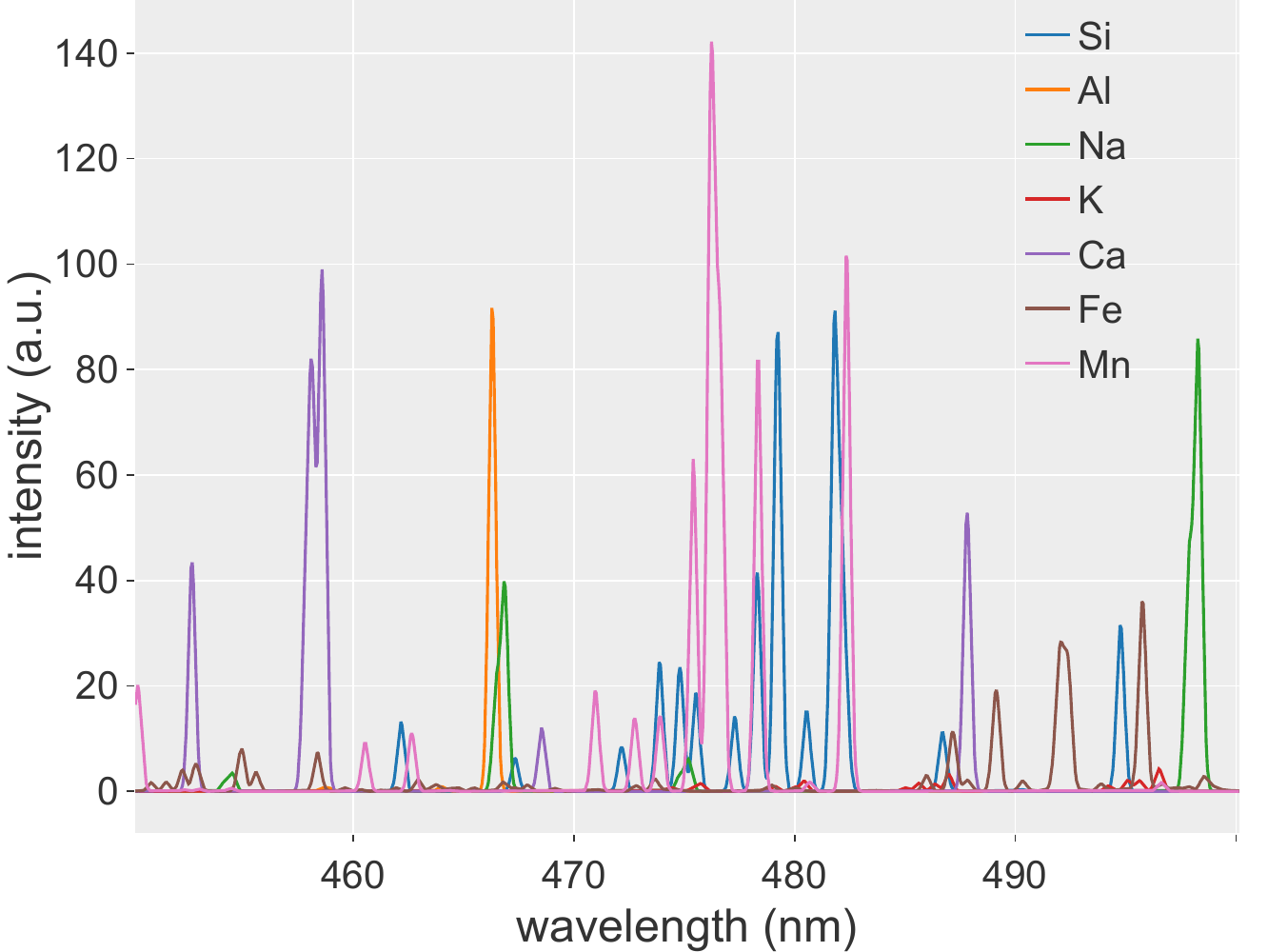}

    \caption{``Granite'' spectra in LTE.}
    \label{fig:granite_spectra_lte}
\end{figure}

\begin{figure}[h]
    \centering

    \includegraphics[width=0.9\linewidth]{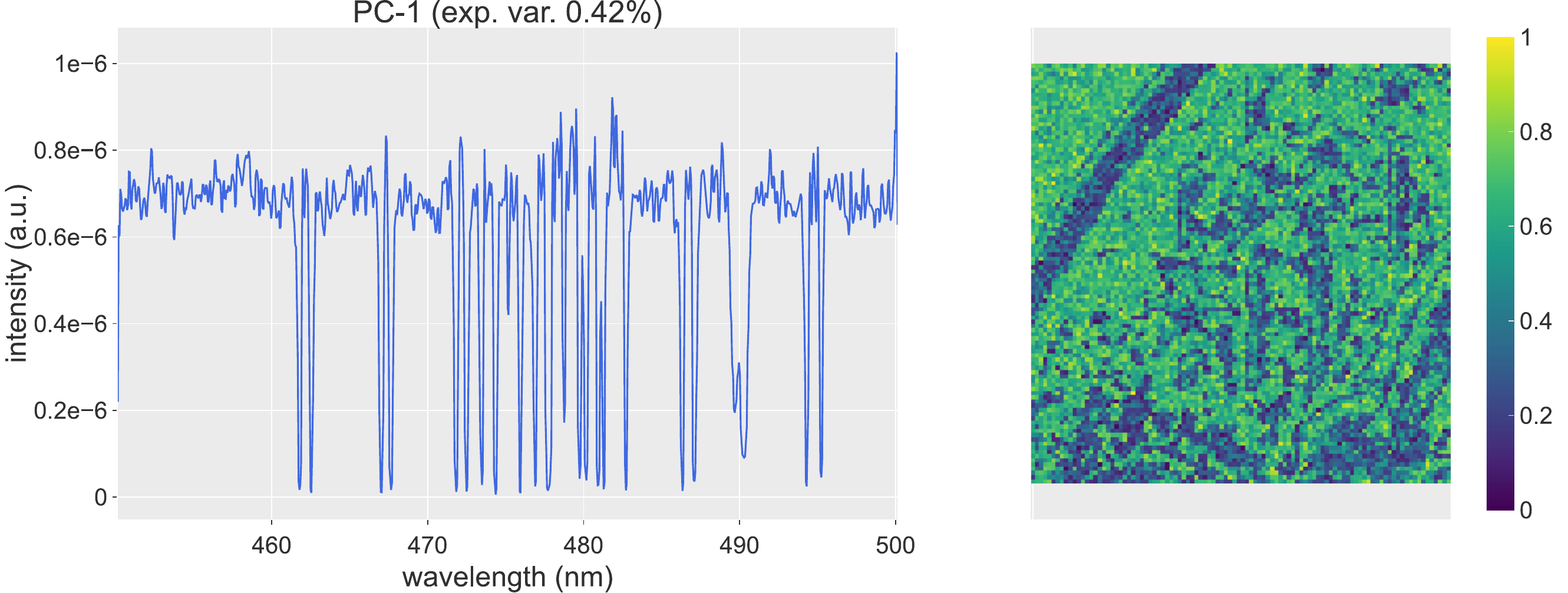}
    \\

    \caption{First \pc of the ``granite'' dataset computed using \hyperpca: the noise present in the dataset deeply spoils the first \pc of both \pca and \hyperpca rendering it mostly unusable.}
    \label{fig:granite_dwt_kpca_granite_noise}
\end{figure}

\begin{figure}[h]
    \centering

    \includegraphics[width=0.475\linewidth]{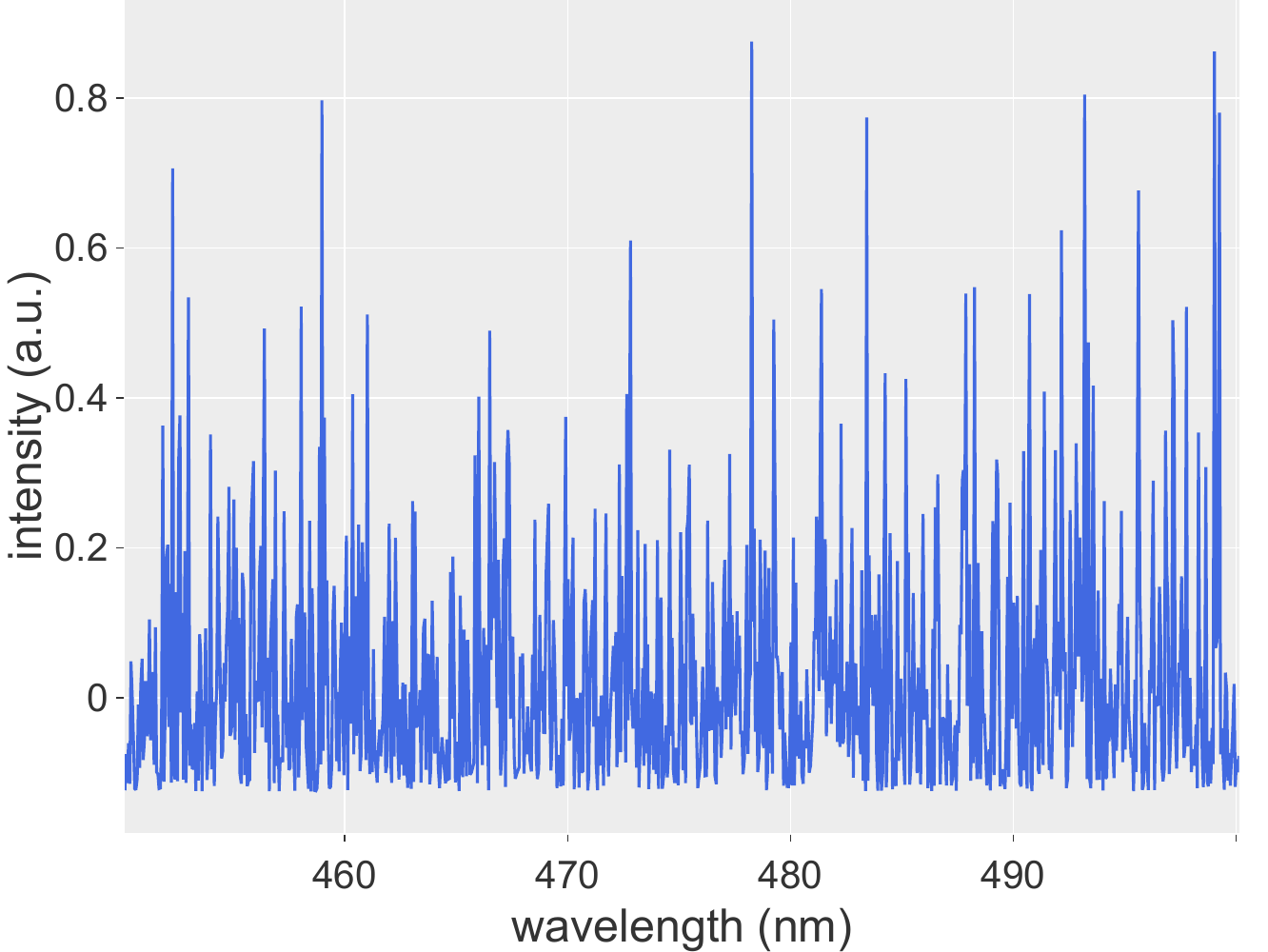}
    \hfill
    \includegraphics[width=0.475\linewidth]{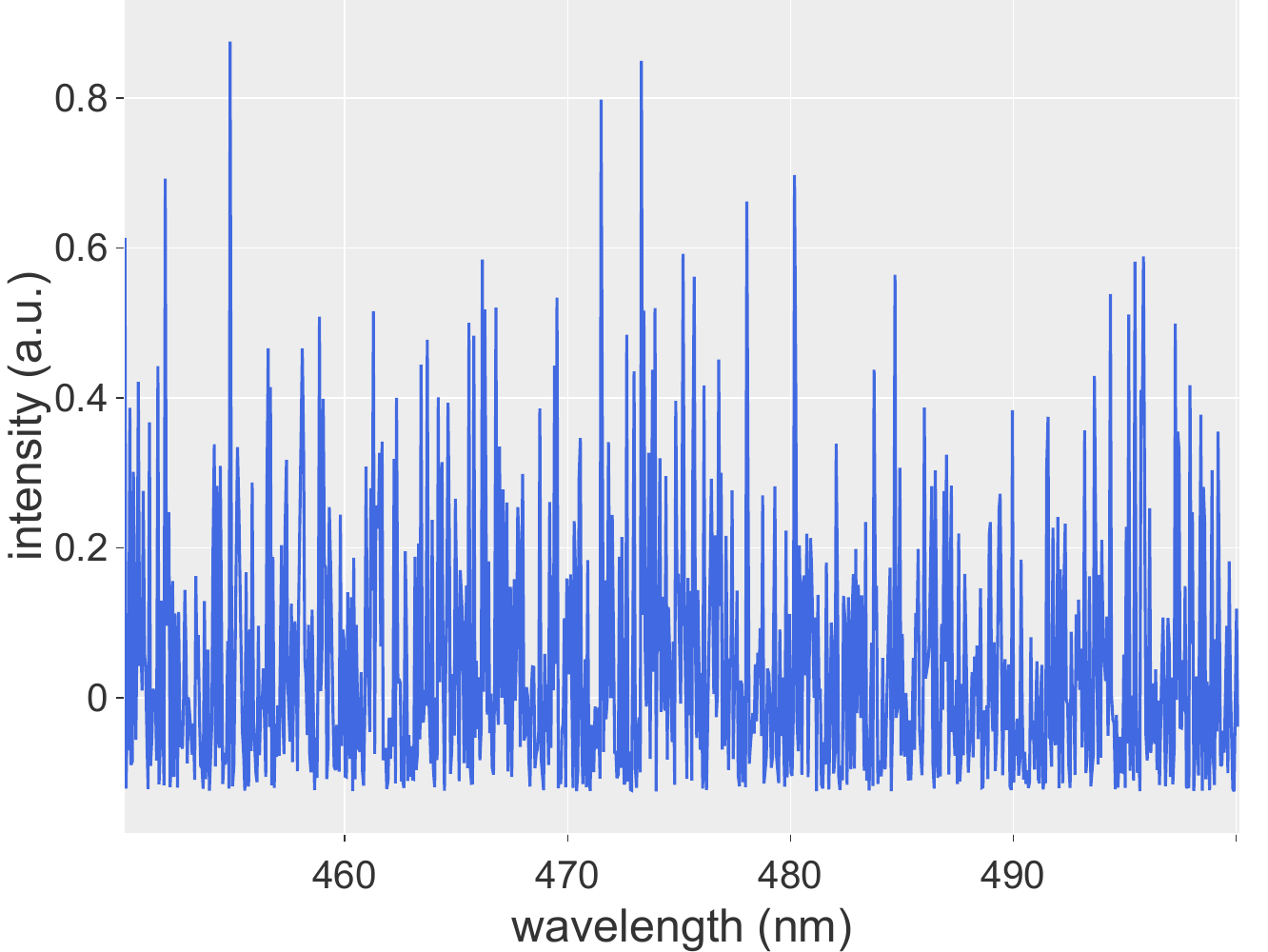}
    \\
    \includegraphics[width=0.475\linewidth]{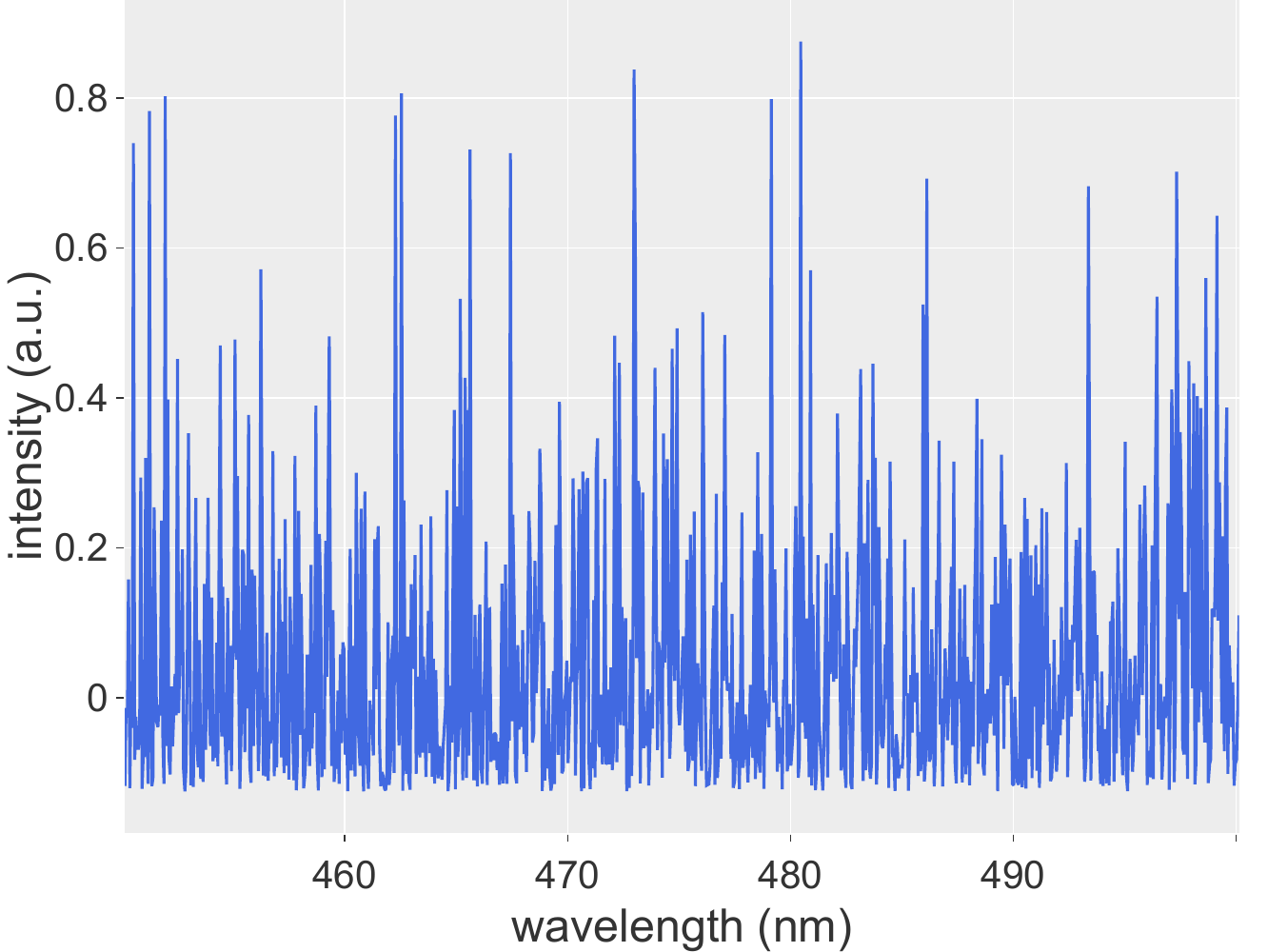}
    \hfill
    \includegraphics[width=0.475\linewidth]{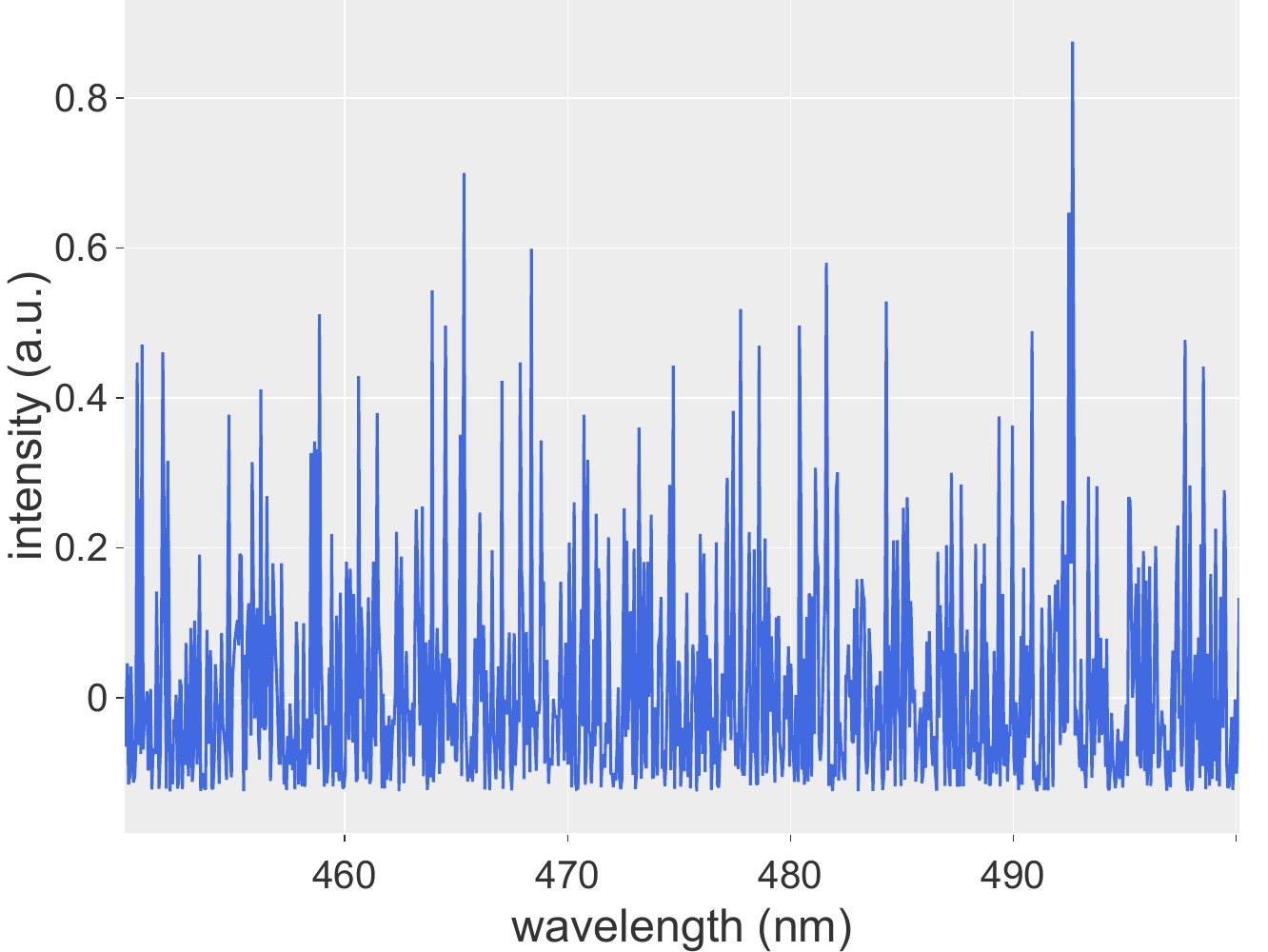}

    \caption{Example spectra in the ``granite'' dataset which illustrate the difficulty in extracting the signal due to the high noise level.}
    \label{fig:granite_examples}
\end{figure}

\begin{figure}[h]
    \centering

    \subfigure[Al]{\includegraphics[width=0.475\linewidth]{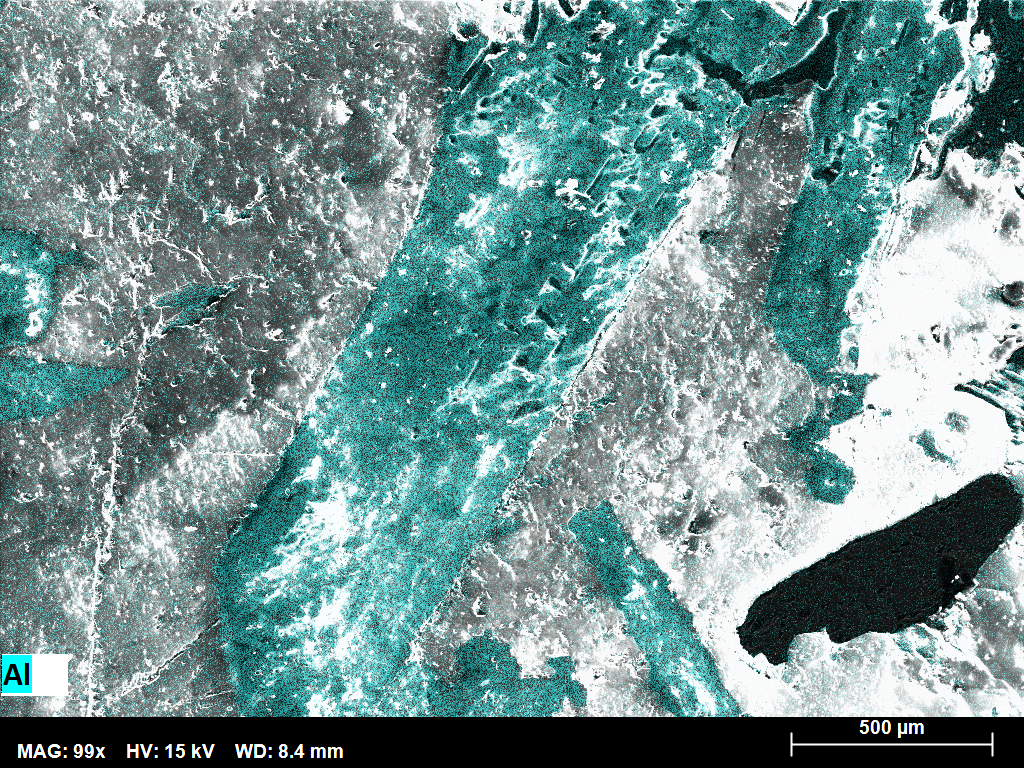}}
    \hfill
    \subfigure[Ca]{\includegraphics[width=0.475\linewidth]{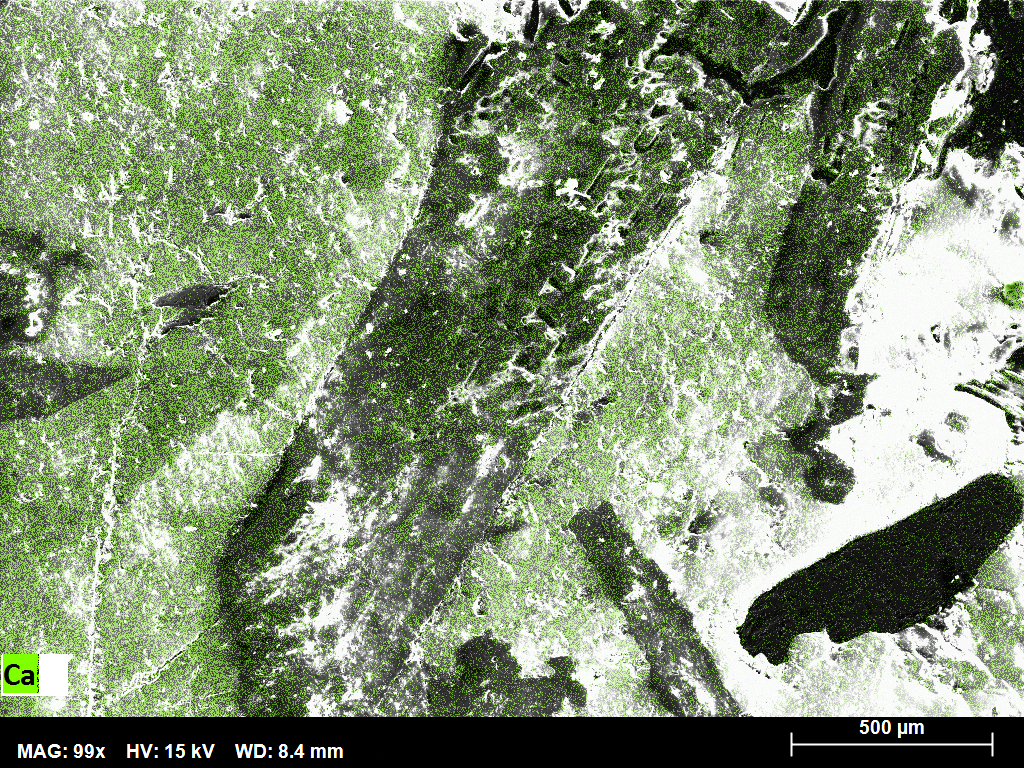}}
    \\
    \subfigure[Fe]{\includegraphics[width=0.475\linewidth]{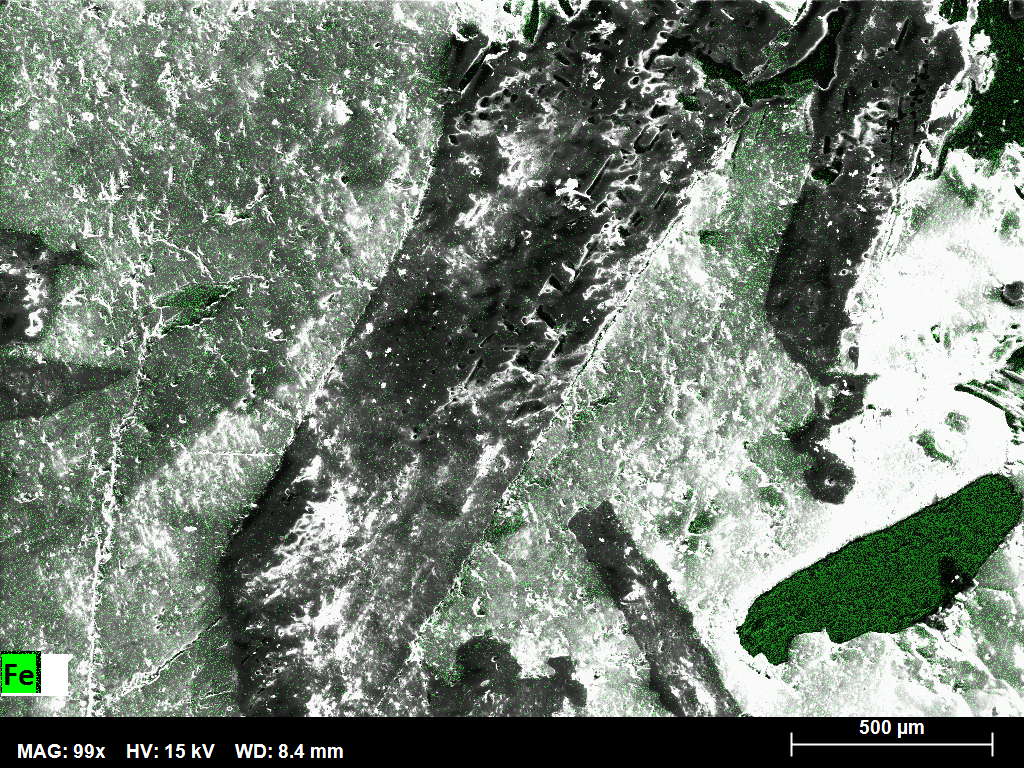}}
    \hfill
    \subfigure[Ti]{\includegraphics[width=0.475\linewidth]{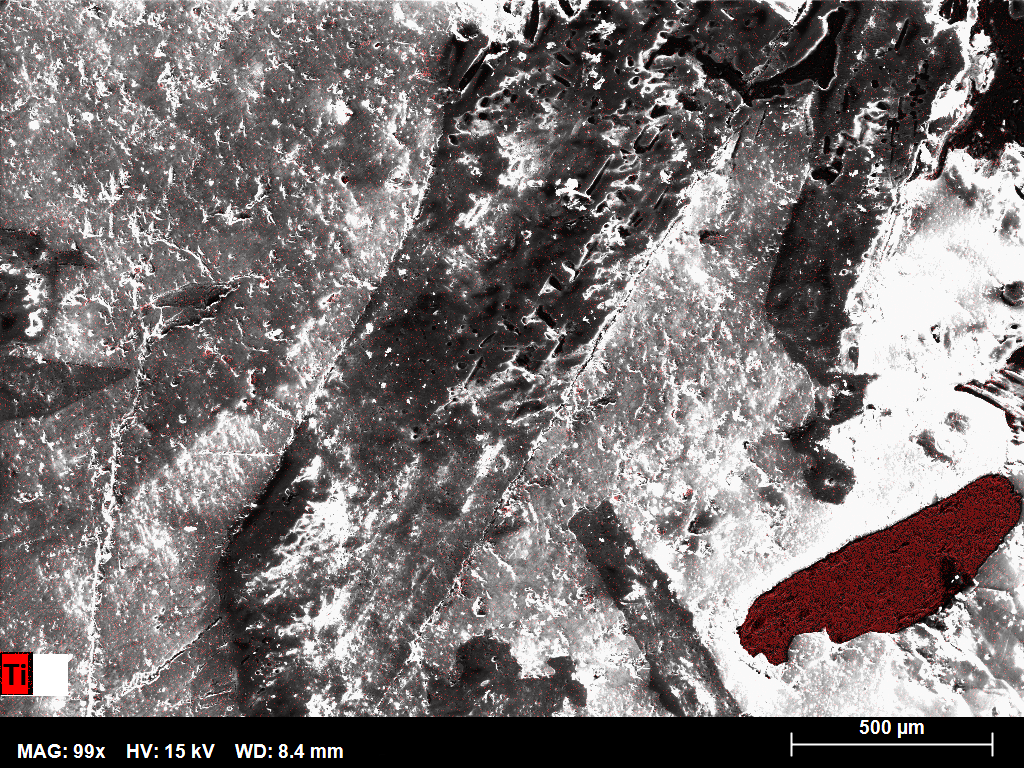}}
    \\

    \caption{Distributions of the main elements in the gabbro specimen via \sem analysis.}
    \label{fig:gabbro_ground_truth}
\end{figure}

\begin{figure}[h]
    \centering

    \includegraphics[width=0.9\linewidth]{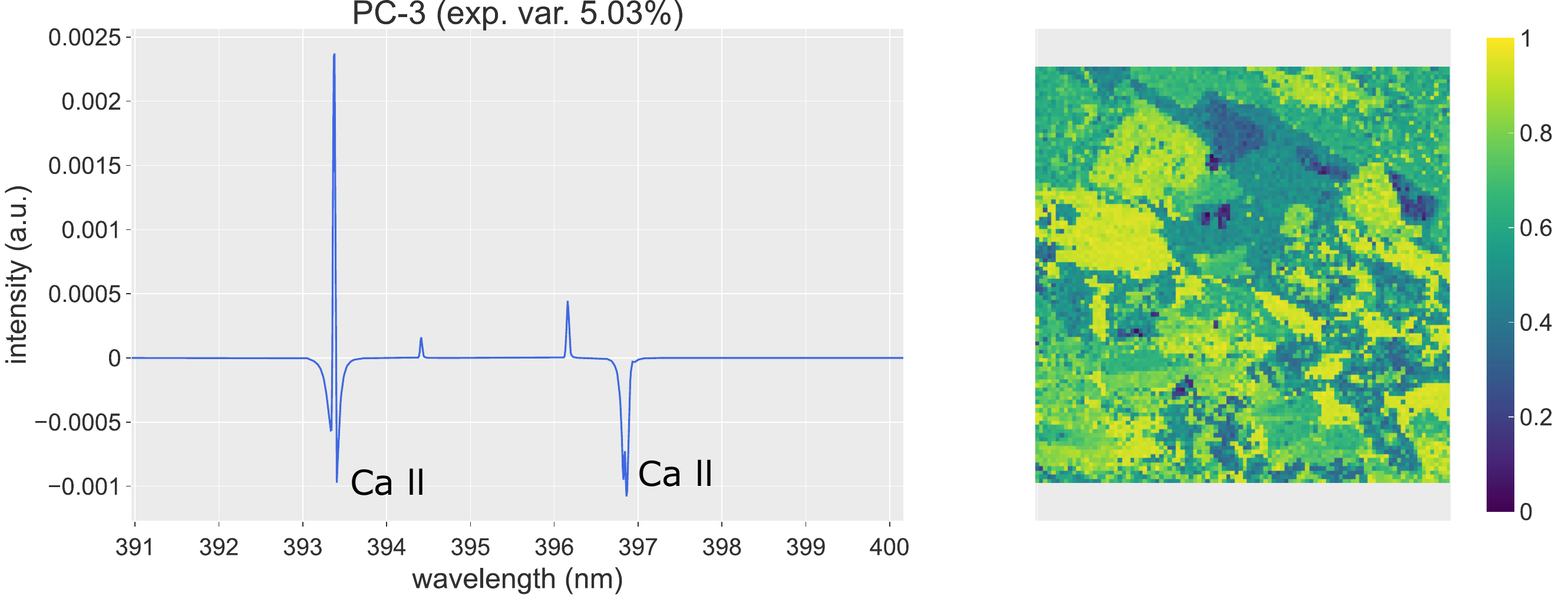}
    \\
    \includegraphics[width=0.9\linewidth]{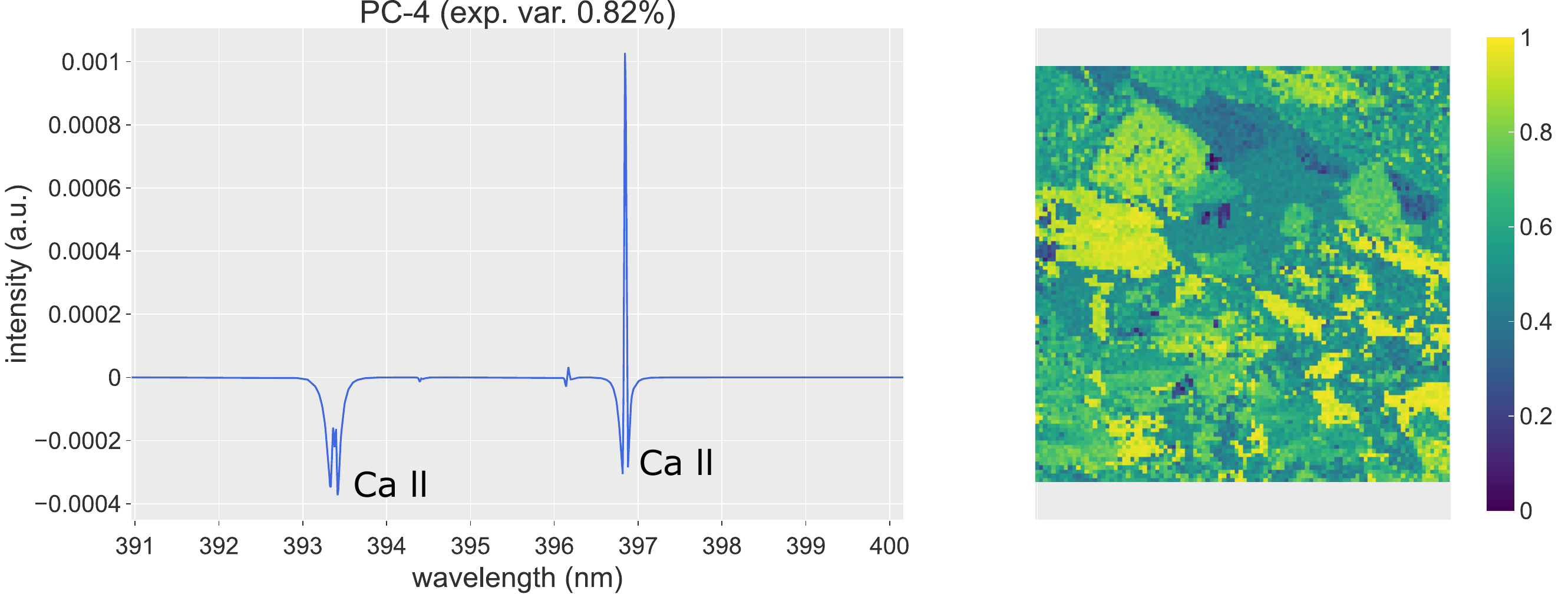}
    \\
    \includegraphics[width=0.9\linewidth]{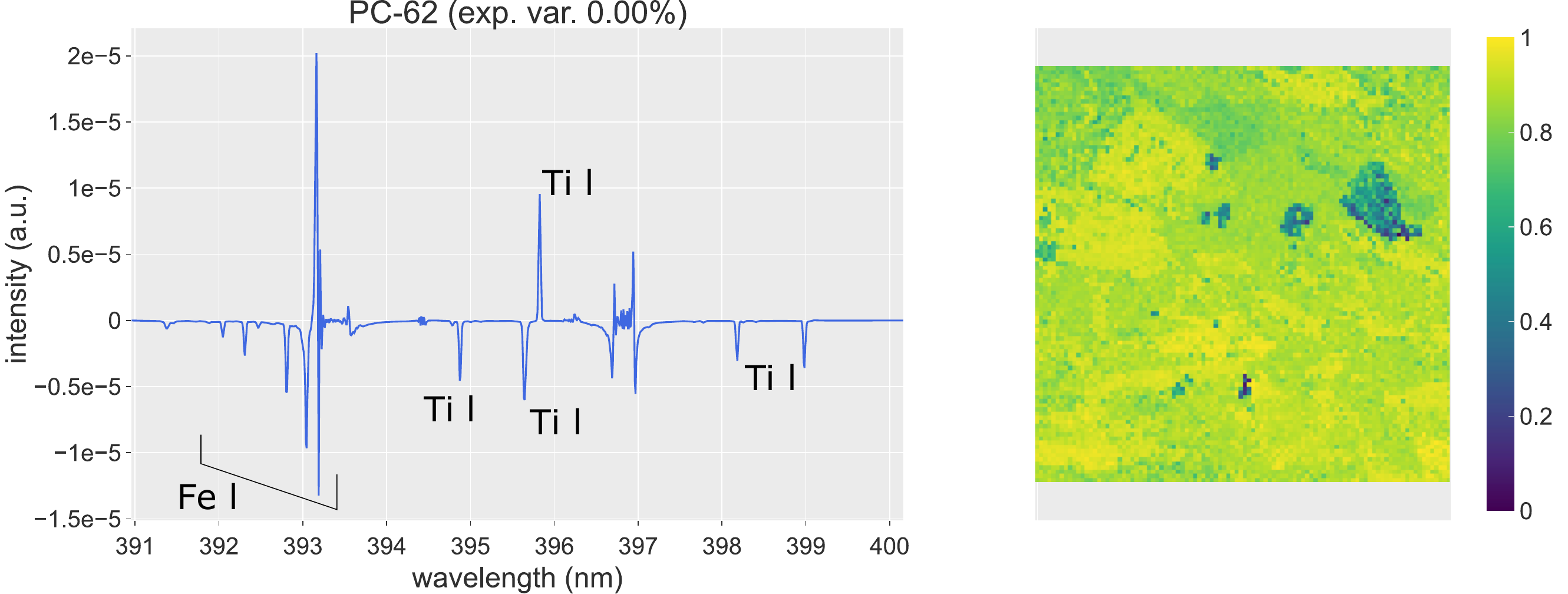}
    \\

    \caption{Complementary \pcs of the gabbro specimen using \hyperpca.}
    \label{fig:gabbro_kpca_less_imp}
\end{figure}

\begin{figure}[h]
    \centering

    \includegraphics[width=0.9\linewidth]{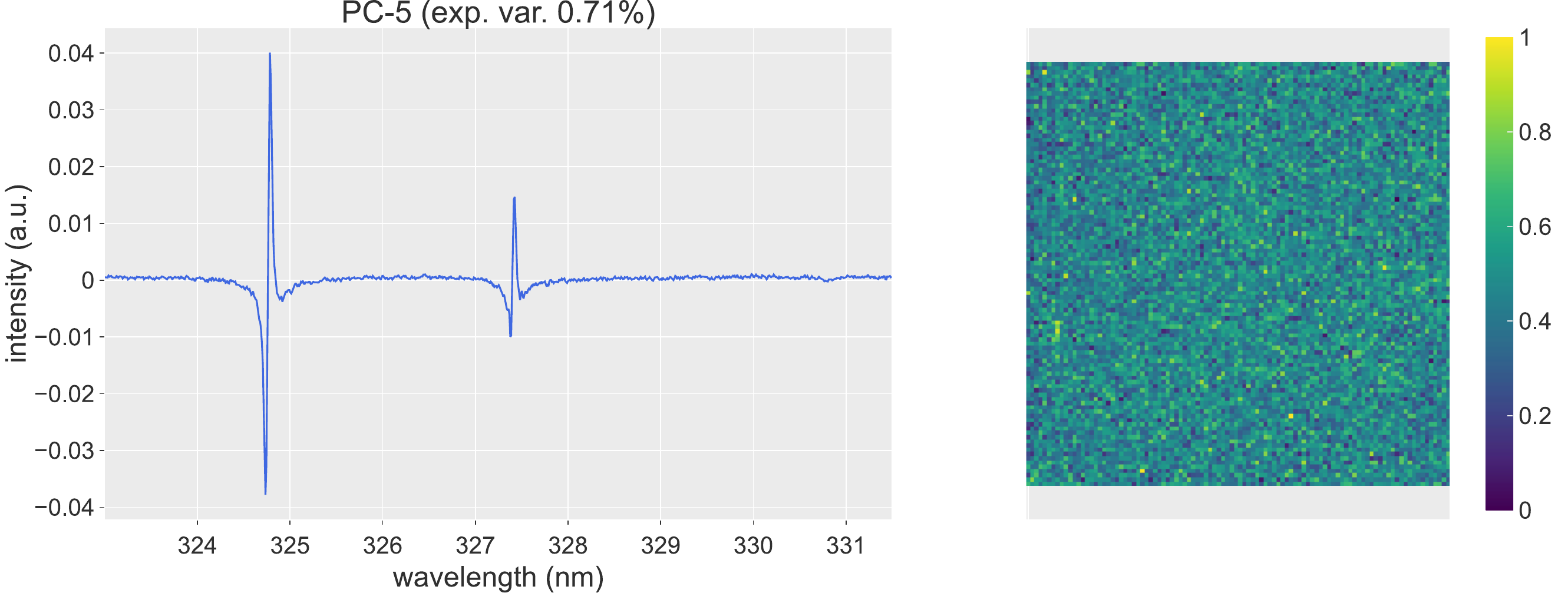}
    \\
    \includegraphics[width=0.9\linewidth]{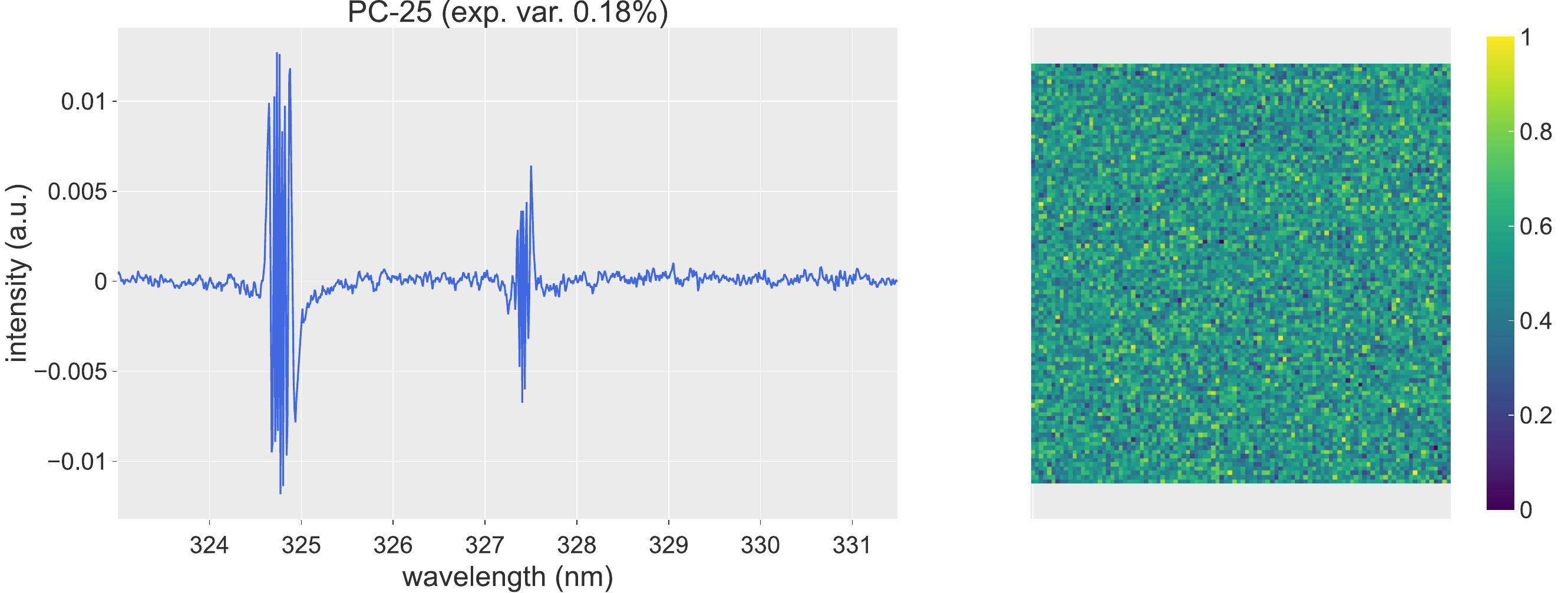}
    \\

    \caption{Complementary \pcs of the AlCu sample using the standard \pca.}
    \label{fig:alcu_pca_less_imp}
\end{figure}

\end{document}